\documentclass[twocolumn]{aastex631}

\usepackage{savesym}
\savesymbol{tablenum}
\usepackage{siunitx}
\restoresymbol{SIX}{tablenum}

\usepackage[version=4]{mhchem}
\usepackage{graphicx,epsf}
\usepackage{subfigure}
\usepackage{graphicx}	
\usepackage{amsmath}	
\usepackage{amssymb}	
\usepackage{newtxtext,newtxmath}
\usepackage{siunitx}
\usepackage{footnote}

\newcommand{\Msun}{{\rm M}_\odot}
\newcommand{\Rsun}{{\rm R}_\odot}
\newcommand{\Lsun}{{\rm L}_\odot}
\newcommand{\kms}{\textrm{km}\,\textrm{s}^{-1}}

\newcommand{\mdot}{M$_{\odot}$~yr$^{-1}$}

\def\arcsec{\hbox{$^{\prime\prime}$}}

\newcommand{\code}[1]{\texttt{#1}}
\def\heracles{{\code{HERACLES}}}
\def\cmfgen{{\code{CMFGEN}}}

\DeclareRobustCommand{\ion}[2]{\relax\ifmmode\ifx\testbx\f@series{\mathbf{#1\,\mathsc{#2}}}\else{\mathrm{#1\,\mathsc{#2}}}\fi\else\textup{#1\,{\mdseries\textsc{#2}}}\fi}

\usepackage[scale=0.94]{tgbonum}
\usepackage[mode=text]{siunitx}

\makeatletter
\DeclareTextCompositeCommand{\r}{OT1}{A}{%
  \leavevmode\vbox{%
    \offinterlineskip
    \ialign{\hfil##\hfil\cr\char23\cr\noalign{\kern-1.15ex}A\cr}%
  }%
}
\makeatother

\shorttitle{Final Moments II}
\shortauthors{Jacobson-Gal\'an et al.}

\begin{document}

\title{Final Moments II: Observational Properties and Physical Modeling of CSM-Interacting Type II Supernovae}

\correspondingauthor{Wynn Jacobson-Gal\'{a}n (he, him, his)}
\email{wynnjg@berkeley.edu}

\author[0000-0002-3934-2644]{W.~V.~Jacobson-Gal\'{a}n}
\affil{Department of Astronomy, University of California, Berkeley, CA 94720-3411, USA}

\author[0000-0003-0599-8407]{L.~Dessart}
\affil{Institut d’Astrophysique de Paris, CNRS-Sorbonne Université, 98 bis boulevard Arago, F-75014 Paris, France}

\author[0000-0002-5680-4660]{K.~W.~Davis}
\affil{Department of Astronomy and Astrophysics, University of California, Santa Cruz, CA 95064, USA}

\author[0000-0002-5740-7747]{C.~D.~Kilpatrick}
\affil{Center for Interdisciplinary Exploration and Research in Astrophysics (CIERA), Northwestern University, Evanston, IL 60202, USA}
\affiliation{Department of Physics and Astronomy, Northwestern University, Evanston, IL 60208, USA}

\author[0000-0003-4768-7586]{R.~Margutti}
\affil{Department of Astronomy, University of California, Berkeley, CA 94720-3411, USA}
\affil{Department of Physics, University of California, Berkeley, CA 94720-7300, USA}

\author[0000-0002-2445-5275]{R.~J.~Foley}
\affiliation{Department of Astronomy and Astrophysics, University of California, Santa Cruz, CA 95064, USA}

\author[0000-0002-7706-5668]{R.~Chornock}
\affil{Department of Astronomy, University of California, Berkeley, CA 94720-3411, USA}

\author[0000-0003-0794-5982]{G.~Terreran}
\affil{Las Cumbres Observatory, 6740 Cortona Dr. Suite 102, Goleta, CA, 93117}

\author[0000-0002-1125-9187]{D.~Hiramatsu}
\affil{Center for Astrophysics \textbar{} Harvard \& Smithsonian, 60 Garden Street, Cambridge, MA 02138-1516, USA}
\affil{The NSF AI Institute for Artificial Intelligence and Fundamental Interactions, USA}

\author[0000-0001-9570-0584]{M.~Newsome}
\affil{Las Cumbres Observatory, 6740 Cortona Dr. Suite 102, Goleta, CA, 93117}
\affil{Department of Physics, University of California, Santa Barbara, Santa Barbara, CA, USA, 93111}

\author[0000-0003-0209-9246]{E.~Padilla~Gonzalez}
\affil{Las Cumbres Observatory, 6740 Cortona Dr. Suite 102, Goleta, CA, 93117}
\affil{Department of Physics, University of California, Santa Barbara, Santa Barbara, CA, USA, 93111}

\author[0000-0002-7472-1279]{C.~Pellegrino}
\affil{Department of Astronomy, University of Virginia, Charlottesville, VA 22904, USA}

\author[0000-0003-4253-656X]{D.~A.~Howell}
\affil{Las Cumbres Observatory, 6740 Cortona Dr. Suite 102, Goleta, CA, 93117}
\affil{Department of Physics, University of California, Santa Barbara, Santa Barbara, CA, USA, 93111}

\author[0000-0003-3460-0103]{A.~V.~Filippenko}
\affil{Department of Astronomy, University of California, Berkeley, CA 94720-3411, USA}


\author[0000-0003-0227-3451]{J.~P.~Anderson}
\affil{European Southern Observatory, Alonso de C\'ordova 3107, Casilla 19, Santiago, Chile }
\affil{Millennium Institute of Astrophysics MAS, Nuncio Monsenor Sotero Sanz 100, Off. 104, Providencia, Santiago, Chile}

\author[0000-0002-4269-7999]{C.~R.~Angus}
\affil{DARK, Niels Bohr Institute, University of Copenhagen, Jagtvej 128, 2200 Copenhagen, Denmark}
\affil{Astrophysics Research Centre, School of Mathematics and Physics, Queen’s University Belfast, Belfast BT7 1NN, UK}

\author[0000-0002-4449-9152]{K.~Auchettl}
\affil{Department of Astronomy and Astrophysics, University of California, Santa Cruz, CA 95064, USA}
\affil{School of Physics, The University of Melbourne, VIC 3010, Australia}

\author[0000-0002-4924-444X]{K.~A.~Bostroem}
\affil{Steward Observatory, University of Arizona, 933 North Cherry Avenue, Tucson, AZ 85721-0065, USA}
\affil{LSST-DA Catalyst Fellow}

\author[0000-0001-5955-2502]{T.~G.~Brink}
\affil{Department of Astronomy, University of California, Berkeley, CA 94720-3411, USA}

\author[0000-0003-4553-4033]{R.~Cartier}
\affil{Gemini Observatory, NSF’s National Optical-Infrared Astronomy Research Laboratory, Casilla 603, La Serena, Chile}

\author[0000-0003-4263-2228]{D.~A.~Coulter}
\affil{Space Telescope Science Institute, Baltimore, MD 21218, USA}

\author[0000-0001-5486-2747]{T.~de~Boer}
\affil{Institute for Astronomy, University of Hawaii, 2680 Woodlawn Drive, Honolulu, HI 96822, USA}

\author[0000-0001-7081-0082]{M.~R.~Drout}
\affil{David A. Dunlap Department of Astronomy and Astrophysics, University of Toronto, 50 St. George Street, Toronto, Ontario, M5S 3H4, Canada}

\author[0000-0003-1714-7415]{N.~Earl}
\affil{Department of Astronomy, University of Illinois at Urbana-Champaign, 1002 W. Green St., Urbana, IL 61801, USA}

\author[0000-0001-7251-8368]{K.~Ertini}
\affil{Facultad de Ciencias Astron\'omicas y Geofísicas, Universidad Nacional de La Plata, Paseo del Bosque S/N, B1900FWA, La Plata, Argentina}
\affil{Instituto de Astrofisica de La Plata (IALP), CCT-CONICET-UNLP, Paseo del Bosque S/N, B1900FWA, La Plata, Argentina}

\author[0000-0003-4914-5625]{J.~R.~Farah}
\affil{Las Cumbres Observatory, 6740 Cortona Dr. Suite 102, Goleta, CA, 93117}
\affil{Department of Physics, University of California, Santa Barbara, Santa Barbara, CA, USA, 93111}

\author[0000-0002-6886-269X]{D.~Farias}
\affiliation{DARK, Niels Bohr Institute, University of Copenhagen, Jagtvej 128, 2200 Copenhagen, Denmark}

\author[0000-0002-8526-3963]{C.~Gall}
\affil{DARK, Niels Bohr Institute, University of Copenhagen, Jagtvej 128, 2200 Copenhagen, Denmark}

\author[0000-0003-1015-5367]{H.~Gao}
\affil{Institute for Astronomy, University of Hawaii, 2680 Woodlawn Drive, Honolulu, HI 96822, USA}

\author{M.~A.~Gerlach}
\affil{Department of Astrophysics, Pontificia Universidad Catolica de Chile, Santiago, Chile}

\author{F.~Guo}
\affil{Department of Physics, Tsinghua University, Shuangqing Road, Beijing, China}

\author[0000-0003-4287-4577]{A.~Haynie}
\affiliation{The Observatories of the Carnegie Institute for Science, 813 Santa Barbara St., Pasadena, CA 91101, USA}
\affiliation{Department of Physics \& Astronomy, University of Southern California, Los Angeles, CA 90089, USA}

\author[0000-0002-0832-2974]{G.~Hosseinzadeh}
\affil{Steward Observatory, University of Arizona, 933 North Cherry Avenue, Tucson, AZ 85721-0065, USA}

\author[0000-0003-2405-2967]{A.~L.~Ibik}
\affiliation{David A. Dunlap Department of Astronomy and Astrophysics, University of Toronto, 50 St. George Street, Toronto, Ontario, M5S 3H4, Canada}

\author[0000-0001-8738-6011]{S.~W.~Jha}
\affil{Department of Physics and Astronomy, Rutgers, the State University of New Jersey, 136 Frelinghuysen Road, Piscataway, NJ 08854, USA}

\author[0000-0002-6230-0151]{D.~O.~Jones}
\affiliation{Gemini Observatory, NSF’s NOIRLab, 670 N. A’ohoku Place, Hilo, HI 96720, USA}

\author[0000-0001-5710-8395]{D.~Langeroodi}
\affiliation{DARK, Niels Bohr Institute, University of Copenhagen, Jagtvej 128, 2200 Copenhagen, Denmark}

\author[0000-0002-2249-0595]{N~LeBaron}
\affil{Department of Astronomy, University of California, Berkeley, CA 94720-3411-3411, USA}

\author[0000-0002-7965-2815]{E.~A.~Magnier}
\affil{Institute for Astronomy, University of Hawaii, 2680 Woodlawn Drive, Honolulu, HI 96822, USA}

\author[0000-0001-6806-0673]{A.~L.~Piro}
\affil{The Observatories of the Carnegie Institute for Science, 813 Santa Barbara St., Pasadena, CA 91101, USA}

\author[0000-0002-6248-398X]{S.~I.~Raimundo}
\affil{DARK, Niels Bohr Institute, University of Copenhagen, Jagtvej 128, 2200 Copenhagen, Denmark}
\affil{Department of Physics and Astronomy, University of Southampton, Highfield, Southampton SO17 1BJ, UK}

\author[0000-0002-4410-5387]{A.~Rest}
\affil{Department of Physics and Astronomy, The Johns Hopkins University, Baltimore, MD 21218, USA}
\affil{Space Telescope Science Institute, Baltimore, MD 21218, USA}

\author[0000-0002-3825-0553]{S.~Rest}
\affil{Department of Physics and Astronomy, The Johns Hopkins University, Baltimore, MD 21218, USA}

\author[0000-0003-0427-8387]{R.~Michael~Rich}
\affil{Department Physics and Astronomy, University of California, Los Angeles, Los Angeles, CA, 90095-1547}

\author[0000-0002-7559-315X]{C.~Rojas-Bravo}
\affiliation{Department of Astronomy and Astrophysics, University of California, Santa Cruz, CA 95064, USA}

\author[0000-0001-8023-4912]{H.~Sears}
\affil{Center for Interdisciplinary Exploration and Research in Astrophysics (CIERA), Northwestern University, Evanston, IL 60202, USA}
\affiliation{Department of Physics and Astronomy, Northwestern University, Evanston, IL 60208, USA}

\author[0000-0002-5748-4558]{K.~Taggart}
\affil{Department of Astronomy and Astrophysics, University of California, Santa Cruz, CA 95064, USA}

\author[0000-0002-1125-9187]{V.~A.~Villar}
\affil{Center for Astrophysics \textbar{} Harvard \& Smithsonian, 60 Garden Street, Cambridge, MA 02138-1516, USA}

\author[0000-0002-1341-0952]{R.~J.~Wainscoat}
\affil{Institute for Astronomy, University of Hawaii, 2680 Woodlawn Drive, Honolulu, HI 96822, USA}

\author[0000-0002-7334-2357]{X-F.~Wang}
\affil{Department of Physics, Tsinghua University, Shuangqing Road, Beijing, China}

\author[0000-0002-4186-6164]{A.~R.~Wasserman}
\affil{Department of Astronomy, University of Illinois at Urbana-Champaign, 1002 W. Green St., IL 61801, USA}
\affil{Center for Astrophysical Surveys, National Center for Supercomputing Applications, Urbana, IL, 61801, USA}

\author[0009-0004-4256-1209]{S.~Yan}
\affil{Department of Physics, Tsinghua University, Shuangqing Road, Beijing, China}

\author{Y.~Yang}
\affil{Department of Astronomy, University of California, Berkeley, CA 94720-3411, USA}

\author[0000-0002-8296-2590]{J.~Zhang}
\affil{Yunnan Observatories (YNAO), Chinese Academy of Sciences, Kunming 650216, China}
\affil{Key Laboratory for the Structure and Evolution of Celestial Objects, CAS, Kunming, 650216, China}

\author{W.~Zheng}
\affil{Department of Astronomy, University of California, Berkeley, CA 94720-3411, USA}

\begin{abstract}

We present ultraviolet/optical/near-infrared observations and modeling of Type II supernovae (SNe II) whose early-time ($\delta t < 2$~days) spectra show transient, narrow emission lines from shock ionization of confined ($r < 10^{15}$~cm) circumstellar material (CSM). The observed electron-scattering broadened line profiles (i.e., IIn-like) of \ion{H}{i}, \ion{He}{i/ii}, \ion{C}{iv}, and \ion{N}{iii/iv/v} from the CSM persist on a characteristic timescale ($t_{\rm IIn}$) that marks a transition to a lower-density CSM and the emergence of Doppler-broadened features from the fast-moving SN ejecta. Our sample, the largest to date, consists of 39 SNe with early-time IIn-like features in addition to 35 ``comparison'' SNe with no evidence of early-time IIn-like features, all with ultraviolet observations. The total sample consists of 50 unpublished objects with a total of 474 previously unpublished spectra and 50 multiband light curves, collected primarily through the Young Supernova Experiment and Global Supernova Project collaborations. For all sample objects, we find a significant correlation between peak ultraviolet brightness and both $t_{\rm IIn}$ and the rise time, as well as evidence for enhanced peak luminosities in SNe~II with IIn-like features. We quantify mass-loss rates and CSM density for the sample through matching of peak multiband absolute magnitudes, rise times, $t_{\rm IIn}$ and optical SN spectra with a grid of radiation hydrodynamics and non-local thermodynamic equilibrium (nLTE) radiative-transfer simulations. For our grid of models, all with the same underlying explosion, there is a trend between the duration of the electron-scattering broadened line profiles and inferred mass-loss rate: $t_{\rm IIn} \approx 3.8[\dot{M}/$(0.01 \mdot)]~days. 



\end{abstract}


\section{Introduction} \label{sec:intro}

Shock breakout (SBO) from a red supergiant (RSG) premieres as a burst of luminous ultraviolet (UV) and X-ray radiation that lasts several hours \citep{waxman17,Goldberg22}. The breakout photons escape from a characteristic optical depth ($\tau \approx c/v_{\rm sh}$, where $c$ is the speed of light and $v_{\rm sh}$ is the shock velocity), which could occur either in the outer RSG envelope or inside of high-density circumstellar material (CSM) surrounding the star at the time of first light \citep{Chevalier11,haynie21}. Following first light at the characteristic optical depth, the photons emitted at SBO will ``flash ionize'' the CSM, leading to narrow emission lines in the early-time spectra of highly ionized elements such as \ion{He}{ii}, \ion{C}{iv}, \ion{O}{vi}, and \ion{N}{iii/iv/v}. However, without the presence of a continuous ionizing source in the CSM after SBO, the CSM will quickly recombine and the ``flash ionization'' phase will conclude within minutes to hours after SBO ($t_{\rm rec} \propto 1/n_e$, where $n_e$ is number density of free electrons) given the densities typical of RSG environments (e.g., $n \approx 10^{7-10}$~cm$^{-3}$, $\rho \approx 10^{-14} - 10^{-17}$~g~cm$^{-3}$ at $r < 2R_{\star}$).  

For Type II supernovae (SNe II) propagating in a low-density environment ($\rho < 10^{-15}$~g~cm$^{-3}$ at $r \approx 10^{14-15}$~cm), the fast-moving SN ejecta will then sweep up low-density, optically thin CSM and the Doppler-broadened spectral features of SN ejecta will be visible within hours-to-days after first light. For higher densities associated with some SN~II environments (e.g., $\rho \gtrsim 10^{-14}$~g~cm$^{-3}$), radiative cooling of the shocked regions will result in the formation of a cold dense shell (CDS) even at early times \citep{Chevalier94, Chevalier17}. Consequently, SNe II in dense CSM ($\rho \gtrsim 10^{-14}$~g~cm$^{-3}$ at $r \approx 10^{14-15}$~cm) present a unique opportunity to probe more extreme RSG mass-loss histories through ultrarapid (``flash'') spectroscopy during the explosion’s first days \citep{galyam14,Khazov16,yaron17,terreran22,wjg23a}. 

Following SN ejecta-CSM interaction, the forward-shock kinetic luminosity goes as $L_{\rm sh} = \dot M v_{\rm sh}^3 / 2v_w$, where $v_{\rm sh}$ is the shock velocity, $v_w$ is the wind velocity, and $\dot M$ is the mass-loss rate (e.g., $\dot M = 4\pi\rho r^2 v_w$). Consequently, in high-density CSM, the SN shock power is quite high ($>10^{41}$~erg~s$^{-1}$ for $\dot M > 10^{-4}$~\mdot) and for typical post-shock temperatures ($T_{\rm sh} \approx 10^{5-8}$~K), the gas will cool primarily via free-free emission, as well as line emission \citep{Chevalier17}. High-energy photons emitted at the shock front will continue to ionize the intervening CSM, prolonging the formation of high-ionization recombination lines present during the ``flash ionization'' phase. During this ``photoionization'' phase, recombination photons inside the CSM will encounter a large number density of free electrons and consequently participate in multiple scatterings before they exit the CSM. Observationally, this manifests as spectral line profiles that contain a combination of a narrow core and Lorentzian wings (i.e., IIn-like), the former tracing the expansion velocities in the wind/CSM while the latter resulting from the photon's frequency shift following electron scattering \citep{Chugai01, Dessart09, HuangES}. In the single-scattering limit, the observed emission line will map the thermal velocity of the free electrons ($v_e \approx 10^3 (T/10^{4.5}~{\rm K})^{1/2}~\kms$), but with sufficiently large electron-scattering optical depths ($\tau \approx 3$--10) the resulting line profiles can extend to thousands of $\kms$. However, as the shock samples lower density CSM at large radii (assuming a wind-like profile or CSM shell), these electron-scattering profiles will vanish within days-to-weeks of first light, with the SN photosphere then revealing the CDS, if present, and subsequently the fast moving SN ejecta \citep{dessart17}. However, departures from CSM spherical symmetry and/or homogeneous density may blur the transition between these three phases; for example, Doppler-broadened line profiles can appear while spectral signatures of unshocked optically-thick CSM are still present in the early-time spectra.

Given the transient nature of these spectral features, high-cadence ``flash'' spectroscopy during the first days post-SBO is essential to map the densities, kinematics, and progenitor chemical composition in the pre-explosion environment at radii of $r<10^{15}$~cm. Consequently, such observations provide a window into the largely unconstrained stages of stellar evolution in the final years-to-months before core collapse. Enabled by the advent of high-cadence surveys in the past decade, the study of SNe II with such photoionization spectral features has revealed enhanced, late-stage mass-loss rates in RSG progenitor systems. Interestingly, one of the first records of this phenomenon was in SN 1983K \citep{Niemela85}, but garnered the most attention through observations of SN~1998S \citep{fassia00, Leonard00}, which showed high-ionization features at early-times ($\delta t < 7$~days) and then transitioned to a Type IIL supernova (SN IIL) at later phases ($\delta t > 7$~days) as the IIn-like features disappeared. Spectroscopic and photometric modeling of SN~1998S suggested significant mass loss of $\dot{M} \approx 10^{-2}$~\mdot for $v_w \approx 50-100~\kms$ \citep{shivvers15, Dessart16}, capable of producing transient IIn-like features and an overluminous light curve, placing it as extreme compared to normal SNe II, but not quite placing it in the Type IIn SN subclass. 

Since SN~1998S, a number of SNe II have been discovered with photoionization spectral features from SN ejecta-CSM interaction at early times. Modeling of the photoionized spectra continues to point toward confined ($r<10^{15}$~cm), high-density ($\dot{M} \approx 10^{-3}$-- $10^{-2}$~\mdot, $v_w\approx50$--100~km~s$^{-1}$) CSM created in the final years before explosion (e.g., PTF11iqb, \citealt{smith15}; SN~2013fs, \citealt{yaron17, dessart17}; SN~2014G, \citealt{Terreran16}; SN~2016bkv, \citealt{Hosseinzadeh18, Nakaoka18}; SN~2017ahn, \citealt{Tartaglia21}; SN~2018zd, \citealt{zhang20, Hiramatsu21}; SN~2020pni, \citealt{terreran22}; SN~2020tlf, \citealt{wjg22}; SN~2023ixf, \citealt{wjg23a, Bostroem23, Smith23, Zimmerman23}; Fig. 11 of \citealt{Brethauer22}). Sample studies have sought to uncover the rates of such events, the most recent estimate being $>40\%$ for all SN II discovered within 2 days of first light \citep{bruch21, bruch23}. Intriguingly, mass-loss rates derived for individual explosions stand in contradiction with observations of weak, steady-state mass loss (e.g., $10^{-6}$~\mdot) in observed RSGs \citep{Beasor20} as well as the quiescent behavior of SN II progenitor stars in pre-explosion imaging \citep{Kochanek17}. However, this could be related to the pre-explosion timescales that each method is probing. Furthermore, beyond H-rich SNe, ``flash spectroscopy'' has aided in significant breakthroughs in our understanding of H-poor SN progenitor identity and late-stage evolution e.g., Type IIb, Type Ibc, and calcium-rich (e.g., \citealt{galyam14, Pastorello15, wjg20, chugai22, davis23, Wang23}). However, it is also evident that some massive stars undergo enhanced mass loss even before their final years (e.g., $t \approx 10$--1000~yr) and therefore X-ray/radio observations as well as long-term UV/optical monitoring is essential to reconstruct a more complete mass-loss history (e.g., \citealt{Chevalier98, Fransson96, Chevalier06, Weiler02, wellons12, Milisavljevic15, Chevalier17, Margutti17, Brethauer22, demarchi22, stroh21, dessart23a, Shahbandeh23, Grefenstette23, Berger23, Panjkov23}). 


In this study, we present observations and modeling of the largest sample to date of SNe II with early-time ($\delta t < 2$~days) spectroscopic signatures of CSM interaction. This sample consists of 27 unpublished SNe with photoionization emission features, which includes 293 new spectra as well as 27 UV/optical/near-infrared (NIR) light curves. In Section 2 we define the sample and present the spectroscopic and photometric observations. Section 3 presents an analysis of the bolometric and multiband light curves as well as early-time and photospheric-phase spectra. In Section 4 we present the \heracles/\cmfgen\ model grid and the derived mass-loss rates and CSM densities based on model comparisons to the sample data. Our results are discussed in Section 5 and our conclusions are in Section 6. 

All phases reported in this paper are with respect to the adopted time of first light (Table \ref{tab:gold_sample_specs}) and are in rest-frame days. The time of first light ($\delta t$) and its uncertainty are calculated from the average phase between the last deep non-detection and the first detection using forced photometry from the survey that initially imaged the SN (e.g., ZTF, ATLAS, YSE, DLT40). However, we note that the first-light phase could be earlier in some instances given a shallow depth of the last non-detection limit. Furthermore, ``first light'' in this case only refers to when photons are first detected from the SN, which is unlikely to reflect the first emission from the explosion. When available for a given sample object, we adopt the time of first light reported in a previously published study and confirm that this phase is consistent with first detection and last non-detection using forced photometry. When possible, we use redshift-independent host-galaxy distances and adopt standard $\Lambda$CDM cosmology ($H_{0}$ = 70~km~s$^{-1}$~Mpc$^{-1}$, $\Omega_M = 0.27$, $\Omega_{\Lambda} = 0.73$) if only redshift information is available for a given object. 

\section{Observations} \label{sec:obs}

\begin{figure*}
\centering
\includegraphics[width=\textwidth]{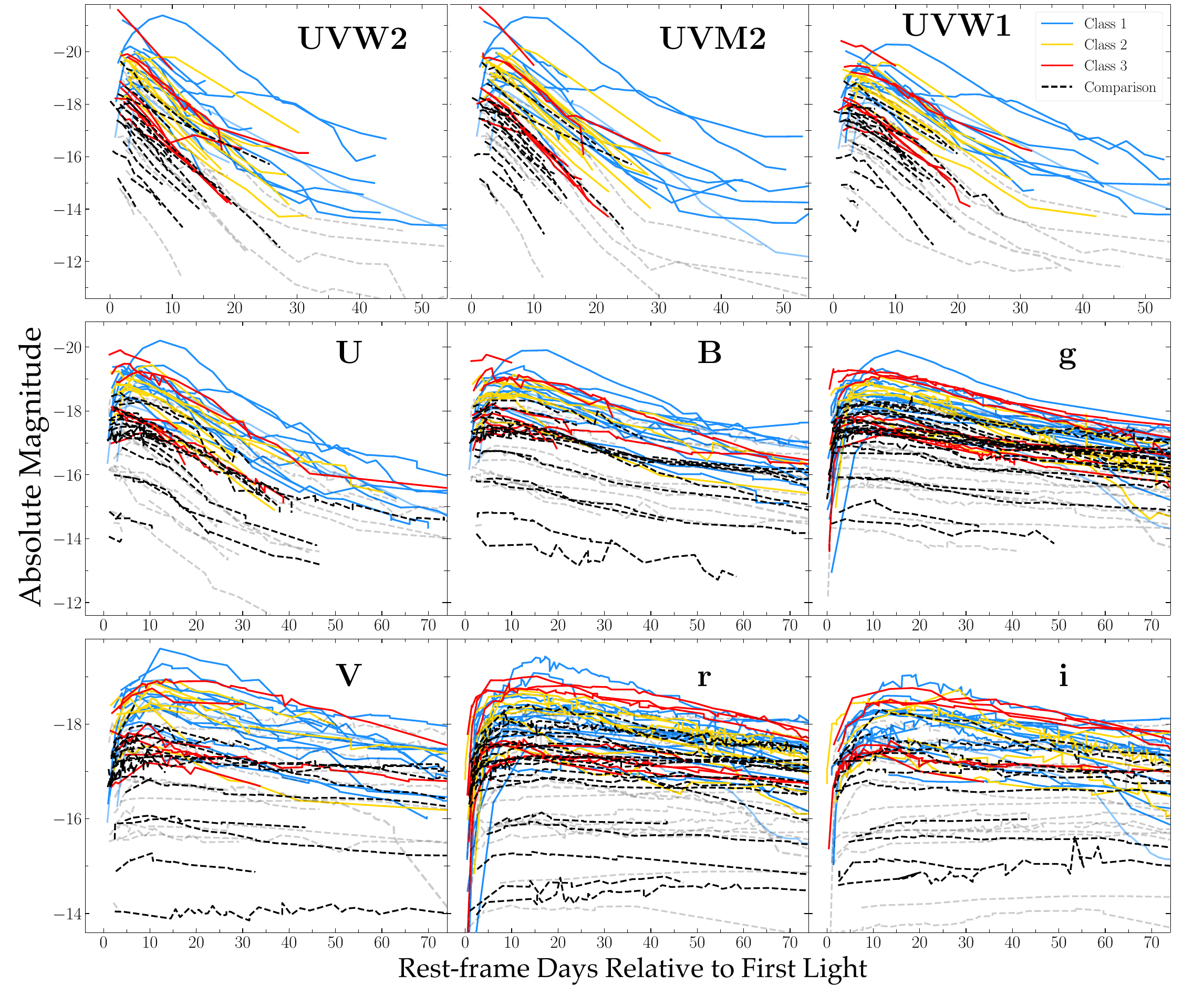}
\caption{ {\it Left to tight, top to bottom:} Early-time, extinction corrected \textit{w2-}, \textit{m2-}, \textit{w1-}, \textit{U/u-}, \textit{B/b-}, \textit{V/v-}, \textit{g-}, \textit{r-}, and \textit{i-}band light curves of SNe~II with IIn-like profiles in their early spectra. No K-corrections have been applied. Gold and silver samples shown in blue, yellow, and red; comparison sample plotted as black dashed lines. Solid colored curves represent the subsample of objects at $D>40$~Mpc. Compared to SNe~II without IIn-like features (i.e., comparison sample), objects with confirmed IIn-like signatures have notably more luminous and longer-lasting UV emission at early times. Furthermore, Class 1 objects that show longer lived IIn-like profiles of \ion{He}{ii} and \ion{N}{iii} are typically brighter than other gold-sample objects with shorter-lived IIn-like features. The variance of the total sample decreases with increasing wavelength, with the least luminous objects being those in the comparison sample. \label{fig:LC_all} }
\end{figure*}

\subsection{Sample Definition}\label{SubSec:Sample}

Our total sample consists of 74 SNe II, 39 of which show spectroscopic evidence for CSM interaction at early times ($\delta t < 10$~days) through the detection of transient IIn-like features. Gold-sample objects have a spectrum obtained at $\delta t<2$~days while silver-sample objects only have spectra obtained at $\delta t>2$~days. Additionally, we include 35 SNe~II with ``flash spectroscopy'' (i.e., spectra at $\delta t < 2$~days) but no detection of IIn-like features (the comparison sample). For the gold and comparison samples, we require that the uncertainty in the time of first light be $<1$~day. To construct the total sample, we first query the Transient Name Server (TNS)\footnote{\url{https://www.wis-tns.org/}} for every transient discovered between 2004-11-20 and 2022-08-01 and then select only objects with Type II-like classification (e.g., SN II, SN IIP, SN IIL, SN IIn, SN II-pec) at redshifts $z < 0.05$, which returns 1697 SNe. For those SNe~II, we keep objects having spectra within 3 days of discovery, which returns 428 objects. Next, we query the {\it Swift}-UVOT data archive and record how many total observations of the SN location exist within 10 days of discovery. We then keep objects with $>2$ {\it Swift}-UVOT observations at $<10$ days post-discovery, which returns 114 total objects, after cutting SNe~IIn. This exercise is repeated using the Weizmann Interactive Supernova Data Repository (WISeREP)\footnote{\url{https://www.wiserep.org/}}, finding 48 total objects, both with and without IAU names, that meet the sample selection criteria listed above. We are then left with 137 total SNe~II after removing duplicate objects. Lastly, we cut all SNe~II with no IIn-like features that do not have a spectrum at $\delta t < 2$~days and/or uncertainty in the time of first light of $>1$~day. Furthermore, we cut all objects that do not have $\Delta m > 1$~mag between last non-detection and first detection, in the same filter, and/or $\Delta M > 3$~mag between first detection and peak brightness. Consequently, our total sample contains 74 objects: 20 gold, 19 silver, and 35 comparison-sample SNe~II. In this data release, we also include multicolor light curves and spectra of five additional SNe~II with IIn-like features: 2018cvn, 2018khh, 2019ofc, 2019nyk, 2021ulv. These objects are not used in our analysis given the lack of UV photometry. 

The gold/silver samples contain 12 previously published objects with a total of 208 spectra and 12 UV/optical light curves, in addition to 27 unpublished objects with a total of 293 spectra and 27 UV/optical light curves. The comparison sample contains 12 previously published and 23 unpublished objects, with a total of 464 spectra. As shown in Figure \ref{fig:selection}, peak absolute magnitude as a function of SN distance reveals a trend consistent with a Malmquist bias i.e., only higher luminosity objects can be detected at further distances. An examination of peak apparent magnitude before extinction corrections are applied shows that the sample extends to low luminosities, with the majority of nearby ($D < 20$~Mpc) events being in the comparison sample. The lack of nearby gold/silver sample objects may be the result of selection effects and/or the intrinsic rarity of SNe~II with IIn-like features. Furthermore, the difference in redshift distribution (top-left panel of Fig. \ref{fig:selection}) implies that the gold and comparison samples may not arise from the same parent distribution. We account for this difference by applying a distance cut in our comparison of observables in each sub-sample in \S\ref{subsec:phot_properties}. Additionally, we note the lack of highly reddened SNe ($A_V > 3$~mag; \citealt{Jencson19}) in our sample, which represents a selection effect in our sample because these objects are unlikely to have associated {\it Swift}-UVOT observations.

Within both subsamples, the color delineation (e.g., Figures \ref{fig:LC_all} and \ref{fig:FS_all_gold}) is as follows: at phases of $t\approx2$~days post-first light, blue-colored objects (``Class 1'') show high-ionization emission lines of \ion{N}{iii}, \ion{He}{ii}, and \ion{C}{iv} (e.g., SNe~1998S, 2017ahn, 2018zd, 2020pni, 2020tlf), yellow-colored objects (``Class 2'') have no \ion{N}{iii} emission but do show \ion{He}{ii} and \ion{C}{iv} (e.g., SNe~2014G, 2022jox), and red-colored objects (``Class 3'') only show weaker, narrow \ion{He}{ii} emission superimposed with a blueshifted, Doppler-broadened \ion{He}{ii} (e.g., SN~2013fs, 2020xua). However, it should be noted that high-ionization lines of \ion{O}{v/vi}, \ion{C}{v}, and \ion{N}{iv} are also present in SN~2013fs at $t < 1$~day owing to a more compact CSM than other CSM-interacting SNe~II \citep{yaron17, dessart17}, thus, the color delineation is epoch dependent.

All targets were selected from private collaborations/surveys as well as all public/published studies on SNe II with prominent or potential IIn-like features in their early-time spectra (Tables \ref{tab:gold_sample_specs} and \ref{tab:silver_sample_specs}). We emphasize that while the SNe in our sample may show IIn-like line profiles at early times, they are not prototypical SNe IIn that show relatively narrow line profiles from CSM interaction for weeks-to-months following explosion (e.g., SNe~2005ip, 2010jl; \citealt{Smith09, Taddia13, Gall14, Fransson14, D15_2n}). The IIn-like profiles in our sample objects fade within days to a week after first light and the explosion proceeds to evolve photometrically and spectroscopically as a seminormal RSG explosion --- a light-curve plateau or linear (in magnitudes) decline where hydrogen recombination mitigates the release of stored radiative energy and the photospheric spectra are dominated by P~Cygni profiles formed from H, He, and Fe-group elements in the SN ejecta.

\subsection{Photometric Observations}\label{SubSec:Phot}

All gold-, silver- and comparison-sample objects were observed during their evolution with the Ultraviolet Optical Telescope (UVOT; \citealt{Roming05}) onboard the {\it Neil Gehrels Swift Observatory} \citep{Gehrels04}. We performed aperture photometry with a 5$\arcsec$ region radius with \texttt{uvotsource} within HEAsoft v6.26\footnote{We used the calibration database (CALDB) version 20201008.}, following the standard guidelines from \cite{Brown14}\footnote{\url{https://github.com/gterreran/Swift_host_subtraction}}. In order to remove contamination from the host galaxy, we employed images acquired at $\delta t > 1$~yr, assuming that the SN contribution is negligible at this phase. This is supported by visual inspection in which we found no flux at the SN location. We subtracted the measured count rate at the location of the SN from the count rates in the SN images and corrected for point-spread-function (PSF) losses following the prescriptions of \cite{Brown14}. We also note that the $w2$ filter has a known red leak \citep{brown2010}, which could impact post-peak observations when the SN is significantly cooler.  

For the total sample, optical/NIR photometry was obtained from a variety of collaborations and telescopes. Pan-STARRS telescope \citep[PS1/2;][]{Kaiser2002, Chambers2017} imaging in the $grizy$ bands was obtained through the Young Supernova Experiment \citep[YSE;][]{Jones2021}. Data storage/visualization and follow-up coordination was done through the YSE-PZ web broker \citep{Coulter22, Coulter23}. The YSE photometric pipeline is based on {\tt photpipe} \citep{Rest+05}, which relies on calibrations from \cite{Magnier20a} and \cite{waters20}. Each image template was taken from stacked PS1 exposures, with most of the input data from the PS1 3$\pi$ survey. All images and templates were resampled and astrometrically aligned to match a skycell in the PS1 sky tessellation. An image zero-point is determined by comparing PSF photometry of the stars to updated stellar catalogs of PS1 observations \citep{flewelling16}. The PS1 templates are convolved with a three-Gaussian kernel to match the PSF of the nightly images, and the convolved templates are subtracted from the nightly images with {\tt HOTPANTS} \citep{becker15}. Finally, a flux-weighted centroid is found for the position of the SN in each image and PSF photometry is performed using ``forced photometry'': the centroid of the PSF is forced to be at the SN position. The nightly zero-point is applied to the photometry to determine the brightness of the SN for that epoch.

\begin{figure}[t!]
\subfigure{\includegraphics[width=0.45\textwidth]{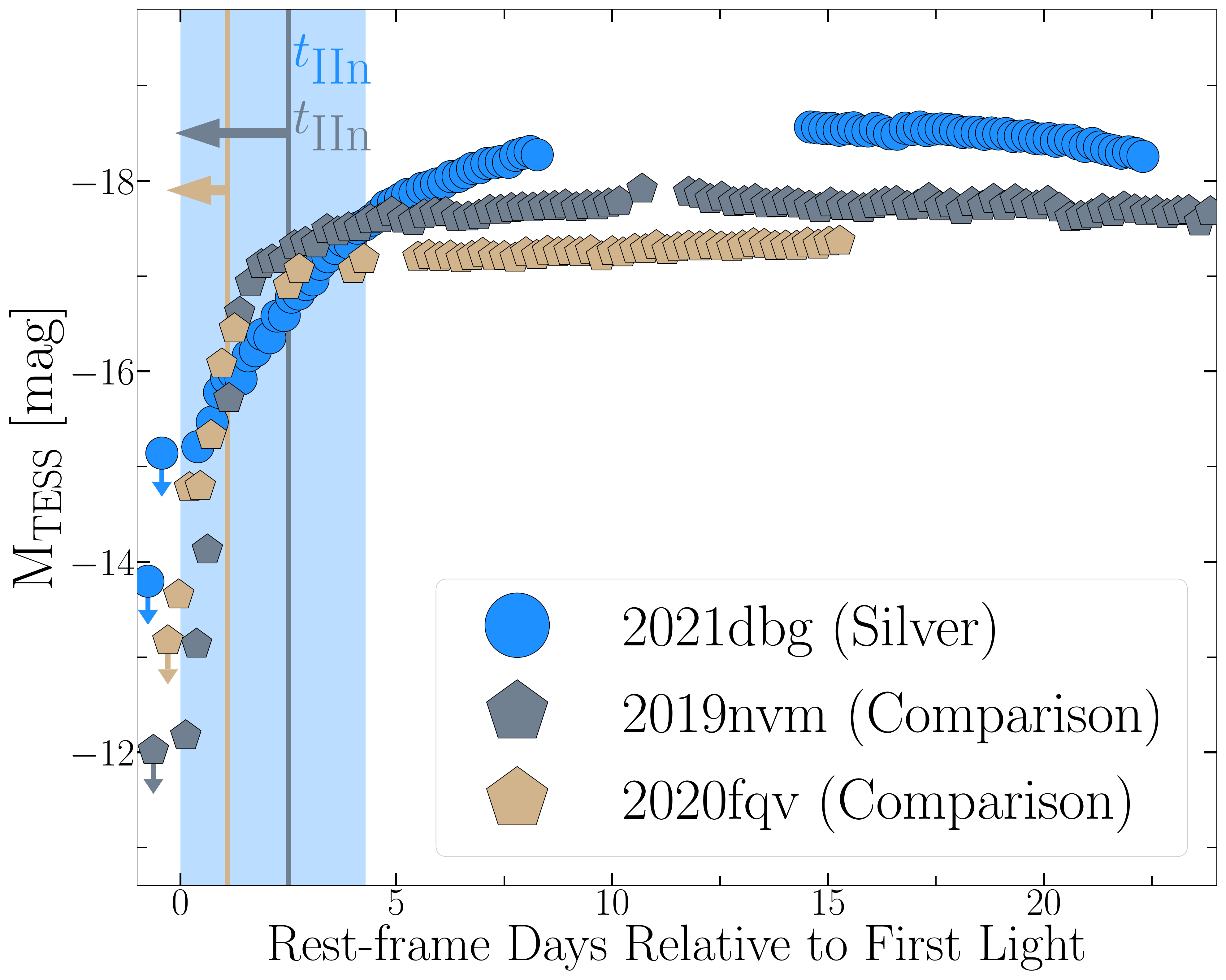}}
\subfigure{\includegraphics[width=0.45\textwidth]{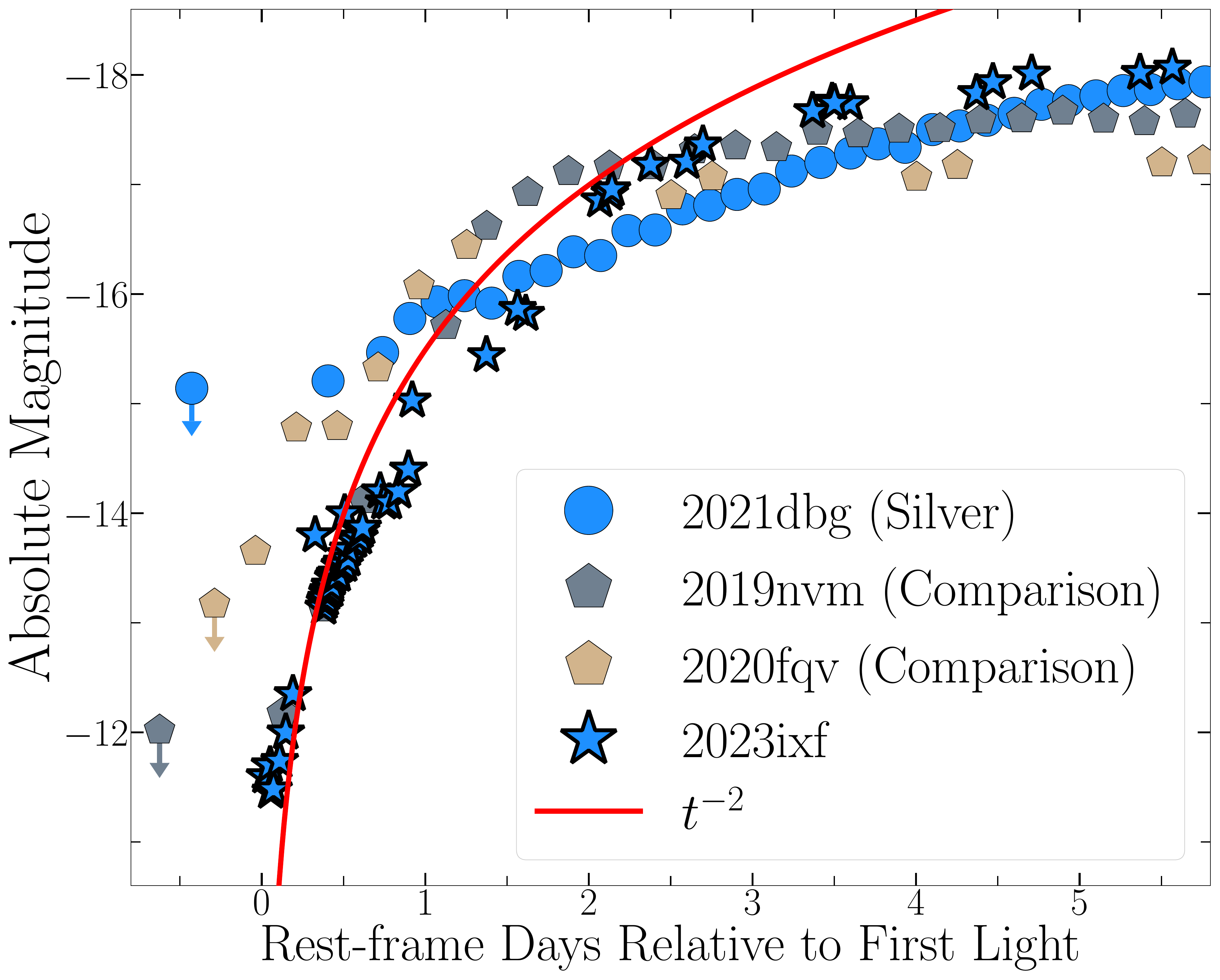}}
\caption{{\it Top:} {\it TESS} ($\lambda_{\rm eff} = 7453$~\AA) light curves (binned) for silver-sample object SN~2021dbg (blue circles) and comparison-sample objects SNe~2019nvm (gray polygons) and 2020fqv (tan polygons). SN~2021dbg shows IIn-like signatures for $\sim 4$~days after first light (blue shaded region), consistent with an increased rise time and peak absolute magnitude. Conversely, the persistence of IIn-like features in SNe~2019nvm and 2020fqv is constrained to $<2.6$ and $<1.1$~days, respectively. These SN light curves are likely consistent with shock-cooling emission from confined ($<2R_{\star}$), high-density stellar material and/or SN ejecta interaction with lower density CSM that extends out to larger distances, neither scenario being able to form IIn-like features. {\it Bottom:} Zoom-in of the first 5 days of the {\it TESS} light curves for SNe~2021dbg, 2020fqv, and 2019nvm compared to ground-based photometry in optical clear- and $r$-band filters of the nearby CSM-interacting SN~II 2023ixf \citep{Hosseinzadeh23}.  \label{fig:TESS} }
\end{figure}

We obtained $uUBVgriz$ imaging with the Las Cumbres Observatory (LCO) 1~m telescopes through the Global Supernova Project (GSP) and YSE. After downloading the {\tt BANZAI}-reduced images from the Las Cumbres Observatory data archive \citep{mccully18}, we used {\tt photpipe} \citep{Rest+05} to perform {\tt DoPhot} PSF photometry \citep{Schechter+93}. All photometry was calibrated using PS1 stellar catalogs described above with additional transformations to the SDSS $u$ band derived from \citet{finkbeiner16}. For additional details on our reductions, see \citet{kilpatrick18}. We also obtained photometry using a 0.7~m Thai Robotic Telescope at Sierra Remote Observatories and the 1~m Nickel telescope at Lick Observatory in the $BVRI$ bands. Images are bias subtracted and field flattened. Absolute photometry is obtained using stars in the $10' \times 10'$ field of view. We also observed objects with the Lulin 1~m telescope in $griz$ bands and the Swope 1~m telescope in $uBVgri$.  Standard calibrations for bias and flat-fielding were performed on the images using {\tt IRAF}, and we reduced the calibrated frames in {\tt photpipe} using the methods described above for the Las Cumbres Observatory images.

Sample objects were also observed with ATLAS, a twin 0.5~m telescope system installed on Haleakala and Maunaloa in the Hawai'ian islands that robotically surveys the sky in cyan (\textit{c}) and orange (\textit{o}) filters \citep[][]{2018PASP..130f4505T}. The survey images are processed as described by \cite{2018PASP..130f4505T} and photometrically and astrometrically calibrated immediately \citep[using the RefCat2 catalogue;][]{2018ApJ...867..105T}. Template generation, image-subtraction procedures, and identification of transient objects are described
by \cite{Smith20}. PSF photometry is carried out on the difference images and all sources greater than 5$\sigma$ are recorded and go through an automatic validation process that removes spurious objects \citep{Smith20}. Photometry on the difference images (both forced and non-forced) is from automated PSF fitting as documented by \cite{2018PASP..130f4505T}. The photometry presented here is weighted averages of the nightly individual 30~s exposures, carried out with forced photometry at the position of each SN. In addition to our observations, we include $gri$-band photometry from the Zwicky Transient Facility (ZTF; \citealt{bellm19, graham19}) forced-photometry service \citep{Masci19}. 

In Figure \ref{fig:TESS}, we present new \textit{Transiting Exoplanet Survey Satellite} ({\it TESS}; \citealt{Ricker15}) light curves of SNe~2019nvm and 2021dbg, reduced using the \texttt{TESSreduce} package \citep{Ridden-Harper21}, compared to previously published {\it TESS} light curve of SN~2020fqv \citep{Tinyanont22}. These observations have been binned to a 6~hr cadence and are able to constrain the uncertainty in the time of first light to a few hours. To our knowledge, SN~2021dbg represents the first SN~II with IIn-like features to have a complete {\it TESS} light curve. 

For all SNe, the Milky Way (MW) $V$-band extinction and color excess along the SN line of site is inferred using a standard \cite{fitzpatrick99} reddening law ($R_V = 3.1$). In addition to the MW color excess, we estimate the contribution of host-galaxy extinction in the local SN environment using \ion{Na}{i}~D absorption lines for all gold-, silver-, and comparison-sample objects. To determine if \ion{Na}{i}~D is detected, we fit the continuum in a region around the transition based on the spectral resolution and calculate the residuals between the continuum fit and the spectral data. We then integrate the residual flux and confirm that it is greater than or equal to three times the residual flux uncertainty in order to claim a ``detection.'' We calculate the \ion{Na}{i}~D equivalent width (EW) and use $A_V^{\rm host} = (0.78\pm0.15)~{\rm mag} \times ({\rm EW_{NaID}}$/\AA) from \cite{Stritzinger18} to convert these EWs to an intrinsic host-galaxy $E(B-V)$, also using the \cite{fitzpatrick99} reddening law. A visualization of this method is shown in Figure \ref{fig:extinction}. For non-detections, we calculate an upper limit on the EW and host reddening using the fitted continuum flux. We present a detailed discussion of the host extinction uncertainties in Appendix Section A. We do not apply alternative methods for estimating host extinction such as using the diffuse interstellar band (DIB) at 5780~\AA\ \citep{Phillips13}, which has been shown to yield consistent extinction values to \ion{Na}{i}~D EW for other SNe \citep{Hosseinzadeh22}. We present cumulative distributions of the gold, silver and comparison sample host extinction in Figure \ref{fig:selection}, which shows consistency across sub-samples. We test this further by applying a logrank test and find a $35\%$ chance probability that the gold and comparison sample reddening come from the same distribution, indicating that the reddening correction is likely applied equally across sub-samples. Furthermore, in Appendix \S\ref{subsec:host_ext}, we discuss the use of colors as a metric for host galaxy reddening.

All adopted extinction (MW and host), redshift, distance, and first light date values are reported for gold-, silver-, and comparison-sample objects in Tables \ref{tab:gold_sample_specs}, \ref{tab:silver_sample_specs}, and \ref{tab:control_sample_specs}, respectively. Complete, multiband light curves are shown in Figure \ref{fig:LC_all}. All photometric data/logs will be publicly available in an online data repository.\footnote{\url{https://github.com/wynnjacobson-galan/Flash_Spectra_Sample}}  

\subsection{Spectroscopic Observations}\label{SubSec:Spec}

\begin{figure*}[t]
\centering
\includegraphics[width=0.99\textwidth]{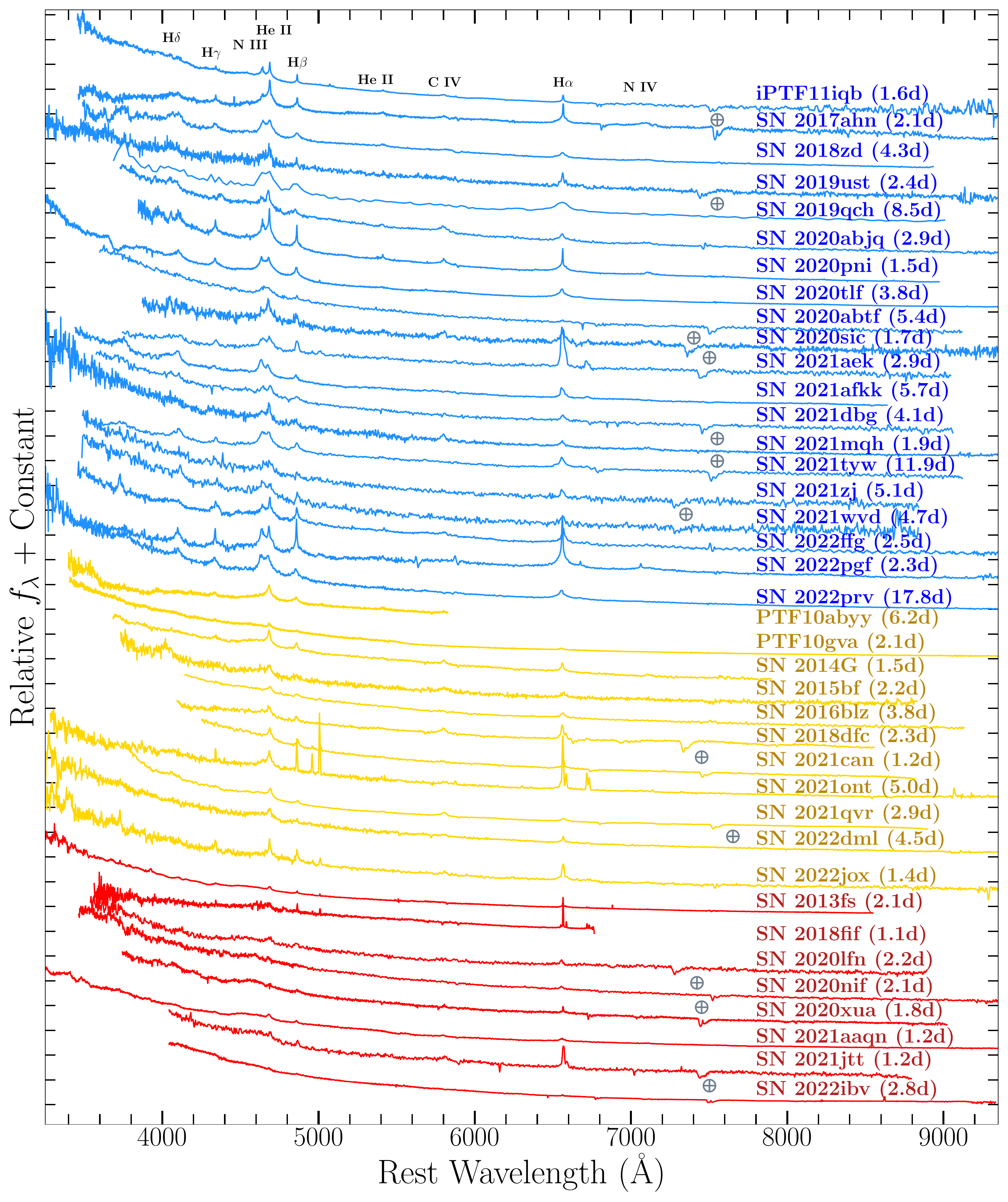}
\caption{Early-time (``flash'') spectra of all gold- and silver-sample SNe II (e.g., \S\ref{SubSec:Sample}); phases are relative to time of first light. All plotted SNe show transient, IIn-like (i.e., electron-scattering broadened) line profiles formed from persistent photoionization of dense, slow, unshocked CSM. Objects in blue (``Class 1'') show prominent \ion{He}{ii} and \ion{N}{iii} emission, objects in yellow (``Class 2'') exhibit only prominent \ion{He}{ii} emission, and objects in red (``Class 3'') have weak \ion{He}{ii} emission. Gray circles with a plus indicate telluric absorption. We note that because a number of spectra were obtained from public databases, there has not been a consistent flux calibration applied and therefore the relative continuum shapes should be interpreted with caution.   \label{fig:FS_all_gold} }
\end{figure*}

We obtained spectra for sample objects with the Kast spectrograph on the 3~m Shane telescope at Lick Observatory \citep{KAST} and Keck/LRIS \citep{oke95}. For all of these spectroscopic observations, standard CCD processing and spectrum extraction were accomplished with \textsc{IRAF}\footnote{\url{https://github.com/msiebert1/UCSC\_spectral\_pipeline}}. The data were extracted using the optimal algorithm of \citet{1986PASP...98..609H}.  Low-order polynomial fits to calibration-lamp spectra were used to establish the wavelength scale and small adjustments derived from night-sky lines in the object frames were applied.

Las Cumbres Observatory optical spectra were taken with the FLOYDS spectrographs \citep{Brown13} mounted on the 2~m Faulkes Telescope North and South at Haleakala (USA) and Siding Spring (Australia), respectively, through the Global Supernova Project. A $2\arcsec$ slit was placed on the target at the parallactic angle \citep{filippenko82}. One-dimensional spectra were extracted, reduced, and calibrated following standard procedures using the FLOYDS pipeline\footnote{\url{https://github.com/svalenti/FLOYDS\_pipeline}} \citep{valenti14b}.


Spectra were also obtained with the Alhambra Faint Object Spectrograph (ALFOSC) on The Nordic Optical Telescope (NOT), the Goodman spectrograph \citep{clemens04} at the Southern Astrophysical Research (SOAR) telescope, Gemini Multi-Object Spectrographs (GMOS), Wide-Field Spectrograph (WiFeS) at Siding Spring, Binospec on the MMT \citep{Fabricant19}, Lijiang 2.4-m telescope (+YFOSC) \citep{fan15}, and SpeX \citep{Rayner03} at the NASA Infrared Telescope Facility (IRTF). All of the spectra were reduced using standard techniques, which included correction for bias, overscan, and flat-field. Spectra of comparison lamps and standard stars acquired during the same night and with the same instrumental setting have been used for the wavelength and flux calibrations, respectively. When possible, we further removed the telluric bands using standard stars. Given the various instruments employed, the data-reduction steps described above have been applied using several instrument-specific routines. We used standard \textsc{IRAF} commands to extract all spectra.

Sample spectral data were also collected using EFOSC2 \citep{buzzoni84} at the 3.58~m ESO New Technology Telescope (NTT) through the ePESSTO+ program \citep{smartt15}. Standard data-reduction processes were performed using the PESSTO pipeline \citep{smartt15}\footnote{\url{https://github.com/svalenti/pessto}}. The reduced spectrum was then extracted, and calibrated in wavelength and flux. In some instances, public classification spectra from TNS as well as published data stored in WISeREP were used in the presented sample. Early-time spectra for the gold and silver samples are presented in Figure \ref{fig:FS_all_gold}, with comparison-sample spectra shown in Figure \ref{fig:FS_all_comp}. In total, this study includes 491 published and 474 previously unpublished spectra of SNe~II. All spectroscopic data/logs will be publicly available in an online data repository.\footnote{\url{https://github.com/wynnjacobson-galan/Flash_Spectra_Sample}}

\section{Analysis}\label{sec:analysis}

\subsection{Photometric Properties}\label{subsec:phot_properties}

\begin{figure*}
\centering
\subfigure{\includegraphics[width=\textwidth]{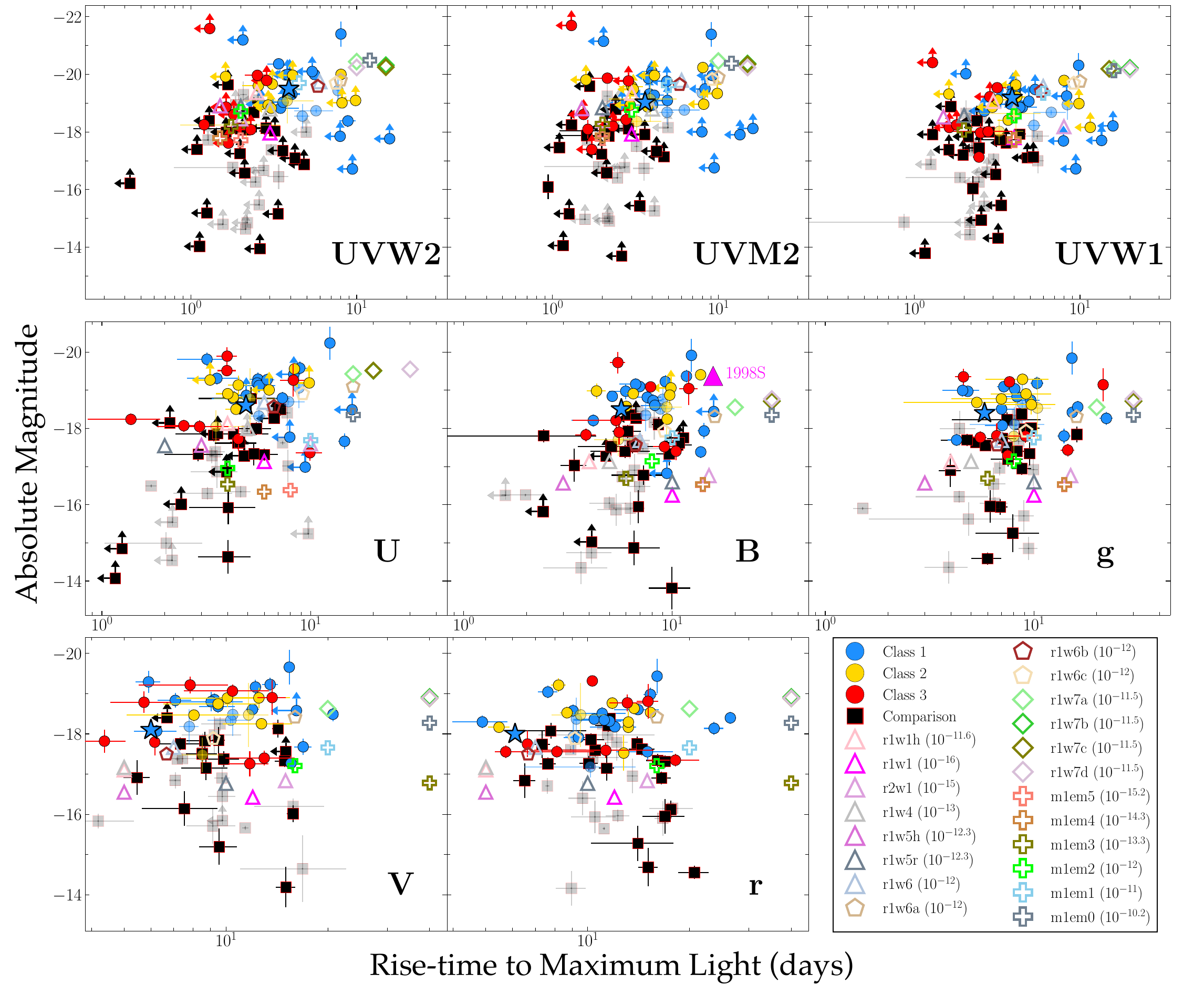}}
\caption{{\it Left to right, top to bottom:} Peak absolute magnitude versus rise time in the \textit{w2}, \textit{m2}, \textit{w1}, \textit{u}, \textit{B/b}, \textit{V/v}, \textit{g}, and \textit{r} bands. Gold/silver samples shown as blue/yellow/red circles and the comparison sample is shown as black squares. Solid colored points represent the subsample of objects at $D>40$~Mpc. Parameters from the \cmfgen\ model grid (\S\ref{subsec:cmfgen}) are plotted as colored stars, polygons, diamonds and plus signs with the CSM densities at $10^{14}$~cm (in g~cm$^{-3}$) for each model displayed in parentheses. SNe~1998S and 2023ixf are shown for reference as a magenta triangle and blue star, respectively. We note that the model parameters do not cover the dynamical range of the observations, which will influence the derivation of CSM properties for some objects (\S\ref{sec:modeling}). Furthermore, in the UV bands, the data show significantly larger variance than the models, which follow a well-defined trend. This likely indicates a dependence on a variable not included in the models.  \label{fig:Max_RT} }
\end{figure*}

\begin{figure*}[t!]
\centering
\subfigure{\includegraphics[width=0.33\textwidth]{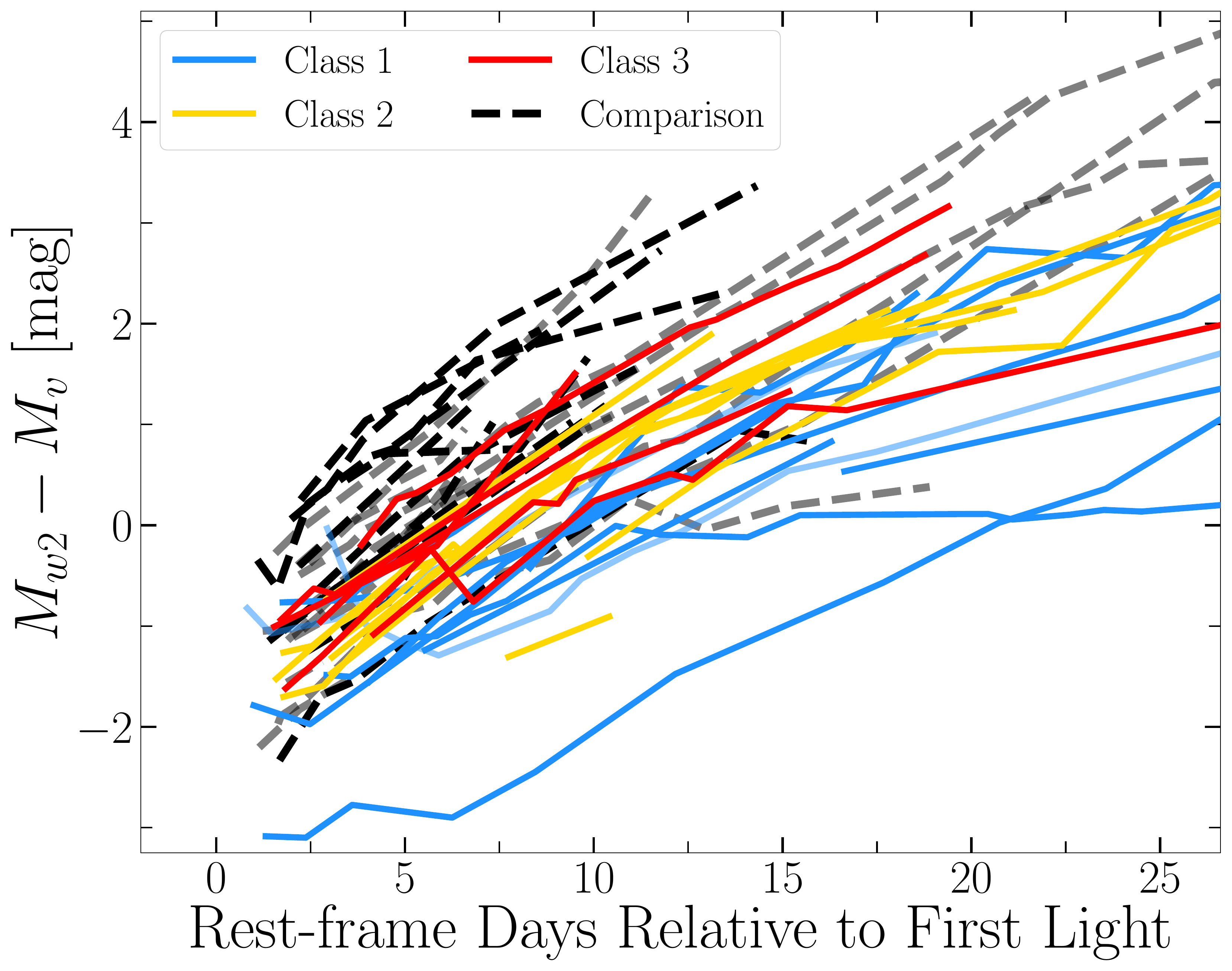}}\subfigure{\includegraphics[width=0.34\textwidth]{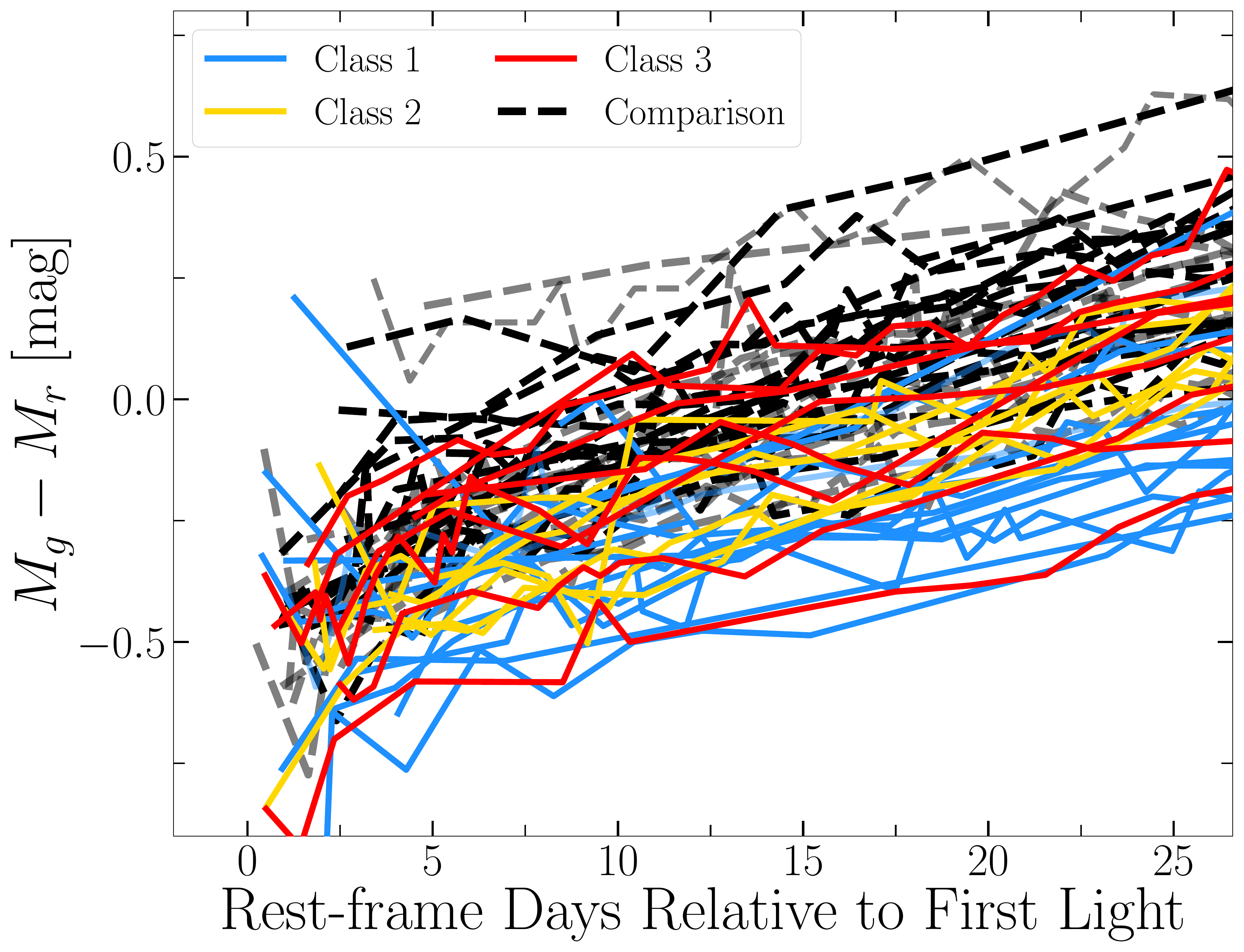}}
\subfigure{\includegraphics[width=0.32\textwidth]{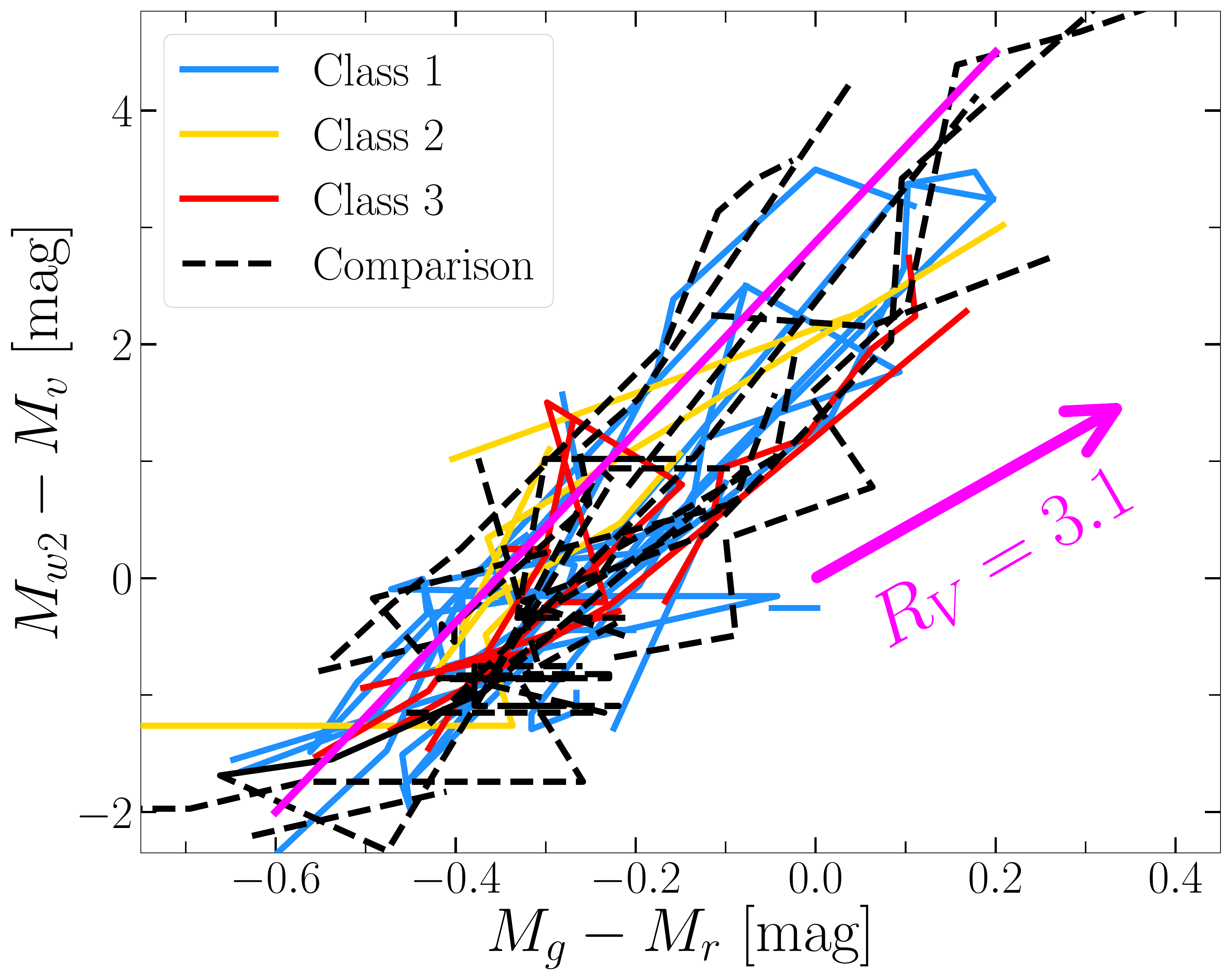}}\\
\caption{{\it Left:} Early-time, reddening-corrected $W2-V$ color plot for gold- and silver-sample objects (red, yellow, blue lines) compared to comparison sample objects (black dashed lines). Solid colored curves represent the subsample of objects at $D>40$~Mpc. Gold- and silver-sample objects, in particular the Class 1 objects, show significantly bluer colors than Class 2/3 or comparison-sample objects, which is indicative of increased temperatures from persistent CSM interaction. {\it Middle:} Early-time, reddening-corrected $g-r$ color plot shows a less clear delineation between objects/classes with varying signatures of CSM interaction, suggesting that the UV colors are the most sensitive metric for confirming ejecta-CSM interaction. {\it Right:} $W2-V$ vs. $g-r$ colors for gold- and comparison-sample objects. The reddening vector for $R_V = 3.1$ using the \cite{fitzpatrick99} reddening law is shown as a magenta arrow.  \label{fig:colors} }
\end{figure*}

\begin{figure*}
\centering
\subfigure{\includegraphics[width=0.51\textwidth]{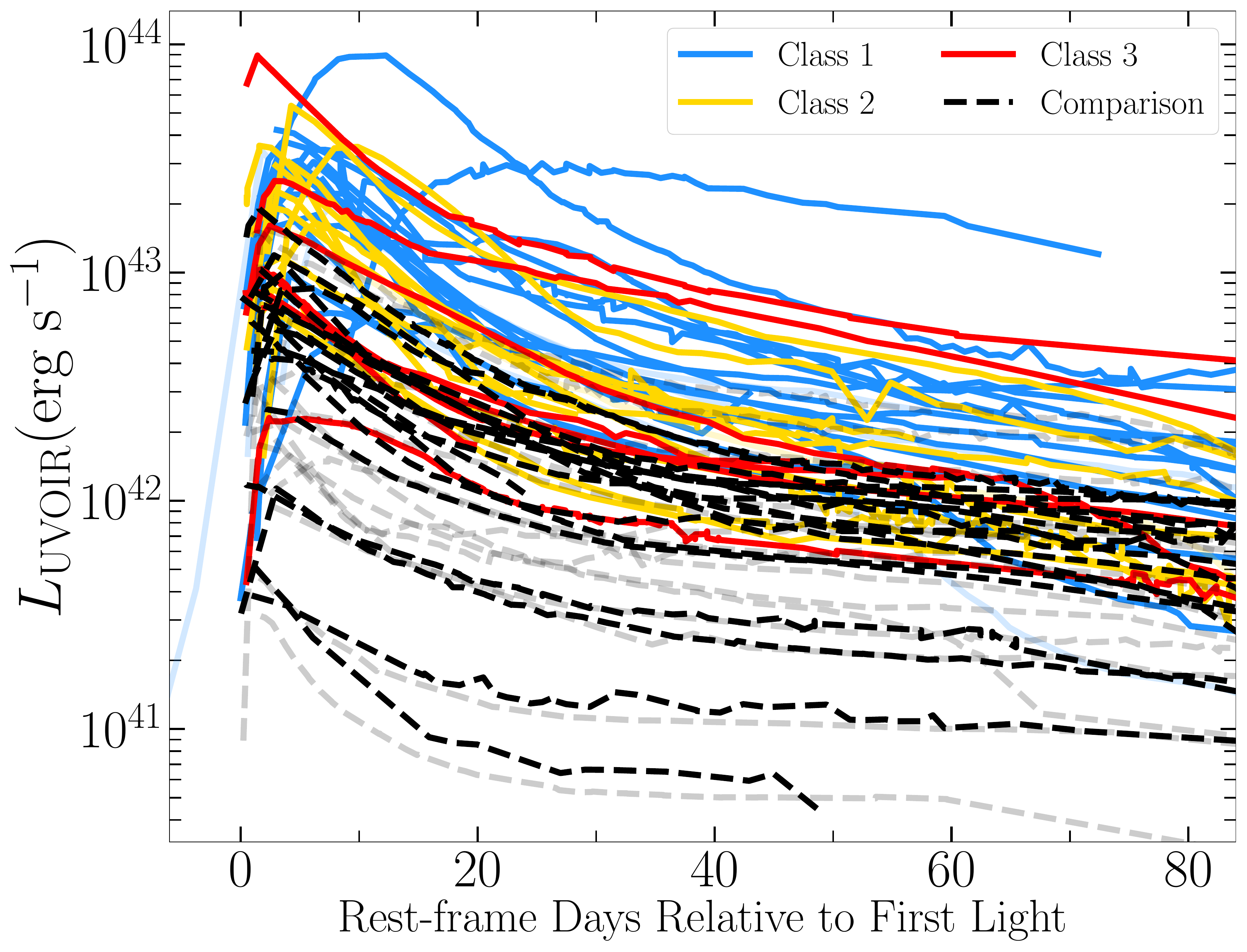}}
\subfigure{\includegraphics[width=0.48\textwidth]{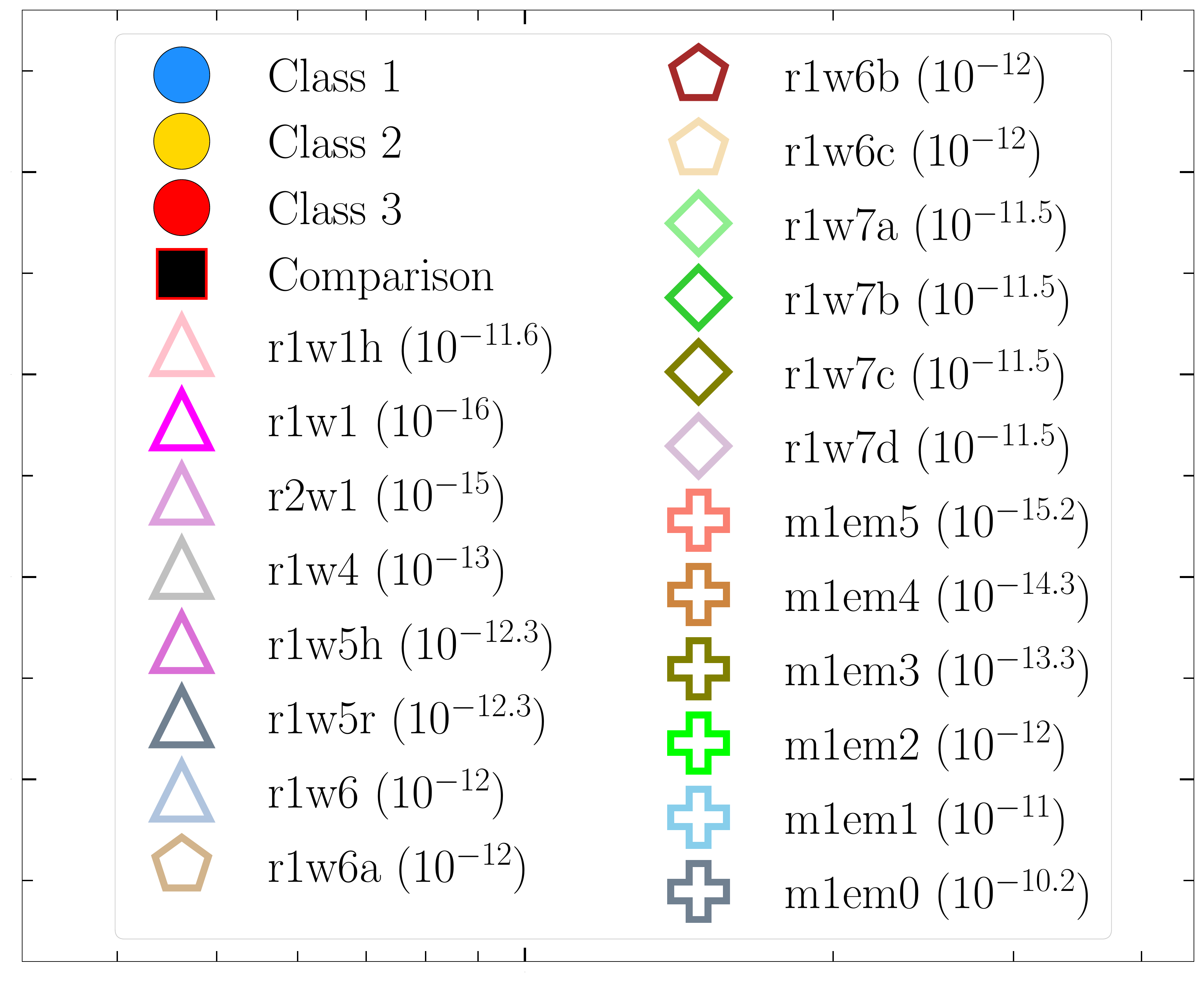}}\\
\subfigure{\includegraphics[width=0.49\textwidth]{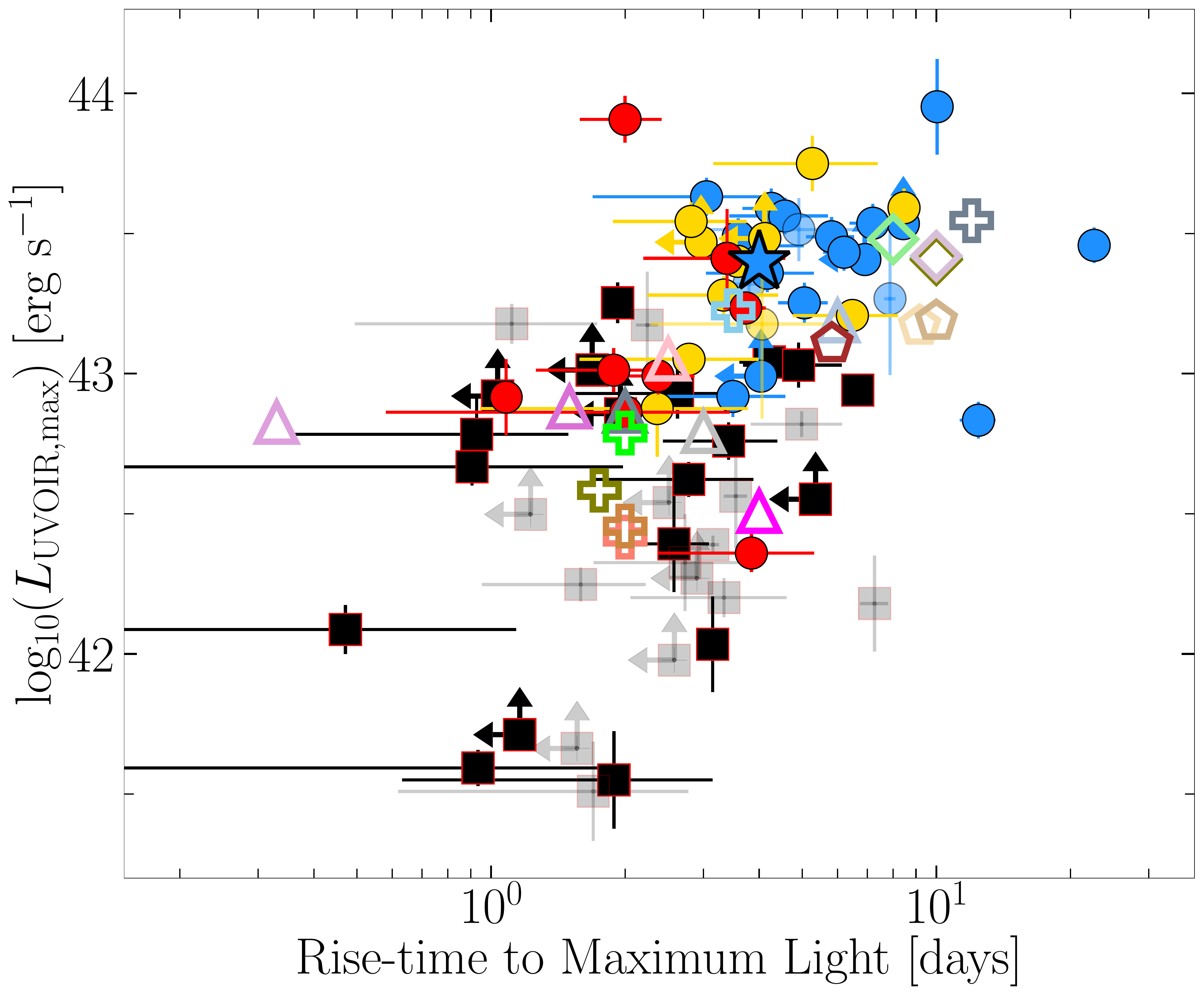}}
\subfigure{\includegraphics[width=0.49\textwidth]{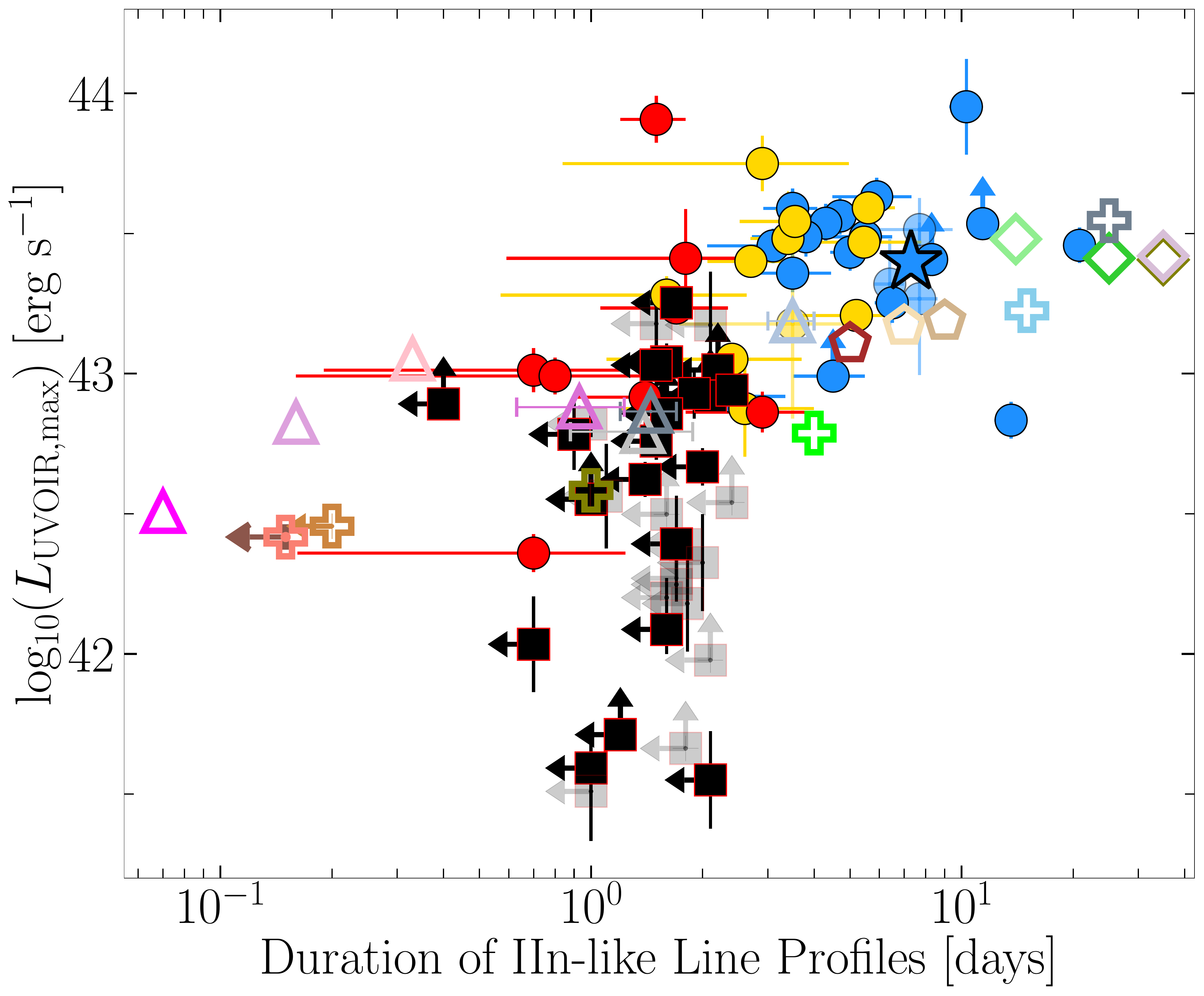}}
\caption{{\it Top left:} Pseudo-bolometric (i.e., UVOIR) light curves of gold/silver samples (blue/yellow/red solid lines) and the comparison sample (dashed black lines). Solid colored points/curves represent the subsample of objects at $D>40$~Mpc. The CSM interaction present in SNe~II with IIn signatures can create more than an order of magnitude luminosity excess beyond SNe~II in low-density CSM. The light curve of gold-sample object SN~2020tlf (blue) extends before first light because of detected precursor emission \citep{wjg22}. {\it Top right:} Legend with all models. {\it Bottom left:} Peak bolometric luminosity versus rise time for gold-, silver-, and comparison-sample objects, compared to \cmfgen\ model grid. {\it Bottom right:} Peak bolometric luminosity versus duration of IIn-like features ($t_{\rm IIn}$) also shows a clear positive trend (\S\ref{subsec:spec_analysis}). SN~2023ixf is shown for reference as a blue star. \label{fig:Bol_RT_TL} }
\end{figure*}

\begin{figure*}
\centering
\subfigure{\includegraphics[width=0.33\textwidth]{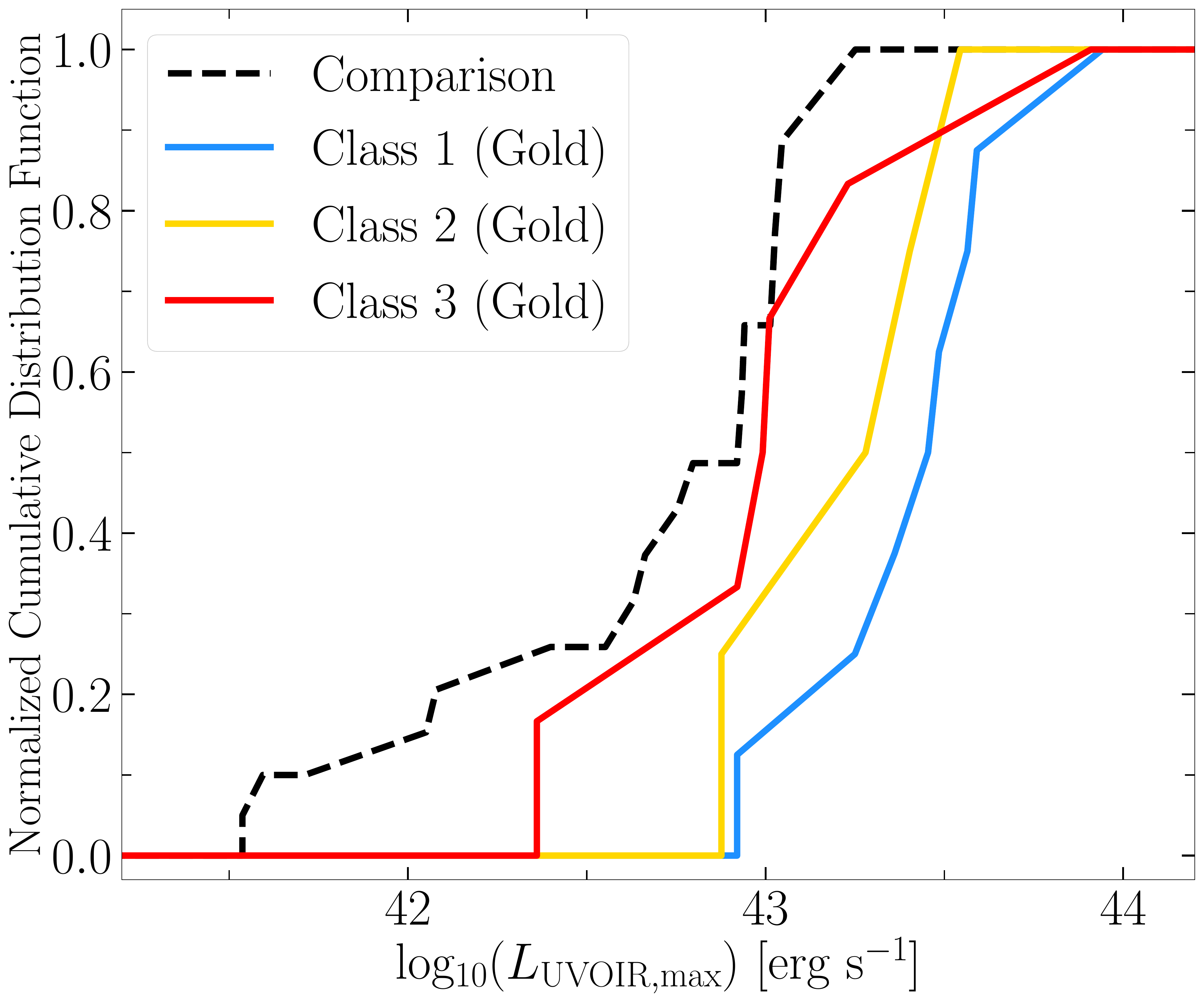}}
\subfigure{\includegraphics[width=0.33\textwidth]{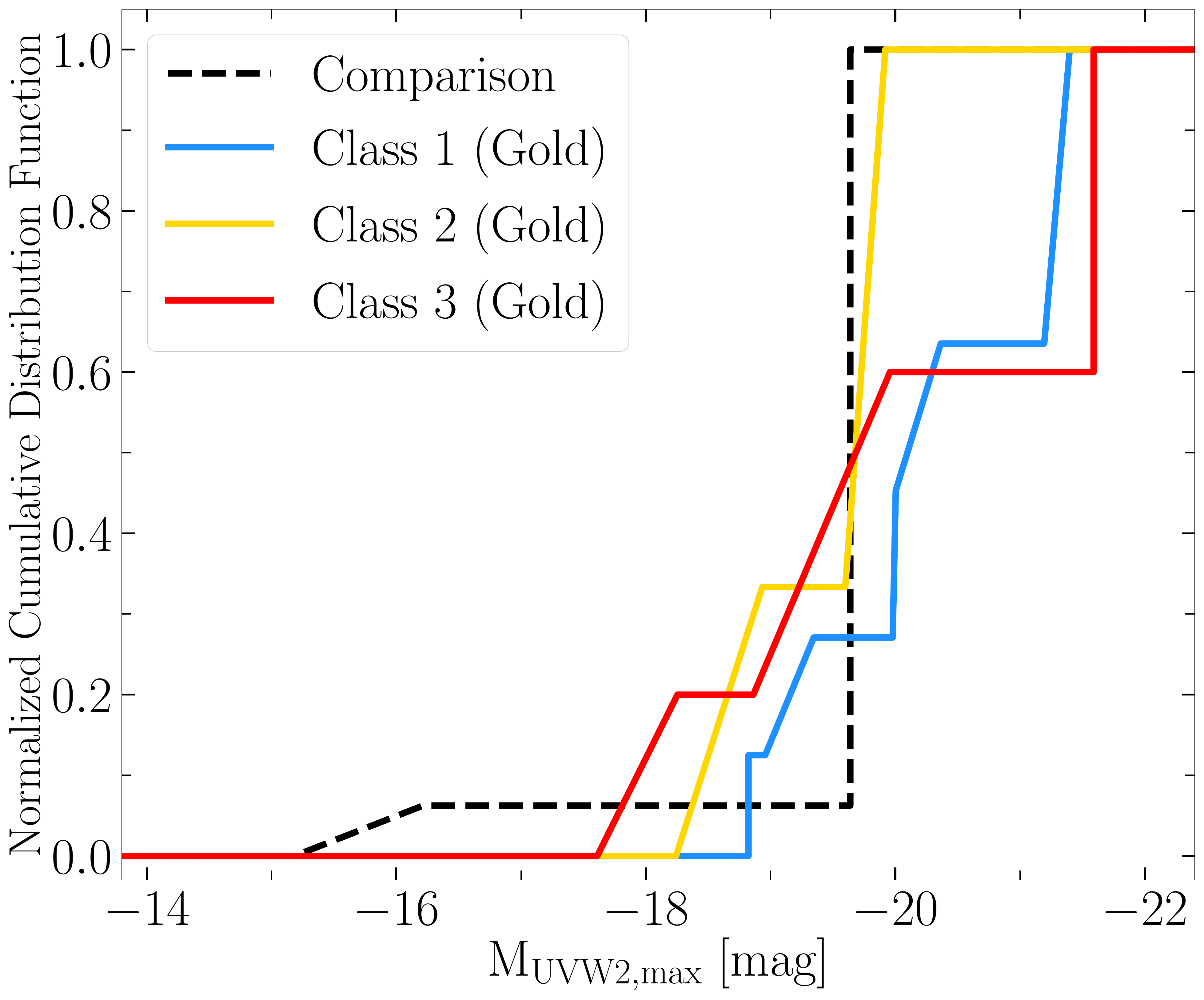}}
\subfigure{\includegraphics[width=0.33\textwidth]{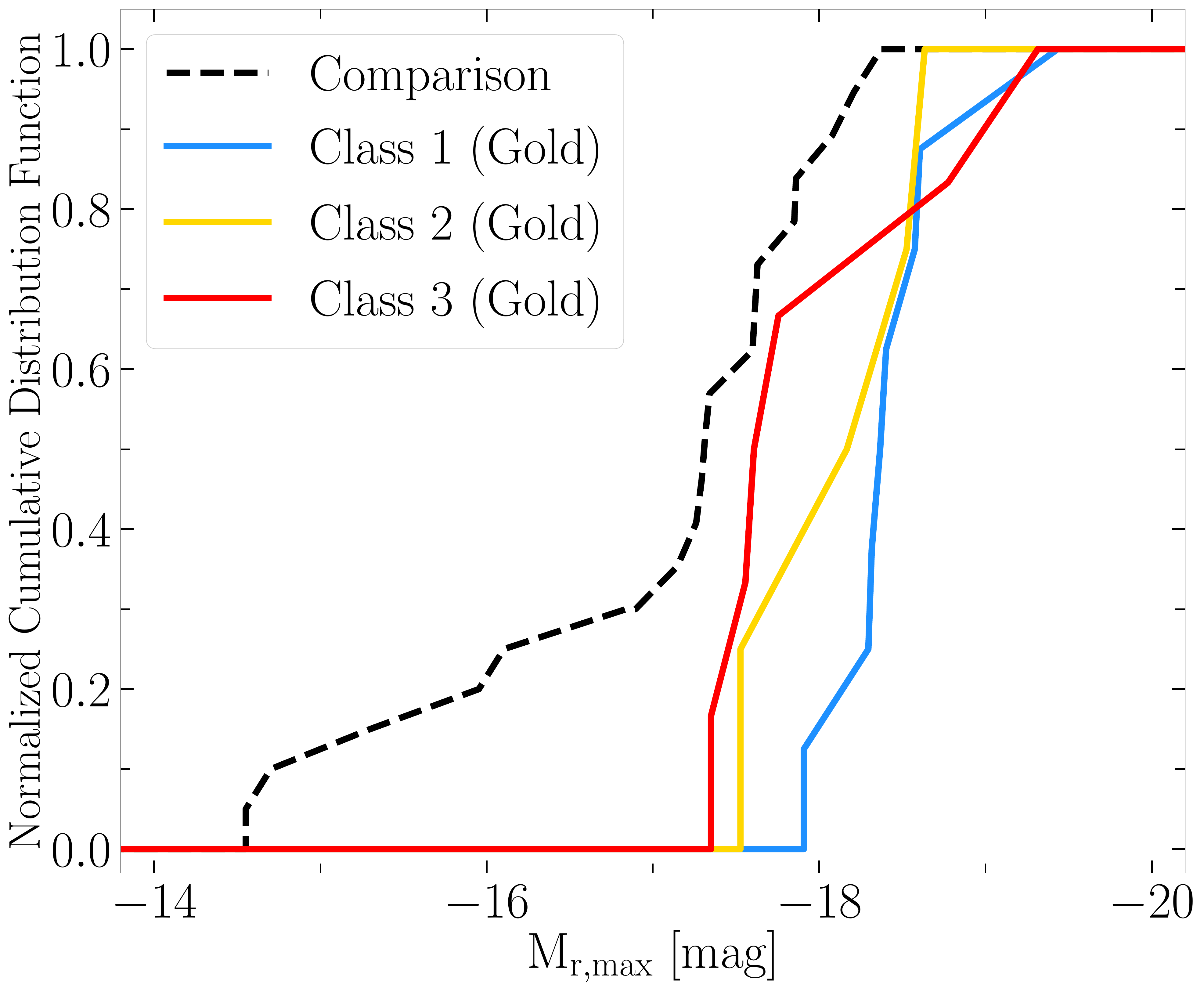}}\\
\subfigure{\includegraphics[width=0.33\textwidth]{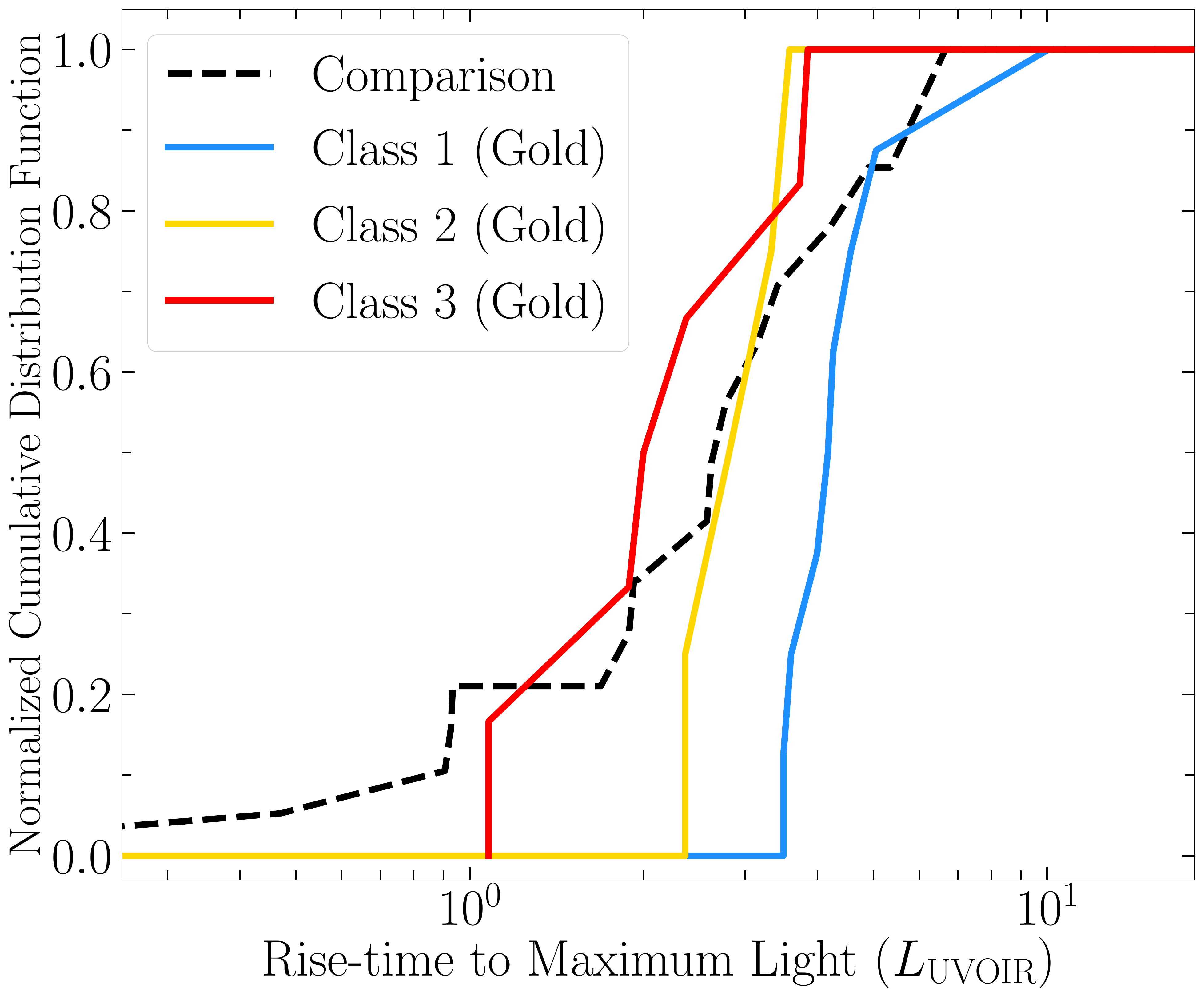}}
\subfigure{\includegraphics[width=0.33\textwidth]{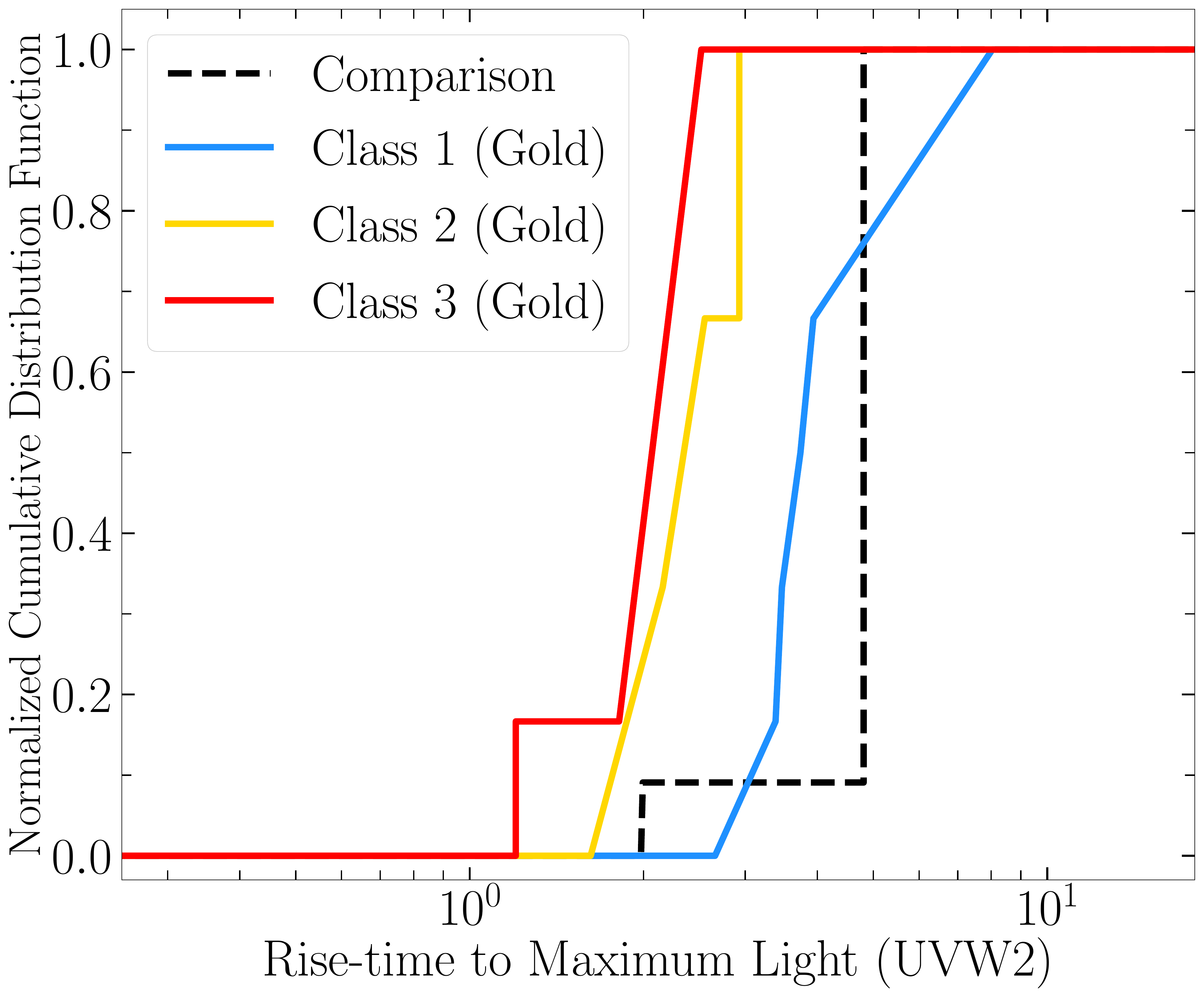}}
\subfigure{\includegraphics[width=0.33\textwidth]{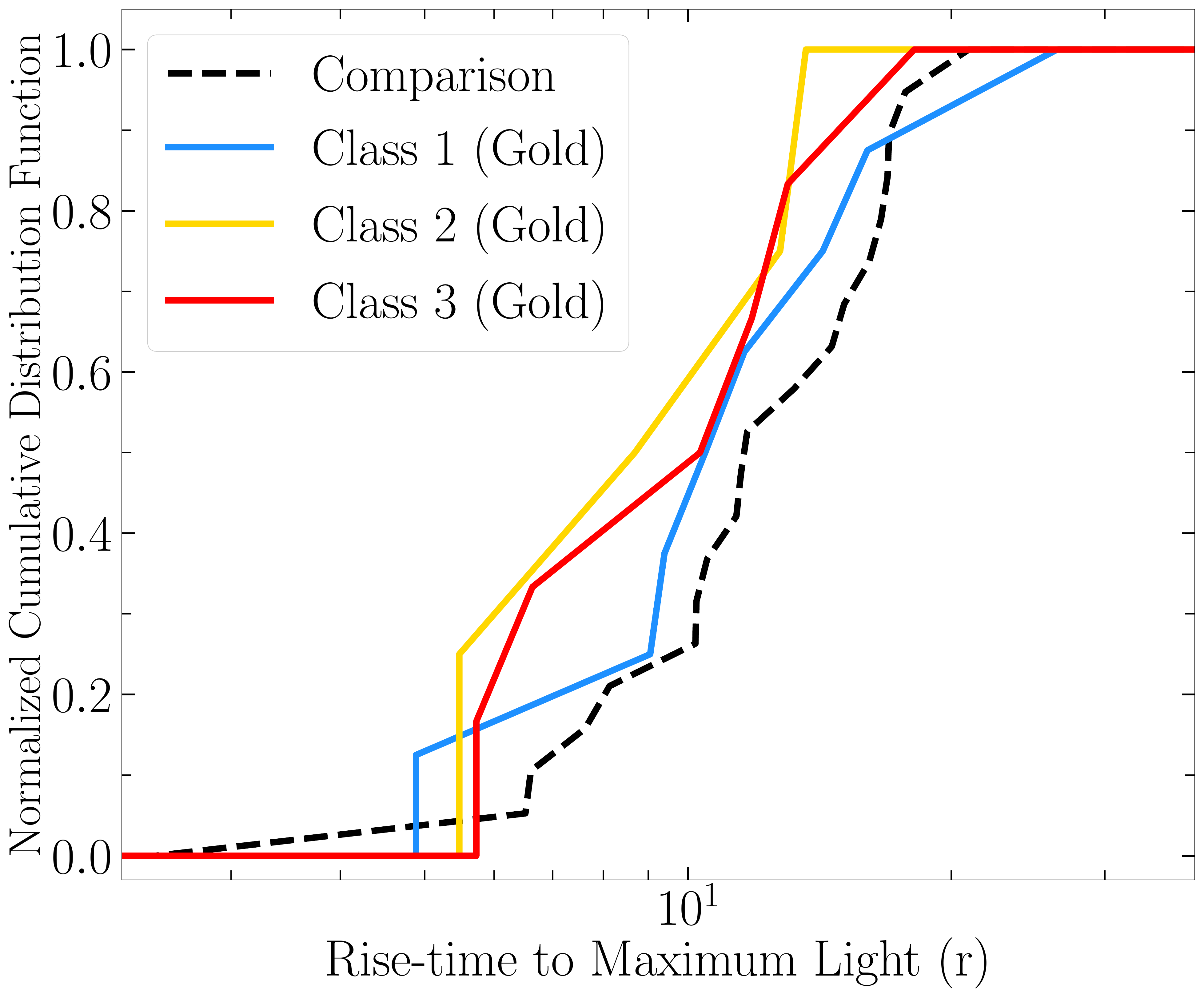}}\\
\caption{{\it Left to right, top to bottom:} Cumulative distributions of peak UVOIR luminosities, peak $w2$-band absolute magnitudes, peak $r$-band absolute magnitudes, UVOIR rise times, $w2$-band rise times, and $r$-band rise times for Class 1, 2, 3 gold-sample (blue, yellow, red lines) and comparison-sample (black dashed lines) objects after a distance cut ($D>40$~Mpc) is applied. Distinct distributions are present in the peak bolometric and optical luminosities for gold-sample objects compared to the comparison-sample SNe, which is most likely due to the effects of CSM interaction on the early-time light curves. \label{fig:hist_all} }
\end{figure*}

\begin{figure*}
\centering
\subfigure{\includegraphics[width=0.32\textwidth]{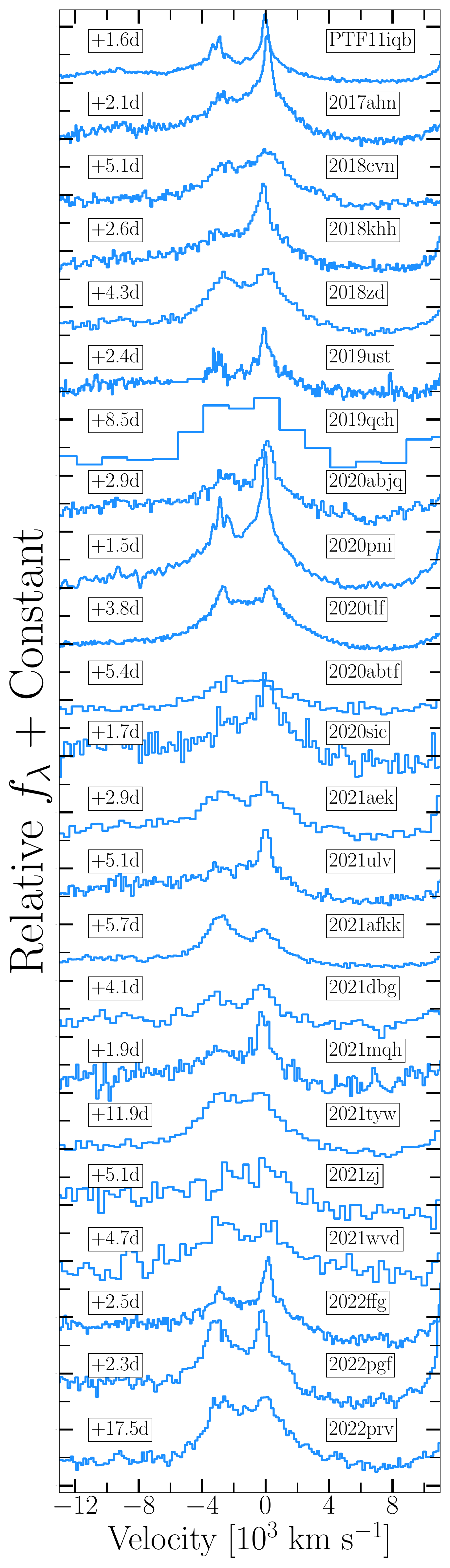}}
\subfigure{\includegraphics[width=0.32\textwidth]{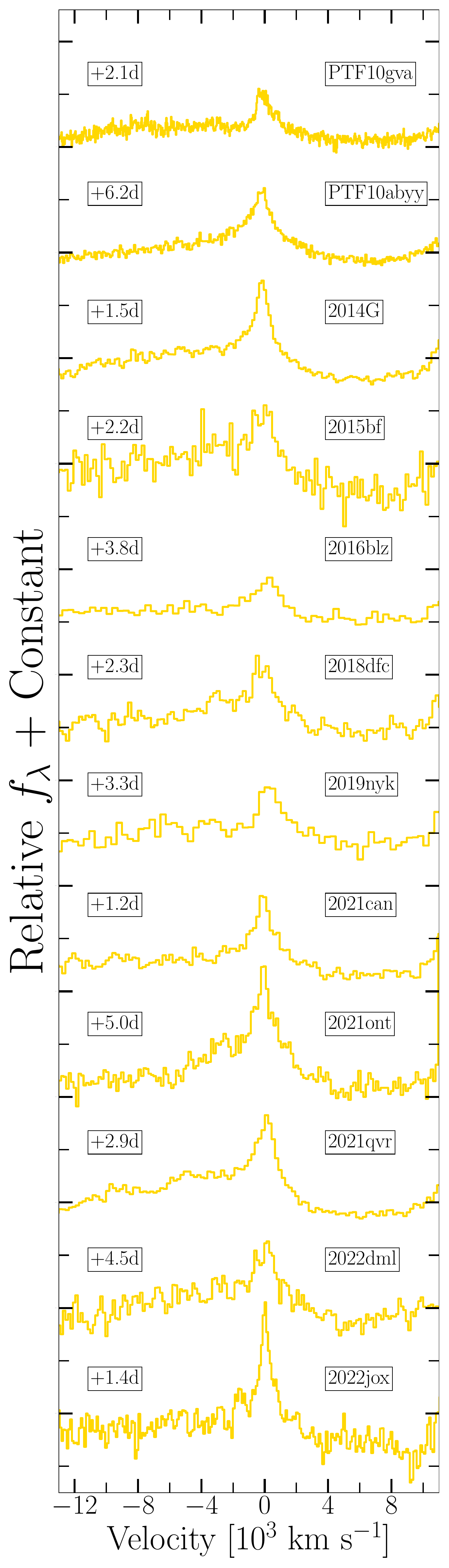}}
\subfigure{\includegraphics[width=0.32\textwidth]{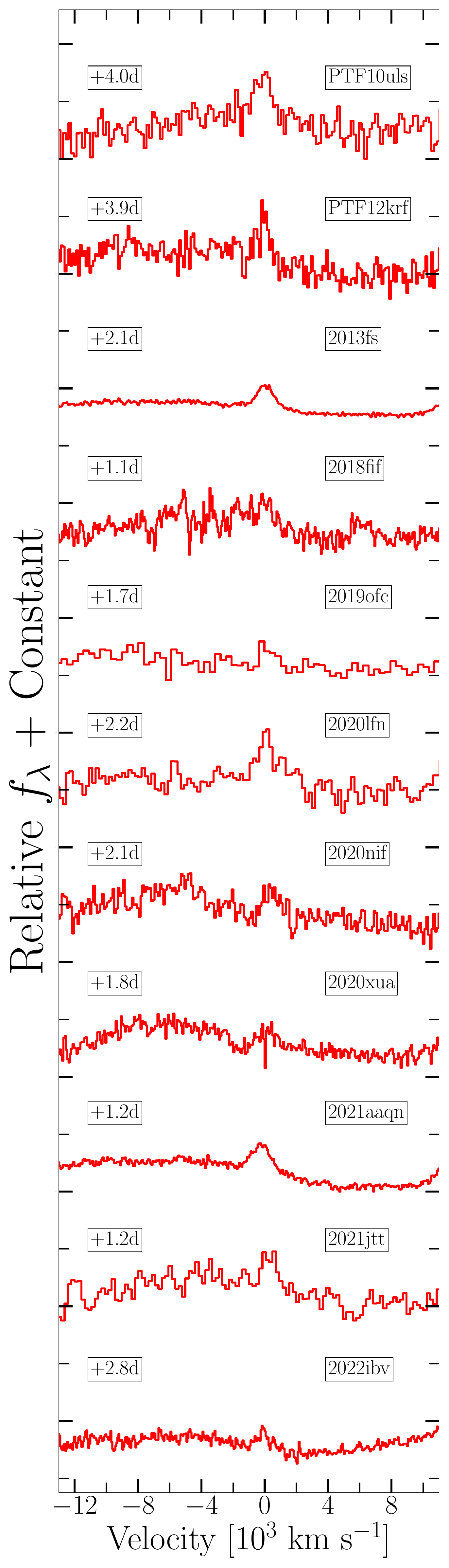}}\\
\caption{\ion{He}{ii} emission-line profiles for gold- and silver-sample objects. {\it Left:} SNe with visible, narrow \ion{N}{iii} emission are shown in blue (Class 1). {\it Middle/right:} Objects plotted in yellow (Class 2) and red (Class 3) show only narrow \ion{He}{ii} emission lines, the latter possessing the weakest emission superimposed on top of the broad \ion{He}{ii} profile from the fastest moving SN ejecta.   \label{fig:vels_lines} }
\end{figure*}

We present extinction-corrected $w2,m2,w1,u,b,v,g,r$ light curves of gold-, silver-, and comparison-sample objects in Figure \ref{fig:LC_all}. Given that the redshift/distance distributions of the gold and comparison samples are not the same, we divide sample objects based on a distance cut of $D>40$~Mpc is applied; this distance being the threshold when the distance distributions of both sub-samples are consistent. In order to quantify the differences between the gold-sample classes and the comparison sample, we fit high-order polynomials to all light curves to derive a peak absolute magnitude and a rise time in all eight filters. These values are reported in Tables \ref{tab:sample_phot_gold}--\ref{tab:sample_phot_control2}, with the uncertainty in peak magnitude being the $1\sigma$ error from the fit and the uncertainty in the peak phase being found from adding the uncertainties in both the time of peak magnitude and the time of first light in quadrature. We note that the pre-peak evolution in the UV filters of some sample objects is unconstrained (e.g., Fig. \ref{fig:LC_all}). For those objects with no constrained rise, we report the peak absolute magnitude and rise time as lower and upper limits, respectively. 

As shown in Figure \ref{fig:Max_RT}, we identify moderate positive trends between $M_{\rm peak}$ and $t_{\rm rise}$ in $w2,m2,w1,u$-band filters, and we find that while such trends are not as significant in $b,g,v,r$ filters, there is still a difference between gold/silver and comparison samples in optical filters. Amongst gold-sample SNe, Class 1 objects display the brightest peak absolute magnitudes and longest rise times compared to Class 3 and comparison-sample objects. On average, gold-sample objects are $>2$~mag brighter in the UV bands than comparison-sample objects (e.g., $M_{\rm avg}^{\rm W2} = -19.5$~mag versus $M_{\rm avg}^{\rm W2} = -17.1$~mag), even after a distance cut is applied, suggesting a significant luminosity boost from CSM interaction at early times. Furthermore, the $w2-v$ and $g-r$ colors plotted in Figure \ref{fig:colors} show that gold-sample objects, in particular Class 1 SNe, are bluer at earlier times than comparison-sample objects. Additionally, most Class 1/2 objects sustain blue colors ($g-r < 0$) longer than the comparison sample, suggesting continued interaction with more distant CSM that is at higher densities than a typical RSG wind. Similarly, the plateau luminosities of Class 1/2 objects remains higher than the control sample, also indicating long-lived interaction power. 

In Figure \ref{fig:Bol_RT_TL} we present pseudo-bolometric UV/optical/NIR (UVOIR) light curves of the gold/silver- and comparison-sample objects generated using the {\tt superbol}\footnote{\url{https://github.com/mnicholl/superbol}} code. For all SNe, we extrapolate between light-curve data points using a low-order polynomial spline in regions without complete color information. Repeating the analysis used for the multiband light curves, we calculate peak pseudo-bolometric luminosities and rise times; these values are presented in Table \ref{tab:sample_phot_gold}. For objects without a constrained rise to peak in all UV filters (i.e., $w2,m2,w1$), we report peak luminosities and rise times as lower and upper limits, respectively. As shown in Figure \ref{fig:Bol_RT_TL}, we find a significant trend between peak UVOIR luminosities and rise time to maximum light; this is similar to UV filters discussed above and indicates that the majority of the flux at early times is focused in the UV bands, especially with the presence of ejecta-CSM interaction. Furthermore, we find that gold/silver-sample objects can be more than an order of magnitude more luminous at peak than comparison-sample SNe (e.g., Table \ref{tab:sample_lum_peak}), also suggesting excess luminosity from CSM interaction. 

In Figure \ref{fig:hist_all}, we present the cumulative distributions of maximum brightness and rise times for the pseudo-bolometric, $w2$-band, and $r$-band light curves of the gold/silver and comparison samples that are constructed using Kaplan-Meier estimation for all objects at $D>40$~Mpc. To test our null hypothesis of whether these sample observables come from the same parent distribution, we apply a logrank test for (i) gold vs. comparison samples, (ii) gold-sample Classes 1 \& 2 vs. 3, and (iii) gold-sample Classes 1 vs. 3. Limits on peak luminosity and rise time are accounted for using survival statistics. For (i), the chance probability that peak-brightness values of the gold and comparison samples come from the same distribution is $0.1\%$ for $L_{\rm max}$, $80.0\%$ for $M_{w2, \rm max}$, and $3\times10^{-3}\%$ for $M_{r, \rm max}$. We find that the pseudo-bolometric, UV, and $r$-band rise times between samples do belong to the same distribution at the 60.6\%, 7.1\%, and 55.6\% levels, respectively. For (ii), the null-hypothesis probability for pseudo-bolometric, UV, and $r$-band peak brightness (rise time) is 23.1(1.67)\%, 73.3(1.9)\%, and 69.4(83.3)\%, respectively. For (iii), the null hypothesis probability for pseudo-bolometric, UV, and $r$-band peak brightness (rise-time) is 17.3(0.24)\%, 92.6(1.51)\%, and 46.6(60.1)\%, respectively. Therefore, we conclude that the gold sample is significantly more luminous than the comparison sample in bolometric and optical light curves, but luminosity differences within the classes of the gold sample are not statistically significant. Given the large number of limits present in the $w2$-band light curves, peak UV luminosity differences between gold and comparison samples cannot be claimed as significant. Furthermore, there is evidence that the differences in bolometric and UV rise-times between Class 1 \& 2 vs 3, as well as Class 1 vs. 3, are statistically significant. However, differences in the rise time between all other groups are not statistically significant.

\subsection{Spectroscopic Properties}\label{subsec:spec_analysis}

We present single epoch, ``flash'' spectroscopy of the gold/silver and comparison samples in Figures \ref{fig:FS_all_gold} and \ref{fig:FS_all_comp}, respectively, with complete spectral series shown for each object in the supplementary, online-only text. As discussed in Section \ref{SubSec:Sample}, the blue (Class 1), yellow (Class 2), and red (Class 3) color delineation is based on the structure of the \ion{He}{ii} $\lambda 4686$ line, which is shown in detail for all gold/silver-sample objects in Figure \ref{fig:vels_lines}. As illustrated in Figure \ref{fig:spec_lor_fit}, the IIn-like features of semi-isolated (i.e., unblended) transitions such as H$\alpha$ can be modeled with a two-component Lorentzian, which includes a narrow component that provides an upper limit on the CSM velocity (due to likely radiative acceleration) and a broad component that forms from electron scattering of recombination-line photons in the optically thick unshocked CSM. The physical origin of the \ion{He}{ii} $\lambda 4686$ profile is slightly more complex and can be modeled with a high-velocity, blueshifted, full width at half-maximum intensity (FWHM) $\approx 10^4~\kms$ component representing fast-moving material in the CDS and/or outer ejecta, plus a narrow, and possibly electron-scattering broadened, emission at the central wavelength for Class 2 and 3 objects (e.g., 2014G and 2013fs; Fig. \ref{fig:spec_lor_fit}). However, Class 1 objects (e.g., 2020pni; \ref{fig:spec_lor_fit}) require multiple narrow and electron-scattering emission components of \ion{He}{ii} and \ion{N}{iii}, which may be superimposed on an underlying, blueshifted \ion{He}{ii} profile, the same as Classes 2 \& 3 (e.g., see \citealt{dessart17})

\begin{figure*}
\centering
\subfigure{\includegraphics[width=0.33\textwidth]{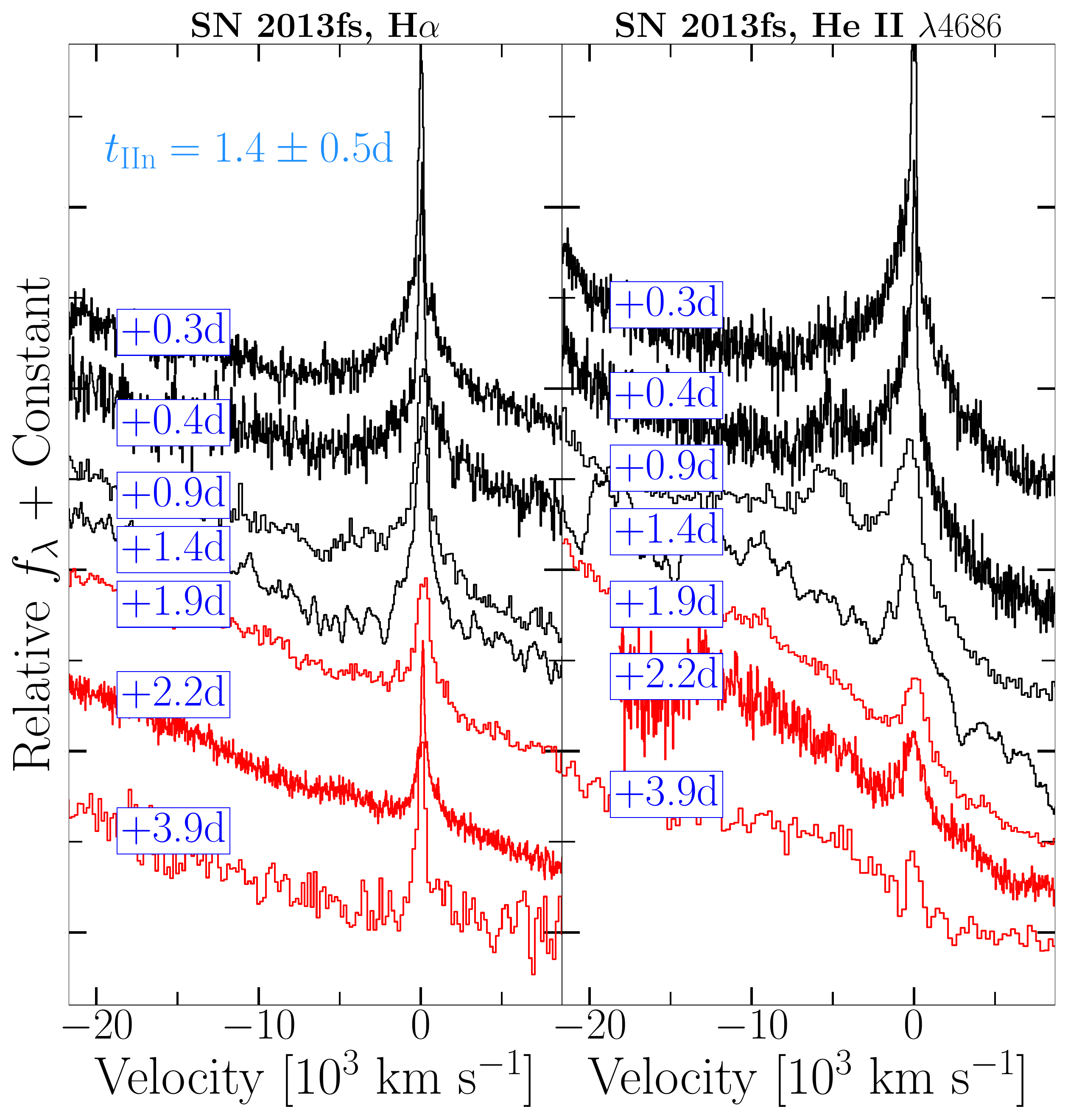}}
\subfigure{\includegraphics[width=0.33\textwidth]{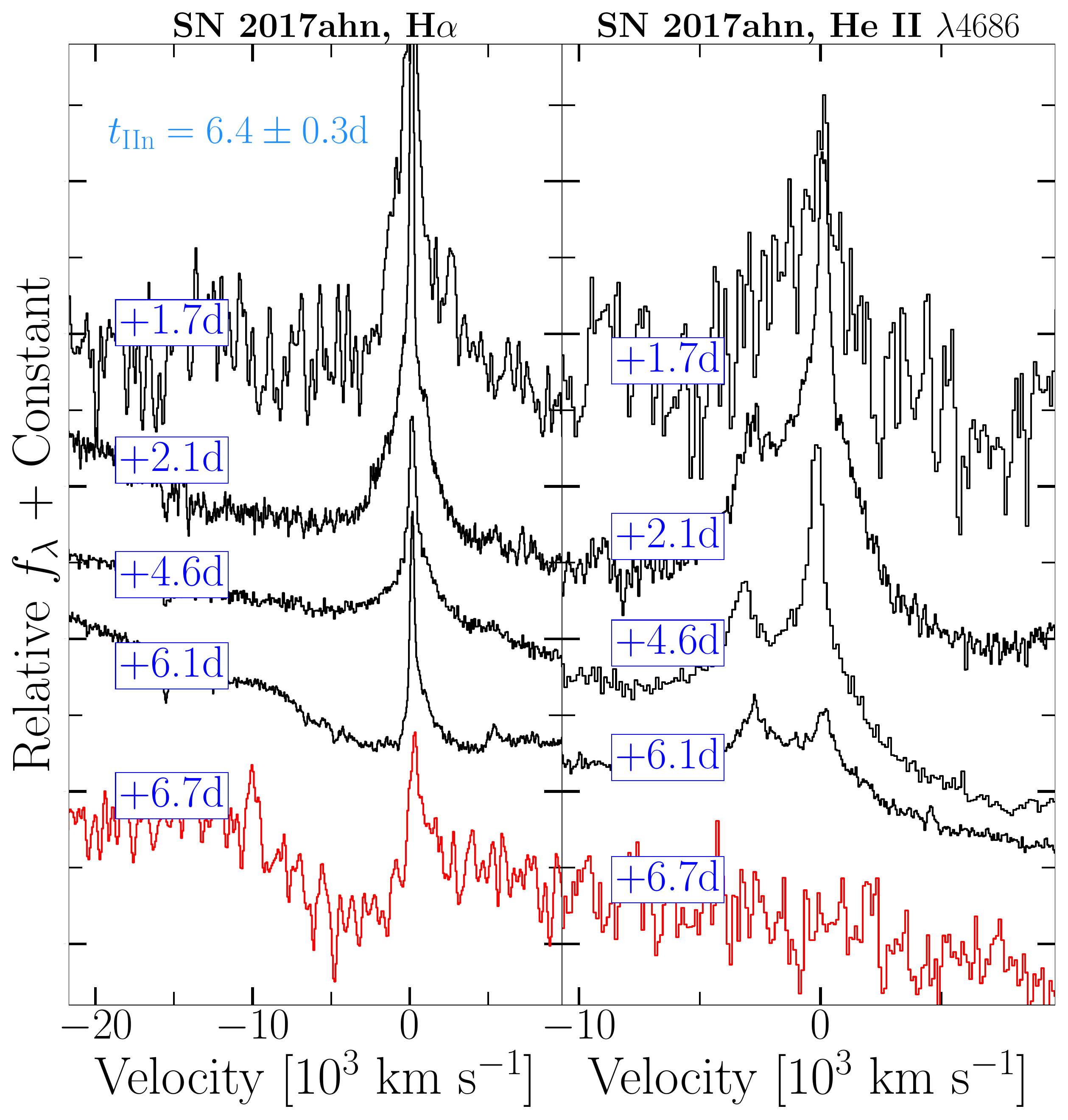}}
\subfigure{\includegraphics[width=0.33\textwidth]{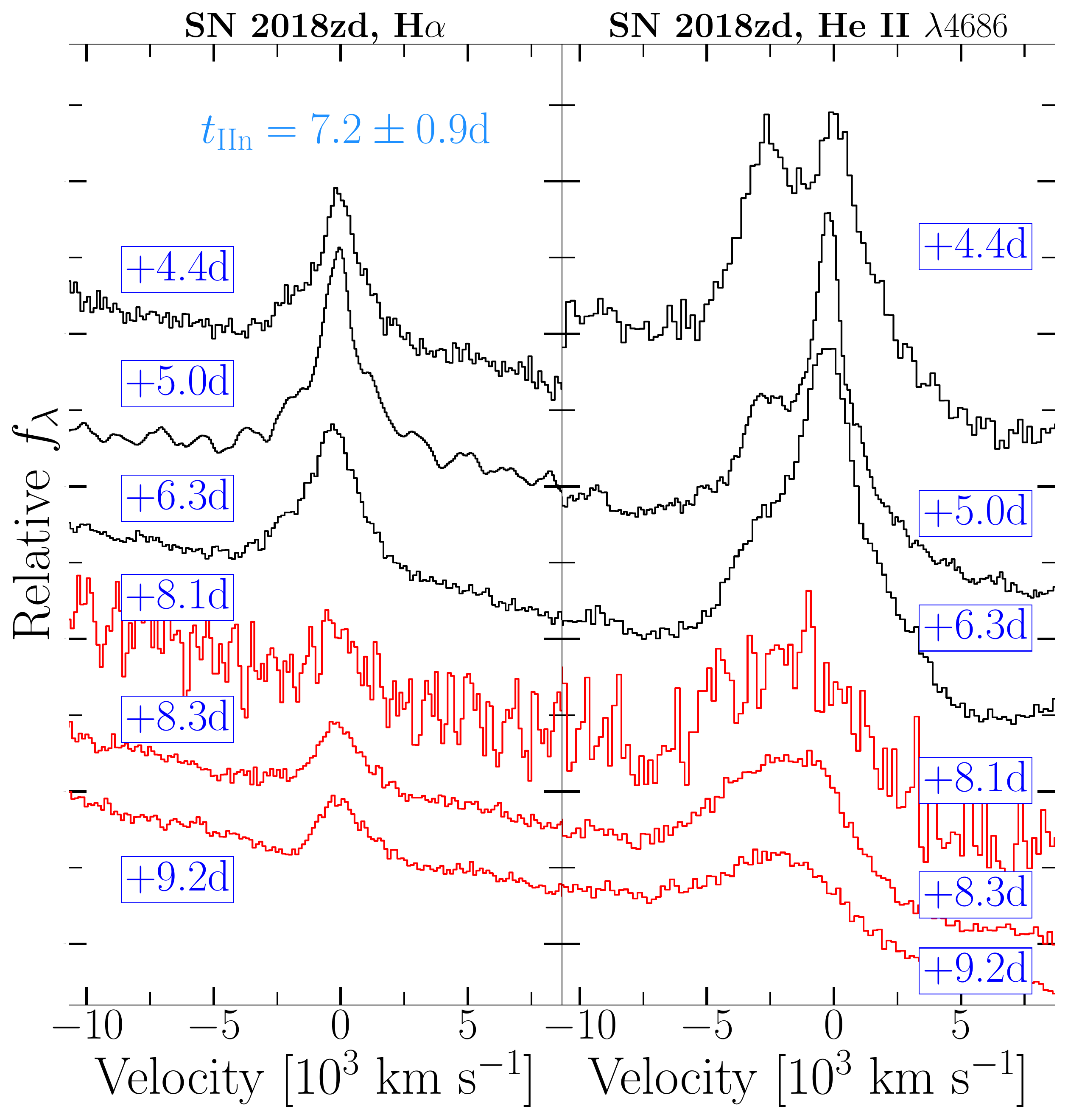}}
\caption{{\it Left:} SN~2013fs spectral series of H$\alpha$ {\it (left panel)} and \ion{He}{ii} $\lambda$4686 {\it (right panel)} velocities during the CSM interaction phase. Spectra in black represent phases when the CSM remains optically thick to electron scattering (e.g., Lorentzian line profiles). The transition shown from black to red lines marks the emergence of broad absorption features derived from the fastest moving SN ejecta. The transition between these two phases is the basis for calculating the $t_{\rm IIn}$ parameter. {\it Middle/right:} Same plot but for SNe~2017ahn and 2018zd, respectively, which show longer-lived IIn profiles.  \label{fig:TL_SNe} }
\end{figure*}

\begin{figure*}
\centering
\subfigure{\includegraphics[width=0.45\textwidth]{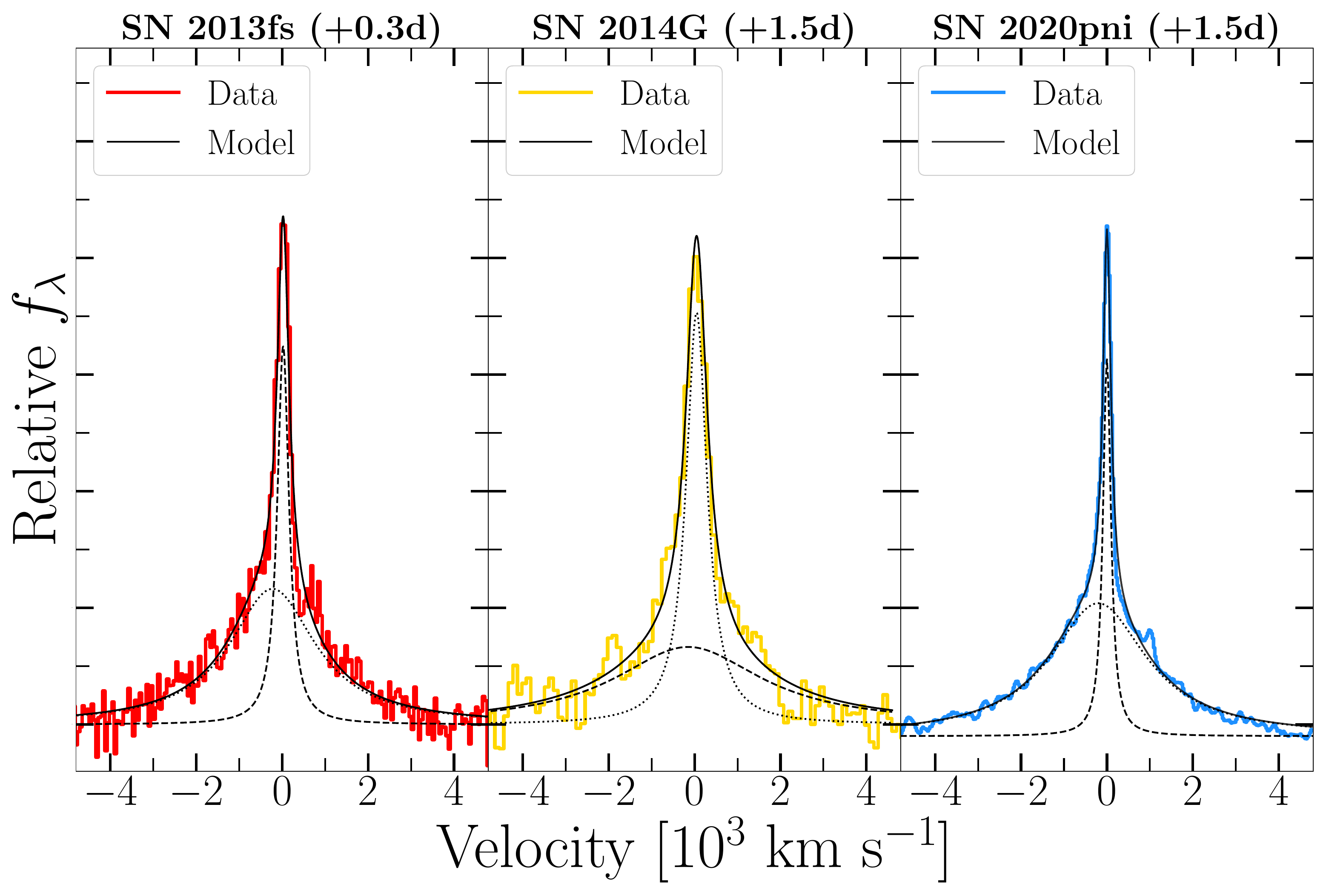}}
\subfigure{\includegraphics[width=0.53\textwidth]{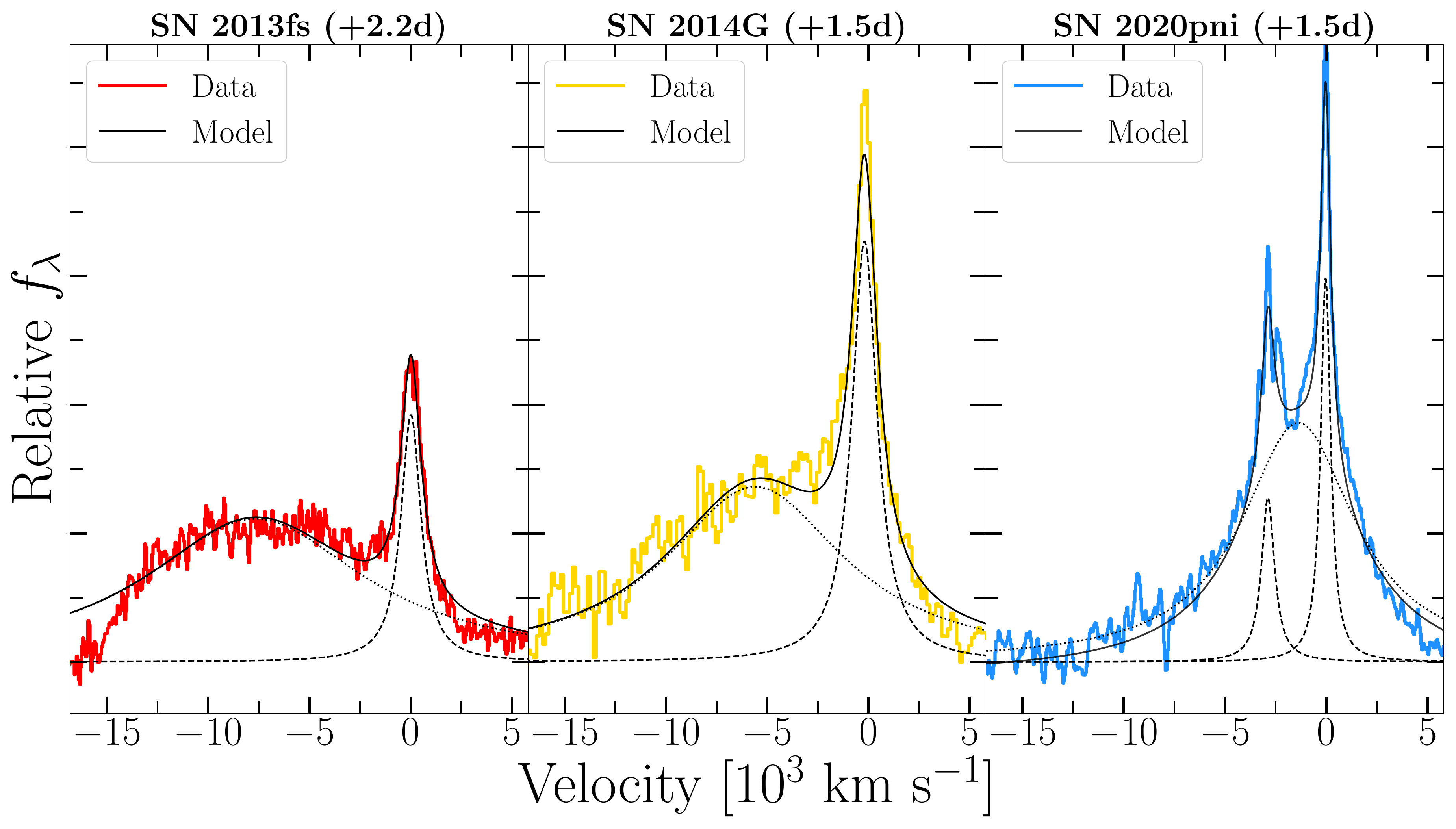}}
\caption{H$\alpha$ ({\it left}) and \ion{He}{ii} $\lambda$4686 ({\it right}) emission lines modeled with multicomponent Lorentzian profiles during the CSM interaction phase. Class 1 objects (shown in blue) possess longer-lived (days-to-weeks) high-ionization species of \ion{He}{ii} and \ion{N}{iii}. Class 2 (shown in yellow) and Class 3 (shown in red) objects show only \ion{He}{ii} emission, with the former having stronger emission lines that last longer. Class 3 objects may represent transitional SNe between the comparison and gold/silver samples given their weak narrow \ion{He}{ii} emission superimposed on a blueshifted \ion{He}{ii} profile, the latter being seen in comparison-sample objects (e.g., Fig. \ref{fig:FS_all_comp}).   \label{fig:spec_lor_fit} }
\end{figure*}

As confirmed by our sample, the narrow, symmetric line profiles with Lorentzian wings caused by electron scattering (i.e., IIn-like) can persist for days after first light. After these phases, the SNe develop broad absorption profiles in all Balmer transitions as a result of the escape of photons from the fast-moving ejecta and a decrease in CSM density. We therefore define the duration of the IIn-like features (i.e., $t_{\rm IIn}$) as the transition point at which the unshocked CSM optical depth to electron scattering has dropped enough to see the emerging fast-moving SN ejecta \citep{Dessart23, wjg23a}. This evolution is shown in Figure \ref{fig:TL_SNe} for gold-sample SNe~2013fs, 2017ahn, and 2018zd, all of which have high enough spectral cadence to allow for a precise observation of the fading of the IIn-like features. We use this transition to calculate $t_{\rm IIn}$ and its uncertainty, which is derived from the cadence of the spectral observations. For gold/silver-sample objects without sufficiently high spectral cadence to confidently estimate $t_{\rm IIn}$, we use spectral comparisons to SNe 2013fs, 2017ahn, and 2018zd to derive a IIn-like feature duration timescale by extrapolating phase measurements and assuming that the spectral evolution is consistent with the SNe used for reference. The uncertainty of $t_{\rm IIn}$ from spectral comparison is added in quadrature with the uncertainty in the time of first light for each sample object. For comparison-sample objects, which do not show IIn-like features, we take the phase of their earliest spectrum to be an upper limit on $t_{\rm IIn}$. All $t_{\rm IIn}$ values are presented in Table \ref{tab:sample_spec_gold}.  

In Figures \ref{fig:Bol_RT_TL} and \ref{fig:Max_lines}, we plot $t_{\rm IIn}$ with respect to peak luminosity for all UV, optical, and pseudo-bolometric light curves. We find a moderate positive trend between between peak luminosity and $t_{\rm IIn}$ in $w2,m2,w1,u,b$-band filters, which is similar to the rise-time trends shown in Figure \ref{fig:Max_RT}. While the peak absolute magnitude in optical $v,g,r$-band filters reveal a more obvious trend with $t_{\rm IIn}$ than $t_{\rm rise}$, their correlation can only be claimed as tentative. Furthermore, as shown in Figure \ref{fig:Bol_RT_TL}, peak pseudo-bolometric luminosities and the duration of IIn-like features are moderately correlated. Amongst the gold/silver samples, Class 1 objects consistently show the highest peak luminosities across wavelengths, coupled with a longer duration of observed IIn-like features, indicating ejecta-CSM interaction with denser, and likely more extended, CSM than Class 2/3 objects (e.g., see Figs. \ref{fig:hist_all} and \ref{fig:TL_hist}). 

As the IIn-like features fade, all gold/silver-sample objects transition into seminormal SNe II with Doppler broadened, blueshifted P~Cygni features of the fast-moving, H-rich ejecta. In Figure \ref{fig:phot_vels}, we present photospheric velocities calculated from the absorption minima of H$\alpha$ and \ion{Fe}{ii} $\lambda 5169$ transitions for gold-, silver-, and comparison-sample objects. Overall, there is some spread in H$\alpha$ velocities amongst gold/silver-sample objects with a few Class 1 SNe displaying slower velocities ($v \approx 5000$--8000~$\kms$) than Classes 2/3 ($v > 10^4~\kms$). However, in general, we find little difference in the H$\alpha$ and \ion{Fe}{ii} velocities found in the absorption minima between gold/silver and comparison sample from $\delta t \approx 10$--100~days. 

\section{Modeling}\label{sec:modeling}

\subsection{\heracles/\cmfgen\ Model Grid}\label{subsec:cmfgen}

In order to quantify the CSM properties in our gold, silver, and comparison samples, we compared the spectral and photometric properties of all SNe to a model grid of radiation hydrodynamics and non-LTE, radiative-transfer simulations covering a wide range of progenitor mass-loss rates ($\dot{M} = 10^{-6}$--$10^{0}~\Msun$~yr$^{-1}$; $v_w = 50~\kms$), maximum radii of dense CSM ($R = 10^{14}$--$10^{16}$~cm), and CSM densities at $10^{14}$~cm ($\rho_{14} = 10^{-16}$--$7.3 \times 10^{-11}$~g~cm$^{-3}$), all in spherical symmetry. Simulations of the SN ejecta-CSM interaction were performed with the multigroup radiation-hydrodynamics code \heracles\ \citep{gonzalez_heracles_07,vaytet_mg_11,D15_2n}, which consistently computes the radiation field and hydrodynamics. Then, at selected snapshots in time post-explosion, the hydrodynamical variables are imported into the non-LTE radiative-transfer code \cmfgen\ \citep{hillier12, D15_2n} for an accurate calculation of the radiative transfer, which includes a complete model atom, $\sim10^6$ frequency points, a proper handling of the complex, nonmonotonic velocity field, and treatment of continuum and line processes as well as electron scattering. For each model, we adopt an explosion energy of $1.2\times 10^{51}$~erg, a 15~$\Msun$ progenitor with a radius in the range $R_{\star} \approx 500$--700~$\Rsun$, and a CSM composition set to the surface mixture of an RSG progenitor \citep{davies_dessart_19}. 

For the simulations presented in this work, the CSM extent is much greater than $R_{\star}$ ($\sim 500$--1200~$\Rsun$ for an RSG mass range of $\sim 10$--20~$\Msun$), and therefore we have found that the progenitor properties have little impact during phases of ejecta-CSM interaction. The progenitor radius plays a more significant role on the light-curve evolution during the plateau phase (e.g., see \citealt{d13_sn2p,Hiramatsu21, wjg22}), once the interaction phase is over and the emission from the deeper ejecta layers dominate the SN luminosity. However, in scenarios with weak CSM interaction, the explosion energy will greatly influence the total luminosity, which could be contributing to the brighter pseudo-bolometric and UV luminosities in comparison-sample events (e.g., Figs. \ref{fig:LC_all} and \ref{fig:Bol_RT_TL}). Specific methods for each simulation are given by \cite{Dessart16, dessart17}, \cite{wjg22},  \cite{Dessart23}, and \cite{wjg23a}; all CSM properties of each model are presented in Table \ref{tab:models}. CSM densities for all models are shown in Figure \ref{fig:model_rho}, which primarily differ at radii above the stellar surface, $r > 4\times 10^{13}$~cm. 

In order to identify a best-matched $\dot M$ and $\rho_{14}$ for all sample objects, we employ three independent methods of matching observables to the model grid. (1) We use the rise times, peak absolute magnitudes, and $t_{\rm IIn}$ to construct a 3-dimensional root-mean-square (RMS) between each model for all eight UV/optical filters and the pseudo-bolometric light curve. We then select the best-matched model for a given filter (as well as pseudo-bolometric) based on the lowest resulting RMS, $[((M_{\rm data} - M_{\rm model})/M_{\rm model})^2 + ((t_{\rm r, data} - t_{\rm r, model})/t_{\rm r, model})^2 + ((t_{\rm IIn, data} - t_{\rm IIn, model})/t_{\rm IIn, model})^2 ]^{0.5}$. This method results in $N+1$ mass-loss inferences: $N$ filters plus the pseudo-bolometric light curve. The range of mass-loss rates and CSM densities for all filters are presented in Table \ref{tab:model_params_gold} and plotted in the left panels of Figure \ref{fig:mdot_IIn}. For this method, we do not incorporate the relative uncertainties in peak luminosities and rise times, but instead report the range of best-matched model parameters as the uncertainty in the derived $\dot M$ and $\rho_{14}$. However, as discussed in \S\ref{subsec:phot_properties}, the peak absolute magnitude and rise times, especially in UV filters, are unconstrained in some sample objects, which will influence the best-matched model parameters. For such objects, we use upper limit or the ill-constrained peak and rise-time values reported in Table \ref{tab:sample_phot_gold} in the above RMS relation, but note that the output model parameters may only represent limits on the true CSM properties in these SNe. (2) We minimize the residuals between only $t_{\rm IIn}$ estimates for each object in order to find the best-matched model in the grid, which is plotted in the middle panels of Figure \ref{fig:mdot_IIn} with error bars on mass-loss/density estimates coming from uncertainties in the given $t_{\rm IIn}$ values. (3) We perform direct spectral matching of \cmfgen\ synthetic spectra to gold-, silver-, and comparison-sample objects in order to estimate the most consistent mass-loss rates and CSM densities. To do this, we degrade the synthetic spectrum to the resolution of the SN spectrum and scale the average flux of each model spectrum to the observations over the wavelength range of the optical spectrum, and calculate the residuals in flux density between model and data in the wavelength ranges that cover emission lines of the \ion{H}{I} Balmer series, \ion{He}{ii} $\lambda\lambda 4686$, 5412, \ion{N}{iii} $\lambda 4641$, \ion{N}{iv} $\lambda 7112$, and \ion{C}{iv} $\lambda 5801$. For each sample object, we estimate a best-matched mass-loss rate and CSM density (right panels of Figures \ref{fig:mdot_IIn}) by selecting the model with the smallest average residual (i.e., $\overline{\Delta}_{\rm IIn}$) between model and SN spectra in all IIn-like feature wavelength ranges. However, we note that the best-matched model spectrum may not reproduce the intrinsic continuum flux of the SN data despite overall consistency with the observed IIn-like features. Similarly, the best-matched model using method 1 may not match the SN light-curve shape on the rise despite consistency with peak brightness and rise time. We discuss inconsistencies between model-matching methods below as well as future improvements to the grid in \S\ref{subsec:future}.

Below, we discuss the resulting mass-loss rates and CSM densities derived for each model-matching method. We find that gold/silver-sample objects with visible IIn-like features reside in a parameter space of progenitor CSM densities of $\sim 10^{-16}$--$10^{-11}$~g~cm$^{-3}$ ($\dot{M} \approx 10^{-6}$--$10^{-1}$~\mdot, $v_w = 50~\kms$) when comparing rise times, peak absolute magnitudes, and $t_{\rm IIn}$ to the model grid (i.e., Methods 1 and 2). However, this parameter space becomes more constrained to $\sim 5 \times 10^{-14}$--$10^{-11}$~g~cm$^{-3}$ ($\dot{M} \approx 10^{-3}$--$10^{-1}$~\mdot, $v_w = 50~\kms$) when using a direct spectral matching method (i.e., Method 3). With regards to subdivisions of the gold and silver samples, the Class 1 objects show the highest mass-loss rates of $\dot{M} \approx 5\times 10^{-3}$--$10^{-1}$~\mdot, Class 2 objects showing low to intermediate mass-loss rates of $\dot{M} \approx 10^{-6}$--$10^{-2}$~\mdot, and Class 3 objects displaying generally lower mass-loss rates of $\dot{M} \approx 10^{-6}$--$10^{-3}$~\mdot. Furthermore, comparison-sample objects that have no detected IIn-like features at $\delta t < 2$~days are consistent with overall low mass-loss rates of $\dot{M} \approx 10^{-6}$--$10^{-3}$~\mdot. Across all three model-matching methods, the average $\dot{M}$ derived is consistent to within an order of magnitude (e.g., see Fig. \ref{fig:model_fidelity}). However, there are instances where mass-loss rates derived from some peak magnitudes or rise times in Method 1 are inconsistent with what would be inferred from Methods 2 and 3 involving $t_{\rm IIn}$ and direct spectral matching. For example, many of the Class 3 objects have $\dot{M}$ ranges of $\sim 10^{-6}$--$10^{-2.3}$~\mdot\ based on Method 1, but have more constrained estimates of $\sim 10^{-3}$--$10^{-2.3}$~\mdot\ based on Methods 2 and 3 that are inconsistent with the lower $\dot{M}$ values. This is caused by similar peak absolute magnitudes and/or rise times across models in optical filters as well as the low resolution of the model grid in general. With future grids, the incorporation of additional explosion parameters such as a variable kinetic energy will provide more self-consistent results between model-matching methods. 

As shown in Figure \ref{fig:mdot_IIn}, there is a clear trend between the $t_{\rm IIn}$ parameter and derived mass-loss rates or CSM densities for both gold/silver- and comparison-sample objects. We then fit a linear function to the mass-loss rates and $t_{\rm IIn}$ from the model grid and overplot the function as black dashed lines in Figure \ref{fig:mdot_IIn}. This relation between the duration of the electron-scattering line profiles and the inferred mass-loss rate, in units of \mdot, goes as $t_{\rm IIn} \approx 3.8[\dot{M}/$(0.01 \mdot)]~days. We note that this correlation is valid for the chosen explosion and progenitor parameters.

\begin{figure*}
\centering
\subfigure{\includegraphics[width=0.49\textwidth]{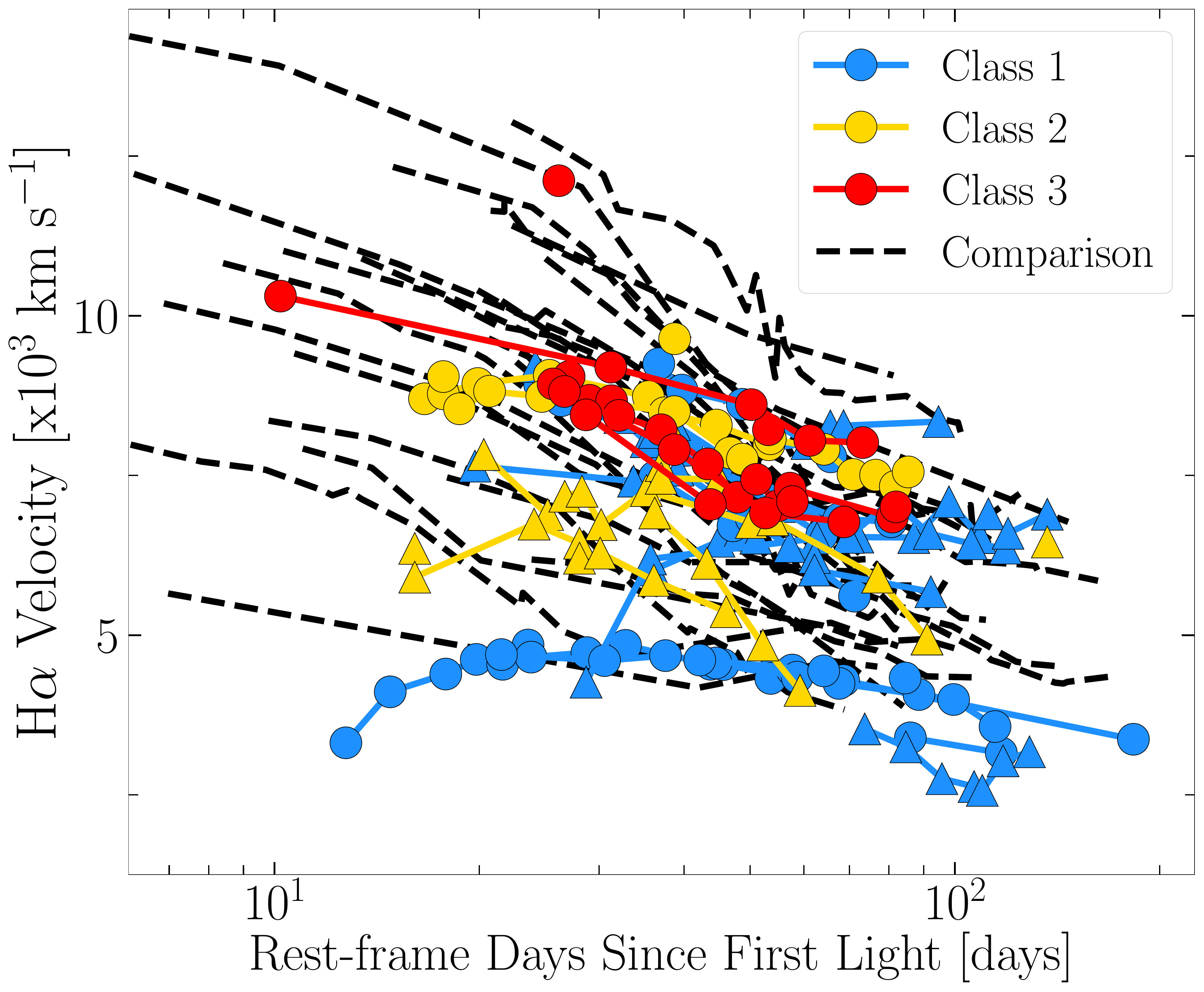}}
\subfigure{\includegraphics[width=0.49\textwidth]{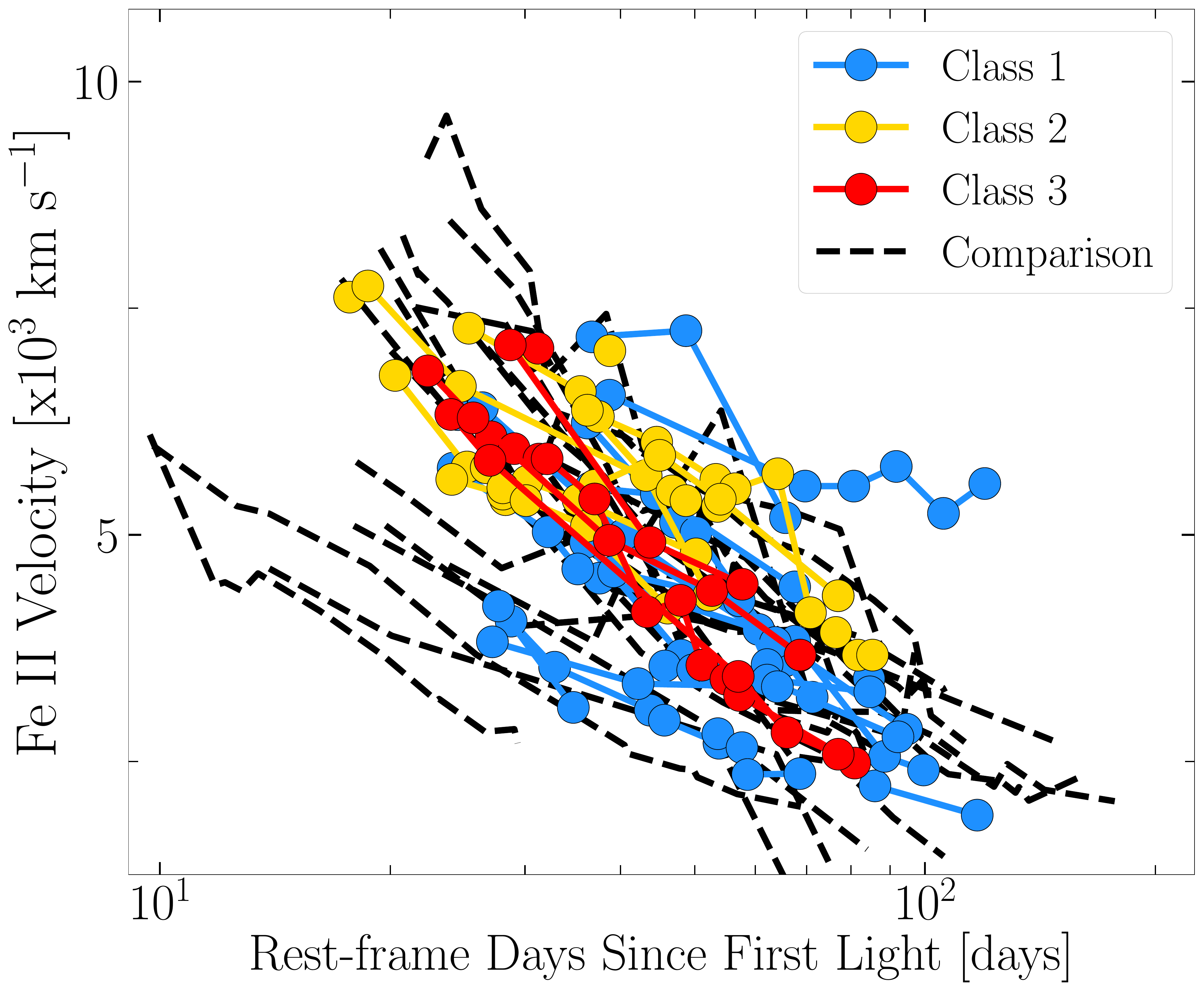}}
\caption{Photospheric-phase velocities for gold/silver- (blue, yellow, red lines) and comparison- (black dashed lines) sample objects calculated from absorption minimum (circles) or emission FWHM (triangles) of H$\alpha$ ({\it left}) and \ion{Fe}{ii} $\lambda 5169$ ({\it right}) line profiles. While some gold-sample objects with more persistent CSM interaction show slower ejecta velocities than the comparison sample, overall both samples possess a consistent evolution in their photospheric velocities.  \label{fig:phot_vels} }
\end{figure*}
\begin{figure*}
\centering
\includegraphics[width=0.99\textwidth]{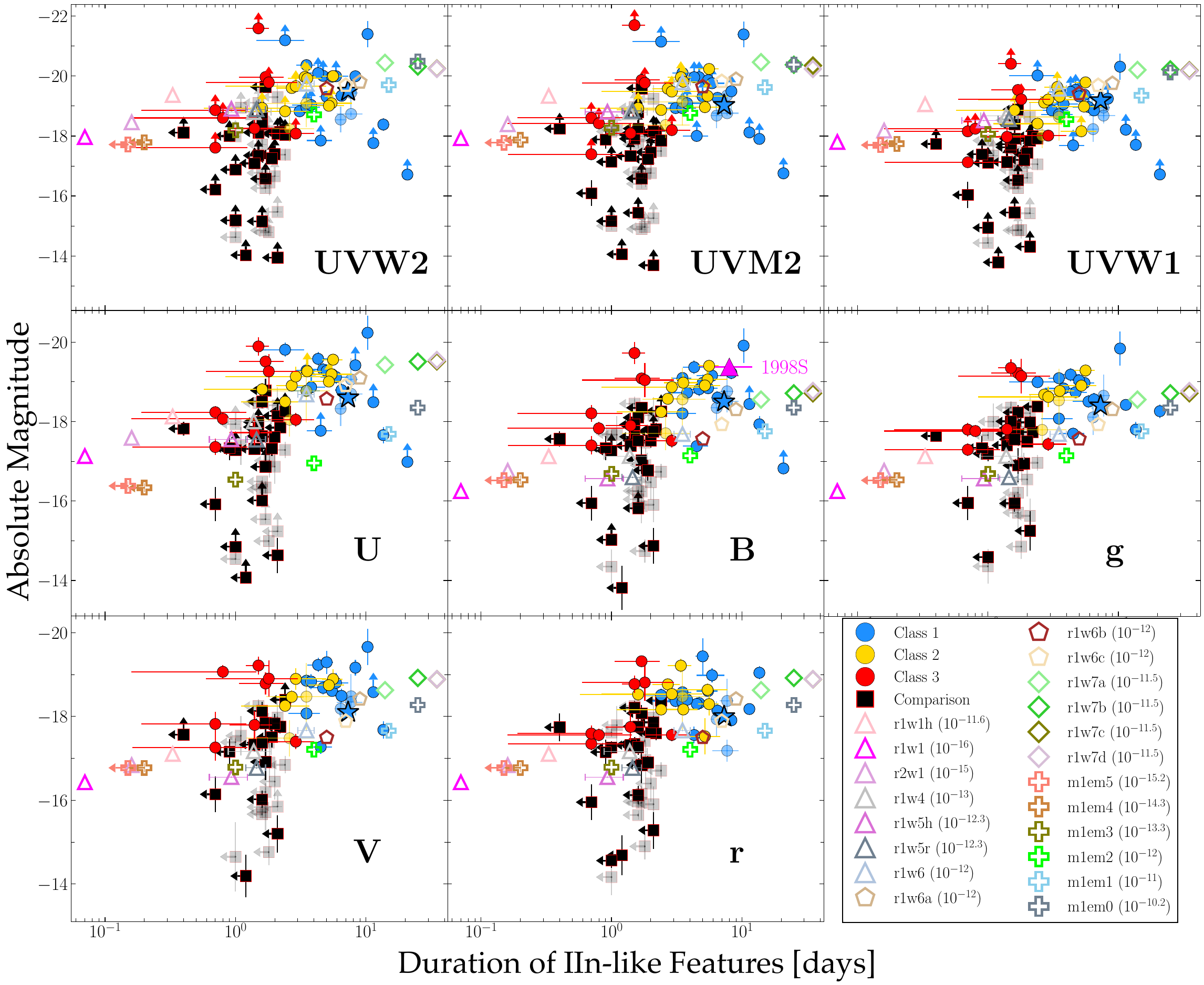}
\caption{{\it Left to right, top to bottom:} Peak absolute magnitude in the \textit{w2}, \textit{m2}, \textit{w1}, \textit{u}, \textit{B/b}, \textit{V/v}, \textit{g}, and \textit{r} bands versus duration of IIn-like features. Gold and silver samples shown as blue/yellow/red circles and comparison sample shown as black squares. Solid colored points represent the subsample of objects at $D>40$~Mpc. Parameters from the \cmfgen\ model grid (\S\ref{subsec:cmfgen}) are plotted as colored stars, polygons, diamonds, and plus signs. SNe~1998S and 2023ixf are shown for reference as a solid magenta triangle and solid blue star, respectively. \label{fig:Max_lines} }
\end{figure*}
\begin{figure}[t!]
\centering
\includegraphics[width=0.47\textwidth]{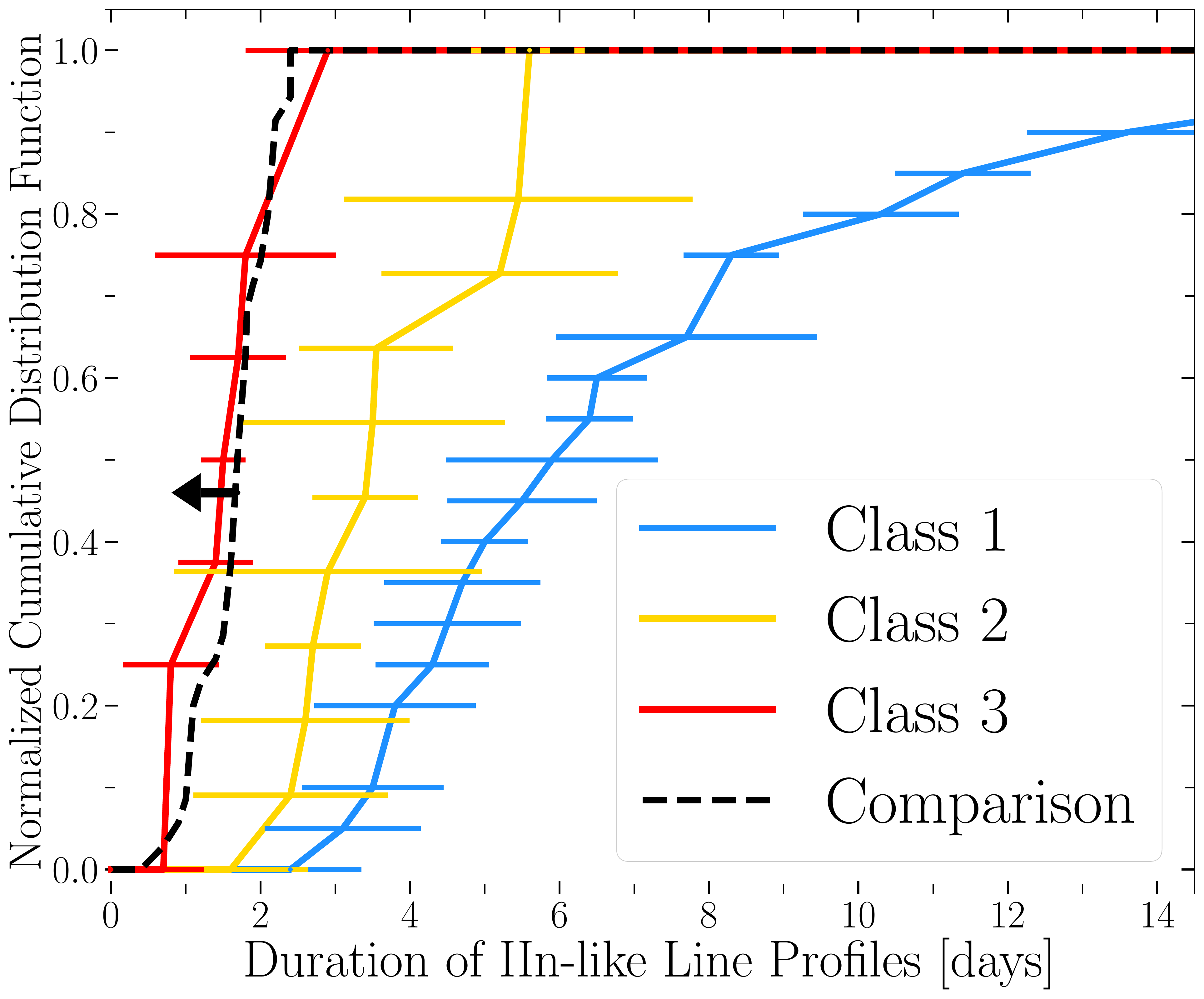}
\caption{Cumulative distribution of $t_{\rm IIn}$ values in Class 1 (blue), 2 (yellow), and 3 (red) gold- and silver-sample objects, as well as upper limits from the comparison sample. Overall, Class 1 objects have longer durations of observed IIn-like features, indicating higher density, and possibly more extended, CSM.  \label{fig:TL_hist} }
\end{figure}
\begin{figure}[t!]
\centering
\includegraphics[width=0.47\textwidth]{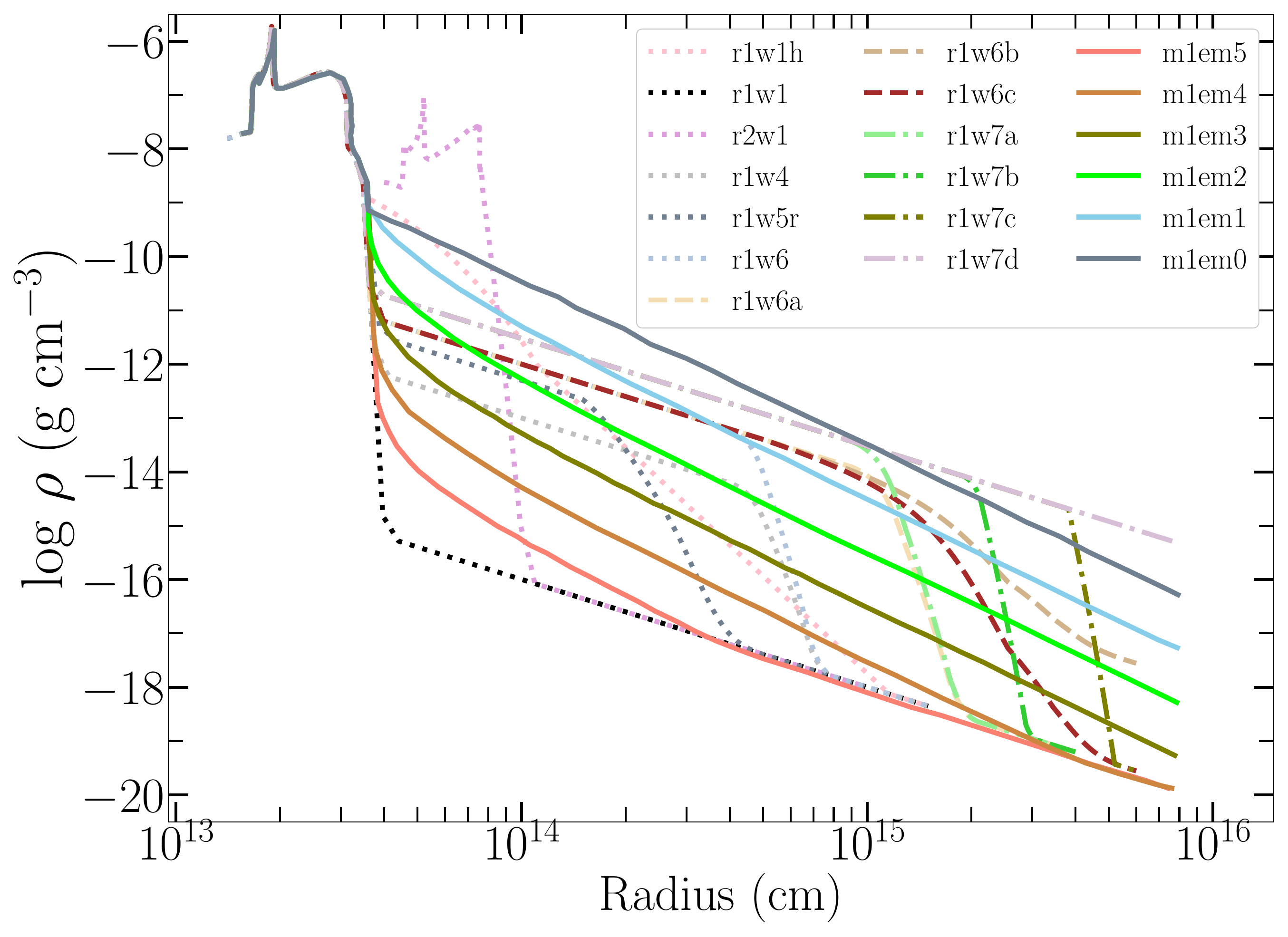}
\caption{CSM densities and radii for complete \cmfgen\ model grid (e.g., Table \ref{tab:models}) used to find the best-matched model for gold-, silver-, and comparison-sample objects. A description of the model setup is provided in \S\ref{subsec:cmfgen}.  \label{fig:model_rho} }
\end{figure}
\begin{figure*}
\centering
\subfigure{\includegraphics[width=0.33\textwidth]{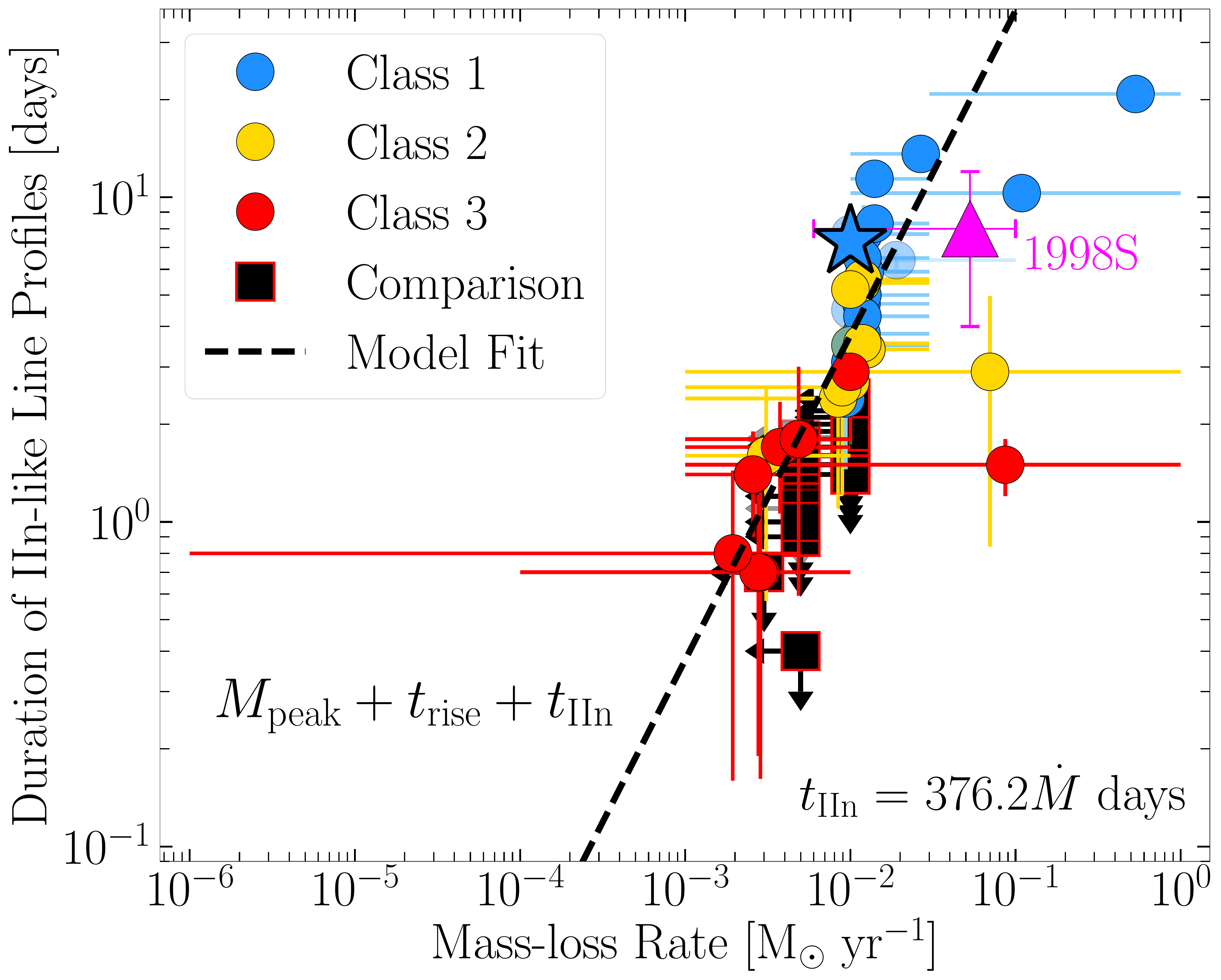}}
\subfigure{\includegraphics[width=0.33\textwidth]{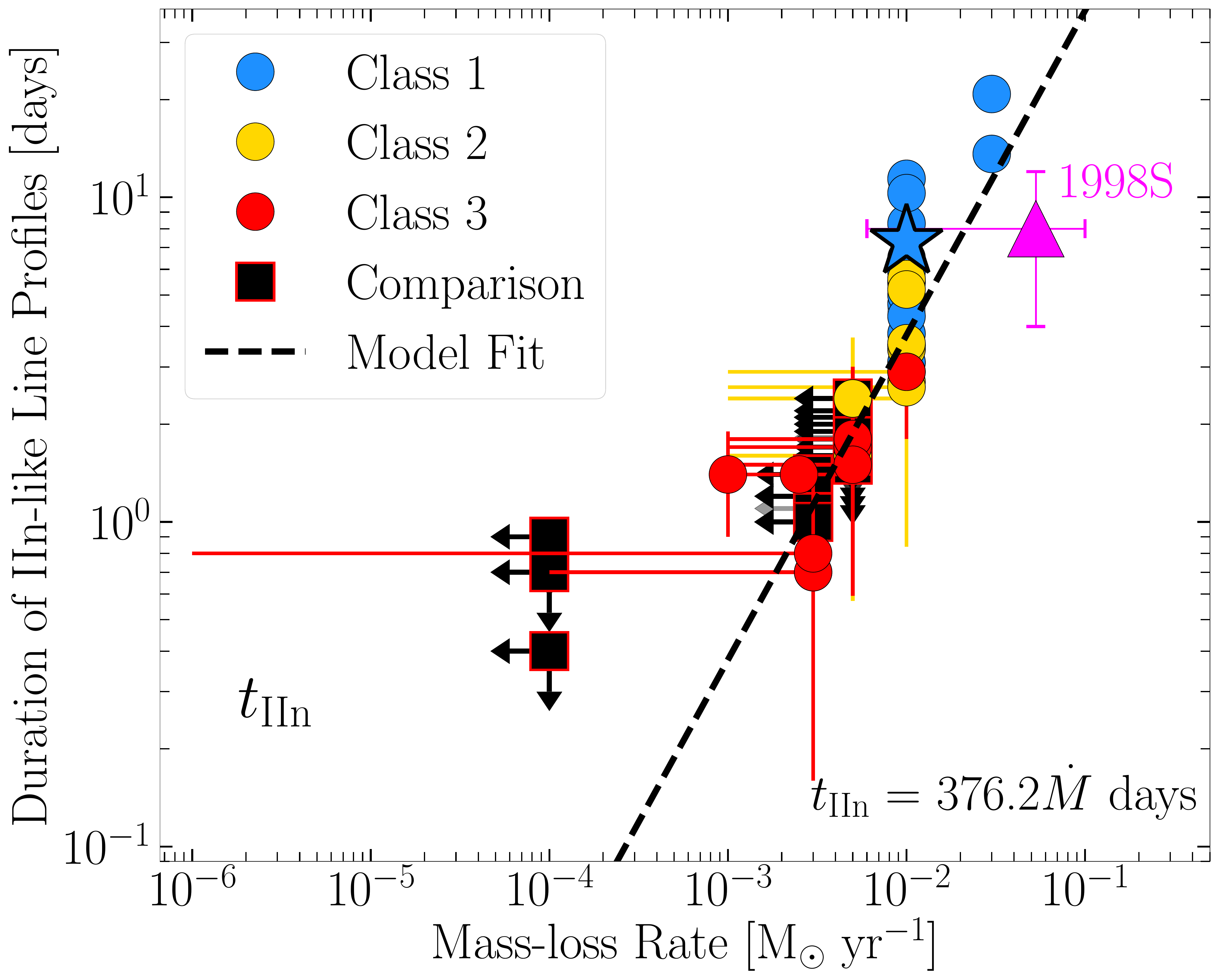}}
\subfigure{\includegraphics[width=0.33\textwidth]{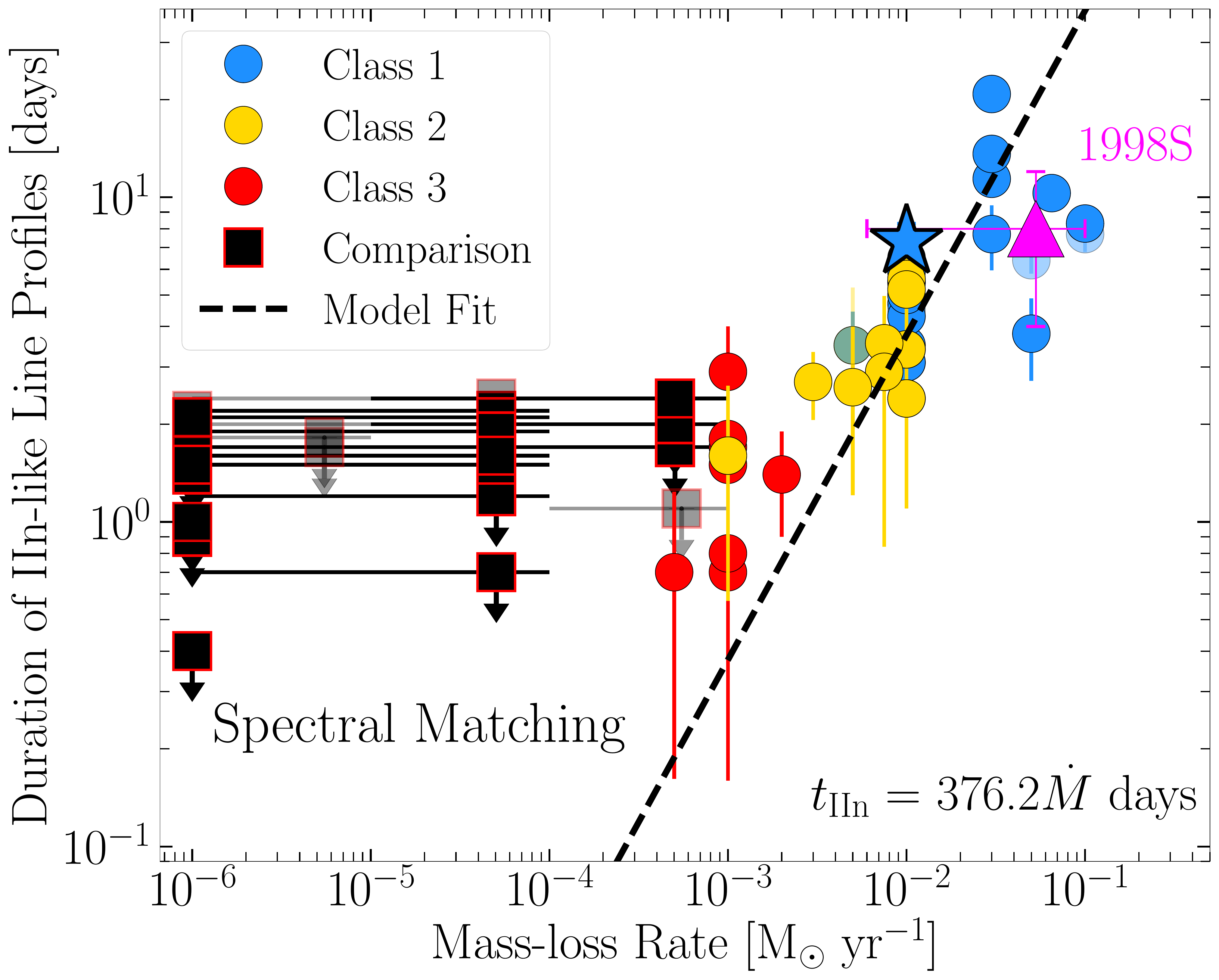}}\\
\subfigure{\includegraphics[width=0.33\textwidth]{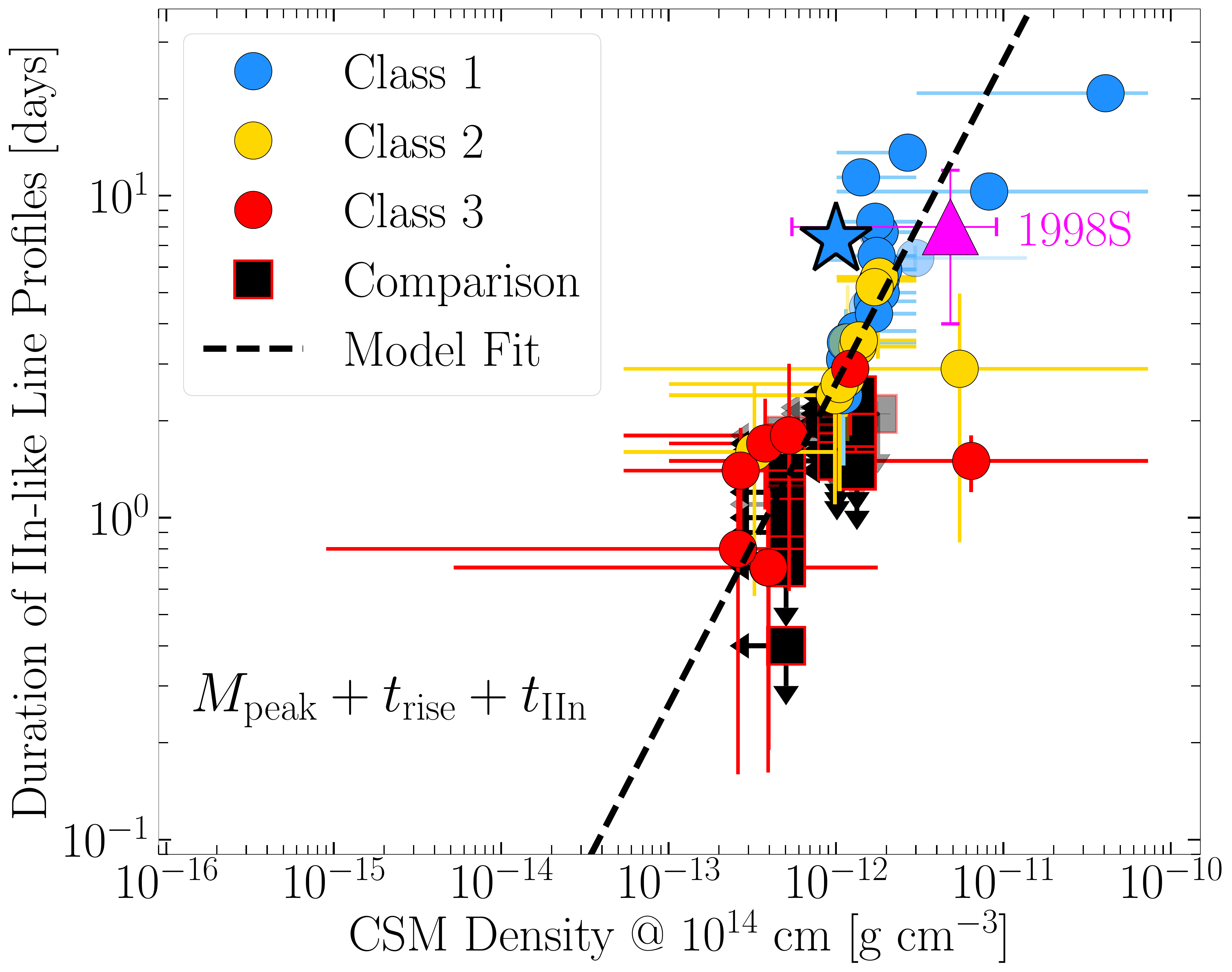}}
\subfigure{\includegraphics[width=0.33\textwidth]{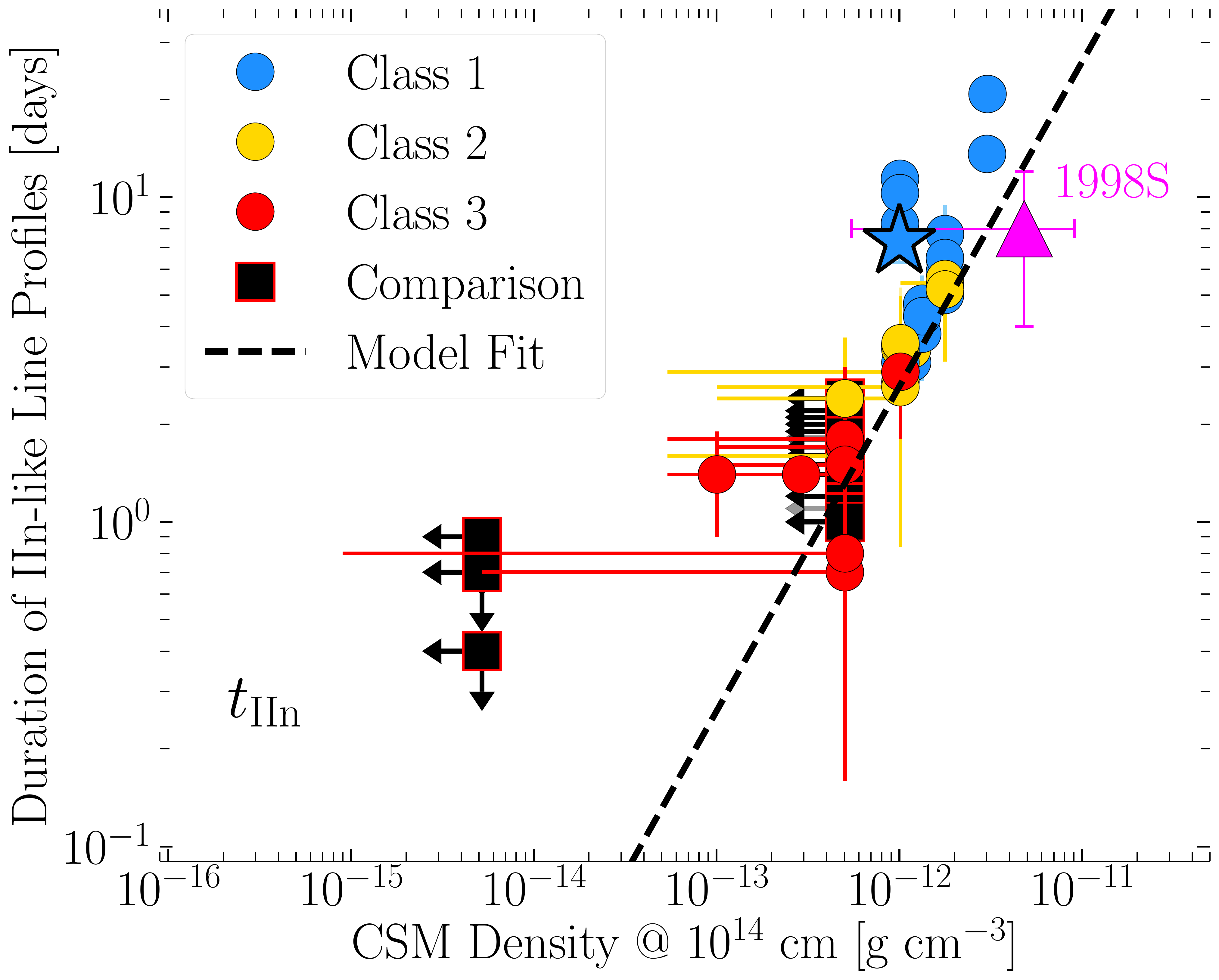}}
\subfigure{\includegraphics[width=0.33\textwidth]{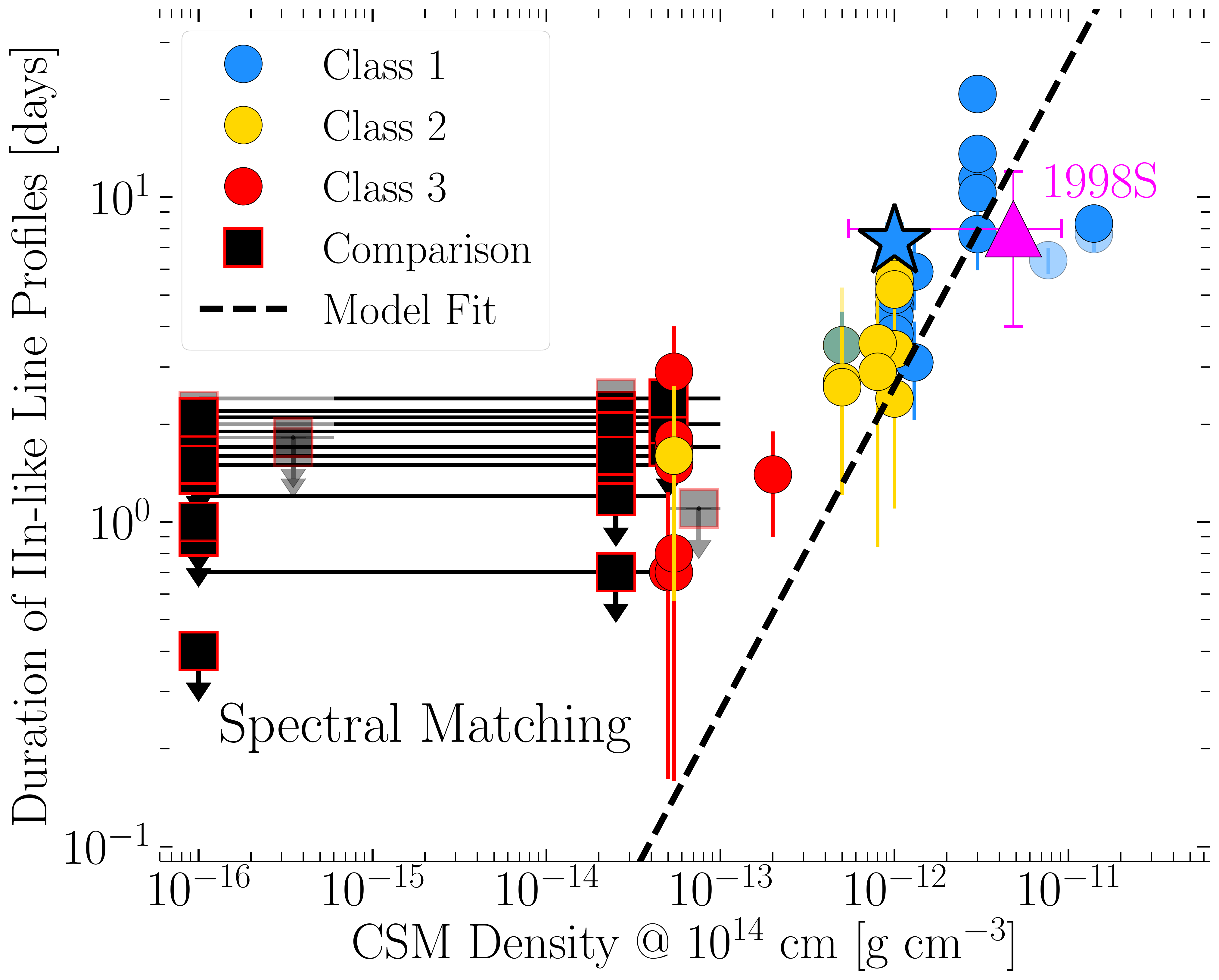}}\\
\caption{Duration of IIn-like features versus best-matched mass-loss rates {\it (top panel)} and CSM densities at $r = 10^{14}$~cm {\it (bottom panel)} for all gold/silver- (blue, yellow and red circles) and comparison- (black squares) sample objects. Solid colored points represent the subsample of objects at $D>40$~Mpc. SNe~1998S and 2023ixf are shown for reference as a magenta triangle and blue star, respectively. Mass-loss rates were estimated for each object based on comparison of ({\it left}) multiband photometry and $t_{\rm IIn}$, ({\it middle}) only $t_{\rm IIn}$, and ($\it right$) early-time spectra, to the \cmfgen\ model grid. Specifics of feature matching and selection of the best model are presented in \S\ref{subsec:cmfgen}. A linear relation between $t_{\rm IIn}$ and $\dot{M}$ (black dashed line) is derived from fitting model parameters used in the \cmfgen\ grid (i.e., the correlations shown are built into our model grid). \label{fig:mdot_IIn} }
\end{figure*}
\begin{figure*}
\centering
\subfigure{\includegraphics[width=0.47\textwidth]{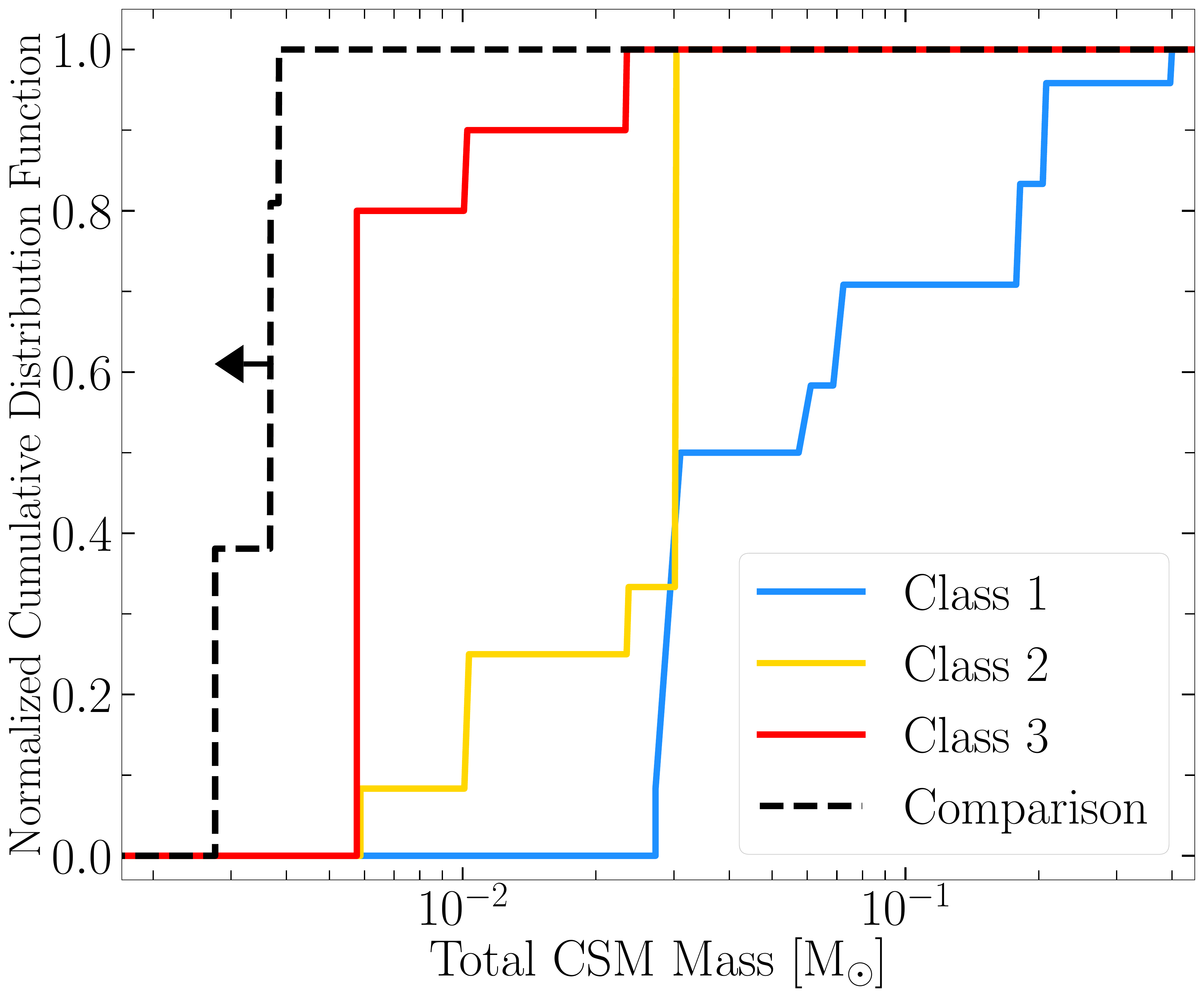}}
\subfigure{\includegraphics[width=0.52\textwidth]{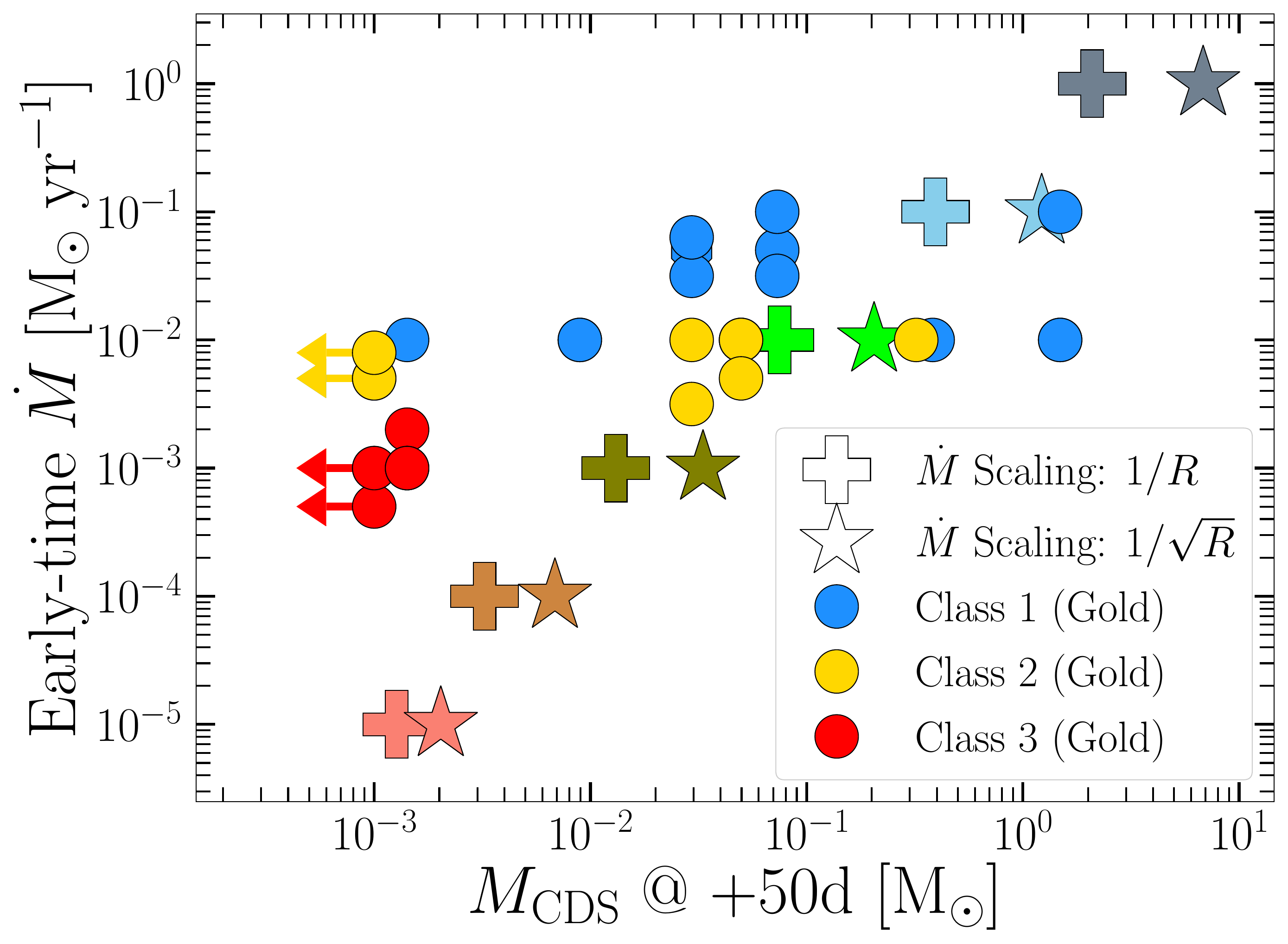}}
\caption{{\it Left:} Histogram of total CSM mass derived from direct spectral matching of the \cmfgen\ grid to Class 1 (blue), 2 (yellow), and 3 (red) gold/silver samples, as well as comparison-sample (black) objects, after a distance cut ($D>40$~Mpc) is applied. {\it Right:} CDS mass (abscissa) derived from the maximum velocity of gold- and silver-sample objects as measured from the bluest edge of the H$\alpha$ absorption profile at $\delta t \approx 50$~days post-first light using the  model trend found by \cite{Dessart23} for \cmfgen\ models of varying mass loss; models shown as plus sign and stars. CDS mass is compared to mass-loss rate (ordinate) derived from comparison of early-time observations to \cmfgen\ model grid.\label{fig:MCSM_hist} }
\end{figure*}

Additionally, as presented in Figure \ref{fig:MCSM_hist}, we calculate the velocities of the fastest moving H-rich ejecta that we can detect at $\delta t = 50$~days by examination of the bluest (reddest) edge of the absorption (emission) profiles in H$\alpha$. However, we note that there is likely faster, optically thin H-rich material that we cannot detect in these spectra and, therefore, these estimates provide a lower limit on velocity of the fastest ejecta. We then compare to model predictions from \cite{Dessart23} for the deceleration of ejecta as a function of total mass in the CDS, which is also connected to the mass-loss rate. From comparison to the models, the slow moving ejecta of some Class 1/2 objects would indicate enhanced mass-loss rates of $\dot{M} = 10^{-3}$--$10^0$~\mdot, while the velocities observed in other Class 1/2 and all Class 3 objects suggest low mass-loss rates of $\dot{M} < 10^{-5}$~\mdot. However, as shown in Figure \ref{fig:phot_vels}, many of the Class 1, as well as all of the Class 2 \& 3, mass-loss rates inferred for gold/silver-sample objects from direct spectral matching are larger than those that are estimated from the fastest moving ejecta. This potentially suggests a degree of CSM asymmetry that would keep some fraction of the ejecta from being decelerated by dense CSM at early times, as is predicted by \cmfgen\ models for spherically symmetric CSM. 

\subsection{Additional Model Grids}\label{subsec:BG_H21}

\begin{figure*}
\centering
\subfigure{\includegraphics[width=0.51\textwidth]{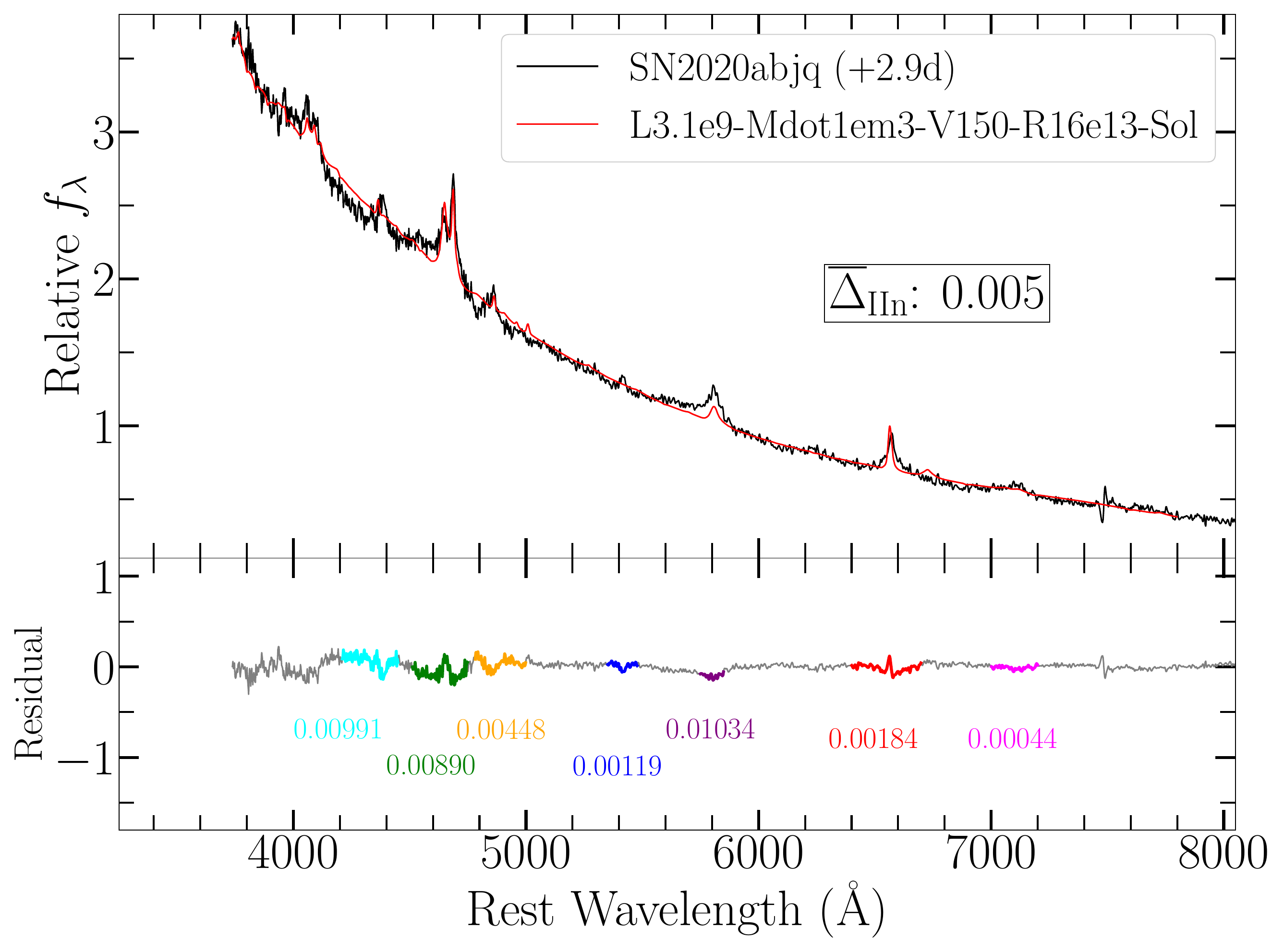}}
\subfigure{\includegraphics[width=0.48\textwidth]{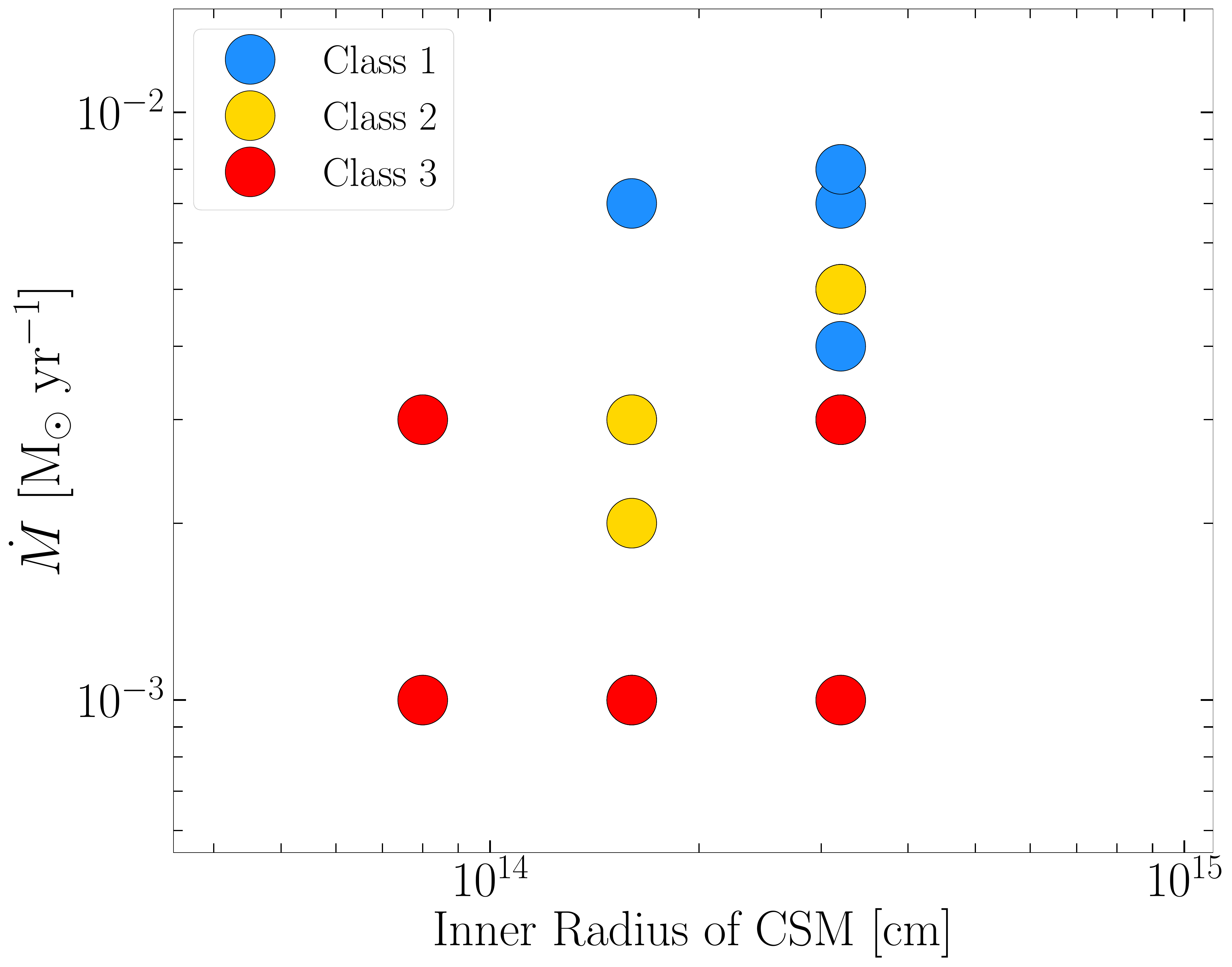}}\\
\caption{{\it Left:} Early-time optical spectra of Class 1 gold-sample object SN~2020abjq is shown with respect to the best-matched \cmfgen\ model from the \cite{Boian20} model grid. Specifics of model matching for the complete sample are presented in \S\ref{subsec:BG_H21}. Numbers in the bottom panel are the residuals between data and model spectra in the wavelength ranges of IIn-like features ($\Delta_{\rm IIn}$). {\it Right:} Best-matching mass-loss rate and inner CSM radius calculated from direct comparison of gold- and silver-sample object spectra to the \cite{Boian20} \cmfgen\ model grid. Some key differences between this grid and that presented in this paper are the lack of spectral time series, multiband photometry, or wider coverage of CSM densities and radii in the former that are present in the latter grid. \label{fig:BG_mdot} }
\end{figure*}

In order to better explore the parameter space of ejecta-CSM interaction in SNe~II, we perform the same spectral matching analysis as above but with the public\footnote{\url{https://www.wiserep.org/object/14764}} grid of \cmfgen\ models presented by \cite{Boian20}. This model grid consists of 137 synthetic spectra with varying CSM compositions (e.g., solar metallicity, CNO-enriched, He-rich), mass-loss rates ($\dot{M} = 10^{-3}$--$10^{-2}$~\mdot), inner radii of the interaction region ($R_{\rm in} = 8\times 10^{13}$--$3.2 \times 10^{14}$~cm), and SN luminosity ($L_{\rm SN} = 1.9\times 10^8$--$2.5 \times 10^{10}~\Lsun$). These models impose an optically thick wind in radiative equilibrium, assume steady state, and have an input luminosity, CSM radius, and mass-loss rate at a given time step. Furthermore, these models contain no radiation hydrodynamics and all of the CSM remains unshocked/unaccelerated at all phases. Similar to our presented model grid, we scale each model spectrum to the observations over the wavelength range of the optical spectrum and calculate the minimum average residual in wavelength regions of IIn-like features (i.e., $\overline{\Delta}_{\rm IIn}$). An example of this matching process is shown for SN~2020abjq in Figure \ref{fig:BG_mdot}, and all best-matched model parameters for gold and silver samples are listed in Tables \ref{tab:model_params_gold} and \ref{tab:model_params_silver} and plotted in Figure \ref{fig:BG_mdot}. 

We find rough agreement between the mass-loss rates derived from the \cite{Boian20} grid and our own: 20/39 objects having mass-loss rates that are consistent to within 50\%. However, the \cite{Boian20} grid does not explore a sufficiently large range of CSM properties (e.g., $\dot{M} > 10^{-2}$~\mdot, $\dot{M} < 10^{-3}$~\mdot, $R_{\rm CSM} > 3 \times 10^{14}$~cm), so these mass-loss estimates may be more biased by the model grid. Furthermore, the \cite{Boian20} model spectra only cover the phases of $\delta t = 1.0$--3.7~days (assuming a shock velocity of $10^4~\kms$) and also do not create synthetic multiband and bolometric light curves to compare with the sample photometry. Nonetheless, the advantage of this model grid is the variety of CSM compositions explored. 

\begin{figure*}
\centering
\subfigure{\includegraphics[width=0.49\textwidth]{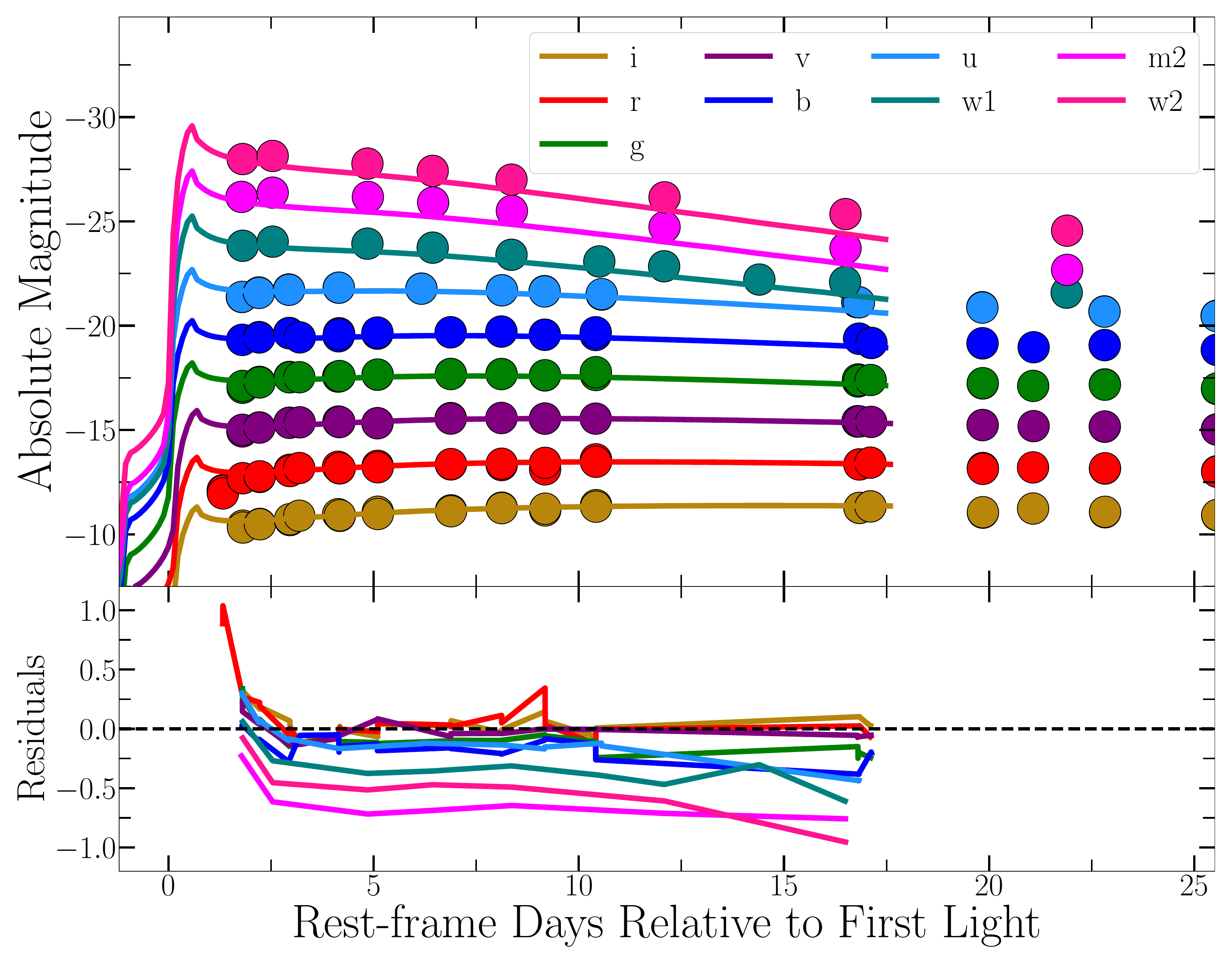}}
\subfigure{\includegraphics[width=0.48\textwidth]{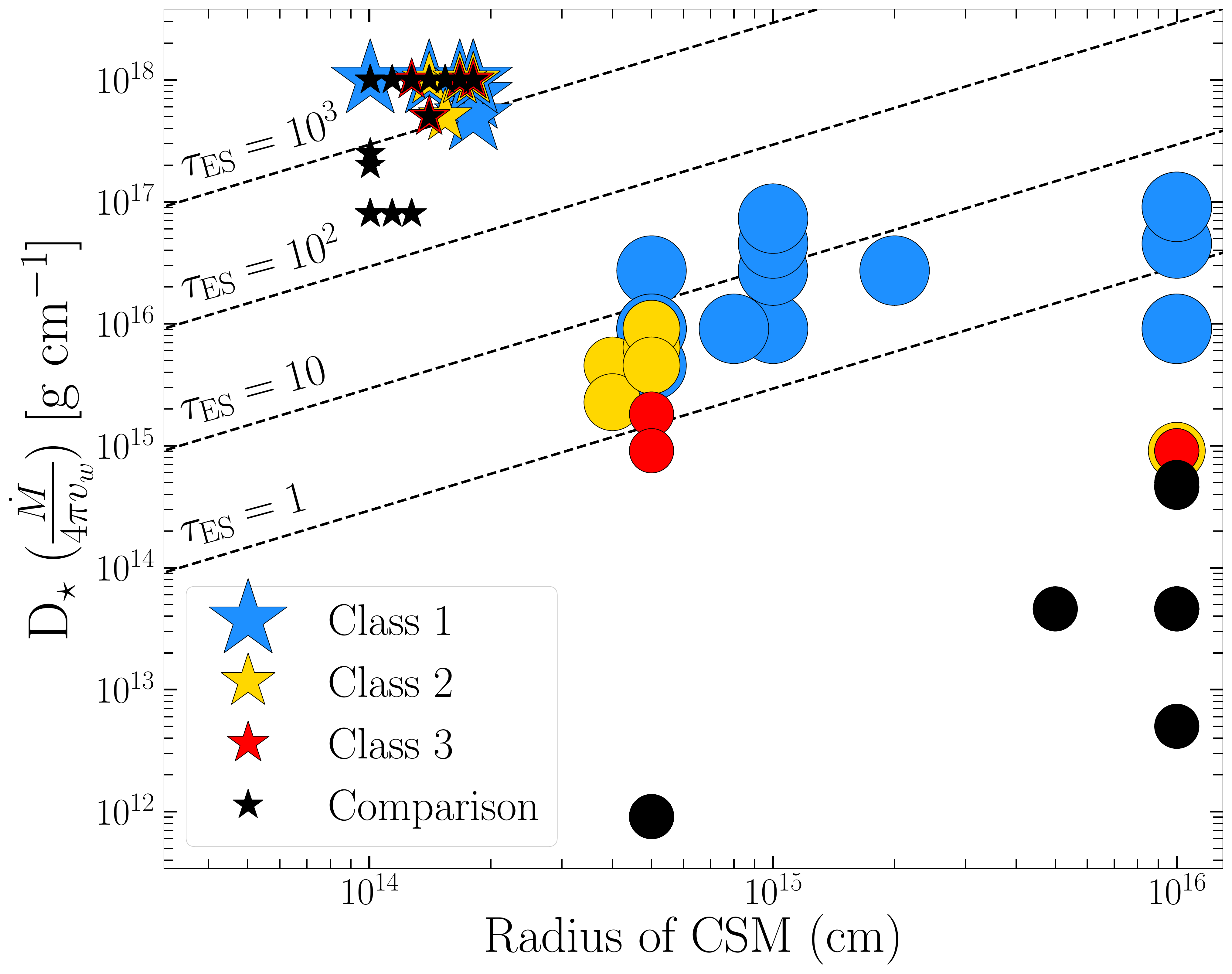}}\\
\subfigure{\includegraphics[width=0.49\textwidth]{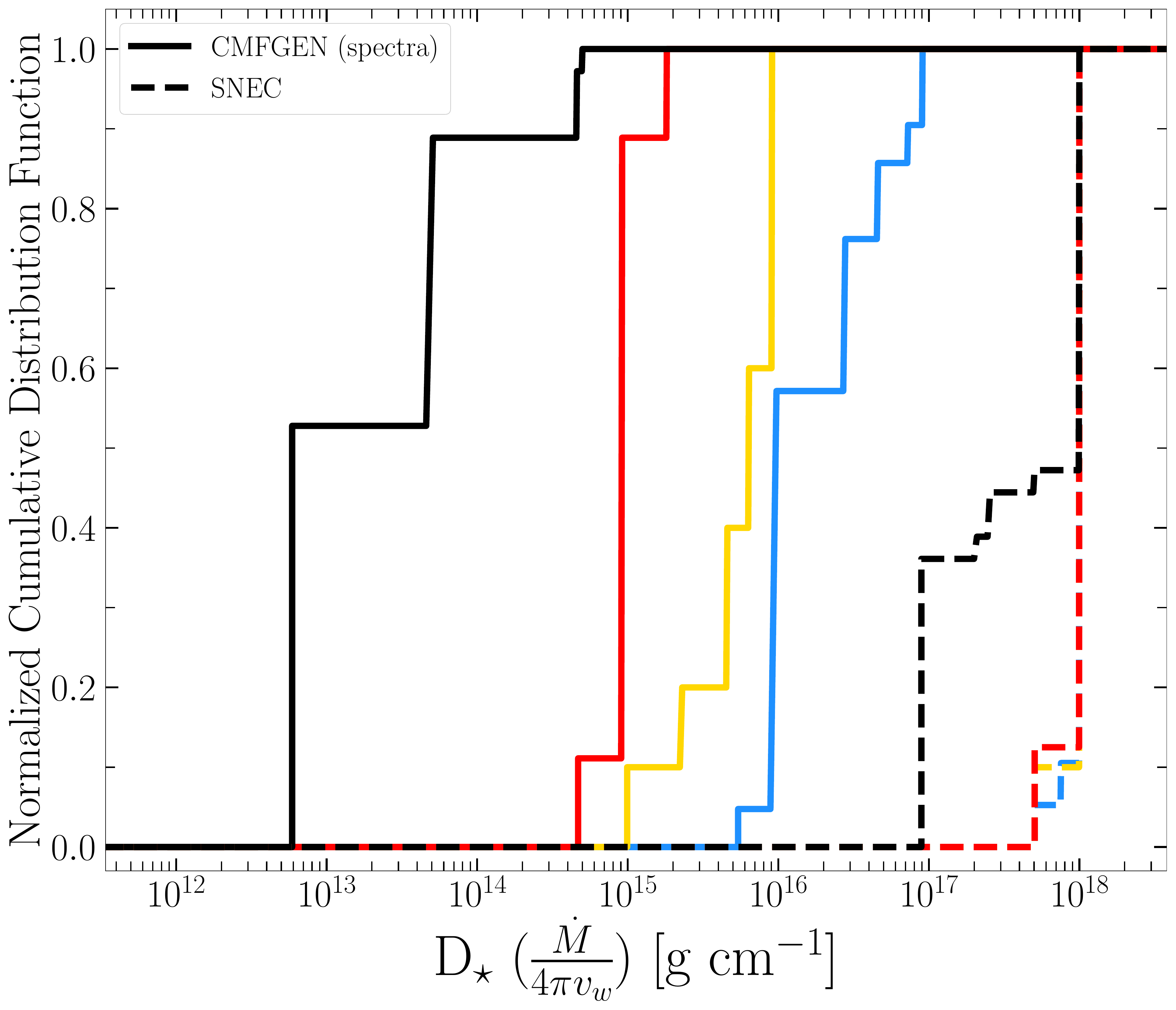}}
\subfigure{\includegraphics[width=0.49\textwidth]{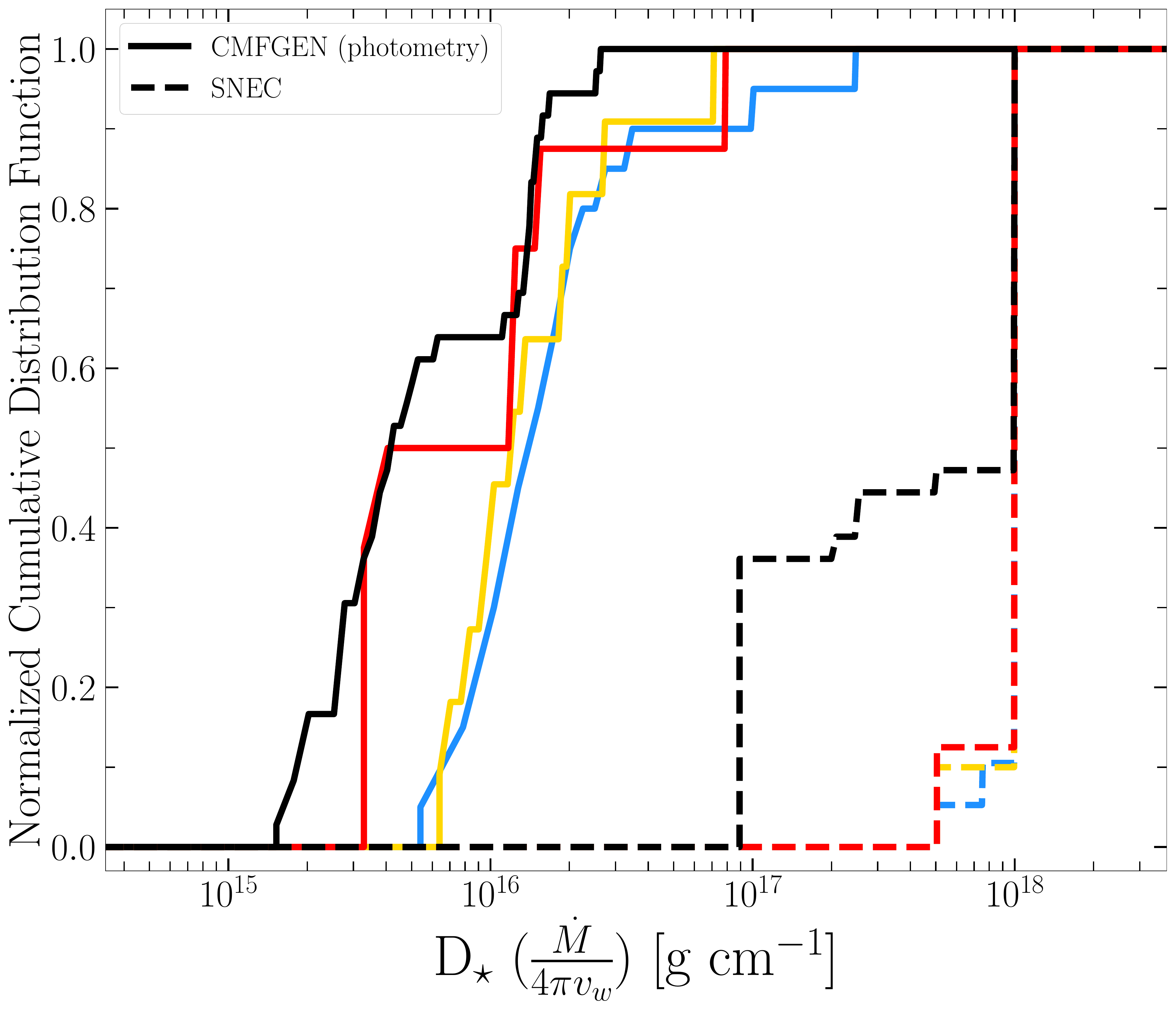}}\\
\caption{{\it Top left:} Multiband photometry of Class 2 gold-sample object SN~2022jox compared to the most consistent CSM interaction model from the grid presented by \cite{haynie21}. Despite the overall match to the photometry, the high CSM densities (e.g., $\rho_{14} \approx 10^{-10}$~g~cm$^{-3}$) required by this model would not allow for the formation of the IIn-like features observed in SN~2022jox. Specifics of model matching for the complete sample are presented in \S\ref{subsec:BG_H21}. {\it Top right:} Mass-loading parameter (D$_{\star}$) versus CSM radius from the best-matched \cite{haynie21} model for all gold/silver- (blue, yellow and red stars) and comparison- (black stars) sample objects. Shown as circles are the best-matching \cmfgen\ models for the gold and comparison samples, which can reproduce both the high peak luminosities and the formation of IIn-like features in the optical spectra. Electron-scattering optical depths shown as dashed lines. {\it Bottom left:} Cumulative distribution of D$_{\star}$ values derived from {\tt SNEC} photometric (dashed lines) and \cmfgen\ spectral (solid lines) model matching. {\it Bottom right:} Cumulative distribution of D$_{\star}$ values derived from {\tt SNEC} (dashed lines) and \cmfgen\ (solid lines) model matching to photometry only. \label{fig:H21_Dstar} }
\end{figure*}

In addition to the \cite{Boian20} spectral models, we also apply a grid of synthetic light curves for shock breakout from dense CSM presented by \cite{haynie21}. The model grid contains 168 multiband light curves created with the LTE, Lagrangian radiative-transfer code {\tt SNEC} \citep{snec} for varying mass-loading parameter D$_{\star} = \dot{M} / (4 \pi v_w) = 8\times 10^{16}$--$10^{18}$~g~cm$^{-1}$, explosion energy ($E_k = (0.3$--3.0) $\times 10^{51}$~erg), and CSM radius ($R_{\rm CSM} = 1500$--$2700~\Rsun$). For all objects in the gold/silver and comparison samples, we find the most consistent model by minimizing the residuals between the synthetic light curves and the observed UV/optical/NIR photometry at $\delta t < 20$~days. First light in these models is assumed to be when the synthetic absolute magnitude rises above $-12$~mag. Furthermore, we note that the uncertainty in the time of first light associated with each sample object could lead to uncertainties in the model parameters derived from the best-matched model light curves. However, these uncertainties are not large enough to impact the overall model trend observed in Figure \ref{fig:H21_Dstar}. An example of a best-match light-curve model to Class 2 gold-sample object SN~2022jox is shown in the left panel of Figure \ref{fig:H21_Dstar}; all derived model parameters are listed in Tables \ref{tab:model_params_gold} and \ref{tab:model_params_silver}.

As shown in Figure \ref{fig:H21_Dstar}, the CSM properties inferred from the best-matched {\tt SNEC} light curves are inconsistent with those derived from both \cmfgen\ model grids. For example, the best-matched light-curve model from \cite{haynie21} implies D$_{\star}[R_{\rm CSM}] = 10^{18}~{\rm g~cm^{-1}}[1900~\Rsun]$ for SN~2013fs, similar to what was found in \cite{Morozova17}, which is several orders of magnitude higher than the most consistent \cmfgen\ model for this SN (e.g., D$_{\star} \approx 10^{15}$~g~cm$^{-1}$). Similarly, the distribution of D$_{\star}$ values derived for the comparison sample is consistent with the distribution of D$_{\star}$ values found by \cite{Morozova18} (e.g., $\sim10^{17-18}~{\rm g~cm^{-1}}$) when modeling the light curves of normal SNe~II with {\tt SNEC}. However, the large densities derived from {\tt SNEC} models ($\rho_{14} \approx 10^{-10}$~g~cm$^{-3}$) would imply mean free paths of $l_{\rm mfp} \approx 3\times 10^{10}$~cm for close-in CSM, $\sim 2R_{\star}$. Such mean free paths are much smaller than the size of extended CSM ($\sim 10^{14}$--$10^{15}$~cm); therefore, electron-scattered photons created from photoionized gas would never escape the CSM to create the IIn-like features observed in the optical spectra while the shock wave is inside of this part of the CSM. Furthermore, at these densities the ionization parameter will be $>10$ (i.e., $\xi = L_{\rm sh} / nr^2$), indicating that the gas will be completely ionized \citep{Lundqvist1996, Chevalier12}. As shown by \cite{Dessart23}, SBO into CSM densities this large will trap the photons stored in the wake of the radiation-dominated shock until the shock has exited the edge of the densest material; the shock front will propagate adiabatically and will not extract kinetic energy that can be used to boost the overall luminosity, as is the case for lower density CSM. Consequently, SBO from such high-density material may provide additional luminosity to early-time light curves, but lower density ($\rho \approx 10^{-12}$--$10^{-14}$~g~cm$^{-3}$) material at larger distances ($r \approx 10^{14-15}$~cm) is needed to create IIn-like features observed in gold/silver-sample objects.

\section{Discussion} \label{sec:discussion}

\subsection{A Continuum of RSG Mass-Loss Rates}\label{subsec:rsg_mdot}
In \S\ref{subsec:cmfgen}, we presented three independent model-matching methods used to derive mass-loss rates and CSM densities for 39 SNe~II (gold/silver samples) with IIn-like features as well as for 35 SNe~II without such spectral signatures. In the total sample, we find significant diversity amongst the mass-loss rates and CSM densities in SNe II, which is intrinsically tied to the distributions of observables between gold/silver and comparison samples such as peak brightness and rise times in their pseudo-bolometric/UV/optical light curves as well as the duration of the IIn-like features. Assuming that all gold-, silver-, and comparison-sample objects arise from the explosion of RSGs, this suggests a continuum of mass-loss histories in the final years-to-months before explosion: Class 1/2 objects (e.g., SNe~20tlf-like, 20pni-like, 98S-like, 14G-like) being associated with RSGs having enhanced mass-loss rates of $\dot{M} \approx 10^{-3}$--$10^{-1}$~\mdot and potentially extended dense CSM ($r \approx 10^{15}$--$10^{16}$~cm), while Class 3 objects (e.g., SN~2013fs-like) may be the result of RSG explosions with lower density ($\dot{M} \approx 10^{-3}$--$10^{-4}$~\mdot), possibly compact ($r < 5 \times 10^{14}$~cm) CSM. Given the lack of IIn-like features at very early-time phases in comparison-sample objects, these SNe need to arise from RSGs with similar or lower mass-loss rates than Class 3 objects ($\dot{M} < 10^{-4}$~\mdot, which may make them more consistent with the weak, steady-state mass-loss rates of Galactic RSGs (e.g., $\dot{M} < 10^{-6}$~\mdot; \citealt{Beasor20}) or highly confined CSM (i.e., $<10^{14}$~cm) at the time of explosion. Nonetheless, the presence of high-density material directly above the RSG surface may be a universal property of SN~II progenitors in order to explain fast-rising light curves (e.g., \citealt{Morozova17}). 

\subsection{Future Improvements to \heracles/\cmfgen\ Grids}\label{subsec:future}

While the differences in $t_{\rm IIn}$, as well as possibly UV $M_{\rm peak}$, are physically linked to differences in CSM density between the gold/silver and comparison samples, the extraction of $\dot{M}$ and $\rho_{14}$ estimates from comparison to the \heracles/\cmfgen\ model grid comes with some assumptions about the physics of the explosion and CSM structure/origin. For the former, this present grid only explores one progenitor mass/radius and explosion energy, which could have an effect on observables such as $t_{\rm rise}$ and $M_{\rm peak}$; future \heracles/\cmfgen\ grids will explore this parameter space in more detail. For the latter, some models in the present grid assume a homogeneous, spherically symmetric CSM with a wind-like density profile, all of which could be potential sources of uncertainty in extracting true mass-loss rates from the present sample. However, some models (e.g., from \citealt{dessart23a}) have varying CSM scale heights as well as different degrees of CSM acceleration. Additionally, the present model grid uses a CSM composition typical of $15~\Msun$ RSGs (\citealt{davies_dessart_19}, which could be varied in future models. 

We are also aware of CSM asymmetries from polarization measurements of SNe II during the photoionization phase (e.g., SN~1998S, \citealt{Leonard00}; SN~2023ixf, \citealt{Vasylyev23}), which suggest that there the CSM is denser along certain lines of sight. Such a physical picture could account for discrepancies between the mass-loss rates inferred from the fastest detectable H$\alpha$ velocities (e.g., Fig. \ref{fig:phot_vels}) and those estimated from the model grid for Class 1/2 objects in the gold/silver samples. In this case, high mass-loss rates (e.g., $\sim 10^{-2}$~\mdot) could still be inferred from electron scattering of recombination photons in dense parts of CSM, while lower density material along different lines of sight would still allow typical ejecta velocities of $\sim 10^4~\kms$, with little to no deceleration by dense CSM. This physical picture may also be able to explain the discrepancies in the derived mass-loss rates between UV/optical vs. X-ray/radio observations of SN~2023ixf (\citealt{Berger23, Chandra23, Grefenstette23, wjg23a, Matthews23}; Nayana et al. 2024, in prep.). Furthermore, a deviation from a steady-state CSM density profile ($\rho \propto r^{-2}$) in these models may be necessary to adequately match the early-time light-curve slope (e.g., SN~2023ixf; \citealt{wjg23a, Hiramatsu23}). For example, SBO from close-in ($r < 10^{14}$~cm) high-density CSM as in the {\tt SNEC} model grid (e.g., \S\ref{subsec:BG_H21}) followed by interaction with lower density material would yield both the fast-rising, luminous light curves and the observation of IIn-like features in some SNe~II.

\subsection{Implications of Photometry-Only Modeling}\label{subsec:phot_only}

As shown in Figure \ref{fig:H21_Dstar}, the extraction of SN~II mass-loss-rate information can yield discrepant results if photometric information is used independently from early-time spectroscopic observations. Here, CSM densities inferred from light-curve matching using a grid of {\tt SNEC} models \citep{haynie21} are too high to allow for the escape of recombination-line photons in the CSM and the formation of IIn-like features. Consequently, without early-time spectroscopy, calculated mass-loss rates and densities close-in to the RSG progenitor (e.g., $< 10^{15}$~cm) may be inconsistent with the presence of narrow emission lines in CSM-interacting SNe~II. Similarly, some studies invoke large CSM masses of $\sim 0.1$--$0.5~\Msun$, confined to $<10^{14}$~cm, in order to match model light curves to early-time SNe~II observations \citep{Morozova17,Tinyanont22,Subrayan23}. However, as shown by \cite{Dessart23}, reproducing the enhanced peak UV/optical luminosity in some early-time SN~II light curves can also be accomplished with $\sim$10\% of these CSM masses. Nevertheless, the early-time light curves of some SNe~II may be influenced by high-density, extended mass, but such explosions can only display IIn-like features during these phases if there are also regions of lower density material via CSM asymmetry or inhomogeneity. It is likely that there is a combination of effects present: (1) SBO from extended envelope and/or high-density CSM located at $<2R_{\star}$ (e.g., \citealt{haynie21}), and (2) interaction with lower density CSM that results in the formation of IIn-like features and increased luminosity. Furthermore, it is worth noting that large amounts of spherically symmetric CSM will cause significant deceleration to the fastest moving SN ejecta; this is an observable that could confirm the existence of such CSM properties \citep{hillier19}. Overall, the combination of photometric and spectroscopic modeling is essential in order to probe both high- and low-density components of CSM in SNe~II. 

\section{Conclusions} \label{sec:conclusion}

In this paper we have presented UV/optical/NIR observations and modeling of the largest sample to date of SNe~II with spectroscopic evidence for CSM interaction. Below we summarize the primary observational findings from our sample analysis.  

\begin{itemize}

\item Our sample consists of 39 SNe II whose early-time (``flash'') spectroscopy shows transient, narrow emission lines with electron-scattering wings (i.e., IIn-like) from the photoionization of dense, confined CSM. The total gold/silver sample contains 39 SNe II, 27 of which are unpublished, and includes 501 total spectra (293 previously unpublished) and 39 UV/optical/NIR light curves (27 previously unpublished). The IIn-like features persist on a characteristic timescale ($t_{\rm IIn}$), which signals a transition in CSM density and the emergence of Doppler-broadened features from the fast-moving SN ejecta.

\item Within the total 74 objects, the ``gold'' sample contains 20 SNe with both early-time IIn-like features, complete UV coverage with {\it Swift}-UVOT and spectral observations at $\delta t < 2$~days. The ``silver'' sample contains 19 SNe that have detectable IIn-like features, complete UV coverage with {\it Swift}-UVOT, and spectral observations only at $\delta t > 2$~days. We divide the gold/silver samples into three classes based on their early-time ($t < 3$~days) spectra: Class 1 shows high-ionization lines of \ion{He}{ii}, \ion{N}{iii}, and \ion{C}{iv} (e.g., SNe~1998S, 2020pni, 2020tlf, 2023ixf, etc), Class 2 shows high-ionization lines \ion{He}{ii} and \ion{C}{iv} but not \ion{N}{iii} (e.g., SNe~2014G, 2022jox), and Class 3 shows only weak \ion{He}{ii} (e.g., SN~2013fs). Additionally, we include a ``comparison'' sample of 35 SNe~II that have optical spectra at $t<2$~days with no IIn-like features as well as a complete UV/optical light curve. Furthermore, Class 1 objects show the longest IIn-like feature timescales (i.e., $t_{\rm IIn} \approx 2$--14~days), while Class 2 and 3 objects displayed shorter-lived emission lines of $t_{\rm IIn} < 4$~days and $t_{\rm IIn} < 2$~days, respectively. We interpret this diversity as arising from variations in CSM extent and density: Class 1 objects arise from RSGs with more extended, higher density CSM than Class 2/3 or the comparison samples. 

\item We find a significant contrast between the peak optical and pseudo-bolometric luminosities in the gold versus comparison samples. We also identify clear correlations between peak UV/optical luminosity and both rise time and $t_{\rm IIn}$. Furthermore, as discussed in \S\ref{subsec:phot_properties}, logrank tests on these observables reveal that the peak pseudo-bolometric and optical luminosities of both samples are likely derived from separate distributions. The difference between sub-samples remains statistically significant after a distance cut ($D>40$~Mpc) is applied.  

\item We apply a grid of ejecta/CSM interaction models, generated with the \cmfgen\ and \heracles\ codes, to extract best-matching mass-loss rates and CSM densities for the gold, silver, and comparison samples. Based on three independent model-matching procedures, we find a continuum of RSG mass-loss rates that extends from $\sim 10^{-6}$ to $10^{-1}$~\mdot. From this model set, we derive an approximate relation between the duration of the electron-scattering broadened line profiles and inferred mass-loss rate: $t_{\rm IIn} \approx 3.8[\dot{M}/$(0.01 \mdot)]~days.

\end{itemize}

Beyond the early-time data presented in this work, future studies (e.g., ``Final Moments III-'') will explore the progenitor and explosion properties of this sample through modeling of their late-time photometric and spectroscopic evolution, as well as multi-wavelength (e.g., X-ray/radio) observations. Now that a sample of SNe~II with IIn-like features has been compiled and examined in detail, it is essential to create new, high-resolution grids of \heracles/\cmfgen\ simulations that can be used together to constrain the CSM properties of such events. Future model grids will provide a more accurate coverage of the CSM interaction parameter space and uncover deficiencies in our model approach (e.g., asymmetries, multidimensional effects, etc.). Furthermore, it is important to build spectroscopically complete, volume-limited surveys that will systematically discover and classify SNe~II within days of first light, therefore reducing biases in follow-up observations and subsequent modeling of certain events. Such discovery efforts will enable volumetric rate measurements of enhanced mass loss in the final years of RSG evolution.

\section{Acknowledgements} \label{Sec:ack}

Research at UC Berkeley is conducted on the territory of Huichin, the ancestral and unceded land of the Chochenyo speaking Ohlone people, the successors of the sovereign Verona Band of Alameda County. Keck I/II, ATLAS, and PS1 observations were conducted on the stolen land of the k\={a}naka `\={o}iwi people. We stand in solidarity with the Pu'uhonua o Pu'uhuluhulu Maunakea in their effort to preserve these sacred spaces for native Hawai`ians. MMT observations were conducted on the stolen land of the Tohono O'odham and Hia-Ced O'odham nations; the Ak-Chin Indian Community, and Hohokam people. ZTF observations were conducted on the stolen land of the Pauma and Cupe\~{n}o tribes; the Kumeyaay Nation and the Pay\'{o}mkawichum (Luise\~{n}o) people. Shane 3~m observations were conducted on the stolen land of the Ohlone (Costanoans), Tamyen and Muwekma Ohlone tribes.

We thank Nathan Smith, David Sand and Avishay Gal-Yam for valuable discussions, and Viktoriya Morozova for providing the initial \texttt{SNEC} models. 
IRAF is distributed by NOAO, which is operated by AURA, Inc., under cooperative agreement with the NSF.

The Young Supernova Experiment and its research infrastructure is supported by the European Research Council under the European Union's Horizon 2020 research and innovation programme (ERC Grant Agreement No.\ 101002652, PI K.\ Mandel), the Heising-Simons Foundation (2018-0913, PI R.\ Foley; 2018-0911, PI R.\ Margutti), NASA (NNG17PX03C, PI R.\ Foley), NSF (AST-1720756, AST-1815935, PI R.\ Foley), the David \& Lucille Packard Foundation (PI R.\ Foley), VILLUM FONDEN (project number 16599, PI J.\ Hjorth), and the Center for AstroPhysical Surveys (CAPS) at NCSA and the University of Illinois Urbana-Champaign.

W.J.-G. is supported by the National Science Foundation (NSF) Graduate Research Fellowship Program under grant DGE-1842165. W.J.-G. acknowledges NASA grants in support of {\it Hubble Space Telescope} programs GO-16075 and GO-16500. This research was supported in part by the NSF under grant PHY-1748958.  The Margutti team at UC Berkeley is partially funded by the Heising-Simons Foundation under grants \#2018-0911 and \#2021-3248 (PI R. Margutti). R.C. acknowledges support from NASA {\it Swift} grant 80NSSC22K0946.

C.D.K. is partly supported by a CIERA postdoctoral fellowship. A. Haynie is supported by the USC-Carnegie Graduate Fellowship. D.L. was supported by a VILLUM FONDEN Investigator grant (project number 16599). C.G. is supported by a VILLUM FONDEN Young Investigator Grant (project number 25501).
This work was funded by ANID, Millennium Science Initiative, ICN12\_009.
The work of X.W. is supported by the National Natural Science Foundation of China (NSFC grants 12288102 and 12033003) and the New Cornerstone Science Foundation through the XPLORER PRIZE.
This work was granted access to the HPC resources of TGCC under the allocation 2021 -- A0110410554  and 2022 -- A0130410554 made by GENCI, France.
This research was supported by the Munich Institute for Astro-, Particle and BioPhysics (MIAPbP) which is funded by the Deutsche Forschungsgemeinschaft (DFG, German Research Foundation) under Germany's Excellence Strategy – EXC-2094 – 390783311.
K.A.B. is supported by an LSSTC Catalyst Fellowship; this publication was thus made possible through the support of Grant 62192 from the John Templeton Foundation to LSSTC. The opinions expressed in this publication are those of the authors and do not necessarily reflect the views of LSSTC or the John Templeton Foundation.

A.V.F.'s research group at UC Berkeley acknowledges financial assistance from the Christopher R. Redlich Fund, as well as donations from Gary and Cynthia Bengier, Clark and Sharon Winslow, Alan Eustace, William Draper, Timothy and Melissa Draper, Briggs and Kathleen Wood, and Sanford Robertson (W.Z. is a Bengier-Winslow-Eustace Specialist in Astronomy, T.G.B. is a Draper-Wood-Robertson Specialist in Astronomy, Y.Y. was a Bengier-Winslow-Robertson Fellow in Astronomy). Numerous other donors to his group and/or research at Lick Observatory include Michael and Evelyn Antin, Shawn Atkisson, Charles Baxter and Jinee Tao, Duncan and Catherine Beardsley, Marc and Cristina Bensadoun, Frank and Roberta Bliss, Ann and Gordon Brown, Tina and Greg Butler, Alan and Jane Chew, Curt Covey, Byron and Allison Deeter, Arthur and Cindy Folker, Peter and Robin Frazier, Ellen Fujikawa, Heidi Gerster, Harvey Glasser, John Gnuse, George and Allison Good, Charles and Gretchen Gooding, Thomas and Dana Grogan, Alan Gould and Diane Tokugawa, Timothy and Judi Hachman, Michael and Virginia Halloran, Robert and Tina Hinckley, Alan and Gladys Hoefer, Jeff and Allison Holland, Jerry and Patti Hume, the Hugh Stuart Center Charitable Trust, James and Zem Joaquin, Joel Krajweski, Walter and Karen Loewenstern, Gregory Losito and Veronica Bayduza, Art and Rita Levinson, Jesse Levinson, Herbert Masters III, Bruce and Judith Moorad, Rand Morimoto and Ana Henderson, James and Marie O'Brient, Douglas and Emily Ogden, Jim Ostendorf, Garry Parton, Edward and Ellin Purdom, Jonathan and Susan Reiter, Margaret Renn, Paul Robinson, Catherine Rondeau, Eric Rudney, Stanley and Miriam Schiffman, Thomas and Alison Schneider, Ajay Shah and Lata Krishnan, Bruce and Debby Smith, Hans Spiller Justin and Seana Stephens, Charles and Darla Stevens, David and Joanne Turner, Rolf Weber, Gerald and Virginia Weiss, Byron and Nancy Wood, Weldon Wood, Richard Wylie, David and Angie Yancey, and Thomas Zdeblick.  

The TReX team at UC Berkeley is supported in part by the NSF under grants AST-2221789 and AST-2224255, and by the Heising-Simons Foundation under grant \#2021-3248 (PI R. Margutti). 

M.R.D. acknowledges support from the NSERC through grant RGPIN-2019-06186, the Canada Research Chairs Program, and the Dunlap Institute at the University of Toronto.
This research was supported by the Munich Institute for Astro-, Particle and BioPhysics (MIAPbP) which is funded by the Deutsche Forschungsgemeinschaft (DFG, German Research Foundation) under Germany´s Excellence Strategy – EXC-2094 – 390783311.
V.A.V. acknowledges support by the NSF under grant AST-2108676. C.R.A. was supported by grants from VILLUM FONDEN (project numbers 16599 and 25501). Parts of this research were supported by the Australian Research Council Centre of Excellence for All Sky Astrophysics in 3 Dimensions (ASTRO 3D), through project number CE170100013.
The UCSC team is supported in part by NASA grant 80NSSC20K0953, NSF grant AST--1815935, the Gordon \& Betty Moore Foundation, the Heising-Simons Foundation, and by a fellowship from the David and Lucile Packard Foundation to R.J.F.

Based in part on observations made with the Nordic Optical Telescope, owned in collaboration by the University of Turku and Aarhus University, and operated jointly by Aarhus University, the University of Turku and the University of Oslo, representing Denmark, Finland and Norway, the University of Iceland and Stockholm University at the Observatorio del Roque de los Muchachos, La Palma, Spain, of the Instituto de Astrofisica de Canarias. Observations were obtained under program P62-507 (PI: Angus).

This work includes data obtained with the Swope telescope at Las Campanas Observatory, Chile, as part of the Swope Time Domain Key Project (PI A. Piro; CoIs Coulter, Drout, Phillips, Holoien, French, Cowperthwaite, Burns, Madore, Foley, Kilpatrick, Rojas-Bravo, Dimitriadis, Hsiao). We thank Abdo Campillay, Yilin Kong-Riveros, Piera Soto-King, and Natalie Ulloa for observations on the Swope telescope.

Some of the data presented herein were obtained at the W. M. Keck Observatory, which is operated as a scientific partnership among the California Institute of Technology, the University of California, and NASA. The Observatory was made possible by the generous financial support of the W. M. Keck Foundation. The authors wish to recognize and acknowledge the very significant cultural role and reverence that the summit of Maunakea has always had within the indigenous Hawaiian community. We are most fortunate to have the opportunity to conduct observations from this mountain.
A major upgrade of the Kast spectrograph on the Shane 3~m telescope at Lick Observatory, led by Brad Holden, was made possible through generous gifts from the Heising-Simons Foundation, William and Marina Kast, and the University of California Observatories. Research at Lick Observatory is partially supported by a generous gift from Google.

Based in part on observations obtained with the Samuel Oschin 48-inch Telescope at the Palomar Observatory as part of the Zwicky Transient Facility project. ZTF is supported by the NSF under grant AST-1440341 and a collaboration including Caltech, IPAC, the Weizmann Institute for Science, the Oskar Klein Center at Stockholm University, the University of Maryland, the University of Washington, Deutsches Elektronen-Synchrotron and Humboldt University, Los Alamos National Laboratories, the TANGO Consortium of Taiwan, the University of Wisconsin at Milwaukee, and the Lawrence Berkeley National Laboratory. Operations are conducted by the Caltech Optical Observatories (COO), the Infrared Processing and Analysis Center (IPAC), and the University of Washington (UW).

The Pan-STARRS1 Surveys (PS1) and the PS1 public science archive have been made possible through contributions by the Institute for Astronomy, the University of Hawaii, the Pan-STARRS Project Office, the Max-Planck Society and its participating institutes, the Max Planck Institute for Astronomy, Heidelberg and the Max Planck Institute for Extraterrestrial Physics, Garching, The Johns Hopkins University, Durham University, the University of Edinburgh, the Queen's University Belfast, the Harvard-Smithsonian Center for Astrophysics, the Las Cumbres Observatory Global Telescope Network Incorporated, the National Central University of Taiwan, STScI, NASA under grant NNX08AR22G issued through the Planetary Science Division of the NASA Science Mission Directorate, NSF grant AST-1238877, the University of Maryland, Eotvos Lorand University (ELTE), the Los Alamos National Laboratory, and the Gordon and Betty Moore Foundation.

This work makes use of observations taken by the Las Cumbres Observatory global telescope network. The Las Cumbres Observatory Group is funded by NSF grants AST-1911225 and AST-1911151.
The new SALT data presented here were obtained through Rutgers University program 2022-1-MLT-004 (PI S. Jha).
Funding for the Lijiang 2.4~m telescope has been provided by the CAS and the People's Government of Yunnan Province.

We are grateful to the staffs at the various observatories where data were obtained. We thank S. Bradley Cenko, Thomas de Jaeger, Ori Fox, Melissa Graham, Goni Halevi, Michael Kandrashoff, Patrick Kelly, Io Kleiser, Jon Mauerhan, Adam Miller, Sarafina Nance, Kishore Patra, Neil Pichay, Anthony Rodriguez, Isaac Shivvers, Jeffrey Silverman, Benjamin Stahl, Erika Strasburger, Heechan Yuk, and Sameen Yunus for assistance with some of the Lick/Shane/Kast observations or reductions. The following U.C. Berkeley undergraduate students helped with the Lick/Nickel observations:
Raphael   Baer-Way,
Sanyum    Channa,
Teagan    Chapman,
Nick      Choksi,    
Maxime    de Kouchkovsky,
Nachiket  Girish,
Goni      Halevy,
Andrew    Halle, 
Romain    Hardy,
Andrew    Hoffman,
Benjamin       Jeffers,
Connor    Jennings,
Sahana    Kumar,
Evelyn    Liu,
Emma      McGinness,
Jeffrey   Molloy,
Yukei     Murakami,     
Andrew    Rikhter,
Timothy   Ross,
Jackson   Sipple,
Samantha  Stegman,
Haynes    Stephens,
James     Sunseri,
Kevin     Tang, and
Sameen    Yunus.

\facilities{\emph{Neil Gehrels Swift Observatory}, Zwicky Transient Facility, ATLAS, YSE/PS1, Lick/Shane (Kast), Lick/Nickel, MMT (Binospec), Keck I/II (LRIS, DEIMOS), Las Cumbres Observatory, {\it TESS}}

\software{IRAF (Tody 1986, Tody 1993),  photpipe \citep{Rest+05}, DoPhot \citep{Schechter+93}, HOTPANTS \citep{becker15}, HEAsoft (v6.22; HEASARC 2014), YSE-PZ \citep{Coulter22, Coulter23}, \cmfgen\ \citep{hillier12, D15_2n}, \heracles\ \citep{gonzalez_heracles_07,vaytet_mg_11,D15_2n} }

\bibliographystyle{aasjournal} 
\bibliography{references} 


\clearpage
\appendix
\counterwithin{figure}{section}

Here we present SN properties for all gold-, silver-, and comparison-sample objects in Tables \ref{tab:gold_sample_specs}, \ref{tab:silver_sample_specs}, and \ref{tab:control_sample_specs}, respectively. Model properties for all \heracles/\cmfgen\ simulations are listed in Table \ref{tab:models}. In Tables \ref{tab:sample_phot_gold}, \ref{tab:sample_phot_gold2}, \ref{tab:sample_phot_control}, and \ref{tab:sample_phot_control2}, we present photometric properties of all gold- and comparison-sample objects after correcting for MW and host reddening. Table \ref{tab:sample_lum_peak} gives gold- and comparison-sample peak luminosity and rise-time distributions. Spectroscopic properties of the gold sample are listed in Table \ref{tab:sample_spec_gold}. In Tables \ref{tab:model_params_gold} and \ref{tab:model_params_silver} we present best-matching model parameters for all gold-, silver-, and comparison-sample objects.  Logs of optical/NIR spectroscopic observations of all unpublished gold-, silver-, and comparison-sample objects are provided in Tables \ref{tab:spec_all_one} - \ref{tab:spec_all_last}. All multicolor/bolometric light curves, spectral sequences, and best-matching light-curve and spectral models are shown for each gold-, silver-, and comparison-sample object in the supplementary pages.\footnote{\url{https://github.com/wynnjacobson-galan/Flash_Spectra_Sample}} \\ 

\section{Host-Galaxy Extinction Uncertainty}\label{subsec:host_ext}

Host-galaxy extinction for sample objects is estimated by measuring the EW of the \ion{Na}{i}~D line and converting it to a host $E(B-V)$ using the relation derived by \cite{Stritzinger18}. We also test the relations between EW and host extinction from \cite{Poznanski12} and find that for the total sample, that relation returns average[min,max] $E(B-V)_{\rm host}$ values of 2.3[0.014, 87.0] compared to 0.19[0.018, 0.81] when using \cite{Stritzinger18}. We choose to adopt the \cite{Stritzinger18} relation given the large scatter associated with the \cite{Poznanski12} relations (e.g., see \citealt{Phillips13}) and inaccuracy of the latter at large EWs due to limited number of objects used in their fitting procedure. In Figure \ref{fig:extinction}, we present the cumulative distributions of the host $E(B-V)$ values as well as the observed $g-r$ color versus \ion{Na}{i}~D EW. For the latter, we note that there is a large scatter relative to the \cite{Stritzinger18} relation i.e., gold/silver sample objects are bluer than comparison sample objects for similar EW. Consequently, it appears that \ion{Na}{i}~D and/or colors are likely limited measures of reddening in SNe~II, especially for large EWs and reddened colors. Additionally, in the top panel of Figure \ref{fig:extinction2}, we compare peak UV/optical magnitudes to host extinction derived from \ion{Na}{i}~D for all subsamples. We note that there is clearly a lack of highly-reddened objects in the sample (e.g., lower-right panel of Fig. \ref{fig:extinction}). Also, there appears to be a correlation present in this host extinction correction method that traces the reddening vector at larger reddening values (e.g., $>0.3$~mag) indicating inaccuracy in using \ion{Na}{i}~D as a tracer of reddening. Nonetheless, when looking at the distribution of peak UV magnitudes for objects without large host reddening, there remains a contrast in absolute magnitude between gold- and comparison-sample objects, most likely the result of CSM interaction. Furthermore, we note that using \ion{Na}{i}~D absorption as a probe of host extinction is dependent on the resolution of the spectrograph used to observe each SN in our sample. However, most of the spectra obtained for this study have resolutions of $R > 500$, which corresponds to $\Delta \lambda \lesssim 12$~\AA\ for a combination of both the \ion{Na}{i}~D1 \& D2 transitions. Reliable detections of this transition only become problematic with very low resolution (e.g., $R<100$, $\Delta \lambda > 60$~\AA) spectrographs for the typical signal-to-noise ratio of our SN spectra.

In Figure \ref{fig:extinction}, we compare our host reddening distribution to SN~II samples from \cite{Anderson14} and \cite{Irani23b}, where the former derives host extinction from the \ion{Na}{i}~D EW using \cite{Poznanski12} and the latter derives it using shock-cooling modeling. Overall, our host reddening distributions contain larger values than both the \cite{Anderson14} and \cite{Irani23b} samples. We note that for some objects that are in both our sample and that of \cite{Irani23b} (e.g., SNe~2020pni, 2019nvm, 2018dfc, 2019ust), the derived host-galaxy $E(B-V)$ is larger by $\sim 0.1$--0.2~mag when using the \ion{Na}{I}~D EW. However, \cite{Irani23b} also fit for an $R_V$ value while we apply a consistent $R_V = 3.1$ with a \cite{fitzpatrick99} reddening law; the choice of both the $R_V$ and the reddening law could lead to bias in the host extinction correction.

We test whether the enhanced UV/optical luminosities observed in the gold sample are a product of the explosion and not uncertainty in the host extinction by first comparing the reddening vector for $R_V = 3.1$ in the \cite{fitzpatrick99} reddening law to the $w2-v$ vs. $g-r$ color evolution, as shown in Figure \ref{fig:colors}. The reddening vector has a slope of $\sim4.3$, which is inconsistent with a slope of $\sim8.1$ measured in the color-color evolution of the gold and comparison samples. This implies that extinction correction alone is not able to make all of these SNe have the sample peak absolute magnitude. Additionally, we apply a synthetic host extinction correction to the $g-r$ colors of the gold/comparison samples until the colors of each object are consistent with the bluest object in the sample at $\delta t = 5$~days, prior to any host reddening correction (e.g., see Figures \ref{fig:colors} and \ref{fig:colors_obs}). We find that an average of $0.21$~mag of host reddening is needed, which translates to $\sim 1.9$~mag of UV extinction. However, even this amount of reddening cannot account for an average difference $>3$~mag observed between gold and comparison-sample UV luminosities, further indicating that this observed phenomenon is not a result of host-galaxy extinction. Furthermore, even after this relative host reddening is applied based on colors, there remains a difference between the peak UV/optical luminosities of many comparison objects relative to those in the gold sample (e.g., see Fig. \ref{fig:extinction2}).

\renewcommand\thetable{A\arabic{table}} 
\setcounter{table}{0}

\begin{deluxetable*}{cccccccc}[h!]
\tablecaption{Gold-Sample Objects \label{tab:gold_sample_specs}}
\tablecolumns{8}
\tablewidth{0pt}
\tablehead{\colhead{Name} & \colhead{Time of First Light} & \colhead{Host Galaxy} & \colhead{Redshift} & \colhead{Distance} &  \colhead{$E(B-V)_{\textrm{MW}}$} &  \colhead{$E(B-V)_{\textrm{host}}$} & {References} \\
\colhead{} & \colhead{(MJD)} & \colhead{} & \colhead{$z$} & \colhead{(Mpc)} & \colhead{(mag)} & \colhead{(mag)} & \colhead{} }
\startdata
PTF10gva & $55320.3 \pm 0.9$ & SDSS J122355.39+103448.9 & 0.028 & $114.5\pm 8.0^{c}$ & 0.026 & $<0.007$ & \cite{Khazov16}\\%
PTF11iqb & $55764.4 \pm 1.0$ &  NGC~151 & 0.013 & $50.4\pm 3.5^{d}$ & 0.028 & $<0.003$ &  \cite{smith15}\\
2013fs & $56571.11 \pm 0.02$ &  NGC~7610 & 0.012 & $50.6 \pm 0.9^{c}$ & 0.035 & $0.02 \pm 0.004$ & \cite{yaron17}\\
2017ahn & $57791.8 \pm 0.5$ & NGC~3318 & 0.0090 & $33.0 \pm 6.5^{d}$ & 0.068 & $0.21 \pm 0.04$ &  \cite{Tartaglia21}\\
2018zd & $58178.4 \pm 0.2$ & NGC~2146 & 0.0030 & $18.4 \pm 4.5^{d}$ & 0.085 & $0.17 \pm 0.03$  & \tablenotemark{a}\\
2018fif & $58350.3 \pm 0.2$ & UGC~00085 & 0.017 & $73.8\pm 5.2^{c}$ & 0.094 & $0.18 \pm 0.04$ & \cite{Soumagnac20}\\
2018dfc & $58302.3 \pm 0.9$ & SDSS J164807.68+241815.4 &  0.037 & $153.5\pm 10.7^{c}$ & 0.072 & $0.14 \pm 0.03$ & \cite{bruch21}\\%
2019ust & $58799.8 \pm 0.5$ & UGC~00548 & 0.022 & $92.2\pm 6.5^{c}$ & 0.053 & $0.28 \pm 0.06$ & \cite{Bruch2019ust} \\
\enddata
\tablenotetext{a}{\cite{zhang20, Hiramatsu21, Callis21}.}
\tablenotetext{c}{Distance derived from redshift with peculiar-velocity uncertainties included.}
\tablenotetext{d}{Redshift-independent distance.}
\end{deluxetable*}

\begin{deluxetable*}{cccccccc}[h!]
\tablecaption{Silver-Sample Objects \label{tab:silver_sample_specs}}
\tablecolumns{8}
\tablewidth{0pt}
\tablehead{\colhead{Name} & \colhead{Time of First Light} & \colhead{Host Galaxy} & \colhead{Redshift} & \colhead{Distance} &  \colhead{$E(B-V)_{\textrm{MW}}$} &  \colhead{$E(B-V)_{\textrm{host}}$} & {References} \\
\colhead{} & \colhead{(MJD)} & \colhead{} & \colhead{$z$} & \colhead{(Mpc)} & \colhead{(mag)} & \colhead{(mag)} & \colhead{} }
\startdata
PTF10abyy & $55532.0 \pm 0.5$ & WISEA J051639.83+064740.8 & 0.030 & $123.6\pm8.6^{c}$ & 0.150 & $<0.09$ & \cite{Khazov16}\\
2014G & $56669.6 \pm 1.7$ &  NGC~3448 & 0.0043 & $24.5 \pm 9.0^{d}$ & 0.01 & $0.29 \pm 0.06$ & \cite{Terreran16}\\
2015bf & $57367.8 \pm 1.2$ & NGC~7653 & 0.014 & $60.1 \pm 1.4^{d}$ & 0.059 & $0.19 \pm 0.04$ &  \cite{Lin21}\\
2016blz & $57485.5 \pm 3.0$ & ECO 0866 & 0.012 & $51.4 \pm 3.6^{c}$ & 0.09 & $0.12 \pm 0.024$ & \cite{Johansson2016blz} \\
2019qch & $58739.0 \pm 0.1$ & WISEA J182914.07+410228.6 & 0.024 & $105.7\pm 4.5^{c}$ & 0.1 & $<0.1$ & \cite{Fremling2019qch} \\
2020abtf & $59184.9 \pm 0.4$ & SDSS J075934.54+151759.6 & 0.014 & $61.2\pm 4.4^{c}$ & 0.029 & $0.09 \pm 0.02$ & \cite{Anderson2020abtf} \\%
2020xua & $59142.3 \pm 1.1$ & UGC~12218 & 0.021 & $90.4\pm 6.3^{c}$ & 0.075 & $0.08 \pm 0.02$ & \cite{Terreran2020xua} \\
2020tlf & $59098.7 \pm 1.5^{b}$ & NGC~5731 & 0.0084 & $36.8\pm 1.3^{c}$ & 0.014 & $0.02 \pm 0.01$ & \cite{wjg22}\\
\enddata
\tablenotetext{a}{\cite{zhang20, Hiramatsu21, Callis21}.}
\tablenotetext{b}{Based on photometric detection. Time of first light used throughout is MJD 59108 based on modeling.}
\tablenotetext{c}{Distance derived from redshift with peculiar-velocity uncertainties included.}
\tablenotetext{d}{Redshift-independent distance.}
\end{deluxetable*}

\begin{deluxetable*}{cccccccc}[h!]
\tablecaption{Comparison-Sample Objects \label{tab:control_sample_specs}}
\tablecolumns{8}
\tablewidth{0pt}
\tablehead{\colhead{Name} & \colhead{Time of First Light} & \colhead{Host Galaxy} & \colhead{Redshift} & \colhead{Distance} &  \colhead{$E(B-V)_{\textrm{MW}}$} &  \colhead{$E(B-V)_{\textrm{host}}$} & {References} \\
\colhead{} & \colhead{(MJD)} & \colhead{} & \colhead{$z$} & \colhead{(Mpc)} & \colhead{(mag)} & \colhead{(mag)} & \colhead{} }
\startdata
2013ft & $56546.8 \pm 1.0$ & NGC~7732 & 0.0096 & $34.2 \pm 7.1^g$ & 0.047 & $<$0.037 & \cite{Khazov16} \\
2013ab & $56339.5 \pm 1.0$ & NGC~5669 & 0.0046 & $24.0\pm0.9^g$ & 0.023 & $0.11 \pm 0.020$ & \cite{Bose15} \\
2013am & $56371.5 \pm 1.0$ & M65 & 0.0030 & $12.8\pm 2.4^g$ & 0.02 & $0.49 \pm 0.090$ & \tablenotemark{a} \\
2016X & $57405.9 \pm 0.5$ & UGC~08041 & 0.0044 & $15.2\pm 3.0^g$ & 0.02 & $<0.007$ & \cite{Huang18} \\
2016aqf & $57443.6 \pm 1.0$ & NGC~2101 & 0.004 & $10.8\pm1.4^g$ & 0.047 & $0.03 \pm 0.006$ & \cite{MullerBravo20} \\
2017eaw & $57885.7 \pm 0.1$ & NGC~6946 & 0.0013 & $6.70\pm 0.15^g$ & 0.30 & $0.04 \pm 0.01$ & \tablenotemark{b} \\
2017gmr & $57999.1 \pm 0.5$ &  NGC~988 & 0.0050 & $19.6 \pm 1.4^g$ & 0.024 & $0.32 \pm 0.070$ & \tablenotemark{c}\\
2018lab & $58480.4\pm0.1$ & IC~2163 & 0.0089 & $32.8 \pm 0.4^g$ & 0.075 & $0.39\pm0.09$ & \cite{Pearson23} \\
\enddata
\tablenotetext{a}{\cite{Zhang14, Tomasella18}.}
\tablenotetext{b}{\cite{kilpatrick18,Szalai19}.}
\tablenotetext{c}{\cite{Andrews19, Utrobin21}.}
\tablenotetext{d}{\cite{Sollerman21, Teja22, Ailawadhi23, Kilpatrick23b}.}
\tablenotetext{e}{\cite{Murai24, Meza24}.}
\tablenotetext{h}{\cite{Hosseinzadeh22,Vasylyev22,Kozyreva22}.}
\tablenotetext{f}{Distance derived from redshift with peculiar-velocity uncertainties included.}
\tablenotetext{g}{Redshift-independent distance.}
\end{deluxetable*}

\begin{deluxetable*}{ccccccc}
\tablecaption{Model Properties \label{tab:models}}
\tablecolumns{7}
\tablewidth{0pt}
\tablehead{\colhead{Name} & {$t_{\rm IIn}$} & {log$_{10}(\dot M$)} & {log$_{10}(\rho_{\rm CSM,14}$)$^a$} & {log$_{10}(R_{\rm CSM}$)} & {$M_{\rm CSM}$} & {Reference} \\
\colhead{} & \colhead{(days)} & \colhead{(M$_{\odot}$~yr$^{-1}$)} & \colhead{(g cm$^{-3}$)} & \colhead{(cm)} & \colhead{($\Msun$)} & \colhead{}  }
\startdata
r1w1h & $<0.3$ & $-6.0$ & $-11.6$ & 14.5 & 0.016 & \cite{dessart17} \\
r1w1 & $<0.1$ & $-6.0$ & $-16.0$ & 15.0 & 0.0028 & \cite{dessart17} \\
r2w1 & $<0.2$ & $-6.0$ & $-15.0$ & 14.0 & 0.061 & \cite{dessart17} \\
r1w4 & $1.4$ & $-3.0$ & $-13.0$ & 14.7 & 0.0056 & \cite{dessart17} \\
r1w5h & $0.9$ & $-2.5$ & $-12.3$ & 14.5 & 0.036 & \cite{dessart17} \\
r1w5r & $1.4$ & $-2.3$ & $-12.3$ & 14.6 & 0.010 & \cite{dessart17} \\
r1w6 & $3.5$ & $-2.0$ & $-12.0$ & 14.7 & 0.030 & \cite{dessart17} \\
r1w6a & $5.5$ & $-2.0$ & $-12.0$ & 14.8 & 0.044 & \cite{wjg23a} \\
r1w6b & $7.0$ & $-2.0$ & $-12.0$ & 14.9 & 0.059 & \cite{wjg23a} \\
r1w6c & $9.0$ & $-2.0$ & $-12.0$ & 15.0 & 0.072 & \cite{wjg22} \\
r1w7a & $14.0$ & $-1.5$ & $-12.5$ & 15.0 & 0.21 & \cite{wjg22} \\
r1w7b & $25.0$ & $-1.5$ & $-12.5$ & 15.3 & 0.40 & \cite{wjg22} \\
r1w7c & $35.0$ & $-1.5$ & $-12.5$ & 15.6 & 0.75 & \cite{wjg22} \\
r1w7d & $35.0$ & $-1.5$ & $-12.5$ & 15.9 & 1.5 & \cite{wjg22} \\
m1em5 & $<0.1$ & $-5.0$ & $-15.2$ & 16.0 & 0.0037 & \cite{Dessart23} \\
m1em4 & $<0.2$ & $-4.0$ & $-14.3$ & 16.0 & 0.0039 & \cite{Dessart23} \\
m1em3 & $1.0$ & $-3.0$ & $-13.3$ & 16.0 & 0.0056 & \cite{Dessart23} \\
m1em2 & $4.0$ & $-2.0$ & $-11.9$ & 16.0 & 0.024 & \cite{Dessart23} \\
m1em1 & $15.0$ & $-1.0$ & $-10.9$ & 16.0 & 0.18 & \cite{Dessart23} \\
m1em0 & $25.0$ & 1.0 & $-10.1$ & 16.0 & 1.21 & \cite{Dessart23} \\
\enddata
\tablenotetext{a}{Density at $10^{14}$~cm.}
\end{deluxetable*}

\begin{deluxetable*}{cccccccccc}[h!]
\tablecaption{Gold/Silver-Sample Photometric Properties \label{tab:sample_phot_gold}}
\tablecolumns{10}
\tablewidth{0pt}
\tablehead{\colhead{Name} & \colhead{log$_{10}(L_{\rm bol,max}$)} & \colhead{$M_{\rm W2,max}$} & \colhead{$M_{\rm M2,max}$} & \colhead{$M_{\rm W1,max}$} & \colhead{$M_{\rm U,max}$} & \colhead{$M_{\rm B,max}$} & \colhead{$M_{\rm V,max}$} & \colhead{$M_{\rm g,max}$} & \colhead{$M_{\rm r,max}$}\\
\colhead{} & \colhead{(erg s$^{-1}$)} & \colhead{(mag)} & \colhead{(mag)} & \colhead{(mag)} & \colhead{(mag)} & \colhead{(mag)} & \colhead{(mag)} & \colhead{(mag)} & \colhead{(mag)} }
\startdata
PTF10gva & $>43.28$ & $<-18.90$ & $<-18.93$ & $<-18.87$ & $-18.81\pm0.25$ & -- & -- & -- & $-18.53\pm0.15$ \\
PTF10abyy & $>43.49$ & $<-19.80$ & $<-19.96$ & $<-19.67$ & -- & -- & -- & -- & $-19.21\pm0.17$ \\
PTF11iqb & $43.46\pm0.06$ & $-18.80\pm0.16$ & $-18.94\pm0.16$ & $-18.86\pm0.15$ & -- & -- & -- & -- & $-18.37\pm0.15$ \\
2013fs & $42.93\pm0.14$ & $-18.19\pm0.32$ & $-18.10\pm0.32$ & $-17.99\pm0.32$ & $-17.72\pm0.32$ & $-17.90\pm0.32$ & $-17.80\pm0.33$ & $-17.82\pm0.32$ & $-17.75\pm0.32$ \\
2014G & $43.15\pm0.32$ & $-18.75\pm0.80$ & $-19.14\pm0.80$ & $-18.72\pm0.80$ & $-18.77\pm0.80$ & $-18.56\pm0.80$ & $-18.47\pm0.80$ & $-18.54\pm0.80$ & $-18.47\pm0.80$ \\
2015bf & $43.05\pm0.02$ & $<-18.97$ & $<-19.29$ & $-18.77\pm0.08$ & $-18.72\pm0.08$ & $-18.43\pm0.09$ & $-18.39\pm0.10$ & -- & $-18.27\pm0.07$ \\
2016blz & $>43.47$ & $<-19.09$ & $<-19.33$ & $<-18.98$ & $<-19.19$ & $<-19.06$ & $-19.07\pm0.16$ & -- & -- \\
2017ahn & $43.33\pm0.17$ & $-18.23\pm0.43$ & $-18.68\pm0.43$ & $-18.30\pm0.43$ & $-18.32\pm0.43$ & $-18.35\pm0.43$ & $-18.18\pm0.43$ & $-18.11\pm0.45$ & $-17.94\pm0.43$ \\
\enddata
\end{deluxetable*}

\begin{deluxetable*}{cccccccccc}[h!]
\tablecaption{Gold/Silver-Sample Photometric Properties \label{tab:sample_phot_gold2}}
\tablecolumns{10}
\tablewidth{0pt}
\tablehead{\colhead{Name} & \colhead{$t_{\rm bol,rise}$} & \colhead{$t_{\rm W2,rise}$} & \colhead{$t_{\rm M2,rise}$} & \colhead{$t_{\rm W1,rise}$} & \colhead{$t_{\rm U,rise}$} & \colhead{$t_{\rm B,rise}$} & \colhead{$t_{\rm V,rise}$} & \colhead{$t_{\rm g,rise}$} & \colhead{$t_{\rm r,rise}$}\\
\colhead{} & \colhead{(days)} & \colhead{(days)} & \colhead{(days)} & \colhead{(days)} & \colhead{(days)} & \colhead{(days)} & \colhead{(days)} & \colhead{(days)} & \colhead{(days)} }
\startdata
PTF10gva & $<3.33$ & $<2.93$ & $<2.94$ & $<2.96$ & $2.93\pm1.05$ & $--$ & $--$ & $--$ & $8.70\pm0.90$ \\
PTF10abyy & $<4.12$ & $<4.11$ & $<4.13$ & $<4.10$ & $--$ & $--$ & $--$ & $--$ & $8.21\pm0.50$ \\
PTF11iqb & $4.00\pm1.15$ & $3.48\pm1.00$ & $4.01\pm1.00$ & $4.49\pm1.00$ & $--$ & $--$ & $--$ & $--$ & $10.46\pm1.00$ \\
2013fs & $1.08\pm0.09$ & $1.20\pm0.02$ & $1.64\pm0.02$ & $2.51\pm0.02$ & $4.51\pm0.02$ & $5.56\pm0.02$ & $6.15\pm0.12$ & $6.64\pm0.02$ & $6.64\pm0.02$ \\
2014G & $4.06\pm2.09$ & $3.81\pm1.70$ & $3.88\pm1.70$ & $4.10\pm1.70$ & $5.13\pm1.70$ & $9.40\pm1.70$ & $11.67\pm1.70$ & $10.40\pm1.70$ & $9.53\pm1.71$ \\
2015bf & $2.78\pm1.48$ & $<3.05$ & $<3.75$ & $3.69\pm1.20$ & $4.39\pm1.20$ & $9.45\pm1.20$ & $12.73\pm1.20$ & $--$ & $13.19\pm1.20$ \\
2016blz & $<2.96$ & $<9.86$ & $<9.86$ & $<9.86$ & $<9.86$ & $<9.86$ & $10.48\pm2.96$ & $--$ & $--$ \\
2017ahn & $3.80\pm0.58$ & $4.70\pm0.50$ & $4.90\pm0.50$ & $5.27\pm0.50$ & $5.65\pm0.50$ & $7.45\pm0.50$ & $7.78\pm0.51$ & $7.91\pm2.26$ & $9.08\pm0.50$ \\
\enddata
\end{deluxetable*}

\begin{deluxetable*}{cccccccccc}[h!]
\tablecaption{Comparison-Sample Photometric Properties \label{tab:sample_phot_control}}
\tablecolumns{10}
\tablewidth{0pt}
\tablehead{\colhead{Name} & \colhead{log$_{10}(L_{\rm bol,max}$)} & \colhead{$M_{\rm W2,max}$} & \colhead{$M_{\rm M2,max}$} & \colhead{$M_{\rm W1,max}$} & \colhead{$M_{\rm U,max}$} & \colhead{$M_{\rm B,max}$} & \colhead{$M_{\rm V,max}$} & \colhead{$M_{\rm g,max}$} & \colhead{$M_{\rm r,max}$}\\
\colhead{} & \colhead{(erg s$^{-1}$)} & \colhead{(mag)} & \colhead{(mag)} & \colhead{(mag)} & \colhead{(mag)} & \colhead{(mag)} & \colhead{(mag)} & \colhead{(mag)} & \colhead{(mag)} }
\startdata
2013ft & $>41.71$ & $<-14.03$ & $<-14.05$ & $<-13.79$ & $<-14.07$ & $-13.78\pm0.58$ & $-14.22\pm0.52$ & -- & $-14.69\pm0.48$ \\
2013am & $>42.27$ & $<-14.85$ & $<-14.90$ & $<-14.86$ & $<-15.54$ & $-15.86\pm0.42$ & $-15.74\pm0.43$ & $-15.71\pm0.41$ & $-15.94\pm0.41$ \\
2013ab & $42.82\pm0.04$ & $<-17.99$ & $<-18.15$ & $-17.85\pm0.09$ & $<-17.93$ & $-17.68\pm0.08$ & $-17.39\pm0.08$ & $-17.66\pm0.08$ & $-17.61\pm0.09$ \\
2016X & $42.18\pm0.17$ & $-16.44\pm0.43$ & $-16.73\pm0.43$ & $-16.99\pm0.43$ & $-17.01\pm0.43$ & $-16.90\pm0.43$ & $-16.90\pm0.43$ & $-16.92\pm0.43$ & $-16.95\pm0.43$ \\
2016aqf & $>41.66$ & $<-14.79$ & $<-14.97$ & $<-14.86$ & $-14.99\pm0.28$ & $-14.74\pm0.29$ & $-14.75\pm0.31$ & $-14.83\pm0.30$ & $-14.90\pm0.31$ \\
2017eaw & $42.39\pm0.03$ & $<-18.16$ & $-17.99\pm0.05$ & $-17.48\pm0.07$ & $-17.62\pm0.05$ & $-17.48\pm0.05$ & $-17.37\pm0.05$ & $-17.75\pm0.05$ & $-17.41\pm0.05$ \\
2017gmr & $43.18\pm0.07$ & $<-19.11$ & $<-18.96$ & $<-18.53$ & $-18.47\pm0.16$ & $-18.44\pm0.18$ & $-18.26\pm0.17$ & $-18.35\pm0.16$ & $-18.25\pm0.16$ \\
2018lab & $>42.50$ & $<-17.72$ & $-17.72\pm0.11$ & $<-16.87$ & $-16.50\pm0.04$ & $-16.25\pm0.04$ & $-15.67\pm0.06$ & $-15.90\pm0.04$ & $-15.65\pm0.04$ \\
\enddata
\end{deluxetable*}

\begin{deluxetable*}{cccccccccc}[h!]
\tablecaption{Comparison-Sample Photometric Properties \label{tab:sample_phot_control2}}
\tablecolumns{10}
\tablewidth{0pt}
\tablehead{\colhead{Name} & \colhead{$t_{\rm bol,rise}$} & \colhead{$t_{\rm W2,rise}$} & \colhead{$t_{\rm M2,rise}$} & \colhead{$t_{\rm W1,rise}$} & \colhead{$t_{\rm U,rise}$} & \colhead{$t_{\rm B,rise}$} & \colhead{$t_{\rm V,rise}$} & \colhead{$t_{\rm g,rise}$} & \colhead{$t_{\rm r,rise}$}\\
\colhead{} & \colhead{(days)} & \colhead{(days)} & \colhead{(days)} & \colhead{(days)} & \colhead{(days)} & \colhead{(days)} & \colhead{(days)} & \colhead{(days)} & \colhead{(days)} }
\startdata
2013ft & $<1.16$ & $<1.13$ & $<1.16$ & $<1.16$ & $<1.16$ & $9.70\pm2.43$ & $15.06\pm1.01$ & $--$ & $15.10\pm1.01$ \\
2013am & $<2.90$ & $<2.19$ & $<2.20$ & $<2.17$ & $<2.18$ & $5.38\pm1.00$ & $9.00\pm1.03$ & $8.95\pm1.05$ & $10.55\pm1.57$ \\
2013ab & $4.98\pm1.24$ & $<4.97$ & $<4.97$ & $5.52\pm1.00$ & $<4.97$ & $6.79\pm1.00$ & $7.44\pm1.00$ & $8.46\pm1.00$ & $8.89\pm1.01$ \\
2016X & $7.27\pm0.56$ & $3.39\pm0.50$ & $3.56\pm0.50$ & $5.56\pm0.50$ & $7.80\pm0.50$ & $6.59\pm0.51$ & $9.73\pm0.50$ & $9.96\pm0.52$ & $13.56\pm0.51$ \\
2016aqf & $<1.56$ & $<1.56$ & $<1.57$ & $<0.87$ & $2.03\pm1.00$ & $4.10\pm1.00$ & $62.54\pm1.00$ & $9.43\pm1.00$ & $60.49\pm1.00$ \\
2017eaw & $3.15\pm0.11$ & $<1.89$ & $2.13\pm0.10$ & $2.37\pm0.10$ & $7.36\pm0.18$ & $6.80\pm2.29$ & $7.18\pm0.11$ & $6.92\pm0.12$ & $10.65\pm2.15$ \\
2017gmr & $1.11\pm0.57$ & $<2.54$ & $<2.52$ & $<2.53$ & $6.58\pm0.50$ & $6.71\pm0.90$ & $9.53\pm0.51$ & $8.35\pm0.50$ & $11.08\pm0.53$ \\
2018lab & $<1.23$ & $<1.27$ & $1.62\pm0.10$ & $<1.26$ & $1.72\pm0.10$ & $1.96\pm0.33$ & $11.43\pm0.23$ & $1.52\pm0.15$ & $11.18\pm0.11$ \\
\enddata
\end{deluxetable*}

\begin{deluxetable*}{ccccccc}[h!]
\tablecaption{Gold/Comparison-Sample Peak Luminosity and Rise-Time Distribution \label{tab:sample_lum_peak}}
\tablecolumns{7}
\tablewidth{0pt}
\tablehead{\colhead{Class} & \colhead{$L_{\rm avg}[L_{\rm min},L_{\rm max}]$} & \colhead{$L_{\rm avg}[L_{\rm min},L_{\rm max}]$} &  \colhead{$M^{\rm W2}_{\rm avg}[M^{\rm W2}_{\rm min},M^{\rm W2}_{\rm max}]$} &  \colhead{$M^{\rm W2}_{\rm avg}[M^{\rm W2}_{\rm min},M^{\rm W2}_{\rm max}]$} & \colhead{$M^{\rm r}_{\rm avg}[M^{\rm r}_{\rm min},M^{\rm r}_{\rm max}]$} & \colhead{$M^{\rm r}_{\rm avg}[M^{\rm r}_{\rm min},M^{\rm r}_{\rm max}]$}\\
\colhead{} & \colhead{(erg s$^{-1}$)} & \colhead{(days)} & \colhead{(mag)} & \colhead{(days)} & \colhead{(mag)} & \colhead{(days)} }
\startdata
1 & $3.1[0.83,9.0]\times10^{43}$ & $5.1[3.5,10.0]$ & $-19.8[-18.5,-21.4]$ & $4.4[2.1,8.0]$ & $-18.4[-17.9,-19.4]$ & $12.0[4.9,26.4]$\\
2 & $2.2[0.77,3.5]\times10^{43}$ & $3.0[2.4,3.6]$ & $-19.2[-18.2,-19.9]$ & $2.3[1.6,2.9]$ & $-18.2[-17.5,-18.6]$ & $10.2[5.5,13.6]$\\
3 & $2.2[0.23,8.2]\times10^{43}$ & $2.5[1.1,3.8]$ & $-19.1[-17.6,-21.6]$ & $1.7[1.2,2.5]$ & $-18.1[-17.3,-19.3]$ & $10.9[5.7,18.1]$\\
Comparison & $3.8[0.36,17.9]\times10^{42}$ & $2.4[0.1,6.7]$ & $-17.1[-13.9,-19.6]$ & $2.2[0.1,4.8]$ & $-17.0[-14.5,-18.4]$ & $12.9[6.5,21.1]$\\
\enddata
\end{deluxetable*}

\begin{deluxetable*}{cccccccccccc}[h!]
\tablecaption{Gold/Silver-Sample Spectroscopic Properties \label{tab:sample_spec_gold}}
\tablecolumns{9}
\tablewidth{0pt}
\tablehead{\colhead{Name} & {$t_{\rm IIn}$} & \colhead{IIn Phase} & \colhead{\ion{H}{i}} & \colhead{\ion{He}{i}} & \colhead{\ion{He}{ii}} & \colhead{\ion{C}{iii}} & \colhead{\ion{C}{iv}} & \colhead{\ion{N}{iii}} & \colhead{\ion{N}{iv}} & \colhead{$V_{\rm max}$\tablenotemark{a}} & \colhead{$M_{\rm CDS}$\tablenotemark{b}}\\ \colhead{} & \colhead{(days)} & \colhead{(days)} & \colhead{} & \colhead{} & \colhead{} & \colhead{} & \colhead{} & \colhead{} & \colhead{} & \colhead{($10^4~\kms$)} & \colhead{($\Msun$)} }
\startdata
PTF11iqb$^1$ & $3.1\pm1.0$ & 1.6 & Y & N & Y & N & Y & Y & N & $1.1$ & 0.005 \\
2017ahn$^1$ & $6.4\pm0.3$ & 2.1 & Y & Y & Y & N & Y & Y & Y & $0.9$ & 0.04 \\
2018zd$^1$ & $7.7\pm0.9$ & 3.8 & Y & N & Y & N & Y & Y & N & 0.5 & 1.5 \\
2019ust$^1$ & $3.5\pm0.6$ & 1.7 & Y & N & Y & N & Y & Y & -- & 1.0 & 0.02\\
2019qch$^1$ & $11.4\pm0.9$ & 8.5 & Y & Y$^c$ & Y & N & Y & Y & -- & -- & -- \\
2020pni$^1$ & $5.5\pm1.0$ & 1.5 & Y & N & Y & Y & Y & Y & Y & 1.0 & 0.01\\
2020sic$^1$ & $2.4\pm0.9$ & 1.7 & Y & N & Y & N & Y & Y & Y$^c$ & -- & -- \\
2020abtf$^1$ & $4.5\pm1.0$ & 5.4 & Y & N & Y & N & N & Y & N & 0.6 & $4.2$\\
\enddata
\tablenotetext{a}{Approximate maximum velocity measured in H$\alpha$ profile during phases of $\delta t = 50 \pm 10$~days.}
\tablenotetext{b}{Calculated using trend in \cite{Dessart23}.}
\tablenotetext{c}{Marginal signal above continuum noise.}
\end{deluxetable*}


\begin{deluxetable*}{ccccccccccc}[h!]
\tablecaption{Best-Matched Model Parameters \label{tab:model_params_gold}}
\tablecolumns{10}
\tablewidth{0pt}
\tablehead{\colhead{Name} & \colhead{log$_{10}$($\dot{M}$)\tablenotemark{a}} & \colhead{log$_{10}$($\rho_{14}$)\tablenotemark{a}} & \colhead{log$_{10}$($\dot{M}$)\tablenotemark{b}} & \colhead{log$_{10}$($\rho_{14}$)\tablenotemark{b}} & \colhead{log$_{10}$($\dot{M}$)\tablenotemark{c}} & \colhead{log$_{10}$($\rho_{14}$)\tablenotemark{c}} & \colhead{log$_{10}$($\dot{M}$[$R_{\rm in}$])\tablenotemark{d}} & \colhead{log$_{10}$(D$_{\star}$)[$R_{\rm CSM}$]\tablenotemark{e}} & \colhead{log$_{10}$($\rho_{14}$)\tablenotemark{e}}\\
\colhead{} & \colhead{(M$_{\odot}$~yr$^{-1}$)} & \colhead{(g cm$^{-3}$)} & \colhead{(M$_{\odot}$~yr$^{-1}$)} & \colhead{(g cm$^{-3}$)} & \colhead{(M$_{\odot}$~yr$^{-1}$)} & \colhead{(g cm$^{-3}$)} & \colhead{(M$_{\odot}$~yr$^{-1}$[cm])} & \colhead{(g cm$^{-1}$[$\Rsun$])} & \colhead{(g cm$^{-3}$)} }
\startdata
PTF11iqb$^{1}$ & -2.0  & -12.0  & -2.0  & -12.0  & -2.0 & -11.9 & -3.0[14.5] & 18.0[2700] & -10.0\\
2017ahn$^{1}$ & -2.0 $-$ -1.0 & -11.8 $-$ -10.9 & -2.0 & -11.8 & -1.3 & -11.1 &-2.4[14.2] & 18.0[1500] & -10.0\\
2018zd$^{1}$ & -2.0  & -11.8  & -2.0  & -11.8  & -1.0 & -10.9 & -2.5[14.5] & 18.0[2700] & -10.0\\
2019ust$^{1}$ & -2.0  $-$ -1.5  & -12.0  $-$ -11.5  & -2.0 & -12.0 & -2.0 & -12.0 & -3.0[14.2] & 18.0[2700] & -10.0\\
2019qch$^{1}$ & -2.0 $-$ -1.5 & -12.0 $-$ -11.5 & -2.0 & -12.0 & -1.5 & -11.5 &-3.0[14.5] & 18.0[2700] & -10.0\\
2020pni$^{1}$ & -2.0  $-$ -1.5  & -11.9  $-$ -11.5  & -2.0 & -11.8 & -2.0 & -12.0 & -3.0[14.2] & 18.0[2500] & -10.0\\
2020sic$^{1}$ & -2.3 $-$ -2.0 & -12.3 $-$ -11.9 & -2.3 & -12.3 & -2.0 & -12.0 &-2.3[14.5] & 18.0[2100] & -10.0\\
2020abtf$^{1}$ & -2.0  & -11.9  & -2.0  & -11.9  & -2.0 & -12.0 & -2.5[14.5] & 17.9[2700] & -10.1\\
\enddata
\tablenotetext{a}{Calculated using multicolor photometry and $t_{\rm IIn}$.}
\tablenotetext{b}{Calculated using only $t_{\rm IIn}$.}
\tablenotetext{c}{Calculated using spectral comparison.}
\tablenotetext{d}{Calculated using spectral models from \cite{Boian20}.}
\tablenotetext{e}{Calculated using multi-color models from \cite{haynie21}.}
\end{deluxetable*}

\begin{deluxetable*}{ccccccccccc}[h!]
\tablecaption{Best-Matched Model Parameters (Cont.) \label{tab:model_params_silver}}
\tablecolumns{10}
\tablewidth{0pt}
\tablehead{\colhead{Name} & \colhead{log$_{10}$($\dot{M}$)\tablenotemark{a}} & \colhead{log$_{10}$($\rho_{14}$)\tablenotemark{a}} & \colhead{log$_{10}$($\dot{M}$)\tablenotemark{b}} & \colhead{log$_{10}$($\rho_{14}$)\tablenotemark{b}} & \colhead{log$_{10}$($\dot{M}$)\tablenotemark{c}} & \colhead{log$_{10}$($\rho_{14}$)\tablenotemark{c}} & \colhead{log$_{10}$($\dot{M}$)[$R_{\rm in}$]\tablenotemark{d}} & \colhead{log$_{10}$(D$_{\star}$)[$R_{\rm CSM}$]\tablenotemark{e}} & \colhead{log$_{10}$($\rho_{14}$)\tablenotemark{e}}\\
\colhead{} & \colhead{(M$_{\odot}$~yr$^{-1}$)} & \colhead{(g cm$^{-3}$)} & \colhead{(M$_{\odot}$~yr$^{-1}$)} & \colhead{(g cm$^{-3}$)} & \colhead{(M$_{\odot}$~yr$^{-1}$)} & \colhead{(g cm$^{-3}$)} & \colhead{(M$_{\odot}$~yr$^{-1}$[cm])} & \colhead{(g cm$^{-1}$[$\Rsun$])} & \colhead{(g cm$^{-3}$)} }
\startdata
\hline
2013ft & $<$-2.3 & $<$-12.3 & $<$-2.5 & $<$-12.3 & -4.3 & -13.6 & -- & 16.9[1500] & -11.1\\
2013am & $<$-2.3 & $<$-12.3 & $<$-2.3 & $<$-12.3 & -5.3 & -15.5 & -- & 17.4[1500] & -10.6\\
2013ab & $<$-2.3 & $<$-12.3 & $<$-2.5 & $<$-12.3 & -6.0 & -16.0 & -- & 18.0[2100] & -10.0\\
2016X & $<$-2.0 & $<$-12.0 & $<$-2.3 & $<$-12.3 & -5.3 & -15.5 & -- & 17.7[2100] & -10.3\\
2016aqf & $<$-2.3 & $<$-12.3 & $<$-2.3 & $<$-12.3 & -6.0 & -16.0 & -- & 16.9[1500] & -11.1\\
2017eaw & $<$-2.0 & $<$-12.0 & $<$-2.3 & $<$-12.3 & -6.0 & -16.0 & -- & 18.0[2500] & -10.0\\
2017gmr & $<$-2.0 & $<$-12.0 & $<$-2.3 & $<$-12.3 & -6.0 & -16.0 & -- & 17.4[1500] & -10.6\\
2018lab & $<$-2.3 & $<$-12.3 & $<$-2.3 & $<$-12.3 & -4.3 & -13.6 & -- & 16.9[1500] & -11.1\\
\enddata
\tablenotetext{a}{Calculated using multicolor photometry and $t_{\rm IIn}$.}
\tablenotetext{b}{Calculated using only $t_{\rm IIn}$.}
\tablenotetext{c}{Calculated using spectral comparison.}
\tablenotetext{d}{Calculated using spectral models from \cite{Boian20}.}
\tablenotetext{e}{Calculated using multicolor models from \cite{haynie21}.}
\end{deluxetable*}

\begin{figure*}
\centering
\includegraphics[width=0.99\textwidth]{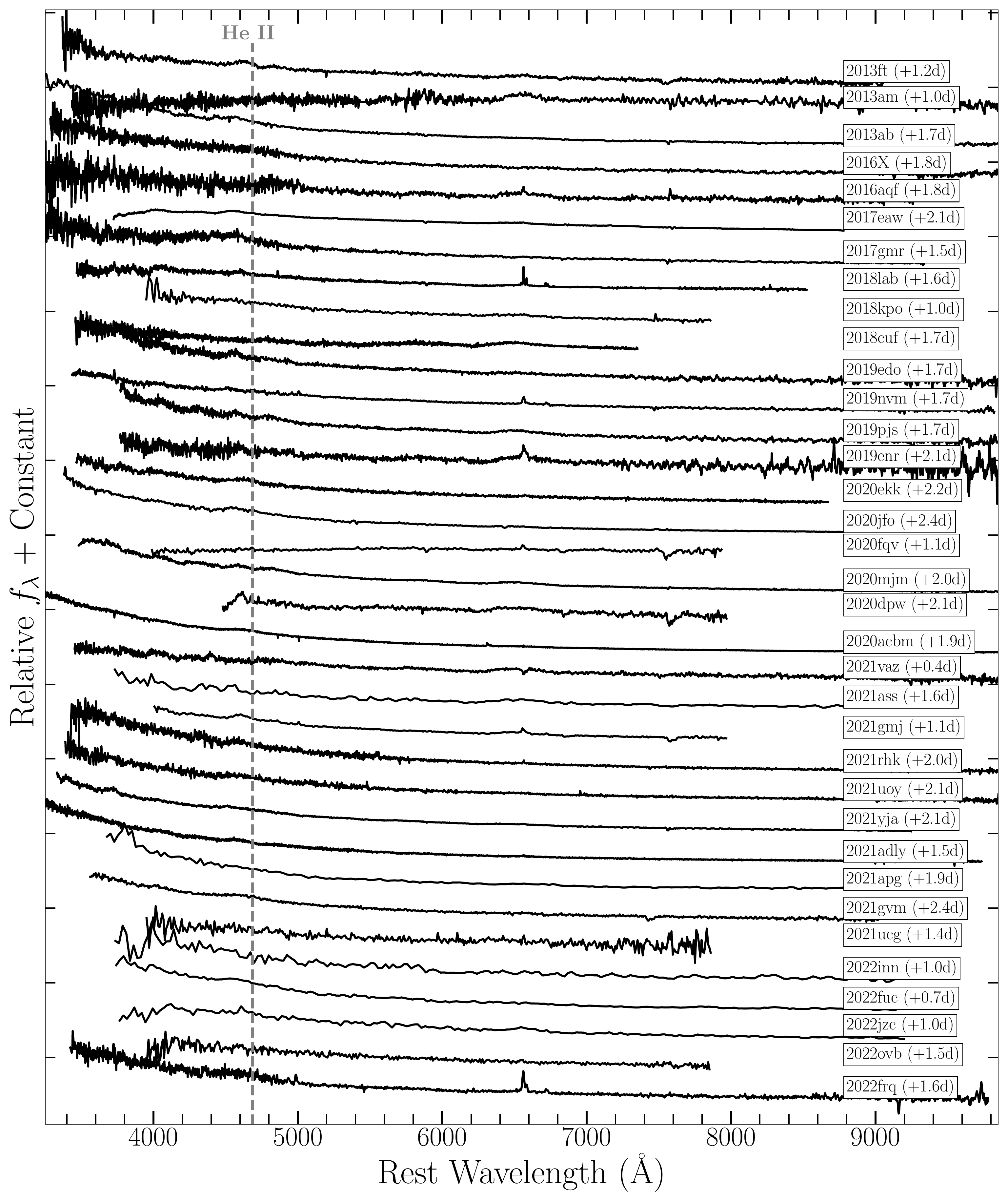}
\caption{Comparison-sample spectra obtained at $t\lesssim2$~days post-first light. These SNe~II do not show prominent spectroscopic evidence for CSM interaction but do have complete UV photometry for comparison to the gold-sample objects.   
\label{fig:FS_all_comp} }
\end{figure*}

\begin{figure*}[t!]
\centering
\subfigure{\includegraphics[width=0.49\textwidth]{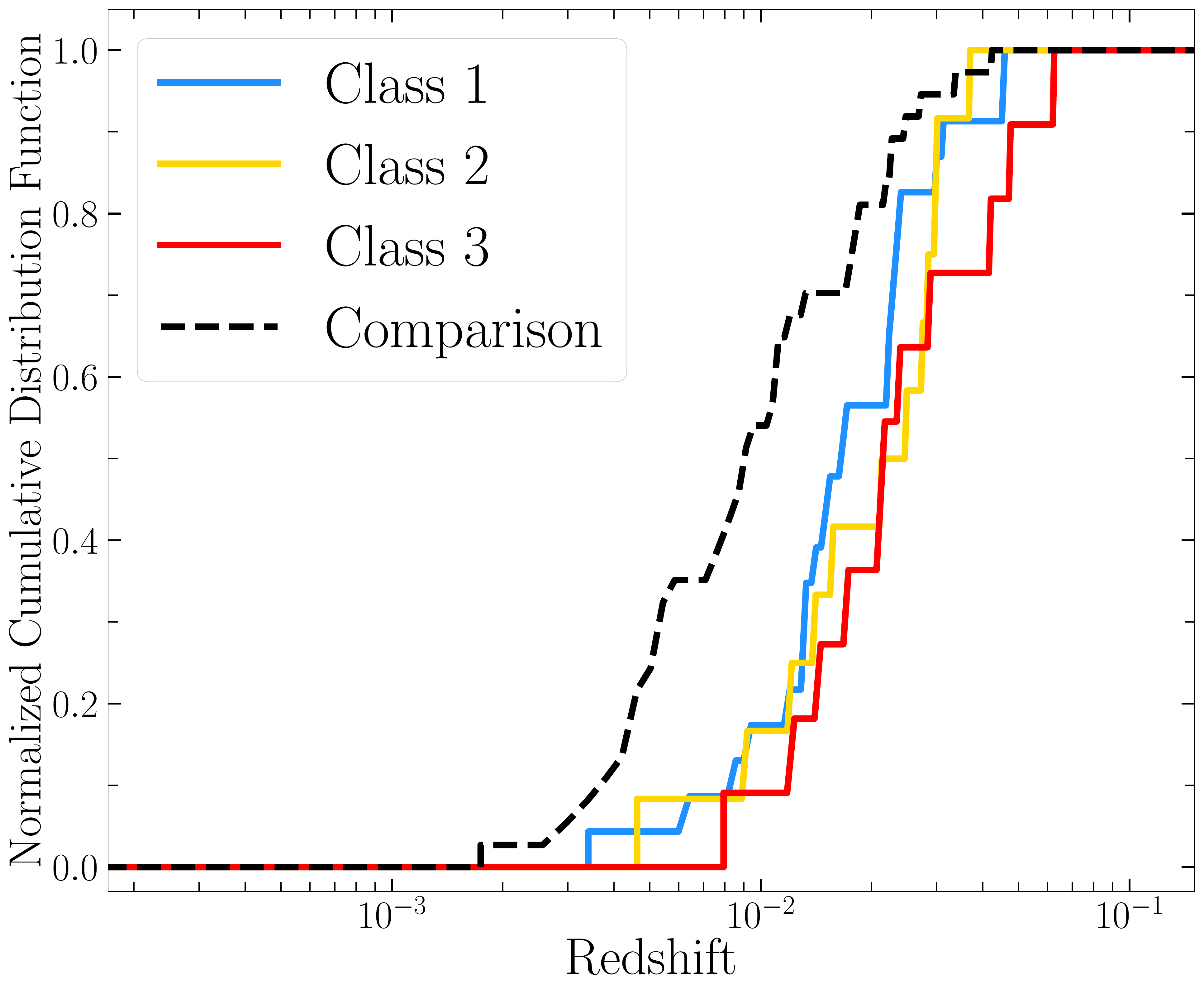}}
\subfigure{\includegraphics[width=0.49\textwidth]{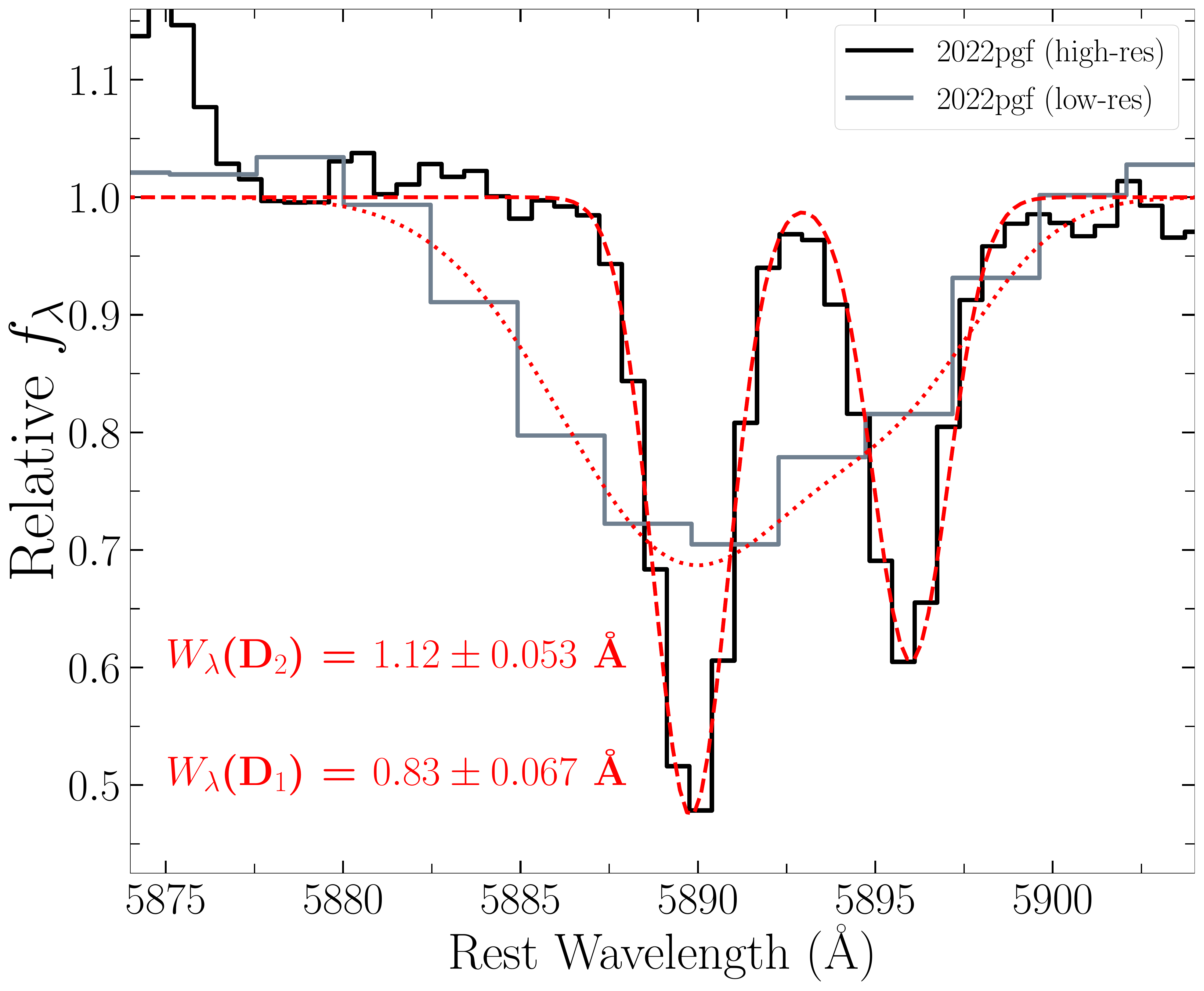}}\\
\subfigure{\includegraphics[width=0.48\textwidth]{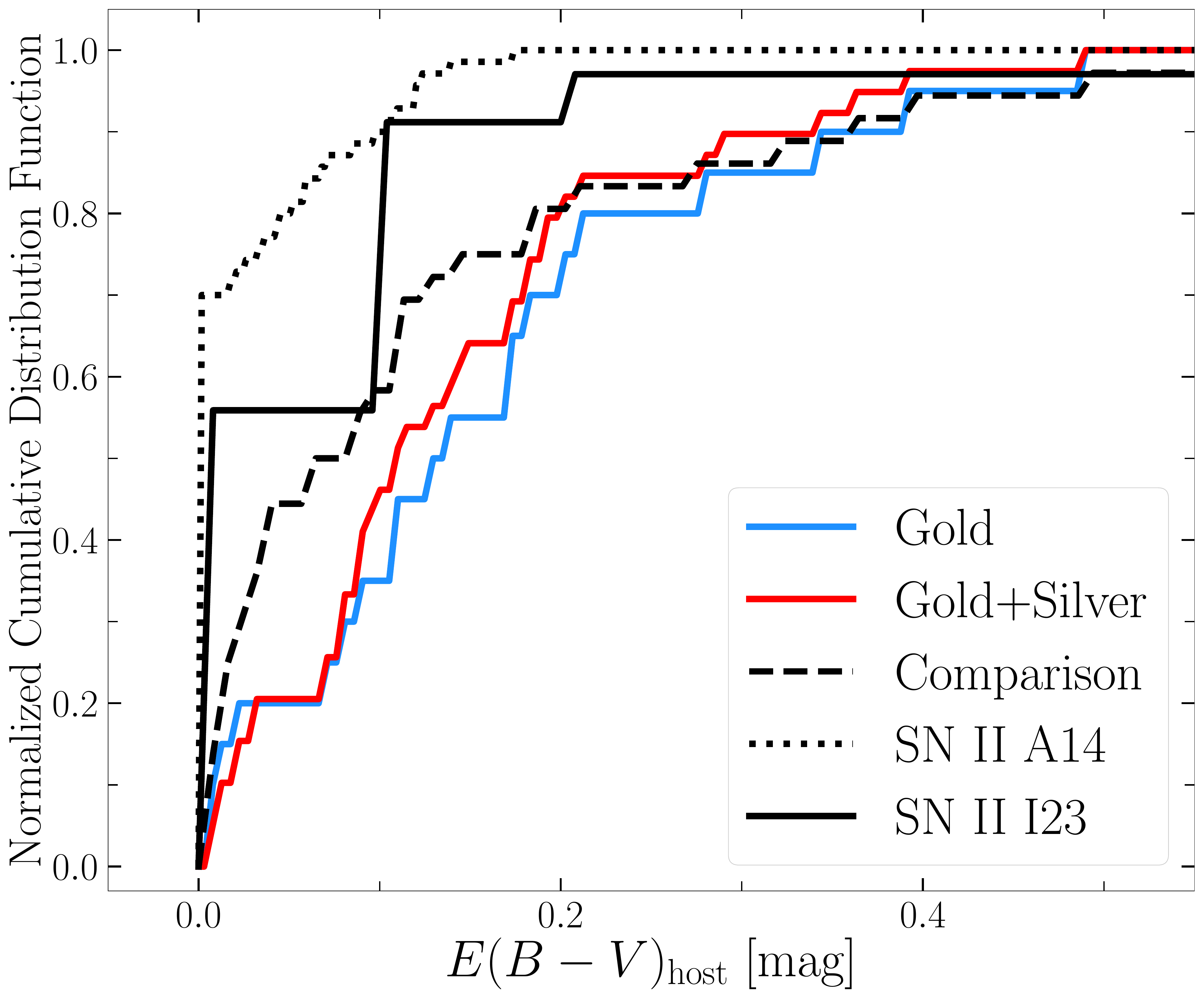}}
\subfigure{\includegraphics[width=0.51\textwidth]{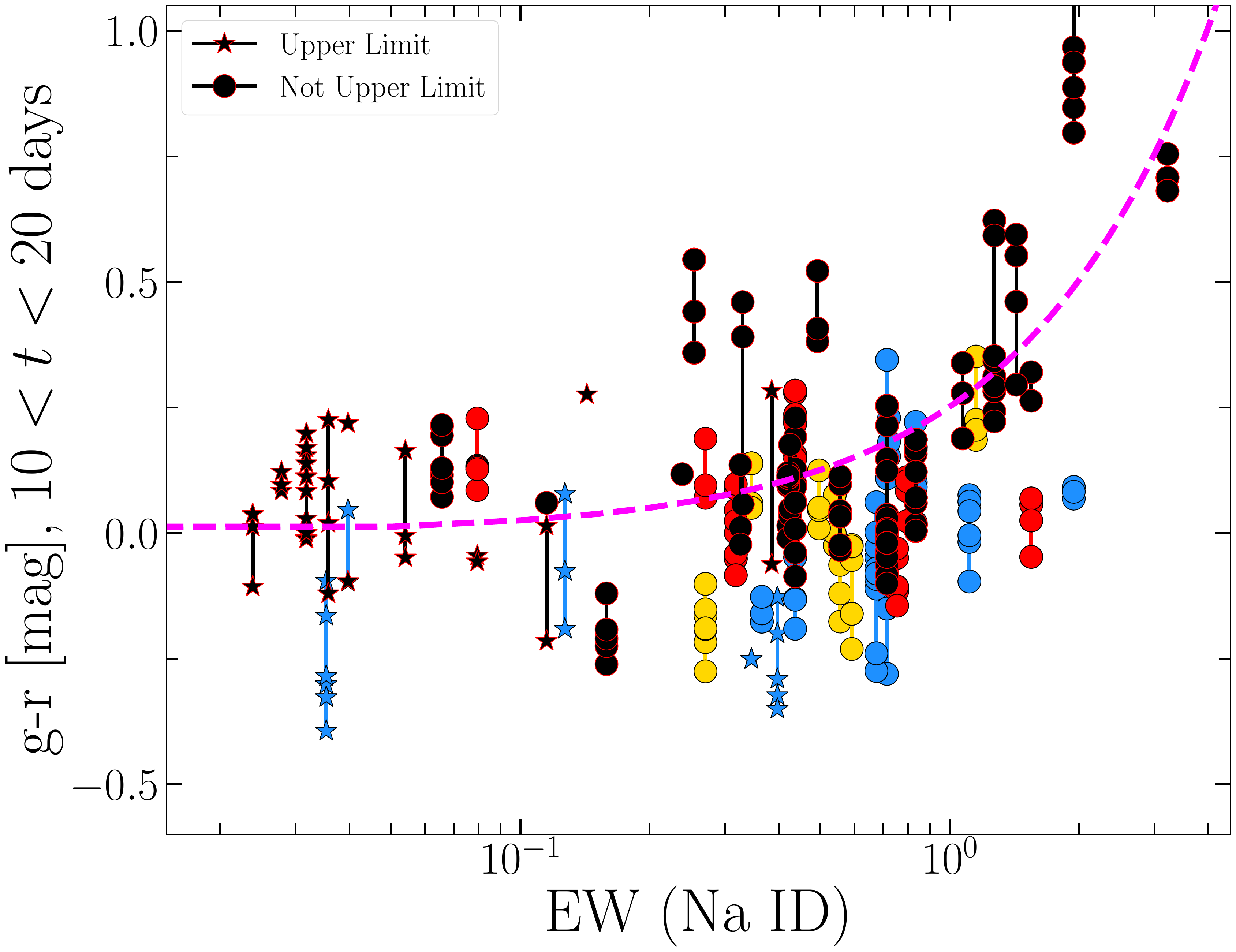}}
\caption{{\it Top left:} Redshift distribution of gold/silver Class 1 (blue), 2 (yellow), 3 (red) sample compared to comparison sample (black dashed lines). {\it Top right:} Visualization of host-galaxy extinction calculation using EWs of \ion{Na}{I}~D1 and D2 transitions and applying the relation from \cite{Stritzinger18}. {\it Bottom left:} Cumulative distribution of host-galaxy extinction for gold-, silver-, and comparison-sample objects. Host-galaxy extinction distribution from SN~II sample by \cite{Anderson14} shown as a black dotted line and by \cite{Irani23b} as a black solid line. {\it Bottom right:} Observed $g-r$ colors at $10 < \delta t < 20$~days versus equivalent width of \ion{Na}{i}~D. Relation from \cite{Stritzinger18} shown as dashed magenta line. Notably, gold-sample objects tend to reside below the reddening relation, implying that they are intrinsically bluer than the inferred host extinction.     
\label{fig:extinction} }
\end{figure*}

\begin{figure*}[t!]
\centering
\subfigure{\includegraphics[width=0.33\textwidth]{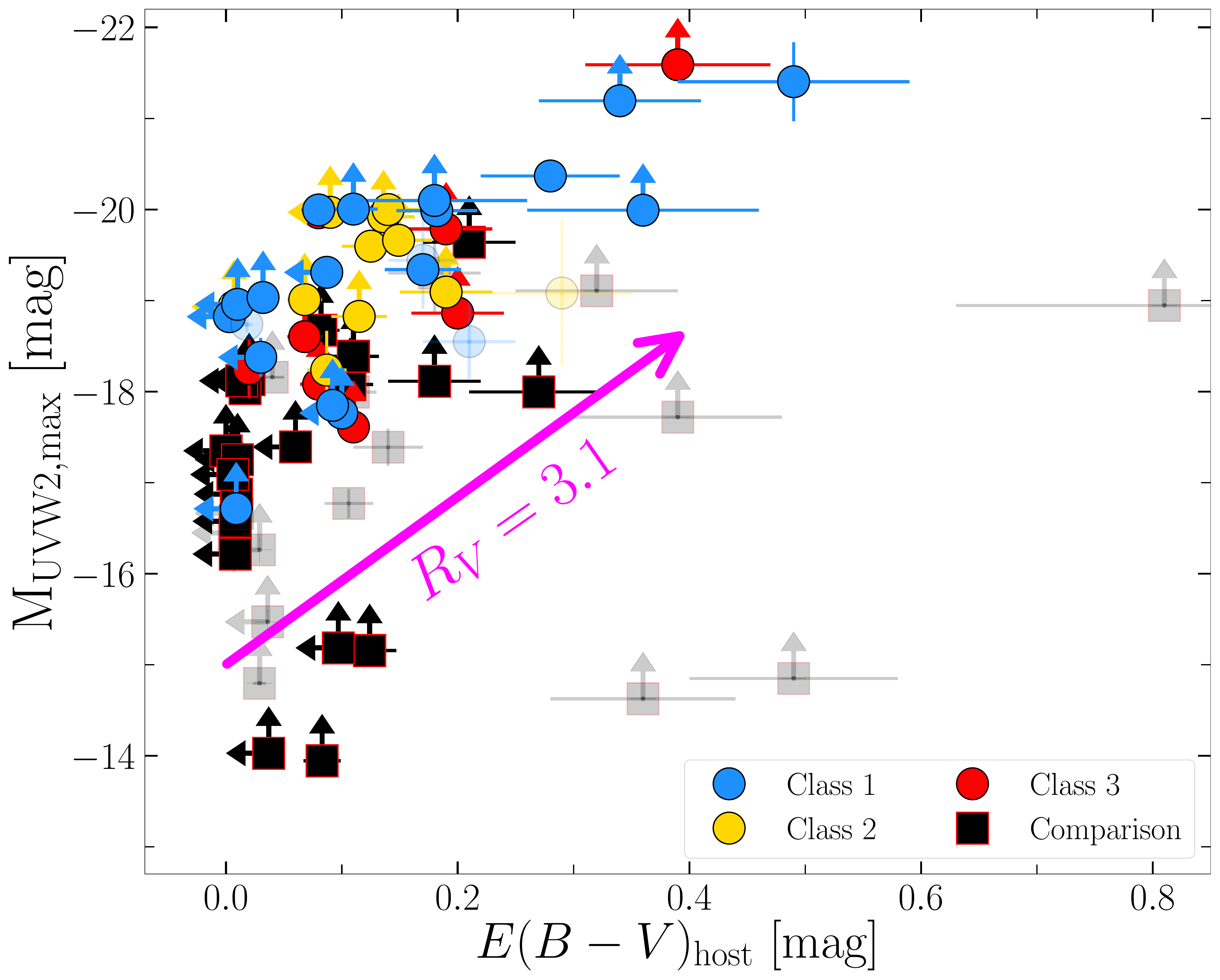}}
\subfigure{\includegraphics[width=0.33\textwidth]{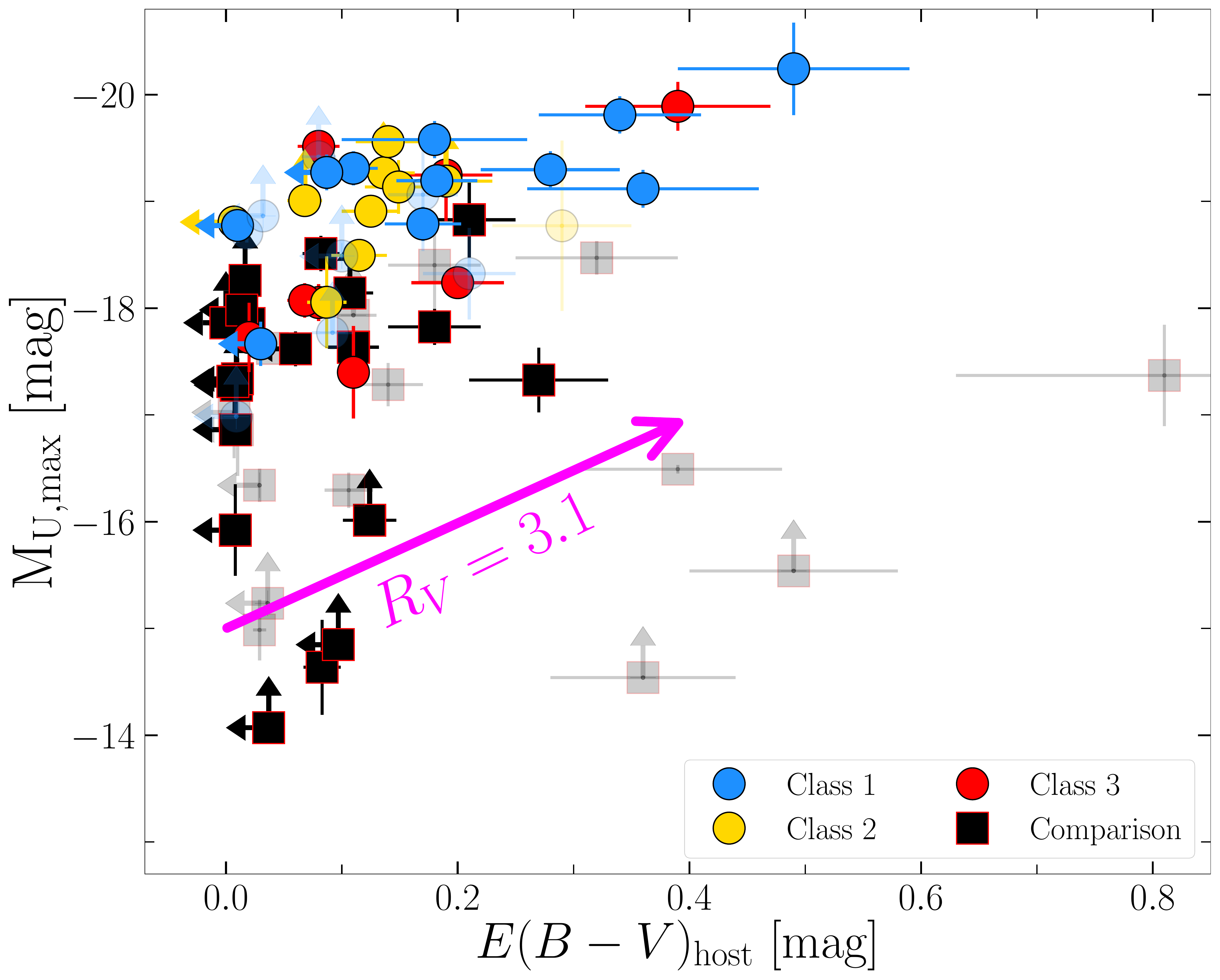}}
\subfigure{\includegraphics[width=0.33\textwidth]{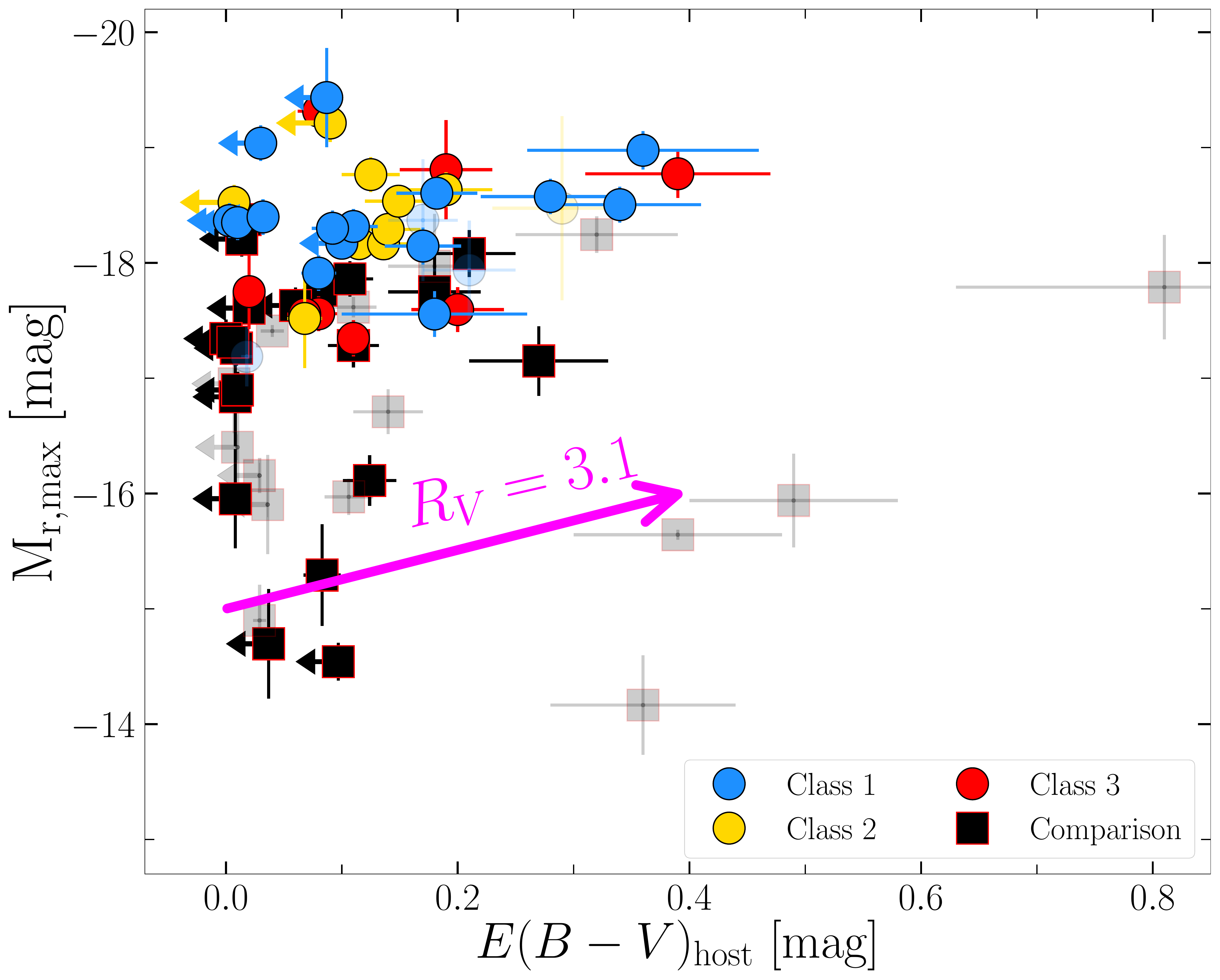}}\\
\subfigure{\includegraphics[width=0.33\textwidth]{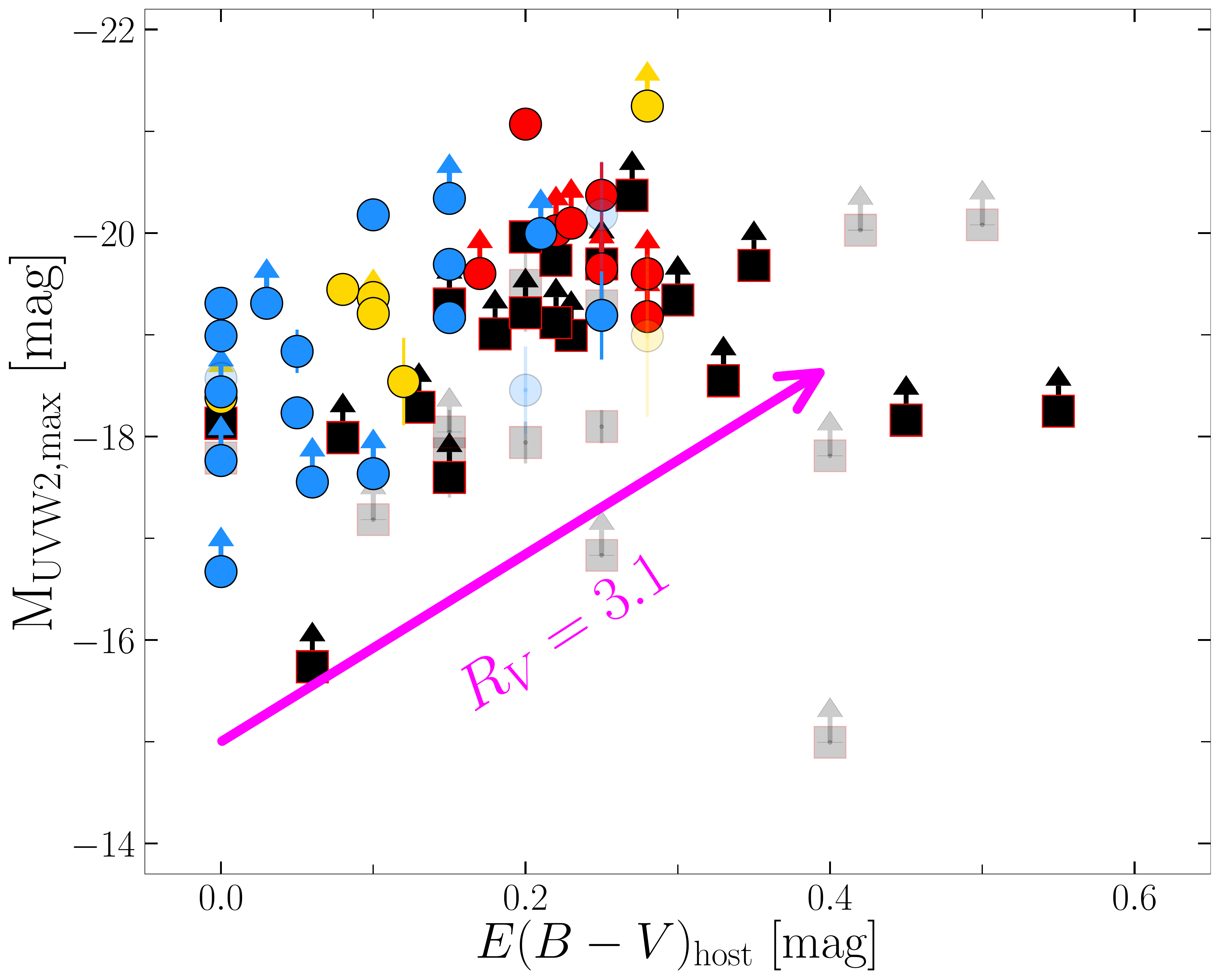}}
\subfigure{\includegraphics[width=0.33\textwidth]{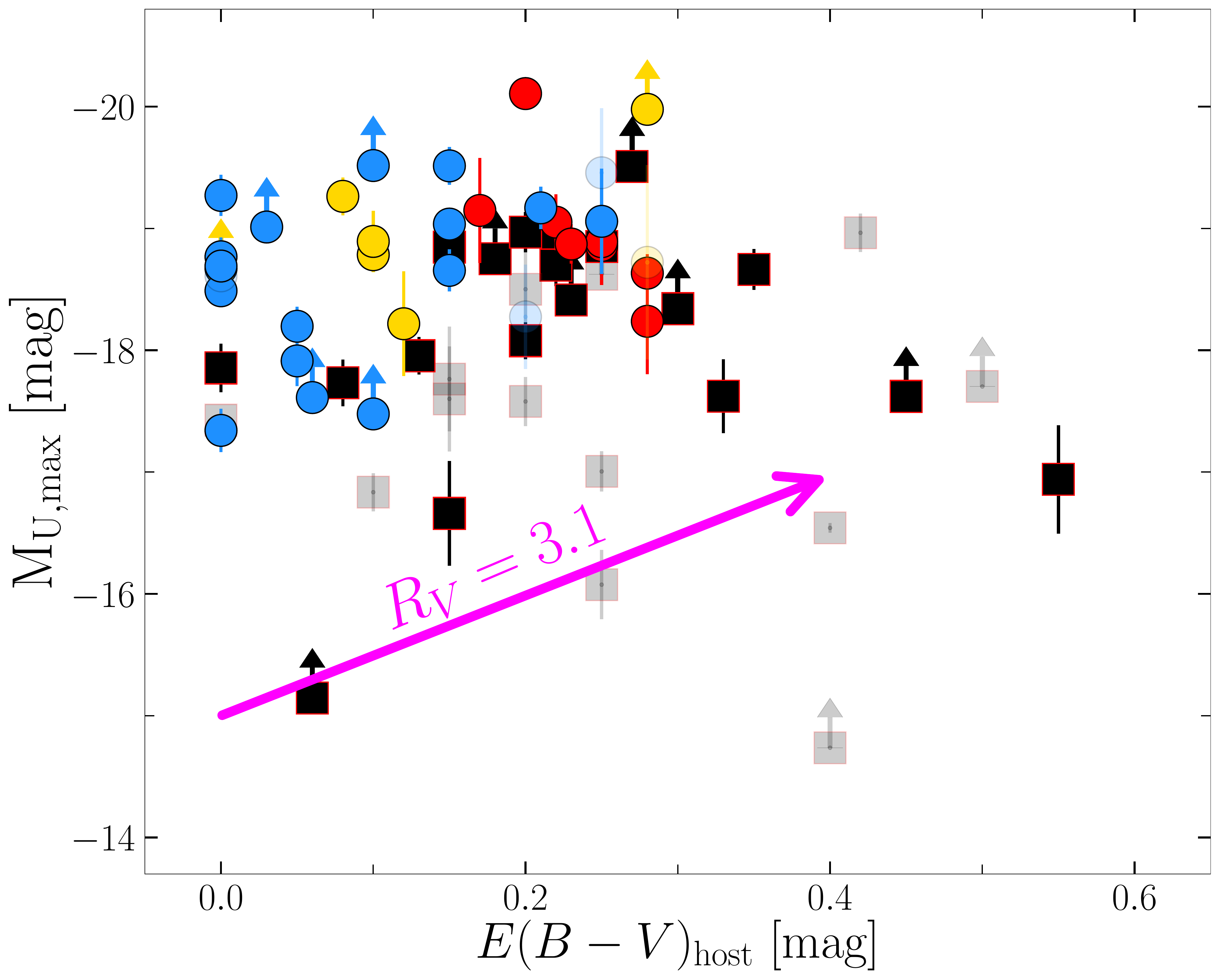}}
\subfigure{\includegraphics[width=0.33\textwidth]{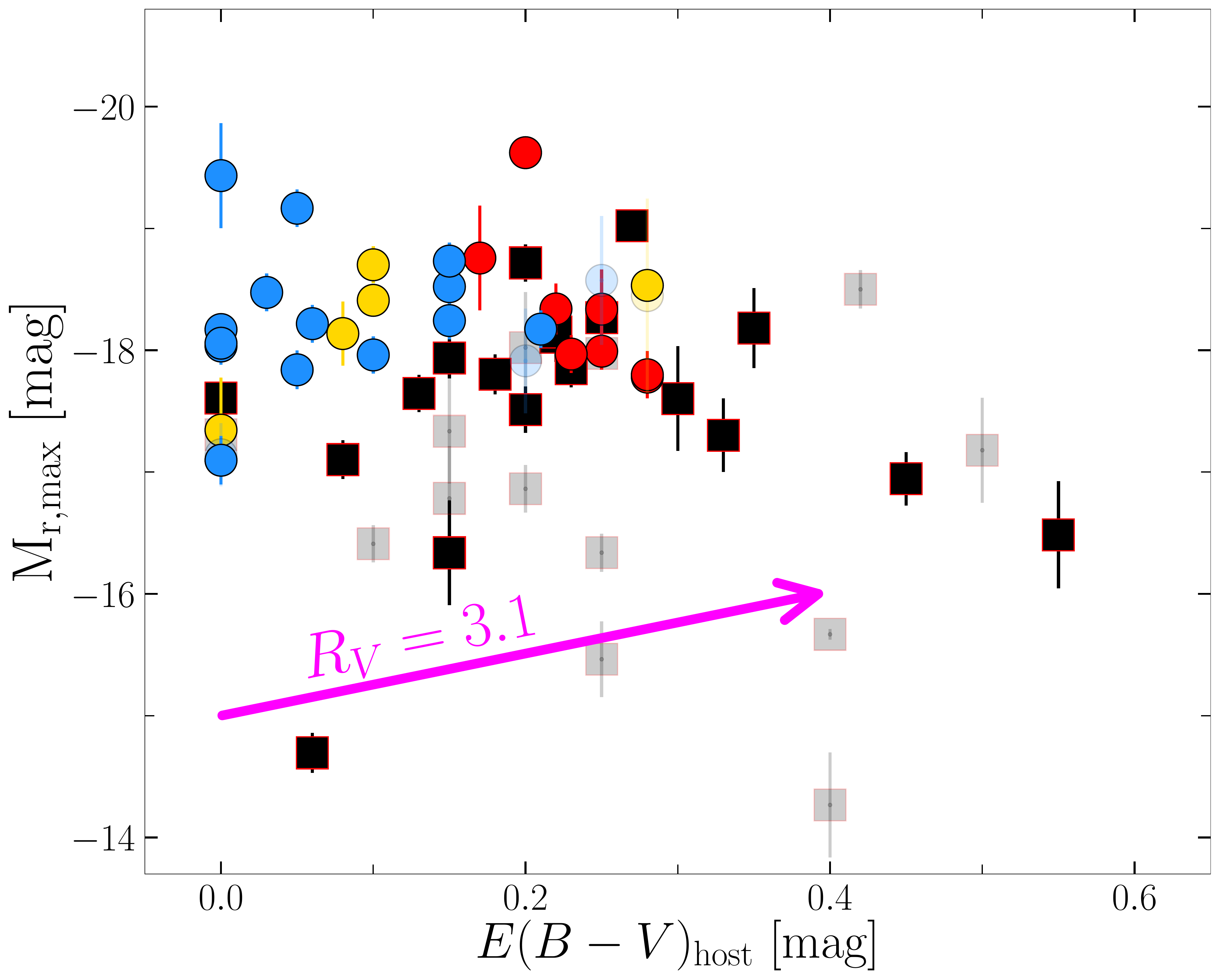}}\\
\caption{{\it Top panel:} Comparison of host-galaxy extinction versus extinction-corrected peak $w2$-, $u$-, and $r$-band absolute magnitudes for all of the sample objects. Host-extinction correction based on \ion{Na}{I}~D EW. Solid colored points represent the subsample of objects at $D>40$~Mpc. We note that the highest luminosity objects ($M_{\rm w2} < -21$~mag) also have some of the largest host extinction values, suggesting that \ion{Na}{I}~D is a limited measure of host reddening. This is further supported by the lack of objects with similarly high luminosities at low $E(B-V)_{\rm host}$ values. {\it Bottom panel:} Peak $w2$-, $u$-, and $r$-band absolute magnitudes versus host-extinction correction using $g-r$ colors.     
\label{fig:extinction2} }
\end{figure*}

\begin{figure*}[t!]
\centering
\subfigure{\includegraphics[width=0.32\textwidth]{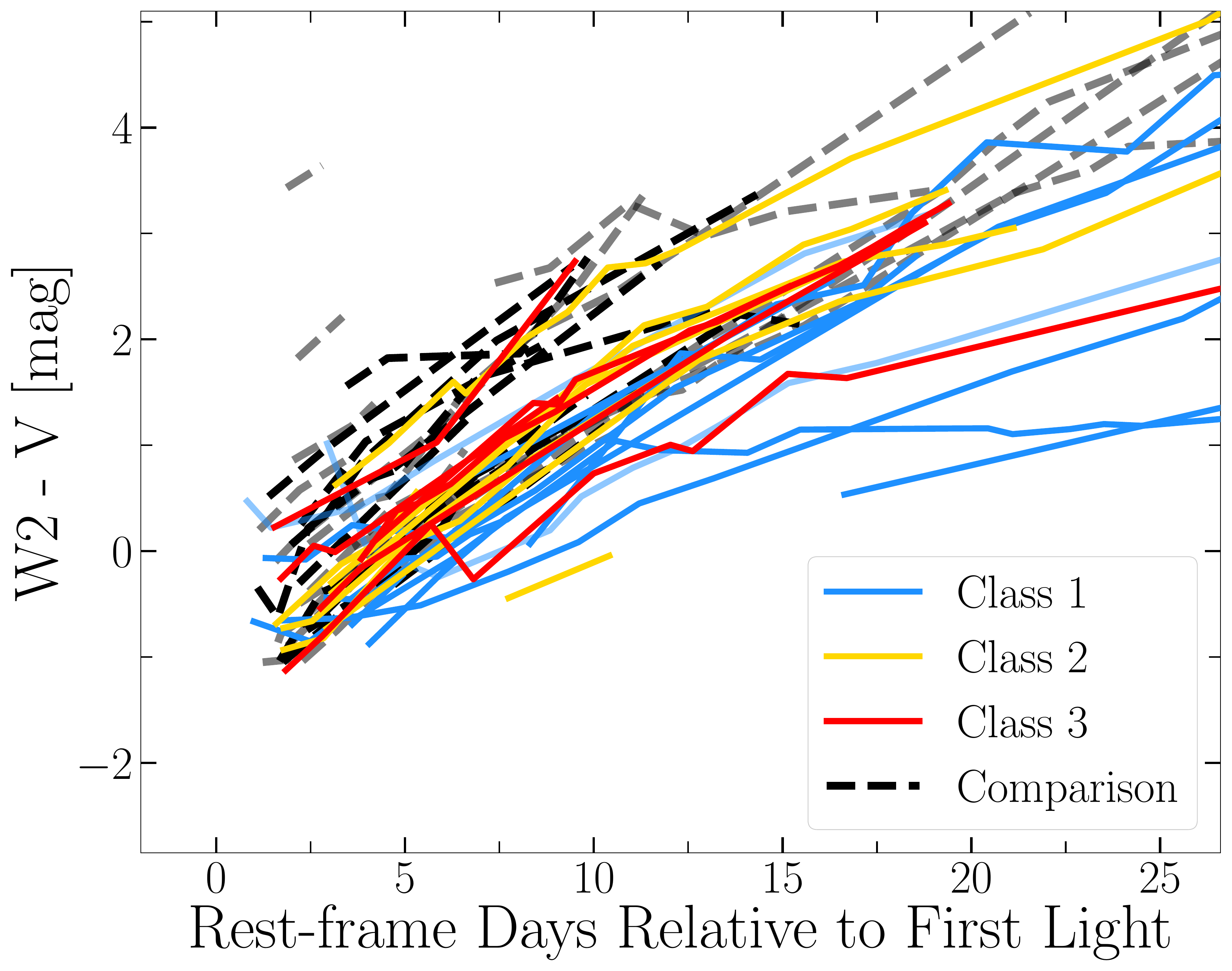}}
\subfigure{\includegraphics[width=0.33\textwidth]{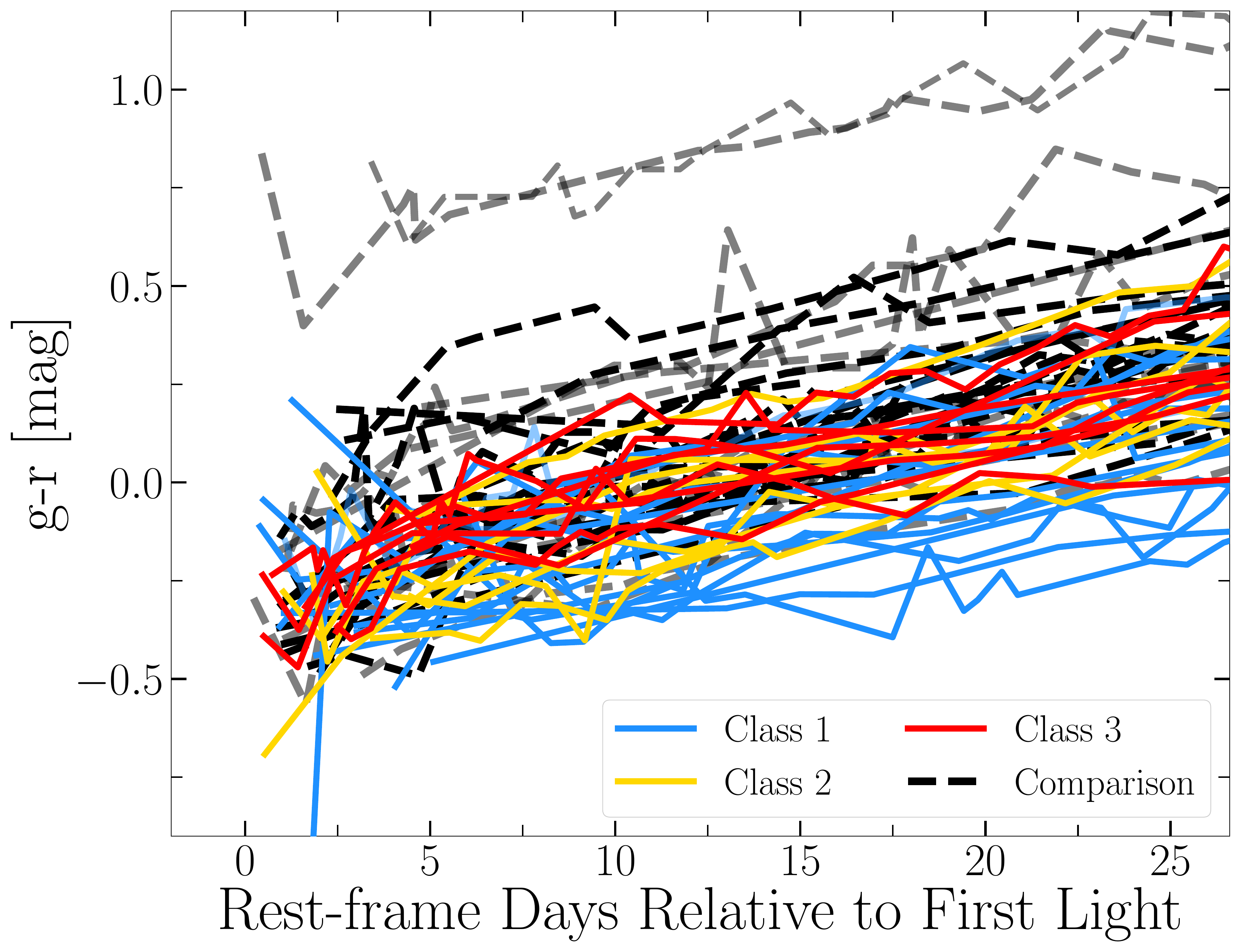}}
\subfigure{\includegraphics[width=0.33\textwidth]{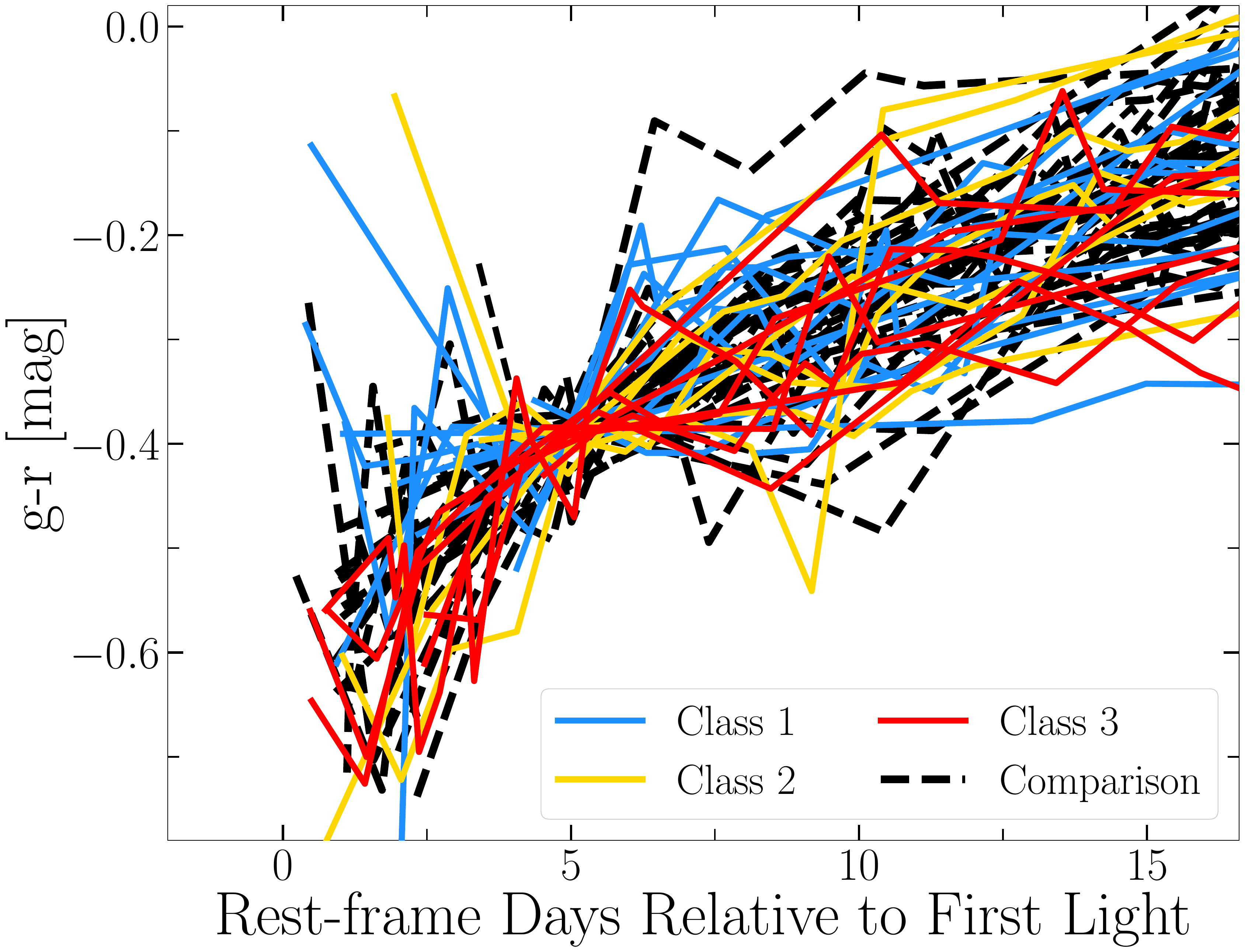}}
\caption{ Observed $W2 - V$ {\it (left)} and $g-r$ {\it (middle)} colors before host-extinction correction is applied. The reddest objects are comparison-sample objects 2013am and 2020fqv. As discussed in Appendix Section A, host reddening is unlikely to cause the contrast observed between the gold and comparison samples. Class 1 objects remain the bluest objects for all phases, suggesting continued CSM interaction. {\it Right:} $g-r$ colors after applying synthetic host extinction correction until all objects have the same color as the bluest object in the sample at $\delta t = 5$~days.
\label{fig:colors_obs} }
\end{figure*}

\begin{figure*}[t!]
\centering
\subfigure{\includegraphics[width=0.33\textwidth]{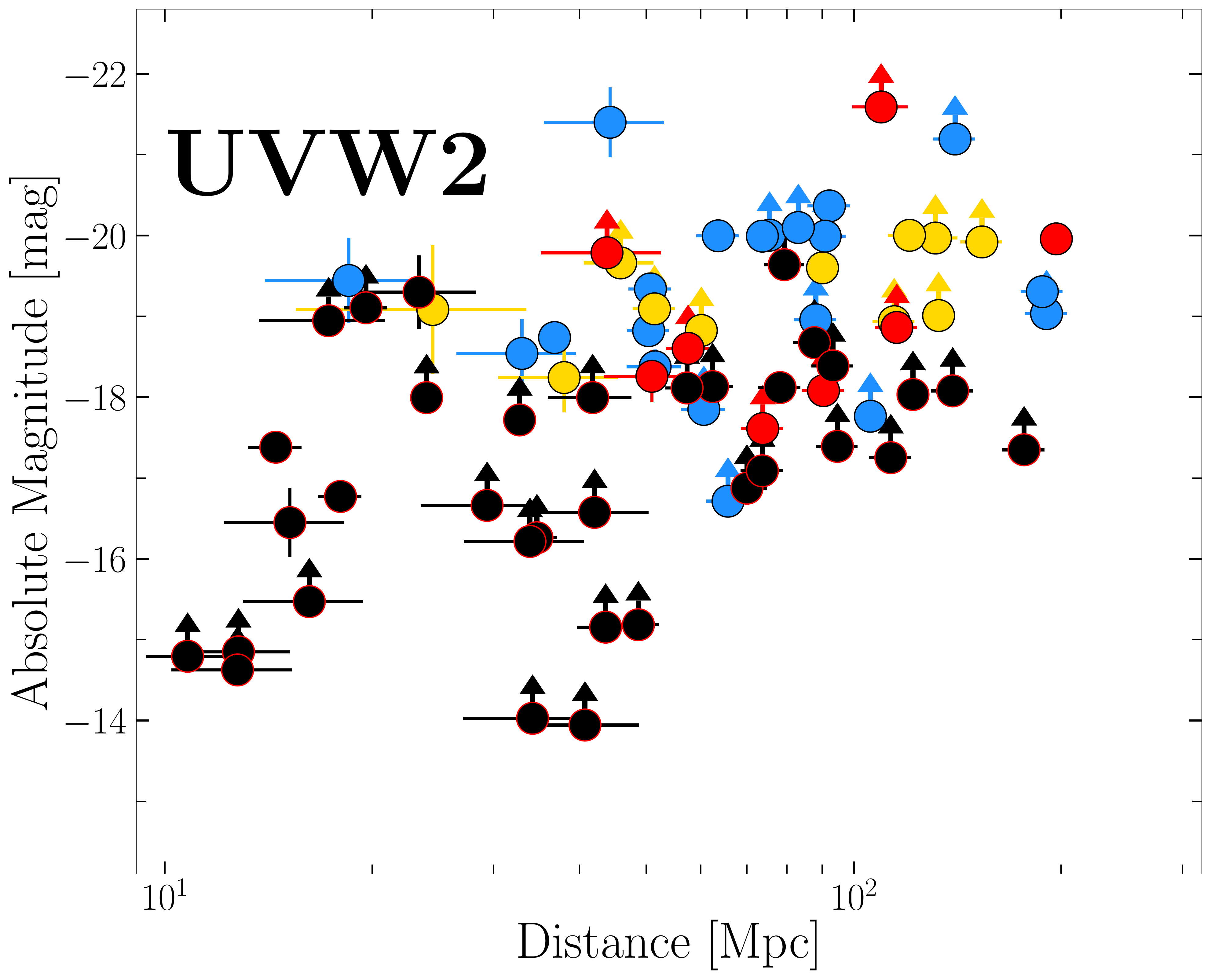}}
\subfigure{\includegraphics[width=0.33\textwidth]{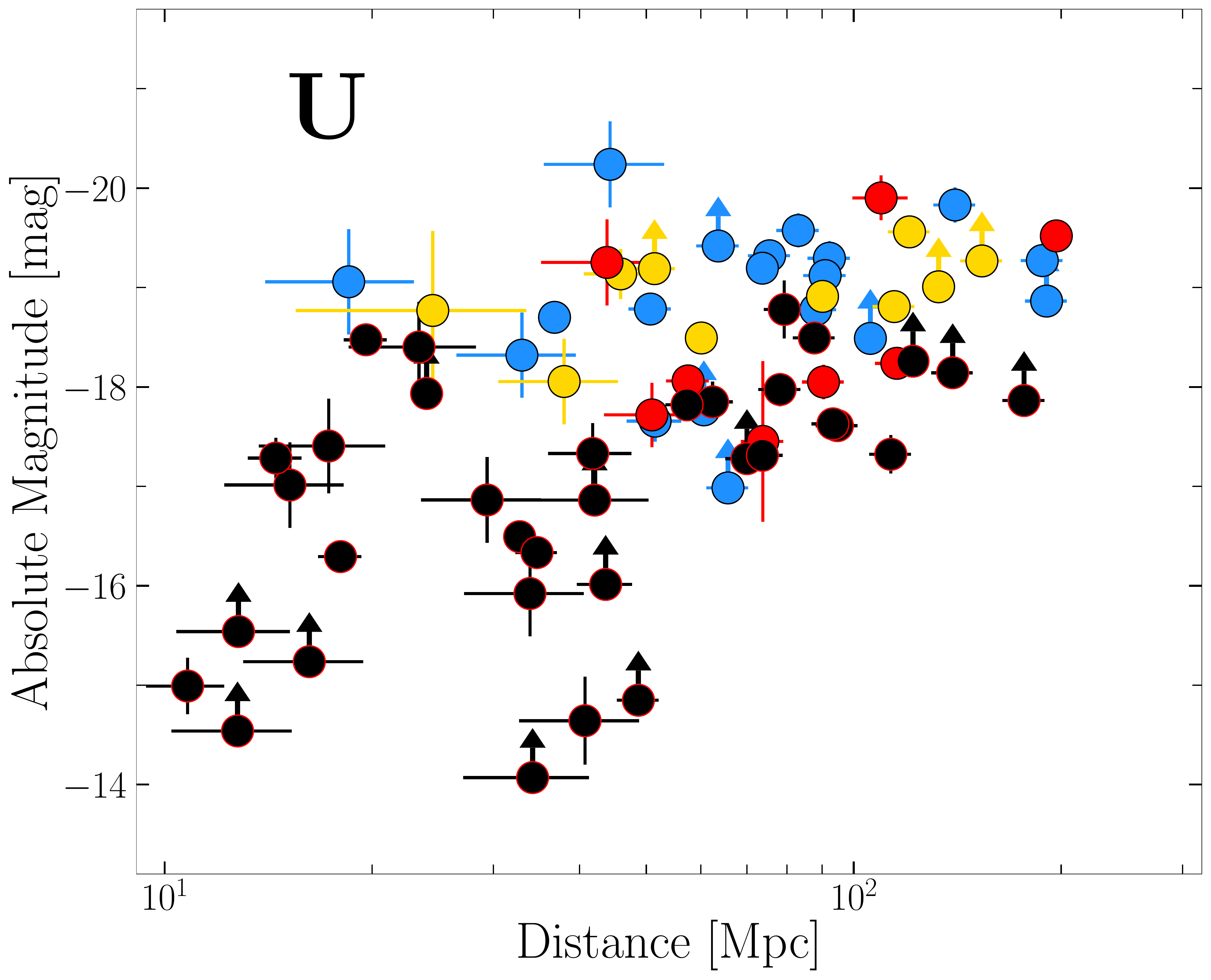}}
\subfigure{\includegraphics[width=0.33\textwidth]{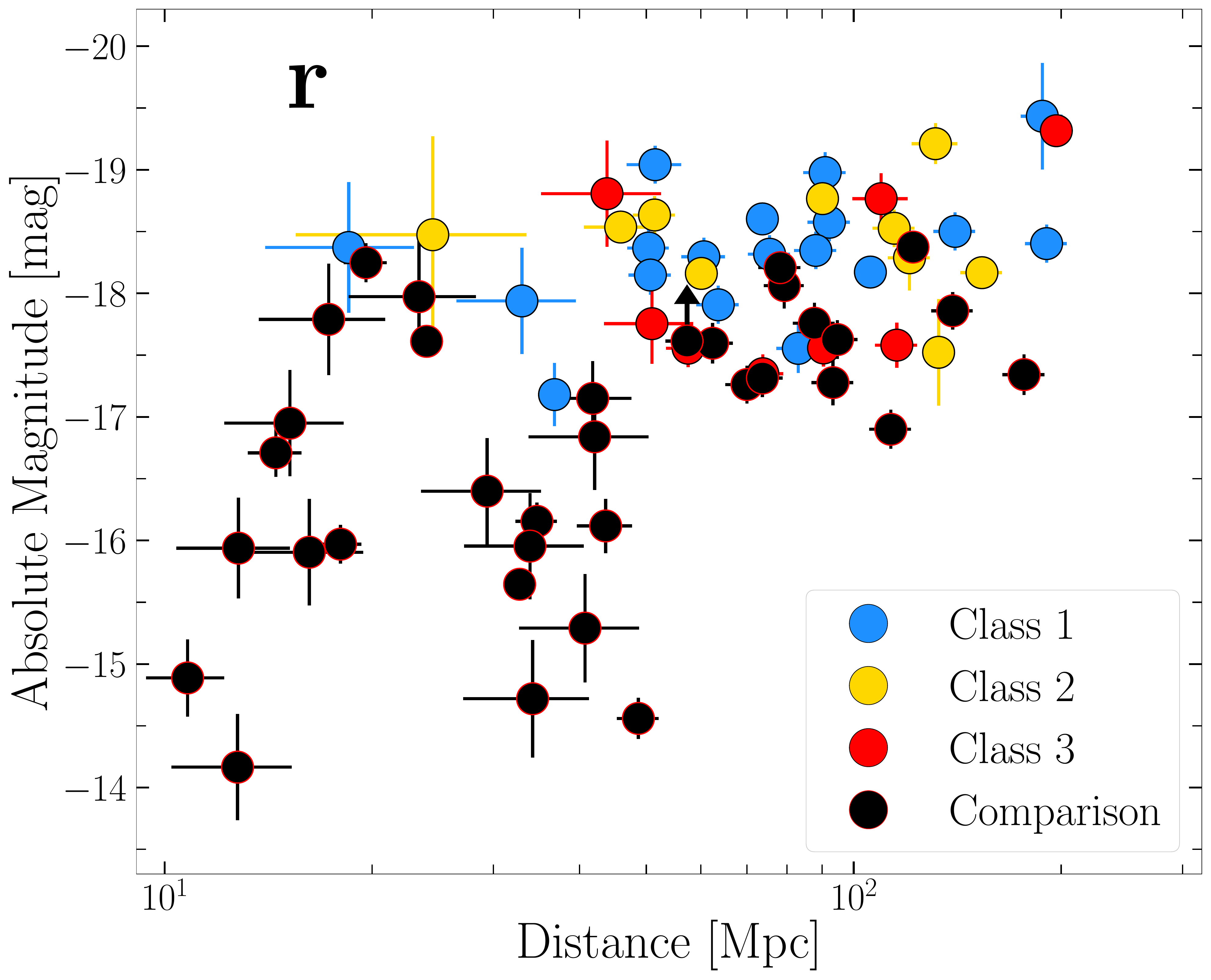}}\\
\subfigure{\includegraphics[width=0.33\textwidth]{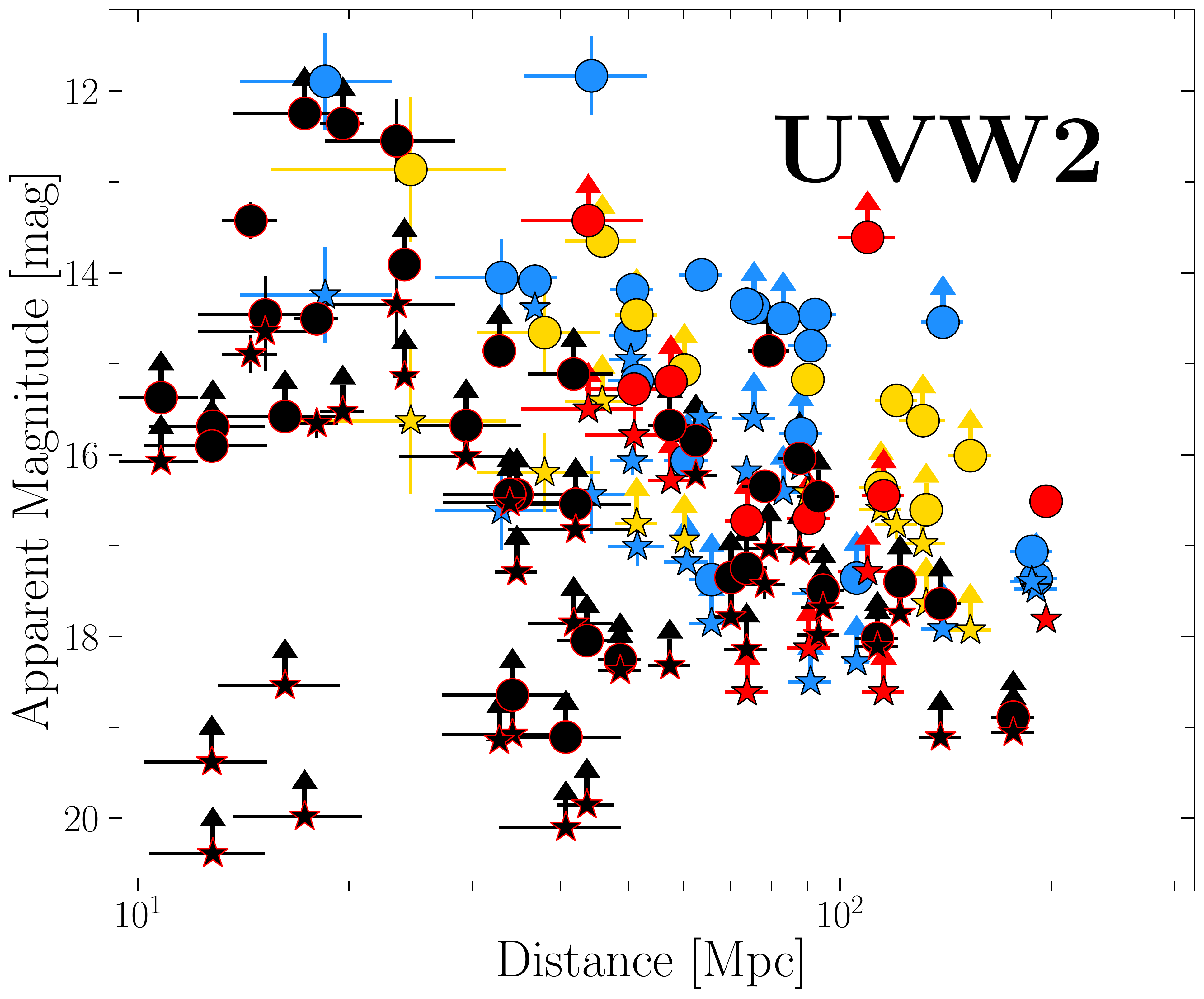}}
\subfigure{\includegraphics[width=0.33\textwidth]{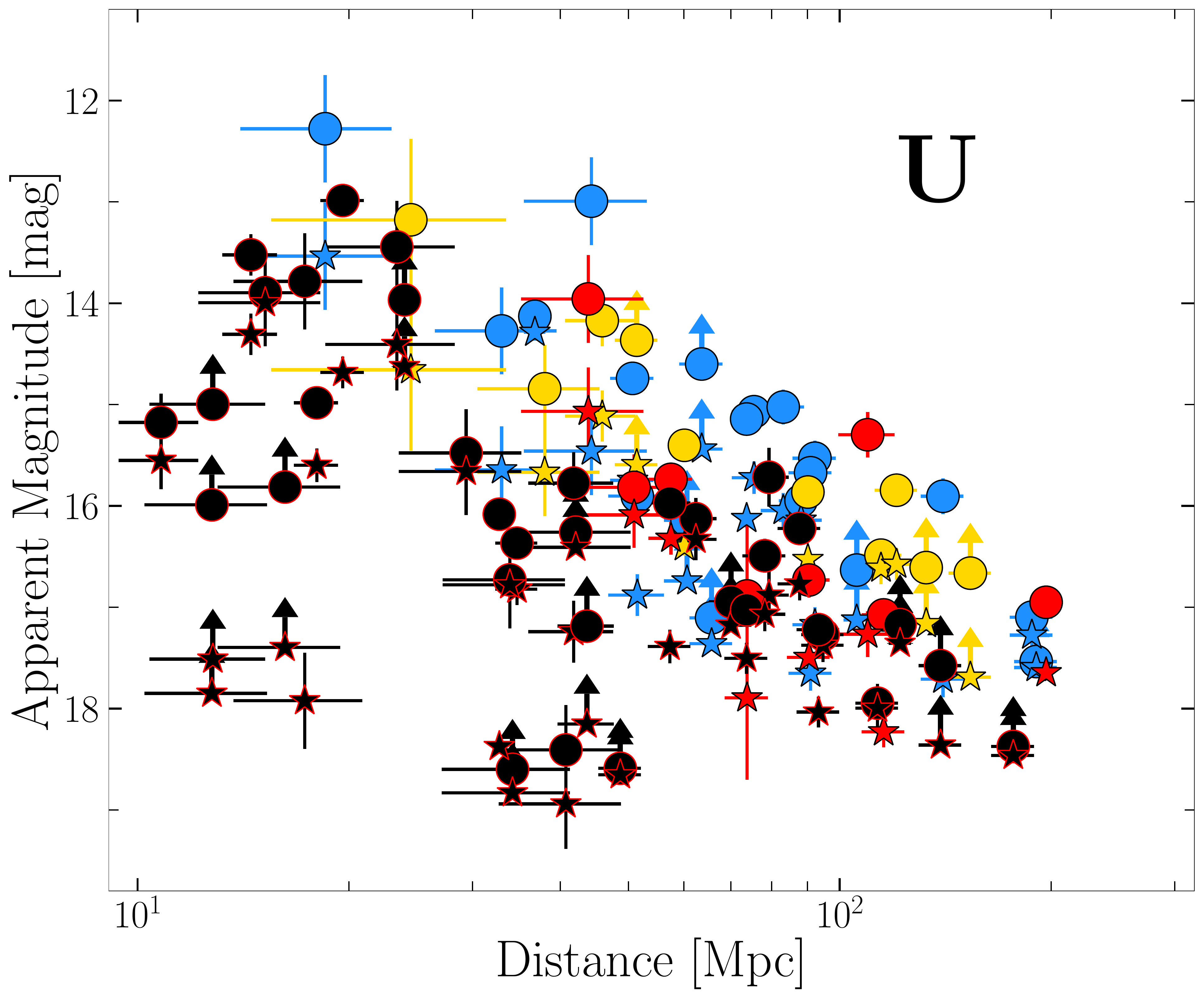}}
\subfigure{\includegraphics[width=0.33\textwidth]{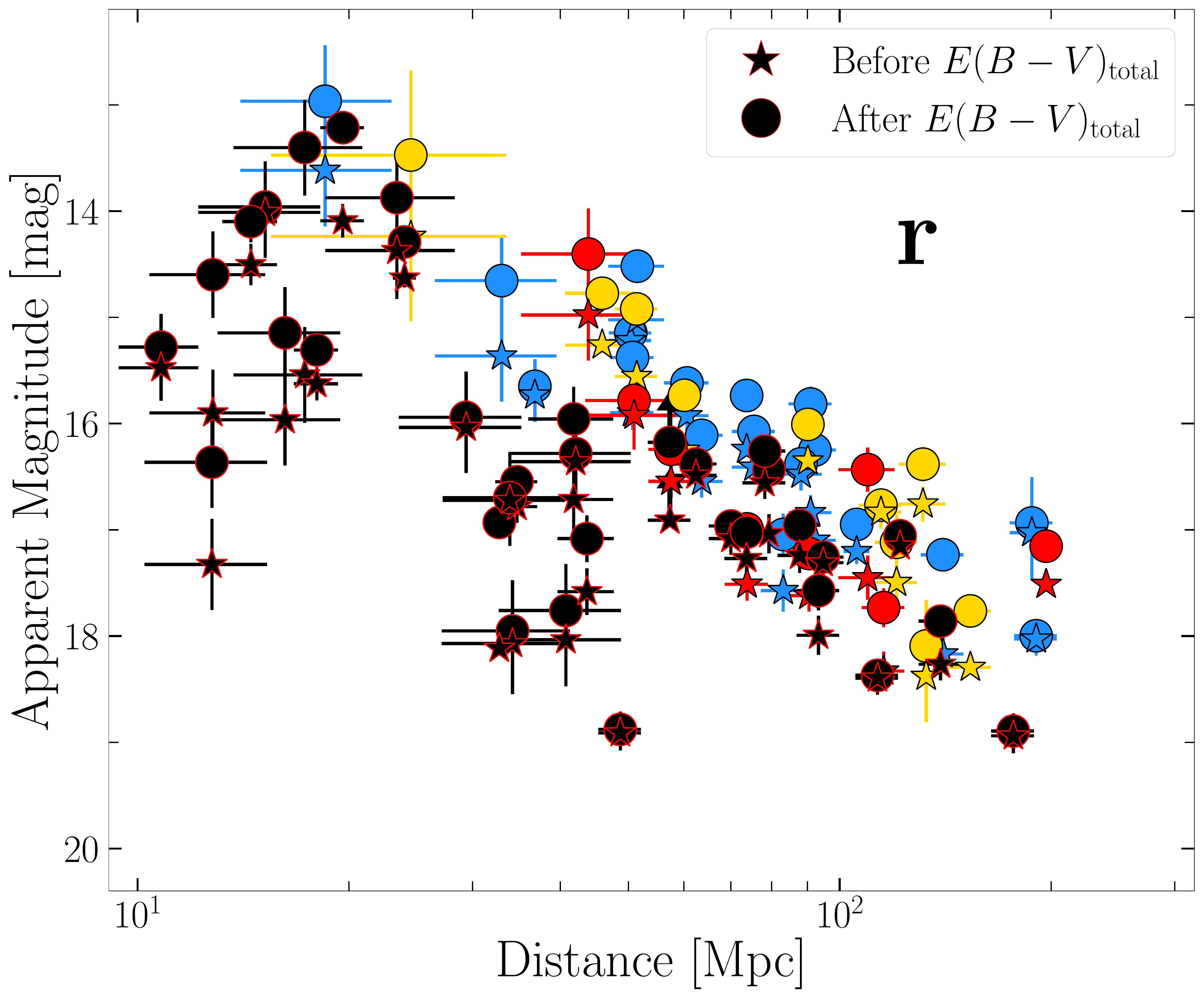}}
\caption{{\it Top:} Peak $w2$- (left), $u$- (middle), and $r$-band (right) absolute magnitudes versus distance for gold/silver (blue, yellow, red circles/stars) and comparison (black circles/stars) samples. Plotted stars (circles) represent peak magnitudes before (after) host extinction is applied using the \ion{Na}{I}~D EW. {\it Bottom:} Peak $w2$- (left), $u$- (middle), and $r$-band (right) apparent magnitude versus distance.  
\label{fig:selection} }
\end{figure*}

\begin{figure*}[t!]
\centering
\subfigure{\includegraphics[width=0.49\textwidth]{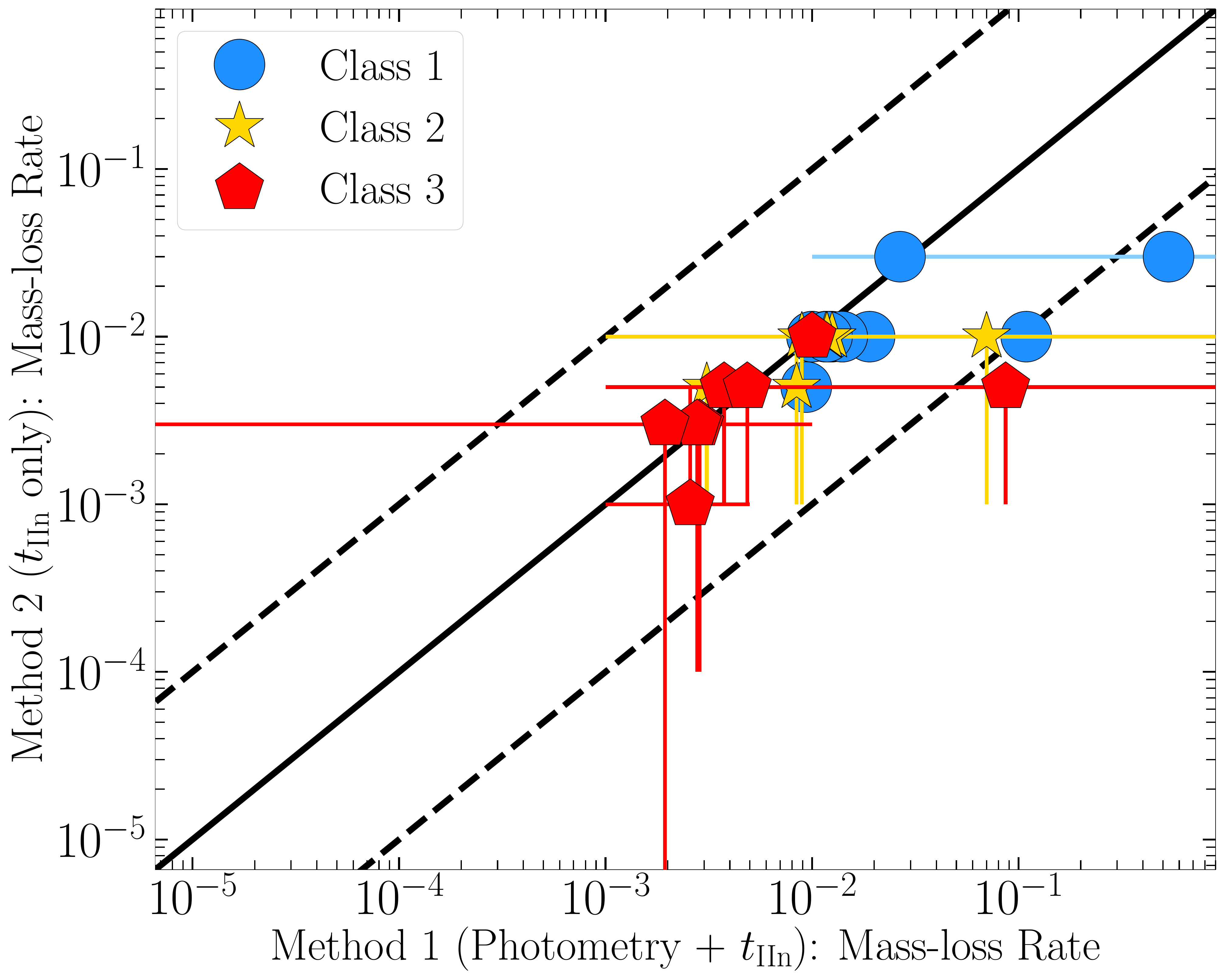}}
\subfigure{\includegraphics[width=0.49\textwidth]{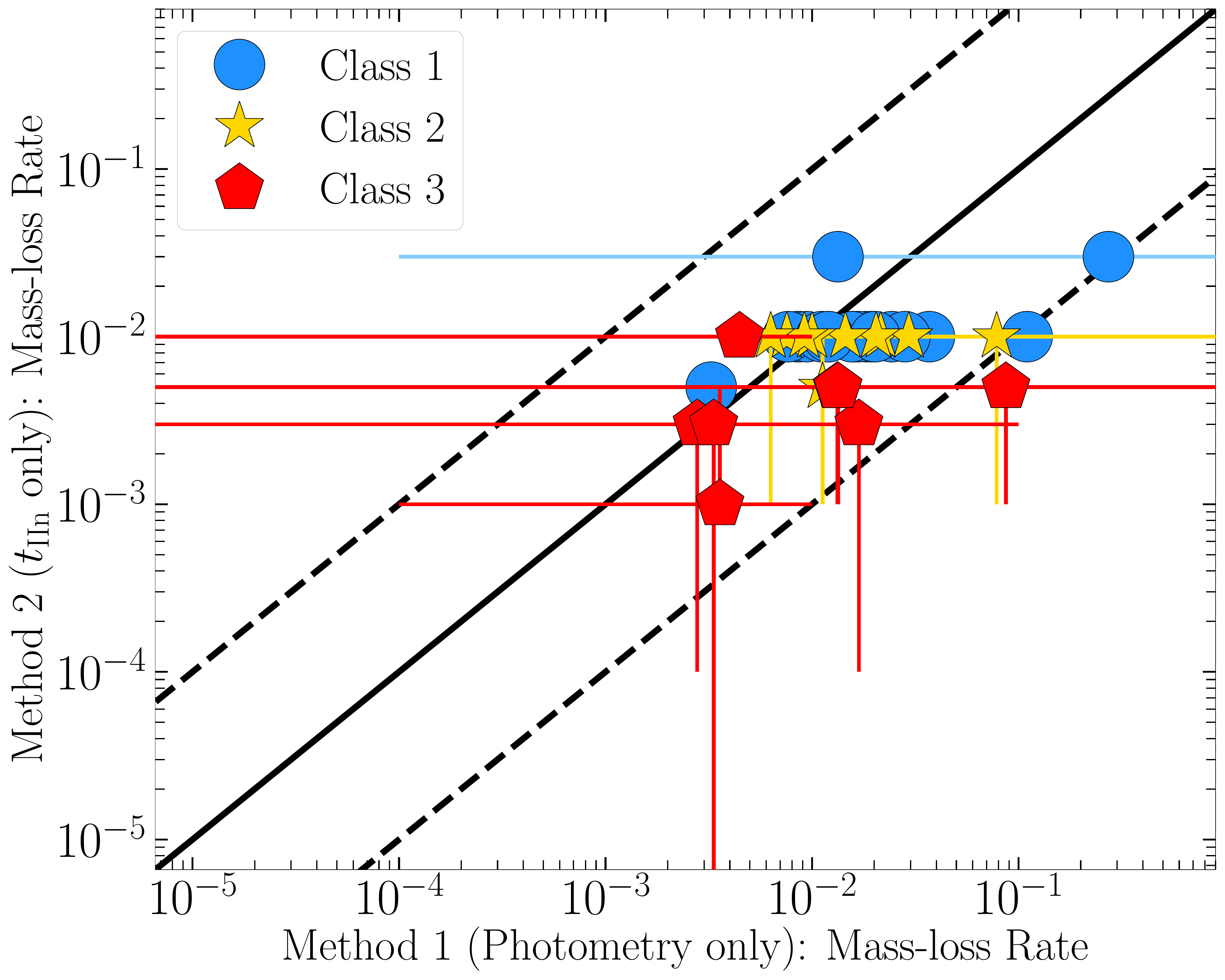}}\\
\subfigure{\includegraphics[width=0.49\textwidth]{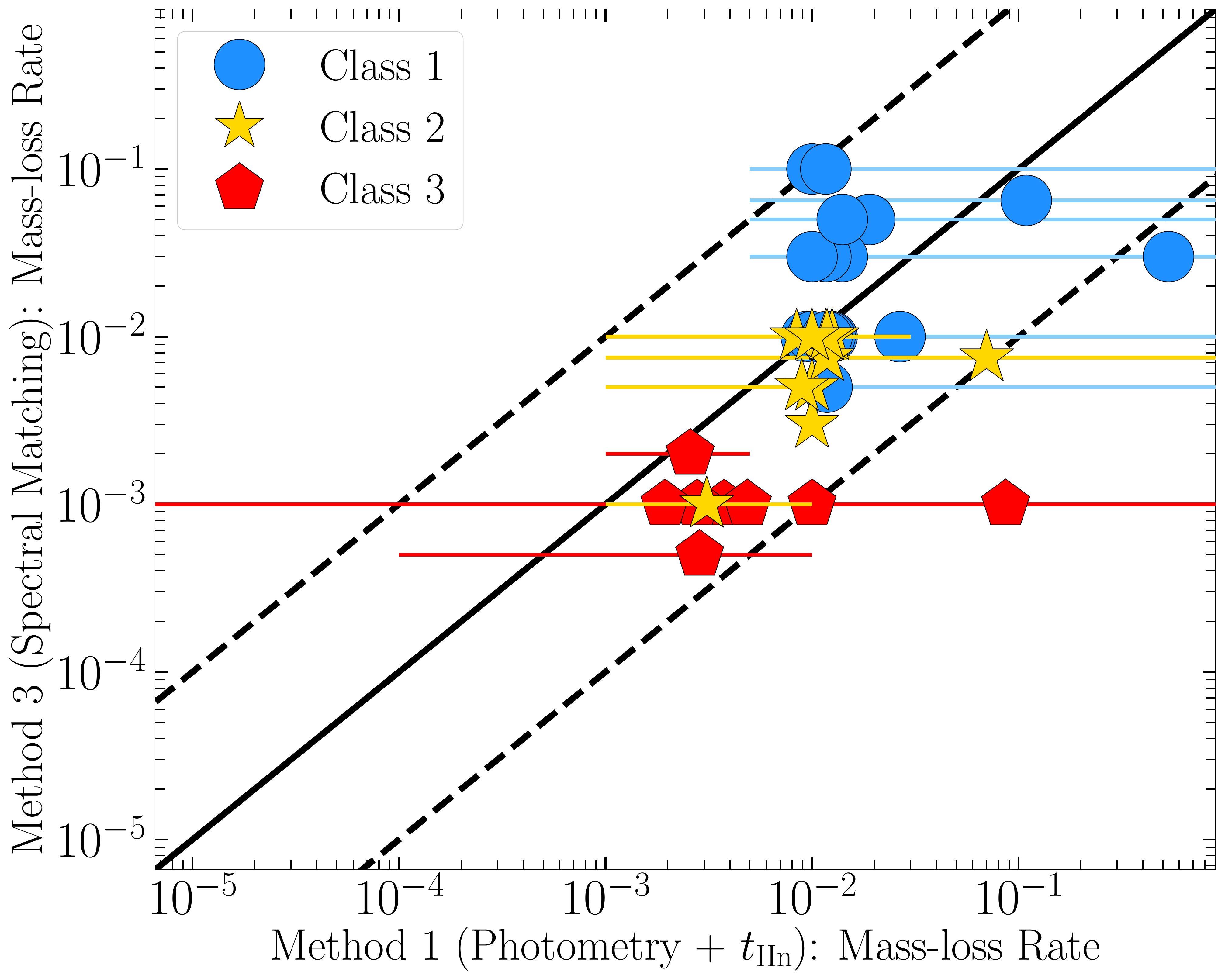}}
\subfigure{\includegraphics[width=0.485\textwidth]{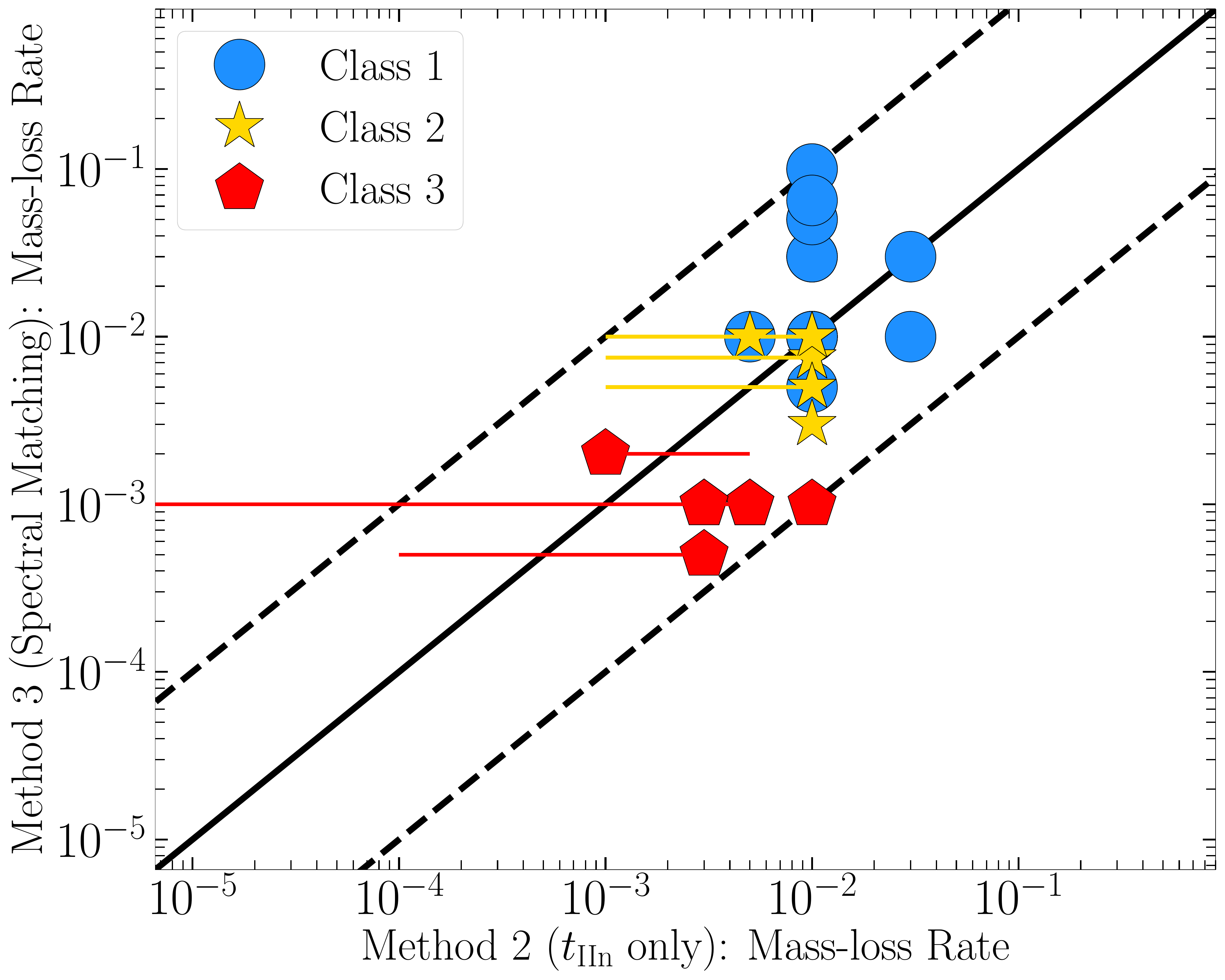}}
\caption{Comparison of model-matching methods (\S \ref{subsec:cmfgen}): {\it Top left:} Multicolor/bolometric light-curve properties plus $t_{\rm IIn}$ versus $t_{\rm IIn}$ only. {\it Top right:} Multicolor/bolometric light-curve properties versus $t_{\rm IIn}$. {\it Bottom left:} multicolor/bolometric light-curve properties plus $t_{\rm IIn}$ versus direct spectral matching. {\it Bottom right:} $t_{\rm IIn}$ versus direct spectral matching. The plotted points represent the average mass-loss rates derived from each method and the error bars represent the range of model parameters that are consistent with the observations (e.g., see \S \ref{subsec:cmfgen}).
\label{fig:model_fidelity} }
\end{figure*}

\begin{deluxetable*}{cccccccc}
\tablecaption{Optical/NIR Spectroscopy \label{tab:spec_all_one}}
\tablecolumns{8}
\tablewidth{0.45\textwidth}
\tablehead{\colhead{SN Name} &
\colhead{UT Date} & \colhead{MJD} &
\colhead{Phase\tablenotemark{a}} &
\colhead{Telescope} & \colhead{Instrument} & \colhead{Wavelength Range} & \colhead{Data Source}\\
\colhead{} & \colhead{} & \colhead{} & \colhead{(days)} & \colhead{} & \colhead{} & \colhead{(\AA)}
}
\startdata
2018cvn & 2018-06-29T09:21:36 & 58298.4 & 5.1 & SOAR & Goodman & 4057-8021 & Cartier \\
2018cvn & 2018-12-22T02:38:24 & 58474.1 & 180.8 & SOAR & Goodman & 3833-7779 & Cartier \\
2018cvn & 2019-01-14T01:40:48 & 58497.1 & 203.8 & SOAR & Goodman & 3899-7765 & Cartier \\
2018khh & 2018-12-21T02:09:36 & 58473.1 & 2.6 & SOAR & Goodman & 3892-7760 & Cartier \\
2018khh & 2018-12-22T01:26:24 & 58474.1 & 3.6 & SOAR & Goodman & 3331-9000 & Cartier \\
2018khh & 2019-01-14T00:43:12 & 58497.0 & 26.5 & SOAR & Goodman & 3793-7756 & Cartier \\
2018khh & 2019-04-18T02:09:36 & 58591.1 & 120.6 & SOAR & Goodman & 3302-9001 & Cartier \\
2018khh & 2019-05-17T08:38:24 & 58620.4 & 149.9 & SOAR & Goodman & 3813-7779 & Cartier \\
\enddata
\tablenotetext{a}{Relative to first light}
\end{deluxetable*}

\begin{deluxetable*}{cccccccc}
\tablecaption{Optical/NIR Spectroscopy \label{tab:spec_all}}
\tablecolumns{8}
\tablewidth{0.45\textwidth}
\tablehead{\colhead{SN Name} &
\colhead{UT Date} & \colhead{MJD} &
\colhead{Phase\tablenotemark{a}} &
\colhead{Telescope} & \colhead{Instrument} & \colhead{Wavelength Range} & \colhead{Data Source}\\
\colhead{} & \colhead{} & \colhead{} & \colhead{(days)} & \colhead{} & \colhead{} & \colhead{(\AA)}
}
\startdata
2019pjs & 2019-09-04T07:38:00 & 58730.3 & 0.7 & Shane & Kast & 3626-10702 & Filippenko \\
2019pjs & 2019-09-05T03:50:00 & 58731.2 & 1.5 & Shane & Kast & 3626-10710 & Filippenko \\
2019pjs & 2019-09-05T09:22:38 & 58731.4 & 1.7 & Faulkes-South & FLOYDS & 3799-9999 & GSP \\
2019pjs & 2019-09-10T07:17:14 & 58736.3 & 6.7 & P60 & SEDM & 3777-9223 & TNS \\
2019pjs & 2019-09-10T09:47:56 & 58736.4 & 6.8 & Faulkes-South & FLOYDS & 3800-10001 & GSP \\
2019pjs & 2019-09-14T08:13:45 & 58740.3 & 10.7 & Faulkes-North & FLOYDS & 3501-10000 & GSP \\
2019pjs & 2019-09-23T08:59:47 & 58749.4 & 19.7 & Faulkes-South & FLOYDS & 3801-10001 & GSP \\
2019pjs & 2019-09-25T11:22:60 & 58751.5 & 21.8 & Xinglong & BFOSC & 3623-8796 & X.Wang \\
\enddata
\tablenotetext{a}{Relative to first light}
\end{deluxetable*}

\begin{deluxetable*}{cccccccc}
\tablecaption{Optical/NIR Spectroscopy (cont.) \label{tab:spec_all2}}
\tablecolumns{8}
\tablewidth{0.45\textwidth}
\tablehead{\colhead{SN Name} &
\colhead{UT Date} & \colhead{MJD} &
\colhead{Phase\tablenotemark{a}} &
\colhead{Telescope} & \colhead{Instrument} & \colhead{Wavelength Range} & \colhead{Data Source}\\
\colhead{} & \colhead{} & \colhead{} & \colhead{(days)} & \colhead{} & \colhead{} & \colhead{(\AA)}
}
\startdata
2020dpw & 2020-02-27T19:12:00 & 58906.8 & 2.1 & Seimei & KOOLS & 4501-8008 & TNS \\
2020dpw & 2020-03-02T13:12:00 & 58910.6 & 5.8 & Shane & Kast & 3614-10708 & Filippenko \\
2020dpw & 2020-03-03T13:08:08 & 58911.5 & 6.8 & Shane & Kast & 3503-9995 & YSE \\
2020dpw & 2020-05-27T11:11:31 & 58996.5 & 91.7 & Shane & Kast & 3504-10494 & YSE \\
2020ekk & 2020-03-12T20:24:00 & 58920.8 & 2.2 & Lijiang & YFOSC & 3504-8767 & TNS \\
2020ekk & 2020-03-17T17:33:26 & 58925.7 & 7.1 & Faulkes-South & FLOYDS & 3501-9999 & GSP \\
2020ekk & 2020-03-21T17:05:40 & 58929.7 & 11.0 & Faulkes-South & FLOYDS & 3499-9999 & GSP \\
2020ekk & 2020-03-23T11:16:24 & 58931.5 & 12.8 & Faulkes-North & FLOYDS & 3501-10001 & GSP \\
\enddata
\tablenotetext{a}{Relative to first light.}
\end{deluxetable*}

\begin{deluxetable*}{cccccccc}
\tablecaption{Optical/NIR Spectroscopy (cont.) \label{tab:spec_all2}}
\tablecolumns{8}
\tablewidth{0.45\textwidth}
\tablehead{\colhead{SN Name} &
\colhead{UT Date} & \colhead{MJD} &
\colhead{Phase\tablenotemark{a}} &
\colhead{Telescope} & \colhead{Instrument} & \colhead{Wavelength Range} & \colhead{Data Source}\\
\colhead{} & \colhead{} & \colhead{} & \colhead{(days)} & \colhead{} & \colhead{} & \colhead{(\AA)}
}
\startdata
2020abjq & 2020-12-03T08:11:34 & 59186.3 & 2.9 & Faulkes-North & FLOYDS & 3799-10000 & GSP \\
2020abjq & 2020-12-04T08:09:04 & 59187.3 & 3.9 & Faulkes-North & FLOYDS & 3501-10000 & GSP \\
2020abjq & 2020-12-07T08:02:13 & 59190.3 & 6.9 & Shane & Kast & 3354-9994 & YSE \\
2020abjq & 2020-12-09T11:37:20 & 59192.5 & 9.0 & Faulkes-South & FLOYDS & 3800-10001 & GSP \\
2020abjq & 2020-12-11T04:31:21 & 59194.2 & 10.7 & Keck & LRIS & 3162-10147 & YSE \\
2020abjq & 2020-12-11T07:43:04 & 59194.3 & 10.9 & Faulkes-North & FLOYDS & 3500-10001 & GSP \\
2020abjq & 2020-12-15T04:42:45 & 59198.2 & 14.7 & Faulkes-North & FLOYDS & 3499-10001 & GSP \\
2020abjq & 2020-12-22T03:36:10 & 59205.2 & 21.7 & Shane & Kast & 3254-10894 & YSE \\
\enddata
\tablenotetext{a}{Relative to first light.}
\end{deluxetable*}

\begin{deluxetable*}{cccccccc}
\tablecaption{Optical/NIR Spectroscopy (cont.) \label{tab:spec_all3}}
\tablecolumns{8}
\tablewidth{0.45\textwidth}
\tablehead{\colhead{SN Name} &
\colhead{UT Date} & \colhead{MJD} &
\colhead{Phase\tablenotemark{a}} &
\colhead{Telescope} & \colhead{Instrument} & \colhead{Wavelength Range} & \colhead{Data Source}\\
\colhead{} & \colhead{} & \colhead{} & \colhead{(days)} & \colhead{} & \colhead{} & \colhead{(\AA)}
}
\startdata
2021zj & 2021-01-13T08:49:37 & 59227.4 & 5.1 & NTT & EFOSC2 & 3650-9245 & TNS \\
2021zj & 2021-01-15T13:27:46 & 59229.6 & 7.3 & Faulkes-North & FLOYDS & 3501-10000 & GSP \\
2021zj & 2021-02-10T11:29:17 & 59255.5 & 33.2 & Faulkes-North & FLOYDS & 3500-9999 & GSP \\
2021zj & 2021-02-16T08:16:56 & 59261.3 & 39.0 & Faulkes-North & FLOYDS & 3500-10000 & GSP \\
2021zj & 2021-03-19T06:36:26 & 59292.3 & 70.0 & Faulkes-North & FLOYDS & 3500-10000 & GSP \\
2021zj & 2021-03-30T09:46:30 & 59303.4 & 81.1 & Faulkes-North & FLOYDS & 4001-10000 & GSP \\
2021zj & 2021-04-10T10:07:26 & 59314.4 & 92.1 & Faulkes-North & FLOYDS & 3500-10001 & GSP \\
2021zj & 2021-04-24T07:26:28 & 59328.3 & 106.0 & Faulkes-North & FLOYDS & 3500-9999 & GSP \\
\enddata
\tablenotetext{a}{Relative to first light.}
\end{deluxetable*}

\begin{deluxetable*}{cccccccc}
\tablecaption{Optical/NIR Spectroscopy (cont.) \label{tab:spec_all3}}
\tablecolumns{8}
\tablewidth{0.45\textwidth}
\tablehead{\colhead{SN Name} &
\colhead{UT Date} & \colhead{MJD} &
\colhead{Phase\tablenotemark{a}} &
\colhead{Telescope} & \colhead{Instrument} & \colhead{Wavelength Range} & \colhead{Data Source}\\
\colhead{} & \colhead{} & \colhead{} & \colhead{(days)} & \colhead{} & \colhead{} & \colhead{(\AA)}
}
\startdata
2021gmj & 2021-03-20T20:35:50 & 59293.9 & 1.6 & LT & SPRAT & 4020-7994 & TNS \\
2021gmj & 2021-03-21T01:22:39 & 59294.1 & 1.8 & NOT & AFOSC & 3383-8174 & TNS \\
2021gmj & 2021-03-21T05:00:00 & 59294.2 & 2.0 & IRTF & SpeX & 6905-25708 & YSE \\
2021gmj & 2021-03-22T00:00:00 & 59295.0 & 2.8 & Shane & Kast & 3404-10497 & YSE \\
2021gmj & 2021-04-06T03:32:00 & 59310.1 & 17.9 & Shane & Kast & 3506-9994 & YSE \\
2021gmj & 2021-04-10T05:00:00 & 59314.2 & 22.0 & IRTF & SpeX & 6906-25709 & YSE \\
2021gmj & 2021-04-19T03:46:53 & 59323.2 & 30.9 & Shane & Kast & 3266-10896 & YSE \\
2021gmj & 2021-05-09T05:26:51 & 59343.2 & 51.0 & IRTF & SpeX & 6905-25708 & YSE \\
\enddata
\tablenotetext{a}{Relative to first light.}
\end{deluxetable*}

\begin{deluxetable*}{cccccccc}
\tablecaption{Optical/NIR Spectroscopy (cont.) \label{tab:spec_all3}}
\tablecolumns{8}
\tablewidth{0.45\textwidth}
\tablehead{\colhead{SN Name} &
\colhead{UT Date} & \colhead{MJD} &
\colhead{Phase\tablenotemark{a}} &
\colhead{Telescope} & \colhead{Instrument} & \colhead{Wavelength Range} & \colhead{Data Source}\\
\colhead{} & \colhead{} & \colhead{} & \colhead{(days)} & \colhead{} & \colhead{} & \colhead{(\AA)}
}
\startdata
2021qvr & 2021-06-23T00:00:00 & 59388.0 & 2.4 & NOT & ALFOSC & 3800-9000 & YSE \\
2021qvr & 2021-06-23T12:31:47 & 59388.5 & 2.9 & Faulkes-North & FLOYDS & 3500-10000 & YSE \\
2021qvr & 2021-06-24T03:21:39 & 59389.1 & 3.5 & LT & SPRAT & 4020-7994 & TNS \\
2021qvr & 2021-06-25T00:00:00 & 59390.0 & 4.4 & NOT & ALFOSC & 3800-9000 & YSE \\
2021qvr & 2021-06-25T13:32:39 & 59390.6 & 5.0 & Faulkes-North & FLOYDS & 3297-10178 & YSE \\
2021qvr & 2021-07-01T08:35:05 & 59396.4 & 10.8 & SOAR & Goodman & 4004-8985 & YSE \\
2021qvr & 2021-07-02T19:34:58 & 59397.8 & 12.2 & Faulkes-South & FLOYDS & 3301-10185 & YSE \\
2021qvr & 2021-07-02T11:19:30 & 59397.5 & 11.9 & Shane & Kast & 3505-10494 & YSE \\
\enddata
\tablenotetext{a}{Relative to first light.}
\end{deluxetable*}

\begin{deluxetable*}{cccccccc}
\tablecaption{Optical/NIR Spectroscopy (cont.) \label{tab:spec_all3}}
\tablecolumns{8}
\tablewidth{0.45\textwidth}
\tablehead{\colhead{SN Name} &
\colhead{UT Date} & \colhead{MJD} &
\colhead{Phase\tablenotemark{a}} &
\colhead{Telescope} & \colhead{Instrument} & \colhead{Wavelength Range} & \colhead{Data Source}\\
\colhead{} & \colhead{} & \colhead{} & \colhead{(days)} & \colhead{} & \colhead{} & \colhead{(\AA)}
}
\startdata
2021tyw & 2021-07-30T06:28:28 & 59425.3 & 11.9 & NTT & EFOSC2 & 3644-9239 & TNS \\
2021tyw & 2021-07-31T12:37:21 & 59426.5 & 13.1 & Faulkes-North & FLOYDS & 3500-9999 & GSP \\
2021tyw & 2021-08-02T11:45:00 & 59428.5 & 15.1 & Shane & Kast & 3624-10760 & Filippenko \\
2021tyw & 2021-08-03T13:45:36 & 59429.6 & 16.2 & Faulkes-North & FLOYDS & 3500-10001 & GSP \\
2021tyw & 2021-08-03T10:09:28 & 59429.4 & 16.0 & Shane & Kast & 3254-10893 & YSE \\
2021tyw & 2021-08-06T13:33:30 & 59432.6 & 19.2 & Faulkes-North & FLOYDS & 3500-10000 & GSP \\
2021tyw & 2021-08-09T13:13:02 & 59435.6 & 22.2 & Faulkes-North & FLOYDS & 3500-10000 & GSP \\
2021tyw & 2021-08-12T11:45:00 & 59438.5 & 25.1 & Shane & Kast & 3622-7857 & Filippenko \\
\enddata
\tablenotetext{a}{Relative to first light.}
\end{deluxetable*}

\begin{deluxetable*}{cccccccc}
\tablecaption{Optical/NIR Spectroscopy (cont.) \label{tab:spec_all3}}
\tablecolumns{8}
\tablewidth{0.45\textwidth}
\tablehead{\colhead{SN Name} &
\colhead{UT Date} & \colhead{MJD} &
\colhead{Phase\tablenotemark{a}} &
\colhead{Telescope} & \colhead{Instrument} & \colhead{Wavelength Range} & \colhead{Data Source}\\
\colhead{} & \colhead{} & \colhead{} & \colhead{(days)} & \colhead{} & \colhead{} & \colhead{(\AA)}
}
\startdata
2021vaz & 2021-08-06T14:34:56 & 59432.6 & 0.4 & Faukes-North & FLOYDS & 3499-10000 & GSP \\
2021vaz & 2021-08-08T14:11:54 & 59434.6 & 2.4 & Faukes-North & FLOYDS & 3500-10000 & GSP \\
2021vaz & 2021-08-09T14:34:17 & 59435.6 & 3.4 & Faukes-North & FLOYDS & 3499-10000 & GSP \\
2021vaz & 2021-08-19T13:58:40 & 59445.6 & 13.4 & Faukes-North & FLOYDS & 3499-10001 & GSP \\
2021vaz & 2021-08-25T13:53:28 & 59451.6 & 19.4 & Faukes-North & FLOYDS & 3499-9999 & GSP \\
2021vaz & 2021-09-05T11:48:54 & 59462.5 & 30.3 & Shane & Kast & 3355-9994 & YSE \\
2021vaz & 2021-09-08T12:26:45 & 59465.5 & 33.4 & Faukes-North & FLOYDS & 3500-9999 & GSP \\
2021vaz & 2021-09-15T12:33:37 & 59472.5 & 40.4 & Faukes-North & FLOYDS & 3500-10000 & GSP \\
\enddata
\tablenotetext{a}{Relative to first light.}
\end{deluxetable*}

\begin{deluxetable*}{cccccccc}
\tablecaption{Optical/NIR Spectroscopy (cont.) \label{tab:spec_all3}}
\tablecolumns{8}
\tablewidth{0.45\textwidth}
\tablehead{\colhead{SN Name} &
\colhead{UT Date} & \colhead{MJD} &
\colhead{Phase\tablenotemark{a}} &
\colhead{Telescope} & \colhead{Instrument} & \colhead{Wavelength Range} & \colhead{Data Source}\\
\colhead{} & \colhead{} & \colhead{} & \colhead{(days)} & \colhead{} & \colhead{} & \colhead{(\AA)}
}
\startdata
2022dml & 2022-02-28T12:21:35 & 59638.5 & 4.5 & Shane & Kast & 3505-10895 & YSE \\
2022dml & 2022-03-01T13:48:54 & 59639.6 & 5.6 & Faulkes-North & FLOYDS & 3500-10000 & GSP \\
2022dml & 2022-03-03T14:31:54 & 59641.6 & 7.6 & Faulkes-North & FLOYDS & 3499-10001 & GSP \\
2022dml & 2022-03-06T12:02:53 & 59644.5 & 10.5 & Faulkes-North & FLOYDS & 3500-10000 & GSP \\
2022dml & 2022-03-10T11:25:15 & 59648.5 & 14.5 & Faulkes-North & FLOYDS & 3500-10001 & GSP \\
2022dml & 2022-03-12T14:31:08 & 59650.6 & 16.6 & Faulkes-North & FLOYDS & 3500-10000 & GSP \\
2022dml & 2022-03-15T12:20:13 & 59653.5 & 19.5 & Faulkes-North & FLOYDS & 3500-10000 & GSP \\
2022dml & 2022-03-19T11:33:50 & 59657.5 & 23.5 & Faulkes-North & FLOYDS & 3500-10000 & GSP \\
\enddata
\tablenotetext{a}{Relative to first light.}
\end{deluxetable*}

\begin{deluxetable*}{cccccccc}
\tablecaption{Optical/NIR Spectroscopy \label{tab:spec_all}}
\tablecolumns{8}
\tablewidth{0.45\textwidth}
\tablehead{\colhead{SN Name} &
\colhead{UT Date} & \colhead{MJD} &
\colhead{Phase\tablenotemark{a}} &
\colhead{Telescope} & \colhead{Instrument} & \colhead{Wavelength Range} & \colhead{Data Source}\\
\colhead{} & \colhead{} & \colhead{} & \colhead{(days)} & \colhead{} & \colhead{} & \colhead{(\AA)}
}
\startdata
2022frq & 2022-04-03T10:48:57 & 59672.5 & 1.6 & Faukes-North & FLOYDS & 3500-10000 & GSP \\
2022frq & 2022-04-03T11:00:00 & 59672.5 & 1.6 & Shane & Kast & 3620-10744 & Filippenko \\
2022frq & 2022-04-04T16:20:12 & 59673.7 & 2.8 & Faukes-South & FLOYDS & 3499-10000 & GSP \\
2022frq & 2022-04-07T09:10:55 & 59676.4 & 5.5 & Faukes-North & FLOYDS & 3499-10001 & GSP \\
2022frq & 2022-04-08T11:37:40 & 59677.5 & 6.6 & Faukes-North & FLOYDS & 3499-10000 & GSP \\
2022frq & 2022-04-11T15:25:02 & 59680.6 & 9.7 & Faukes-South & FLOYDS & 3499-9999 & GSP \\
2022frq & 2022-04-18T14:12:11 & 59687.6 & 16.7 & Faukes-South & FLOYDS & 3500-10001 & GSP \\
2022frq & 2022-04-23T11:10:40 & 59692.5 & 21.6 & Faukes-North & FLOYDS & 3500-10000 & GSP \\
\enddata
\tablenotetext{a}{Relative to first light.}
\end{deluxetable*}

\begin{deluxetable*}{cccccccc}
\tablecaption{Optical/NIR Spectroscopy \label{tab:spec_all_last}}
\tablecolumns{8}
\tablewidth{0.45\textwidth}
\tablehead{\colhead{SN Name} &
\colhead{UT Date} & \colhead{MJD} &
\colhead{Phase\tablenotemark{a}} &
\colhead{Telescope} & \colhead{Instrument} & \colhead{Wavelength Range} & \colhead{Data Source}\\
\colhead{} & \colhead{} & \colhead{} & \colhead{(days)} & \colhead{} & \colhead{} & \colhead{(\AA)}
}
\startdata
2022jzc & 2022-05-16T06:24:59 & 59715.3 & 1.0 & Palomar & SEDM & 3777-9223 & TNS \\
2022jzc & 2022-05-17T10:17:10 & 59716.4 & 2.1 & Faukes-North & FLOYDS & 3500-10001 & GSP \\
2022jzc & 2022-05-17T00:00:00 & 59716.0 & 1.7 & Xinglong & BFOSC & 3871-8833 & X.Wang \\
2022jzc & 2022-05-22T06:21:34 & 59721.3 & 7.0 & Shane & Kast & 3579-10194 & TReX \\
2022jzc & 2022-05-23T00:00:00 & 59722.0 & 7.7 & Xinglong & BFOSC & 3895-8844 & X.Wang \\
2022jzc & 2022-05-23T08:02:24 & 59722.3 & 8.0 & Shane & Kast & 3632-10734 & Filippenko \\
2022jzc & 2022-06-26T05:48:01 & 59756.2 & 42.0 & Shane & Kast & 3603-10194 & TReX \\
2022jzc & 2022-07-04T04:37:24 & 59764.2 & 49.9 & Shane & Kast & 3551-5549 & TReX \\
\enddata
\tablenotetext{a}{Relative to first light.}
\end{deluxetable*}

\end{document}